\documentclass[twoside, openright, 12pt]{book}
\usepackage[a4paper,top=2.5cm,bottom=2.5cm,left=3.5cm,right=3.5cm]{geometry}
\usepackage[parfill]{parskip}
\usepackage[english]{babel}
\usepackage[T1]{fontenc}
\usepackage[utf8]{inputenc}

\usepackage{anyfontsize, comment, graphicx, multicol}
\usepackage{amsfonts, amsthm, amsmath, amssymb, float, mathtools, mathrsfs}
\usepackage{bm, braket, color, csquotes, hyperref, lipsum, soul, titlesec, subcaption, verbatim}
\usepackage[font=small, skip=5pt]{caption}
\usepackage[backend=biber, style=phys, defernumbers=true]{biblatex}
\usepackage[immediate]{silence}

\numberwithin{equation}{section}
\usepackage{sectsty}
\DeactivateWarningFilters[temp]
\usepackage[bottom]{footmisc}
\usepackage{hyperref}
\usepackage{epsfig}
\usepackage{fancyhdr}
\titleformat{\chapter}[display]{\normalfont\huge\bfseries}{\chaptertitlename\ \thechapter}{10pt}{\LARGE}
\titlespacing*{\chapter}{0pt}{0pt}{20pt}\chaptertitlefont{\fontsize{22pt}{30pt}\selectfont}

\hypersetup{colorlinks, citecolor=black, filecolor=black, linkcolor=black, urlcolor=black}

\renewcommand{\footnoterule}{
   \kern -3pt
   \hrule width \textwidth height 1pt
   \kern 2pt
 } 

\begin{document}
\pagenumbering{gobble}
\begin{titlepage}
\begin{figure}[!htb]
    \centering
    \includegraphics[width=5cm]{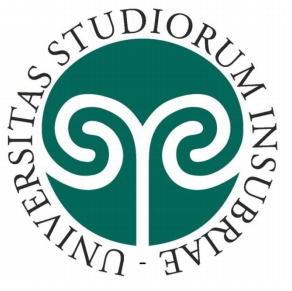}
\end{figure}

\begin{center}
    \Large{\textbf{UNIVERSITÀ DEGLI STUDI DELL'INSUBRIA}}
    \vspace{3mm}
    \\ \normalsize{DIPARTIMENTO DI SCIENZA E ALTA TECNOLOGIA}
    \vspace{6mm}
    \\ \normalsize{DOTTORATO}
    \\  \normalsize{\textbf{FISICA E ASTROFISICA}}
    \vspace{13mm}
\end{center}

\vspace{10mm}
\begin{center}
    \LARGE{\textbf{Boundary and bulk perturbations in vectorial active matter}}
\end{center}

\vspace*{\fill}

\begin{minipage}[t]{1\textwidth}
    \begin{multicols}{2}
    	{\normalsize{\textbf{Supervisor}}{\normalsize\vspace{1mm}
        \\ \normalsize{Prof. Francesco Ginelli }}} \\ 
        
          \columnbreak

         \begin{flushright}
            {\normalsize{\textbf{PhD Candidate}}{\normalsize\vspace{1mm}
            \\ \normalsize{Giuseppe Fava}}} \\
         \end{flushright}
    \end{multicols}
\end{minipage}

\begin{center}
    {\normalsize{\textbf{Academic year}}{\normalsize\vspace{1mm}
    \\ \normalsize{2023/2024}}}  
\end{center}

\end{titlepage}

\tableofcontents
\newpage
\frontmatter
\pagenumbering{roman}
\chapter{Introduction}
Statistical mechanics is a branch of physics that studies how macroscopic properties emerge in systems with a vast number of interacting components. Even in systems with many degrees of freedom, equilibrium conditions allow for a description using only a few key parameters, such as temperature, pressure, and volume that can be expressed as ensemble averages of the microscopic variables, under the assumption of the ergodic hypothesis. In this sense, statistical mechanics links the microscopic dynamics of individual particles and the macroscopic laws of equilibrium thermodynamics.\\
However, most systems in nature are not in equilibrium because they are continuously exposed to flows of matter and energy. At the microscopic level, this means that the transitions between states in a stationary system do not satisfy the detailed balance condition. There are anyhow different ways to be out of equilibrium. Some systems slowly relax towards equilibrium: this is the case, for instance, of isolated systems evolving to a steady state or systems experiencing small external perturbations. In these cases, one can make use of linear response theory which provides a unified framework and strong results like the fluctuation-dissipation theorem \cite{livi2017nonequilibrium}.\\
In contrast, systems far from equilibrium exhibit much more complex behavior that linear theories cannot adequately capture. This is perhaps the case for glasses, which are formed when liquids are cooled rapidly and end up trapped in an energy landscape with an infinity of metastable states. Such materials may take extremely long timescales to relax toward equilibrium, far exceeding the ones accessible experimentally \cite{berthier2011theoretical}. Other examples include systems that are actively driven out of equilibrium, such as interface growth \cite{kardar1986dynamic} or turbulent flows. In turbulent systems, energy is continuously injected at large scales and cascades down to microscopic scales before dissipating \cite{frisch1995turbulence}. These examples represent just a fraction of the rich and diverse phenomena characterizing of far-from-equilibrium systems.\\
This thesis deals with {\it active matter}, a prominent example of far from equilibrium system in which the energy injection takes place at the individual particle level and detailed balance is broken locally.

Active matter was born in 1995 after two seminal papers by Vicsek and collaborators \cite{vicsek1995novel} and by Toner \& Tu \cite{toner1995long} came out. Nowadays with the term active matter we indicate any system capable of transforming some non-thermal (often chemical) energy present in the environment into self-propulsion, or other functional mechanisms (rotate, shrink/expand, etc.).\\
Even before active matter was born \cite{reynolds1987flocks}, self-propelled particles were used to model the {\it flocking} behaviour at the macroscopic scale. Flocking is perhaps one of the most intriguing examples of collective behaviors exhibited by active matter systems, as shown by the aerial displays of starling flocks \cite{cavagna2010scale}. Fish schools \cite{Fish}, mammal herds, bacterial colonies \cite{bacteria1, bacteria2, bacteria3}, cellular migrations \cite{giavazzi2017giant} are just a few examples of the diverse living systems and scales on which flocking is observed. Furthermore, several experimental realizations of synthetic active matter displaying collective motion have been realized in self-propelled colloids \cite{bartolo_prx} and driven granular matter \cite{driven_granular}.\\
The wide range of scales involved and the ubiquity of flocking phenomena, regardless of many different microscopic details, have attracted the interest of the
statistical physics community. Physicists have tackled the problem combining
the analytical study of representative field theories \cite{toner1995long, toner1998flocks} with the numerical study of agent based models \cite{vicsek1995novel, chate2008collective} capturing the essential symmetries and conservation laws of the problem. This approach has revealed that the flocking dynamics is characterized by a strongly fluctuating ordered phase characterized by the presence of long-ranged correlations in both the density and orientation fields. Therefore, flocks exhibit genuinely out-of-equilibrium characteristics even in two spatial dimensions, such as giant number fluctuations, super-diffusive behavior or the development of truly long-ranged order, despite the spontaneous breaking of a continuous symmetry.

While our knowledge of the bulk behaviour of free collective motion is now fairly complete, at least when the surrounding fluid may be safely neglected (the so-called \textit{dry} approximation) much less is known when collective motion is achieved explicit breaking a continuos rotational symmetry, either at the global or local level. This thesis investigate these problems in a series of different but somehow related setups. \\
Symmetry breaking at the global level can perhaps be due to an anisotropic environment, where a favored direction, determines the mean flocking direction: cell motility, for instance, is known to be sensitive to a wide range of external gradients of chemical (chemotaxis), mechanical (durotaxis) and electrical (electrotaxis) origin which often direct cellular migrations. One relevant question regards the signatures of this \textit{directed} motion. How can an experimental observer detect a small anisotropy from the observation of the flocking system without an a priori knowledge of the lack of perfect environmental isotropy?\\
The above example regards an anisotropic perturbation that, in principle,
should be felt by every active particle in the system and breaks the rotational
symmetry globally. A different local source of anisotropy may be due to specific
boundary conditions, often unavoidable in experiments, such as the channel
geometry employed in Ref.\cite{bartolo_prx}. \textit{Confined} flocking not only breaks the translational invariance, but also introduce a local anisotropy between the longitudinal and perpendicular directions w.r.t. the boundary. How does this local symmetry breaking at the boundaries affect the flocking bulk and boundary behavior in large, but finite, systems?\\
In order to find and establish appropriate environments, certain bacteria have developed the ability to respond to light gradients \cite{wilde2017light}. Filamentous cyanobacteria \cite{kurjahn2024collective} for example, are able to respond to light gradients by reverting their direction of motion: they can reverse their direction of motion either when they experience an increase (photophobic) or decrease (scoto-phobic) in illuminance. This {\it virtual} confinement is a more subtle mechanism to control phototactic active matter. By shaping light in order to produce compact light patterns, bacteria with scoto-phobic response are able to accumulate and align their direction of motion with this virtual boundary \cite{kurjahn2024collective}. How does this virtual confinement affect the bulk and boundary properties of a free system?\\
\textbf{Thesis structure}\\ 
In Chapter \ref{dry_am} we will briefly review known results in dry active matter, starting from simple microscopic models of non-aligning particles. We will then focus more specifically on the properties of {\it vectorial} active matter \cite{dadam_benoit}, where aligning torques are explicit, starting with perhaps the simplest model for flocking with discrete symmetry, the active Ising model \cite{Solon2013, 2d_AIM} and then moving to the celebrated Vicsek model \cite{vicsek1995novel, ginelli2016physics}, where the rotational symmetry is continuous. We also introduce a model where torques are an explicit consequence of inter-particle repulsion, the so-called collisional Vicsek model \cite{baconnier2024self}. The final part will be devoted to field theoretical descriptions for flocking obtained both through a phenomenological approach \cite{toner1995long, toner1998flocks, toner2005hydrodynamics, toner2012reanalysis} and from direct coarse graining of the microscopic model at hand \cite{bertin2006boltzmann, bertin2009hydrodynamic}.\\
The second Chapter is dedicate to a brief review of known results concerning bulk and boundary perturbations in dry active matter: we will first show how when interacting with boundaries, active matter systems lack an equation of state \cite{solon2015pressure} as exemplified by the spontaneous compression of a mobile asymmetric partition. We will then move to review some results that show the far-reaching impact that perturbations can have both in the case of {\it scalar} \cite{granek2023inclusions} and vectorial active matter \cite{codina2022small, benvegnen2023metastability}.\\
In Chapter \ref{directed} we will present original results for directed flocking: in the first part we will consider symmetry breaking by a global perturbation and show how it is possible to discriminate between spontaneous and driven  (by an external field) flocking by presenting two different and complementary approaches. The second part is dedicated to explicit symmetry breaking at the boundaries and the effect it has on the bulk properties of flocking, such as the correlation functions and the polar order parameter. Interestingly this two behaviours can be mapped formally mapped one to another. The results presented in this Chapter are published respectively in \cite{BFG} and \cite{Lenzini_2024}.\\
The fourth chapter focuses instead on the boundary effect that confinement by hard mechanical boundaries imposes on a polar active system. We will discuss explicitly how the presence of long range correlations in the ordered phase, results in {\it extensive boundary} layers, showing once again the far-reaching impact that boundaries have in the context of active matter. Moreover, we also report how non-equilibrium long-ranged correlations, present deep in the ordered phase, induce a strong attractive Casimir-like force acting on the parallel walls, which decays slowly and algebraically upon increasing the wall separation. Our results are robust beyond the dilute limit. The main results presented in this chapter are published in \cite{fava2024casimir}, while some more technical aspects are going to be the subject of a second work \cite{preparation1}.\\
In Chapter \ref{ch_nematic} we will present some preliminary results on the study of nematic polar rods under virtual confinement, inspired by a recent experimental work on phototactic cyanobacteria \cite{kurjahn2024collective}. Interestingly we are able to reproduce key experimental results such as boundary accumulation at and alignment with the virtual boundary with a minimal agent-based model \cite{patelli2019understanding}. Direct coarse graining allows to study the effect of virtual boundaries at the mesoscopic level possibly allowing to understand the origin of such an effect. While some result presented here are being prepared for publication \cite{preparation2}, the analysis of the mesoscopic equation is preliminary and subject of current research.
\pagenumbering{arabic}
\setcounter{chapter}{0}
\mainmatter
\fancyhead[LO]{{\it Dry active matter}}
\fancyhead[RE]{{\it Dry active matter}}
\chapter{Dry active matter - A brief overview}\label{dry_am}
All active particles move over a dissipative substrate and/or are surrounded by a fluid: birds \cite{cavagna2010scale} fly through the air, fish \cite{Fish} navigate in water, bacterial colonies \cite{bacteria2} glide on surfaces. If the surrounding fluid can be treated as an inert medium, providing only friction, then the motion of active particles that transfer momentum to the fluid can be modeled as a system with overdamped dynamics and no momentum conservation. Such systems are referred to as \textit{dry}. This can be the case when the moving active objects are walking/crawling on a substrate and/or are strongly confined between boundaries that dissipate momentum. In these cases the surrounding fluid is neglected a priori, as it is done for instance the in kinematic descriptions of flocks of birds \cite{cavagna2010scale} and schools of fish \cite{Fish}. It can also be argued that the effect of the surrounding fluid are mitigated at high Reynolds numbers \cite{high_reynolds}. In any case, whenever the fluid is neglected, legitimately or not, we are dealing with dry active matter.\\
On the other hand, when the momentum transfer to the solvent cannot be ignored, the dynamics of the surrounding fluid must be taken into account and one should give a description of the system as a whole (active particles and fluid) where total momentum is conserved, giving rise to long range interactions between active particles mediated by the fluid. Such systems, with momentum conservation, where fluid is important are known as \textit{wet}.

In this thesis we are only concerned with dry active matter. Therefore, in this introductory chapter we will focus on the physical properties of systems that do not conserve momentum. In the first section we will recall a few microscopic models for non-aligning active particles used in the context of dry \textit{scalar} active matter, by which we mean systems where a field theoretical description can be accounted for by only using the scalar density.\\
The second part will be devoted to \textit{vectorial} active matter, where additional slow degrees of freedom (e.g. the orientation, be it polar or nematic) are needed to get a full field theoretical description: we will first present the general properties common to dry aligning dilute active matter (DADAM) \cite{dadam_benoit}, where active particles interact by aligning forces and volume exclusion effects may be ignored (the so-called dilute limit); we will then focus on microscopic models with polar alignment starting with perhaps the simplest one, the active Ising model (AIM) \cite{AIM_PRL, 2d_AIM} (a lattice model for flocking, characterized by a discrete rotational symmetry) and then moving to discuss the celebrated Vicsek model (VM) \cite{vicsek1995, ginelli2016physics}, where rotational symmetry is continuous. While this is of course a fundamental difference with the VM, the AIM is quite instructive in elucidating the nature of the transition to collective motion.\\
The third and final part of this chapter will be devoted to field theoretical descriptions of flocking. In particular, we will introduce the Toner \& Tu theory \cite{toner1998flocks, toner2005hydrodynamics, toner2012reanalysis}, a fluctuating hydrodynamic description for flocking, and briefly review its behavior, both at the onset of collective motion and deep in the ordered phase.  While the Toner \& Tu equations have been originally obtained by a phenomenological approach (based on symmetry arguments), it is interesting to note that they can also be derived by direct coarse-graining methods. These approaches, while not fully rigorous, allow to keep track of the connection with the corresponding microscopic model, at least at the qualitative level, and consequently determine the dependence of the transport coefficients on microscopic parameters. Here we also review one of them, the so-called Boltzmann-Ginzburg-Landau (BGL) \cite{bertin2006boltzmann, bertin2009hydrodynamic} approach. While the method is illustrated for polar systems, possibly the simplest set-up, we will later on also consider a set of mesoscopic equations derived for nematic systems (see Chapter \ref{ch_nematic}).

\section{Dry scalar active matter}
A wide diversity of active particles, individually capable of dissipating energy at the bulk level in order to self-propel, exist across scales both in nature and artificially. These active particles, while differing widely in specifics, taken in isolation are all \textit{persistent random walkers}: in contrast to passive \textit{ideal} random walkers, this introduces a typical scale dividing \textit{ballistic} behavior (at small scales) from \textit{diffusive} behavior (at large scales). This can be quantified by the \textit{persistence time} $\tau$ or the average distance traveled during that time, the \textit{persistence length} $l_p$.

\begin{figure}[hbt!]
    \centering
    \includegraphics[width=0.75\linewidth]{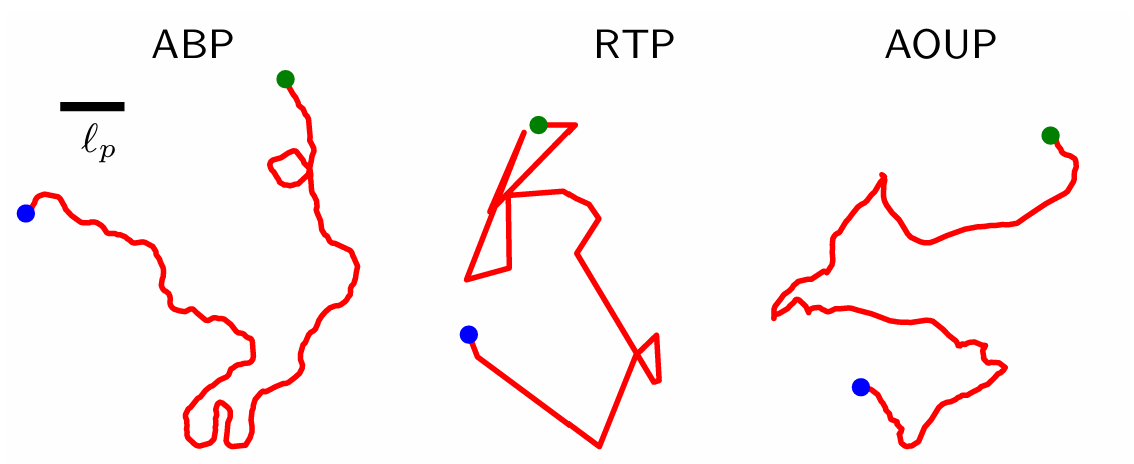}
    \caption{{\bf Schematic trajectories of microscopic models of active particles} Differences of the three basic models (see text) for short times. Figure adapted from \cite{granek2023inclusions}.}
    \label{active_particles}
\end{figure}

With term \textit{scalar} active matter we refer to active systems whose long-time and large-scale behaviors are entirely captured by the stochastic dynamics
of a density field. All non-interacting dry active systems fall in this class, as do a wealth of interacting ones. At the microscopic scale they can be modeled in many different ways: while this, in general does not affect the large scale properties, differences in the microscopic models may lead to relevant differences when interacting with an external potential. For example, in the case of a confining harmonic potential choice of the microscopic model (see Fig.\ref{active_particles}) affects the steady state distribution \cite{szamel2014self}.

Perhaps the simplest idealizations for scalar active matter are active Brownian particles (ABPs) \cite{ABPs_marchetti} and run-and-tumble particles (RTPs) \cite{RTP_1993}. They move with at a constant speed $v_0 = \mu f_a$, where $\mu$ is the particle mobility, and $f_a$ is the self-propelling force's constant magnitude. Because of rotational diffusion, characterized by a rotational diffusivity $D_R$, the self-propulsion orientation of ABPs is constantly changing and the persistence time is simply given by $\tau = 1/[(d-1)D_R]$ . In contrast, RTPs move straight along long stretches and change their self-propulsion direction instantly during \textit{tumbling} events, which happen at a rate $\alpha$, so that their persistence time $\tau = 1/\alpha$.\\
In its most basic form, the spatial dynamics of ABPs and RTPs reads
\begin{equation*}
    \mathbf{\Dot{r}}(t) = \mu \mathbf{f}_a(t)
\end{equation*}
where {\bf r} is the particle position and the self-propulsion force, $\mathbf{f}_a(t)$, can be viewed as a persistent stochastic variable with a fixed magnitude $f_a$.

Both models do not allow fluctuations of the modulo of the propulsion force: a more refined model where particles are allowed to move at different speeds is given by the Active Ornstein-Uhlenbeck particles (AOUPs) where the active force evolves according to an Ornstein-Uhlenbeck process \cite{szamel2014self, sepulveda2013collective}. See Fig. \ref{active_particles} for a sketch of these three classes of stochastic dynamics.\\ 
Despite having distinct dynamics ABPs, RTPs, and AOUPs all result in force auto-correlation functions that decrease exponentially with time:
\begin{equation*}
    \langle f_{a,i}(t)f_{a,j}(0)\rangle = \delta_{ij}\frac{f_a}{d}^2e^{-t/\tau}
\end{equation*}
with $f_{a,k}$ denoting Cartesian components of $\mathbf{f}_a$ and $d$ being the space dimension. Moreover, all result in diffusive dynamics at large scales with an effective diffusion coefficient $D_\text{eff} = \mu^2 f_a^2\tau/d$. For ABPs and RTPs this can also be expressed in terms of the persistence length $l_p=\mu f_a \tau$, so that one has $D_\text{eff} = l_p^2/(d\tau)$, while for AOUPs one gets $D_\text{eff}=\pi l_p^2/(2\tau)$.

Collective behaviors typically encountered in the context of scalar active matter rely on a variety of possible interactions, but for the theory to be a scalar one, interactions must not induce a mutual alignment between individual directions of self-propulsion. This includes a variety of interactions such as: pairwise attraction or repulsion \cite{ABPs_marchetti, structure_dynamics_phase_sep, stenhammar2014phase} or interactions that act only on the magnitude of the self-propulsion velocity and not on its direction (no alignment) \cite{d2017contact, li1991fluctuation}. Such interactions can simply be the effect of volume exclusion forces in spherical particles or, for instance, be mediated by chemical signals as encountered in assemblies of cells interacting via \textit{quorum sensing} \cite{miller2001quorum, cates2015motility}.\\
Many active systems also experience interactions that require more complex mesoscopic descriptions: a typical example is the Vicsek model \cite{vicsek1995novel}, where the proper field theory, for the flocking phase, does not only rely on the density field, but also includes an equation for the orientation field. These systems are the subject of this thesis and will be discussed from section \ref{dadam}.
\newpage

\subsection{Motility induced phase separation - MIPS}
When considering systems whose large-scale behavior is characterized by the conserved dynamics of a density field, possibly the simplest phase transition corresponds to condensation and the breaking of translational uniformity. Condensation has been predicted theoretically \cite{ABPs_marchetti, mognetti2013living, baskaran_prl_2013} and observed experimentally \cite{buttinoni2013dynamical, palacci2013living}, arising from the interplay between attractive, repulsive and self-propulsion forces.\\
With no self-propulsion ($f_a = 0$), an equilibrium system with attracting interactions may undergo equilibrium liquid-gas phase separation. Also if self-propulsion is weak, attractive forces between the particles can easily overcome activity and the expected equilibrium phase separation typically survives. As the self-propulsion force increases, the phase-separated region shrinks until activity overcomes the attractive forces, and the system turns into a homogeneous fluid.\\
Surprisingly, at even larger propulsion forces, a re-entrant phase transition into a phase-separated region is observed \cite{redner2013reentrant}. The mechanism underlying this re-entrant phase transition is now known as motility induced phase separation (MIPS), which is distinct from equilibrium phase separation and does not even require attraction between the particles. Instead, MIPS can be induced by repulsive forces only (at odds with what happens at equilibrium) and results from the interplay between the tendency of active particles to accumulate where they move slower and their slowdown at high density due to collisions and repulsive forces \cite{ABPs_marchetti, cates2015motility}.

MIPS is a consequence of time-reversal symmetry breaking in active matter. In equilibrium, detailed balance ensures that forward and reverse microscopic processes occur with equal probability, ensuring that the dynamics is time-reversible and preventing clustering in the absence of attractive forces. However a persistent self-propulsion in the presence of repulsion violates time-reversal: particles colliding head-on can get stuck for a finite amount of time, while the time-reversed trajectory (by which particles velocities are reversed) are immediately free to move away one from each other \cite{nardini2017entropy}. This leads to irreversible trajectories with the tendency with the tendency to cluster in arrested configurations. This arrested state is inherently out of equilibrium, implicitly sustained by continuous energy input at the particle level, resulting in persistent stresses and anomalous density fluctuations within the clusters. Such non-equilibrium steady states are characteristic of active matter and cannot occur in passive systems.

\subsubsection{A simple linearized argument for MIPS}
To get an intuitive understanding of the origin of MIPS we report here a simple linearized argument, originally given in \cite{cates2015motility}, which shows how the homogeneous solution is unstable in the presence of a position-dependent velocity.

Consider a collection of ABPs moving with position dependent speed $\dot{\bf r}_i=\mathbf{v}_i = v(\mathbf{r}) \hat{\mathbf{n}}_i$, where $\hat{\mathbf{n}}_i$ is a unit vector. In $d=2$ it can be expressed as $\hat{\mathbf{n}}_i = (\cos\theta_i, \sin\theta_i)$ and for ABPs the angle $\theta_i$ will follow
\begin{equation}
    \Dot{\theta}_i(t) = \sqrt{2D_R}\eta_i(t)
\end{equation}
with $D_R$ a rotational diffusivity and $\eta_i$ being a delta-correlated white noise.\\
For ABPs one can then write down the associated Fokker-Planck equation for the probability $p(\mathbf{r}, \theta, t)$ of finding a particle in position $\mathbf{r}$, with orientation $\theta$ at time $t$ as 
\begin{equation}\label{fp_mips}
    \partial_t p(\mathbf{r}, \theta, t) = \nabla_\mathbf{r}\cdot[v(\mathbf{r})p(\mathbf{r}, \theta, t) \hat{\mathbf{n}}] + D_R \partial_\theta^2 p(\mathbf{r}, \theta, t)
\end{equation}
For isotropic processes $p(\mathbf{r}, \theta) = k/v(\mathbf{r})$ is always a stationary solution of \eqref{fp_mips}, with $k$ being some normalization constant, showing indeed that active particles trivially accumulate where they move slower.
The density can then be simply retrieved as $\rho(\mathbf{r}) = \rho_0(2\pi)^{-1}\int p(\mathbf{r},\theta) d\theta$.

Now consider a small perturbation $\delta\rho(\mathbf{r})$ around a initially homogeneous density profile with $\rho(\mathbf{r}) = \rho_0$. This leads to a spatially varying speed $v(\mathbf{r}) \equiv v[\rho(\mathbf{r})]$, that can be expanded as $v[\rho_0+\delta\rho(\mathbf{r})] = v(\rho_0) + \delta\rho(\mathbf{r})v'(\rho_0)$, where $v'(\rho_0)$ is the derivative with respect to the density.
The steady-state density will then be consequently modified, since it depends on the local velocity $v(\mathbf{r})$
\begin{equation}\label{rho_pert}
    \rho(\mathbf{r}) = \frac{k\rho_0}{v[\rho(\mathbf{r})]} \approx \frac{k\rho_0}{v(\rho_0)}\left(1 -\frac{v'(\rho_0)}{v(\rho_0)}\delta\rho(\mathbf{r})\right)
\end{equation}
Mass conservation imposes $\int_V\delta\rho(\mathbf{r}) d^d\mathbf{r} = 0$ and thus determines the value of $k = v(\rho_0)$. Substituting this into Eq.\eqref{rho_pert}, we can express the density as 
\begin{equation}
        \rho(\mathbf{r}) \approx \rho_0\left(1 -\frac{v'(\rho_0)}{v(\rho_0)}\delta\rho(\mathbf{r})\right)\equiv \rho_0 + \widetilde{\delta\rho}(\mathbf{r})
\end{equation}
with the new perturbation $\widetilde{\delta\rho} (\mathbf{r}) = -\rho_0\delta\rho(\mathbf{r})v'(\rho_0)/v(\rho_0)$.\\
These fluctuations, stemming from the perturbed velocity, are amplified when $\widetilde{\delta\rho}>\delta\rho$. This condition is met whenever $ -\rho_0 v'(\rho_0)/v(\rho_0)>1$, which can be conveniently rewritten as 
\begin{equation}\label{condition_mips}
    \frac{d}{d\rho}[\rho v(\rho)]|_{\rho = \rho_0} < 0.
\end{equation}
Whenever \eqref{condition_mips} is satisfied the homogeneous solution is linearly unstable and a feedback mechanism sets in: active particles in regions of high density will move slower which in turn will increase the local density further lowering their velocity, thus leading a positive feedback loop.
\newpage

\section{Dry aligning dilute active matter}\label{dadam}
When \textit{aligning} interactions are at play, one speaks of {\it vectorial} active matter. Perhaps the simplest case is dry aligning dilute active matter (DADAM) \cite{chate2020dry, dadam_benoit}: it is \textit{dry} since there is no fluid surrounding the particles or it may be anyhow ignored (as opposed to the wet case), \textit{aligning} because particles have an explicit aligning mechanism (as in the Ising or XY model) and finally \textit{dilute} since point-wise particles do not occupy any volume (an idealization for systems far away from jamming conditions and where volume exclusion forces may be neglected).
The best known model for collective motion, the Vicsek model (VM) \cite{vicsek1995novel}, falls into the definition of DADAM and is somehow its prototypical example. For this reason, all models that fall under this description are sometimes also called Vicsek-style models.

Three main classes can be identified based on the nature of the alignment and on the symmetry of the particles: the polar/Vicsek class, the self-propelled rods class and active nematics.

\begin{figure}[hbt!]
\centering
\includegraphics[width=1\linewidth]{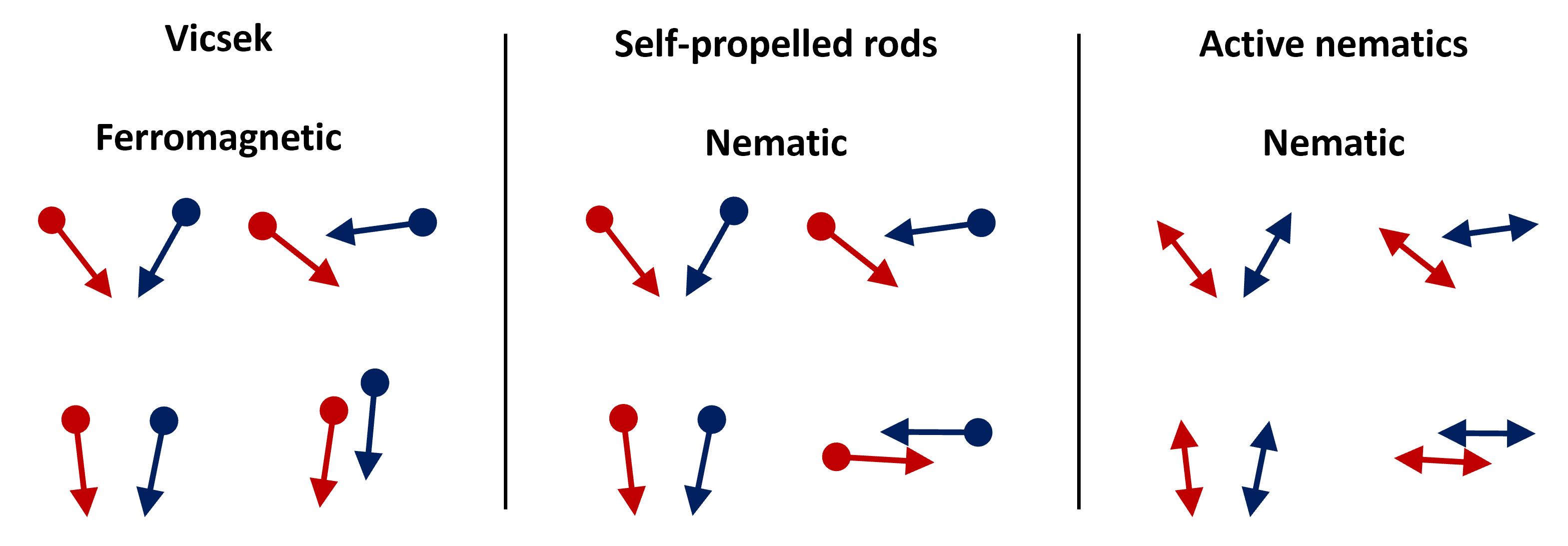}
 \caption{{\bf Three DADAM classes with corresponding alignment} The first row of particles shows their initial state prior to the alignment interaction, while the bottom row shows the effect of alignment on the orientation of particles. Figure adapted from \cite{dadam_benoit}.}
 \label{alignment_dadam}
\end{figure}

As shown in Fig.\ref{alignment_dadam}, the polar class is characterized by \textit{ferromagnetic} alignment, while the two other classes align \textit{nematically} (that is, they are invariant under rotation by $\pi$): self-propelled rods move polarly (they never revert their direction of motion spontaneously) while active nematics do revert their direction of motion with a finite probability rate and have a complete nematic symmetry.\\
The symmetry of the alignment rule directly rules what type of orientational order can emerge locally: ferromagnetic alignment leads to polar local order, while nematic alignment allows for both nematic and polar local order.
Despite these differences, the three classes share many common properties.

\subsection{Phase diagram}
\begin{figure}[hbt!]
\centering
\includegraphics[width=0.5\linewidth]{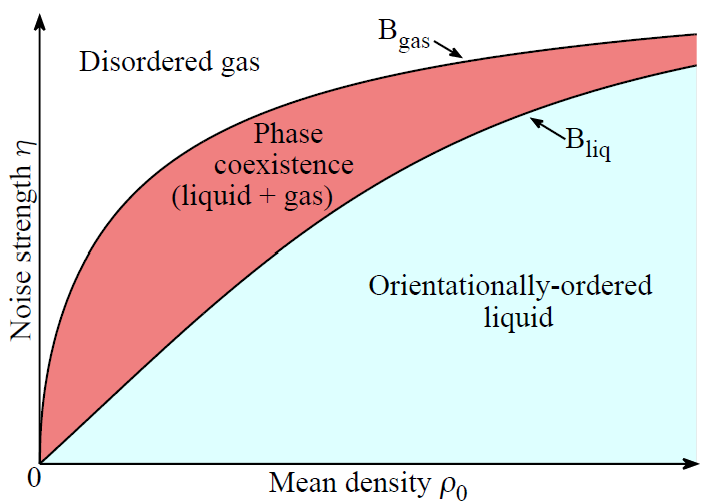}
 \caption{{\bf Schematic phase diagram of all 3 basic DADAM classes in the ($\rho_0, \eta$) plane} The phase coexistence region is characterized by dense ordered bands traveling in a disordered gas and is delimited here by the binodal lines $B_\text{gas}$ and $B_\text{liq}$. Figure adapted from \cite{dadam_benoit}.}
 \label{phase_diagram}
\end{figure}

We start by presenting the phase diagram (see Fig. \ref{phase_diagram}), which holds for all three classes. For high noise and low density the particles' persistent random walk wins over alignment and the system remains in the \textit{disordered gas} phase. When the noise is sufficiently low and/or density sufficiently high, then the alignment mechanism wins over noise and the system undergoes a spontaneous symmetry breaking of its rotational symmetry leading to an ordered state. These two phases are then separated by a coexistence region composed of ordered, dense domains coexisting with a disordered gas: the two lines delimiting this phase meet at the origin and converge again when $\rho_0 \to \infty$. They can be interpreted as the \textit{binodal} lines of a liquid-gas phase separation scenario. At difference with the equilibrium liquid-gas phase separation, here the critical point is sent to infinity. Intuitively, this is due to the fact that the disordered and ordered states have a different symmetry, so that it is impossible to move continuously from one phase to the other as in a supercritical equilibrium fluid. \\
The nature of the coexistence region differs according to the underlying symmetry: in the presence of nematic alignment, for instance, one observes dense bands extending longitudinally w.r.t. nematic order. These objects are however intrinsically unstable to long-wavelength deformations, leading to a mesoscopic chaotic behavior in large enough systems. Ferromagnetic alignment, on the other hand, leads to stable bands which move in a disordered low-density background and extend transversally w.r.t the direction of motion. A more complete discussion of this phase for ferromagnetic alignment can be found in Sec. \ref{liquid_gas_pt} - \ref{physical_vm}.

\subsection{Physical properties}\label{dadam_physical}
In the ordered liquid phase all 3 classes of DADAM display long-range correlations and anomalous fluctuations.
This arises from the spontaneous breaking of the continuous rotational symmetry. Interestingly, since the systems are out of equilibrium, they do not have to obey the Mermin-Wagner-Hohenberg theorem \cite{mermin1966absence, hohenberg1967existence} and long-range order is possible even in $d=2$ for the polar, Vicsek case. Active nematics, on the other hand, only shows quasi-long range order in $d=2$.

Indeed, Toner \& Tu showed \cite{toner1995long, toner1998flocks}, by means of a phenomenological hydrodynamic model, that polar flocks exhibit true long-range order (TLRO) even in $d=2$. They also predicted the anisotropic algebraic decay of the two-point correlation functions of density and velocity fluctuations. For instance, the equal-time velocity correlation function scales as
\begin{equation}
    C_v(\mathbf{r}) = |\mathbf{r}_\perp|^{2\chi} f(r_\parallel/|\mathbf{r}_\perp|^\xi)
\end{equation}
where $\parallel$ and $\perp$ pedices refer respectively to directions longitudinal and transverse the mean flocking direction. The roughness exponent $\chi$ and the anisotropy exponent $\xi$ (and also the scaling function $f$) are universal scaling exponents characteristic of the theory. The existence of this anomalous scaling in the density and velocity fields is equivalent to the existence of giant number fluctuations (GNF), as it will be shown in Sec. \ref{physical_vm}.\\
With the term GNF one usually means that the variance $\Delta n^2 \equiv \langle n^2\rangle - \langle n\rangle ^2$ of the number of particles $n=\rho_0 \ell^2$ contained in a box of linear size $\ell$, much smaller than system size, scales like 
\begin{equation}
    \Delta n^2 \sim \langle n\rangle^\phi \;\;\text{with}\;\; \phi>1,
\end{equation}
as one changes the box size. This is not due to large local fluctuations, but rather to fluctuations that are correlated over large distances, and it is at odds with what happens when density correlations are exponentially decaying and one can use the central limit theorem to predict an exponent $\phi = 1$. GNF are fairly easy to extract from numerical or experimental data and have thus been measured in numerical simulations in the ordered liquid phase of all 3 basic DADAM classes. Numerical estimates of the associated scaling exponent have found that $\phi \simeq 1.6-1.7$ in all 3 cases.

A more specific analysis of the properties of the polar/Vicsek class will be carried out in the next section, while we will discuss the specific properties of self-propelled rods in the introduction of chapter \ref{ch_nematic}.
\newpage

\section{Active Ising model}\label{sec_aim}
The next two sections are dedicated to microscopic models. Before moving to the celebrated Vicsek model, we briefly consider a minimal model for flocking with discrete symmetry, first introduced in \cite{AIM_PRL}. The active Ising model (AIM) drops the continuous rotational symmetry and the off-lattice dynamics of the Vicsek model, for a discrete symmetry on a lattice. It however retains two essential ingredients for flocking: self-propulsion and aligning interactions.
\subsection{The model}
\begin{figure}[hbt!]
    \centering
    \includegraphics[width=0.75\linewidth]{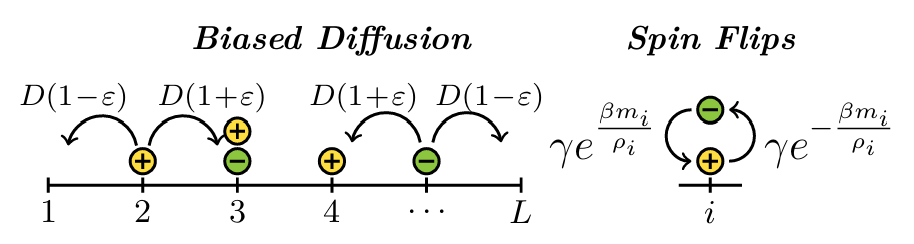}
    \caption{{\bf Sketch of the two possible actions and their rates of occurrence}, adapted from \cite{2d_AIM}. The ferromagnetic interaction between particles is purely on-site and particles diffuse freely. Beyond the biased diffusion shown here, particles also hop symmetrically up or down, with equal rates $D$ in both directions.}
    \label{interactions_aim}
\end{figure}

Consider $N$ particles moving on a $2D$ lattice of $L\times L$ sites with periodic boundary conditions. Each particle carries a discrete orientation, the spin $s = \pm 1$, and there are no excluded volume interactions: a single site $i$ can accommodate an arbitrary number $n_i^\pm$ of particles with spin $\pm 1$. The local density $\rho_i$ is then simply defined as $\rho_i\equiv n_i^+ + n_i^-$ and the magnetization $m_i \equiv n_i^+ - n_i^-$. Particles update their position and magnetization at each time step following two possible actions: spin flip or hopping to a neighboring site (with rates depending on their spin $s$) as sketched in Fig.\ref{interactions_aim}.

A particle with spin $s$ on site $i$ flips its spin at rate $W_{s\to -s} = \exp(-s\beta m_i/\rho_i)$, where $\beta = 1/T$ plays the role of the inverse of a temperature in an equilibrium context. This interaction is purely local since particles only align with other particles on the same site $i$ and without particles hopping, each site will result in an independent fully connected Ising model. \\
Particles also undergo free diffusion on the lattice: the diffusion is unbiased, with rate $D$, in both up and down directions, while their spin induces a bias and thus a net drift in the horizontal direction. A particle carrying spin $s$ will hop to the right with rate $D(1+s\epsilon)$ and to the left with rate $D(1-s\epsilon)$. There is, thus, a mean drift which plays the role of self-propulsion, with particles of spin $\pm1$ traveling along the horizontal axis with mean velocity $\pm2D\epsilon$. \\
The parameter $\epsilon$ allows to interpolate continuously between \textit{totally self-propelled} particles ($\epsilon = 1$), where particles only hop to a neighboring site according to their spin $s$, \textit{self-propelled} when $\epsilon\in(0,1)$ and purely diffusive ($\epsilon = 0$).

\subsection{A liquid gas phase transition to collective motion}\label{liquid_gas_pt}
\begin{figure}[hbt!]
      \centering
      \includegraphics[width=1\linewidth]{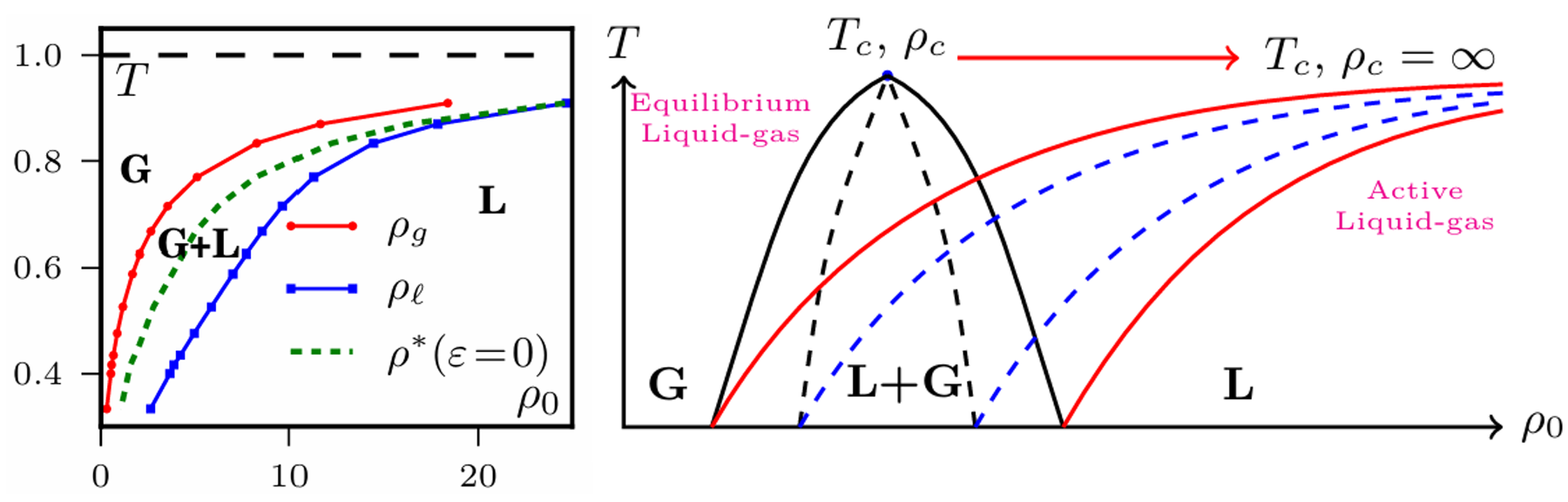}
      \caption{{\bf Phase diagram of the AIM} - Left: Phase diagrams in the $(\rho_0, T)$ plane of the AIM. The red and blue lines delimit the coexistence (liquid and gas) region. The green dashed line indicates the critical points at $\epsilon = 0$. Right: Schematic picture of the differences between the phase diagrams of the passive and active liquid gas transition. In the active case, because the liquid and the gas have different symmetries, the critical point is sent to $\rho = \infty$, thus suppressing the supercritical region. Figures adapted from \cite{2d_AIM}.}
      \label{phase_diagram_aim}
\end{figure}

The phase diagram in the global density $\rho_0 = N/L^2$ and temperature $T = 1/\beta$ plane, at fixed self-propulsion speed $\epsilon$, has been obtained numerically and is reported in Fig.\ref{phase_diagram_aim} (left panel). It roughly resembles that of the DADAM class and, more specifically, of polar particles (see next section). At high temperature and low densities, the particles fail to synchronize their headings and the local magnetization is $\langle m_i\rangle \sim 0$ (\textit{disordered gas} phase). On the contrary, for large enough densities and small temperatures, the particle are able to synchronize and move collectively, forming a \textit{polar liquid} state with $\langle m_i\rangle = m_0\neq 0$. For intermediate densities, when $\rho_0 \in [\rho_g(T), \rho_l(T)]$, the system phase separates into a traveling band of polar liquid and a disordered gaseous background. The two binodal lines $\rho_g(T)$ and $\rho_l(T)$ delimit the domain of existence of phase-separation and are coexistence lines. 

\begin{figure}[hbt!]
    \centering
    \includegraphics[width=1\linewidth]{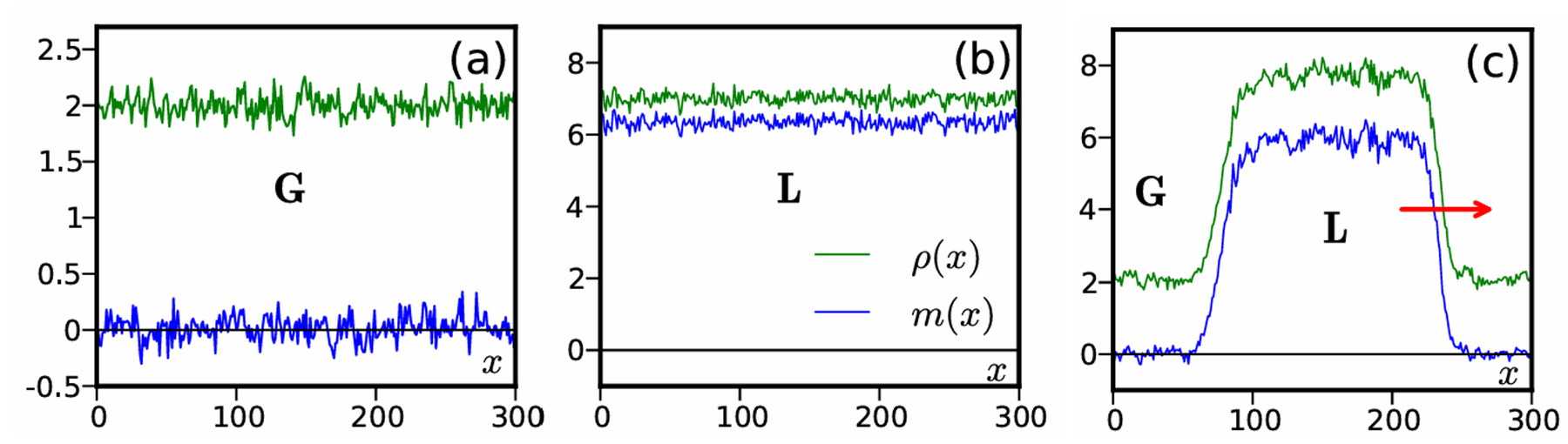}
    \caption{{\bf Examples of density profiles (in green) and magnetization profiles (blue) averaged over the the vertical direction}. (a) Disordered gas. (b) Polar liquid. (c) Liquid-gas coexistence. Figure adapted from \cite{2d_AIM}.}
    \label{fig:phases_aim}
\end{figure}

For all the phase-separated profiles at fixed $T$ the densities in the gas and in the liquid part are $\rho_g$ and $\rho_l$ respectively. Thus, varying the global density $\rho_0$ at constant temperature only varies the width of the band. An example of the density and magnetization profiles is provide in Fig.\ref{fig:phases_aim}.
  
While the general structure of the phase diagram can resemble that of an equilibrium liquid-gas transition (gas, liquid and a coexistence phase), the shapes of the transition lines are quite unusual. This has to deal with the different symmetry that the liquid and gas phase have: indeed the liquid breaks the discrete \textit{rotational} $Z_2$ symmetry to achieve a state of polar order, while the gas phase is rotationally invariant. This implies that the system cannot continuously transform from one homogeneous phase to the other without crossing a transition line. Thus, there is no supercritical region and the critical point is sent to $T_c=1$ and $\rho_c = \infty$, as shown in Fig.\ref{phase_diagram_aim} (right panel).\\
This symmetry argument should hold for all flocking transitions where the symmetry between the disordered gas and the homogeneously ordered phase is broken and phase separation takes place at the onset of order: it surely holds for the Vicsek model and for all 3 DADAM classes (one should then interpret the amplitude of the noise $\eta^2$ as the effective temperature).

Interestingly, the simplicity of the AIM, allows for an exact direct coarse graining. This consist in first deriving the time evolution for the average occupancy on site $i$, $\langle n_i^\pm\rangle$; from these equations one can then derive the evolution for the average density $\langle\rho_i\rangle$ and magnetization $\langle m_i\rangle$. Taking the continuum limit of these equations one obtains the two following equations for the coarse grained density $\langle\rho\rangle$ and magnetization $\langle m \rangle$
\begin{equation}\label{density_aim}
    \partial_t\langle\rho\rangle = D\nabla^2\braket{\rho} - v \partial_x\braket{m},
\end{equation}
\begin{equation}\label{magnetization_aim}
\partial_t \braket{m} = D\nabla^2\braket{m} - v \partial_x \braket{\rho} 
+ \left\langle 2\rho \sinh\left(\frac{\beta m}{\rho}\right) - 2m \cosh\left(\frac{\beta m}{\rho}\right) \right\rangle.
\end{equation}
where $v = 2D\epsilon$. These equations are exact: to solved them explicitly some approximation are needed in order to account for the last term of Eq.\eqref{magnetization_aim}, which describes aligning interactions between particles. \\
A mean-field approximation, which neglects both the correlations between density and magnetization and their fluctuations is not able to correctly account for the nature of the phase transition: this treatment, which amounts to substitute $\langle f(\rho, m)\rangle$ with $ f(\langle\rho\rangle, \langle m \rangle)$ for the last nonlinear term, is nevertheless able to explain a transition to a state with finite magnetization $m$. However, in this simple mean field treatment both the disordered and the ordered solution are found to be linearly stable to spatial perturbations, thus missing a fundamental qualitative difference with the microscopic model since, in the absence of a phase separated regime, the transition is erroneously found to be continuous.\\
A more refined mean-field model, allowing the density and magnetization to fluctuate around their mean value, can be shown \cite{2d_AIM} to introduce a density dependence in the term linear in the magnetization (this is obtained by an expansion of the nonlinear terms up to third order in the magnetization), leading to 
\begin{equation}\label{density_aim_rfmf}
    \partial_t\rho = D\nabla^2\rho - v \partial_x m,
\end{equation}
\begin{equation}\label{magnetization_aim_rfmf}
\partial_t m = D\nabla^2 m - v \partial_x \rho 
+ 2\left(\beta - 1 - \frac{r}{\rho}\right)m - \alpha \frac{m^3}{\rho^2}.
\end{equation}
where we have dropped the brackets $\braket{.}$ for the sake of clarity. This is a key ingredient to describe phase separation at the mesoscopic level as noted already in \cite{bertin2006boltzmann, bertin2009hydrodynamic} in the case of the particles with continuous rotational symmetry (see Section \ref{pd_bgl_mu_dependence}). \\
Linear stability analysis of the equations obtained in the refined mean-field model show that for a finite region of the phase diagram near the onset of order the homogeneous solution is linearly unstable and phase separated solutions appear, with macroscopic liquid bands traveling in a disordered gas background. Direct numerical simulation of Eqs.\eqref{density_aim_rfmf}-\eqref{magnetization_aim_rfmf} yields a phase diagram analogous to the one of the microscopic AIM, with two binodal lines surrounding the spinodal lines where the homogeneous ordered solutions becomes linearly unstable. Finally, The densities in the gas and liquid parts of the profiles remain constant as the global density $\rho_0$ is varied, which was also found in microscopic simulation and is a signature of the liquid-gas phase transition.

The simplicity of the AIM helps in clarifying the nature of the transition, complementing the order-to disorder picture typical of ferromagnetic systems with a liquid-gas point of view. It can be argued that the same scenario also holds for off-lattice models with a continuous rotational symmetry, such as the Vicsek class discussed in the next section. However, some features specific of off-lattice DADAM systems are absent due to the lack of continuous rotational symmetry. In particular one should note that the homogeneous ordered phase does not exhibit long-ranged correlations, at difference with what happens when a continuous symmetry is spontaneously broken. \\
Indeed, in the AIM the number fluctuation are found to be normal both in the polar liquid and gas phases, where $\Delta n^2 \equiv \langle n^2\rangle - \langle n\rangle^2$ is found to scale as $\Delta n \sim \langle n \rangle$, while they are trivially giant in the phase-separated regime where $\Delta n^2 \sim \langle n\rangle^2$. This last scaling is a simple consequence of phase separation (see \cite{2d_AIM} for simple argument) and one should distinguish this from the anomalous scaling reported for DADAM, which is a signature of long-range correlations.\\ 
Another relevant difference w.r.t. the polar ferromagnetic class exemplified by the Vicsek Model (VM) introduced in the next section, is the shape of the bands in the phase-separated region: the macro-separation of the AIM is then replaced by micro-phase separation (multiple bands) in the VM. This difference is understood to be due to the different nature of fluctuations correlations in the two systems and thus, once again, to the discrete symmetry of the AIM.
\newpage

\section{Vicsek model}
The Vicsek Model (VM) \cite{vicsek1995novel} is without any doubts the simplest microscopic model, with continuous rotational symmetry, describing a transition to collective motion. In the study of active matter it plays a prototypical role, analogous to the one played by the Ising model for equilibrium ferromagnetism.

\begin{figure}[hbt!]
    \centering
    \includegraphics[width=0.5\linewidth]{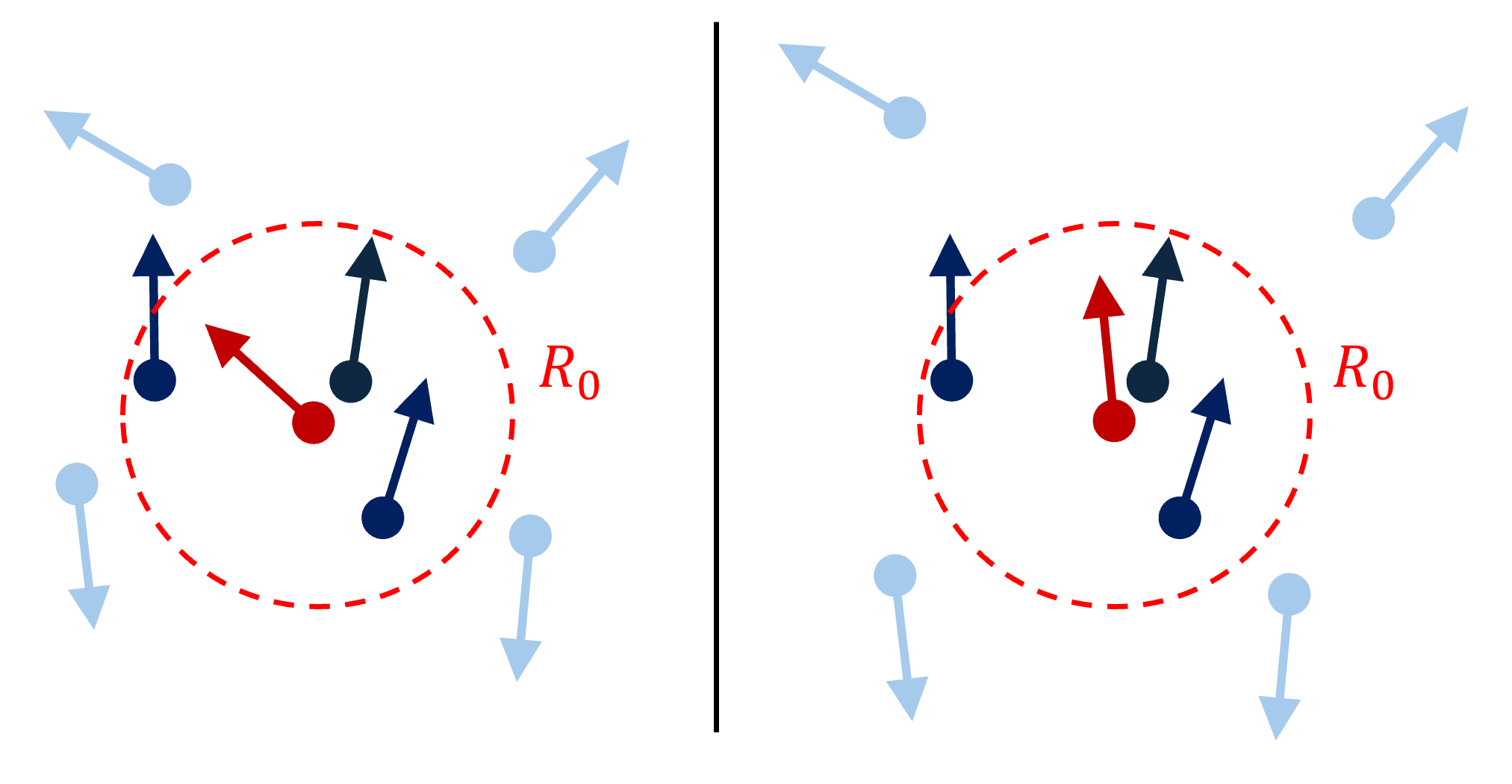}
    \caption{{\bf Schematic representation of Vicsek alignment rule} The red particle aligns only with metric neighbors within a distance $R_0$ in blue, while neighbours outside this metric range do not influence the central particle's orientation. Figure adapted from \cite{ginelli2016physics}.}
    \label{alignment_vicsek}
\end{figure}

The model describes the \textit{overdamped} dynamics of $N$ self-propelled particles, characterized by their position $\mathbf{r}_i^t$ and direction of motion $\mathbf{n}_i^t$, with $|\mathbf{n}_i^t| = 1$. Here $i$ labels the particles ($i = 1, ..., N$) while $t$ indicates time.\\
Since the focus of this thesis will be on $2d$ systems we will discuss explicitly only the bi-dimensional VM, while a more extensive review can be found in \cite{ginelli2016physics}. In $d=2$ spatial dimensions, we can parameterize the orientation $\mathbf{n}_i^t$ with an angle $\theta_i^t$ for which $\mathbf{n}_i^t = (\cos\theta_i^t, \sin\theta_i^t)$. All particles move with constant speed $v_0$ according a simple time-discrete dynamics
\begin{equation}\label{vicsek_theta}
      \theta_i^{t+\Delta t}=\arg(\sum_{i\sim j}\mathbf{n}_i^t) + \eta\xi_i^t
\end{equation}
\begin{equation}\label{position_vicsek}
      \mathbf{r}_i^{t+\Delta t}=\mathbf{r}_i^t + \Delta t v_0 \mathbf{n}_i^{t+\Delta t}
\end{equation}
The first equation describes the elementary mechanism of alignment of the VM: particle $i$ aligns its orientation with all particles ($i$ included) in its spherical metric neighborhood centered in $i$ of radius $R_0$ as schematically reported in Fig.\ref{alignment_vicsek}. Local interaction rules based on topological interactions have been proposed in the literature \cite{ginelli2010relevance}, but for simplicity we will not discuss them here.

The alignment is contrasted by a noise term $\eta\xi_i^t$, where $\xi_i^t$ is a zero average, delta-correlated scalar white noise ($\langle \xi_i^t\rangle = 0, \langle\xi_i^t\xi_j^k\rangle \sim \delta_{ij} \delta_{tk}$) uniformly distributed in $[-\pi, \pi]$. The amplitude $\eta^2$ plays a role analogous to the one played by the temperature $T$ at equilibrium. For the way it is implemented, the maximum possible value is $\eta=1$ and it corresponds to completely randomizing the particles' headings at each time step. 

Finally, the dynamics just described is \textit{synchronous}, meaning that all particles orientations and positions get updated at the same time. Note also that without loss of generality one can rescale space and time to set the interaction radius $R_0$ and the time-step $\Delta t$ equal to one. This choice will be followed in the rest of this thesis.
\subsection{The Vicsek class}
The \textit{Vicsek model} is characterized by certain symmetries and conservation laws that are rather general and define the Vicsek universality class. 
 
\textbf{Continuous rotational symmetry}\\
The time-discrete dynamics just described is \textit{isotropic} in space, since there is no preferential direction for the particles to choose. But in Eq.\eqref{vicsek_theta} there is an explicit polar alignment rule that, analogously to equilibrium ferromagnetism, makes it possible to achieve a state of global orientational order and collective motion by \textit{spontaneously} breaking the \textit{continuous rotational} symmetry. 
 
One can define a \textit{global polar order parameter} simply as $\mathbf{m}(t) = \frac{1}{N}\sum_{i=1}^N \mathbf{n}_i^t$, similarly to the total magnetization in spin systems. We thus expect the stationary time average $ m = \langle\mathbf{m}(t)\rangle_t$ to be close to 1 when in the ordered phase, while $m\sim 0$ in the disordered one (more precisely $m\sim 1/\sqrt{N}$, as one expects for a sum of random variables). 
 
\textbf{Activity and local alignment interactions}\\
The particles self-propel following the rule \eqref{position_vicsek} and, given that the radius of interaction $R_0 \ll L$, with $L$ being the linear system size, they will change neighbors on a regular basis, unlike the equilibrium $XY$ model where the neighbors are fixed a priori, given a lattice. The \textit{exchange of neighbors}, dictated by alignment fluctuations w.r.t. the mean velocities, is what really makes the system out of equilibrium and admits several non-trivial results (LRO in $d=2$, GNF, super-diffusive behavior...).
 
\begin{figure}[hbt!]
    \centering
    \includegraphics[width=0.5\linewidth]{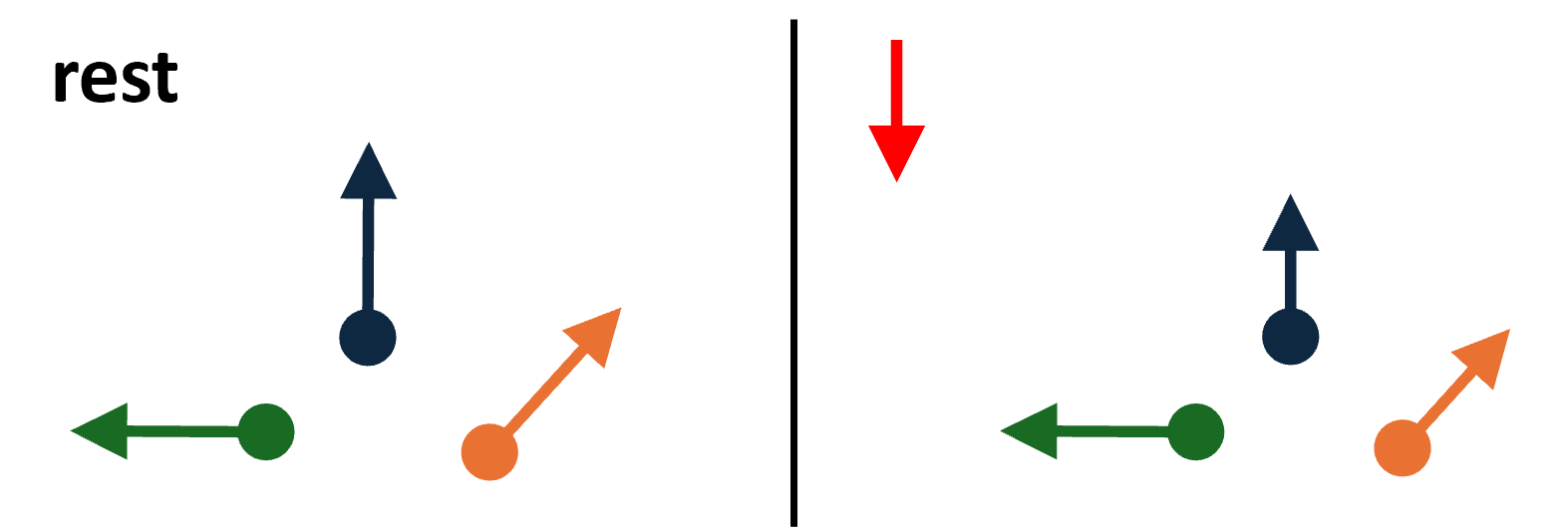}
    \caption{{\bf Schematic representation of the absence of Galilean invariance} The effect of a Galilean boost (indicated by a red arrow) on three different particles shows that in the case of the Vicsek model such an invariance is absent, since the particles' velocity is affected in different ways. In particular note that particles prior the boost (left side) have the same modulo, while after the boost (right side) the three particles do not have the same modulo anymore.}
    \label{absence_galileian}
\end{figure}

\textbf{Conservation laws} \\
The total number of particles $N$ is fixed: particles do not get created, neither die. 
When one allows particles to fluctuate around a fixed global density, as  it was done in \cite{toner_malthusian}, one has to deal with a different class of flocks called \textit{Malthusian flocks}. As one might imagine, relaxing this fundamental conservation law changes the universality class in a non-trivial way.\\ 
It should also be noted that no momentum is conserved since our particles are thought to move on a dissipative substrate which acts as a momentum sink. A notable consequence of the lack of momentum conservation is the absence of Galilean invariance: there is a \textit{preferential frame} that is the one where the dissipative substrate is at rest, as shown in Fig.\ref{absence_galileian}.

\subsection{Physical properties}\label{physical_vm}
The general phase diagram sketched for the 3 DADAM classes in Fig.\ref{phase_diagram} obviously holds also for the VM model. 
The mechanism describing the transition to collective motion is essentially captured by the simplest AIM described in the previous chapter, with a (weakly) first order transition. Here I will first recall some phenomenological results on the coexistence phase and the focus on fluctuations in the homogeneously ordered phase. 
 
\subsubsection{Phase diagram and phase coexistence} 

\begin{figure}[hbt!]
    \centering
    \includegraphics[width=1\linewidth]{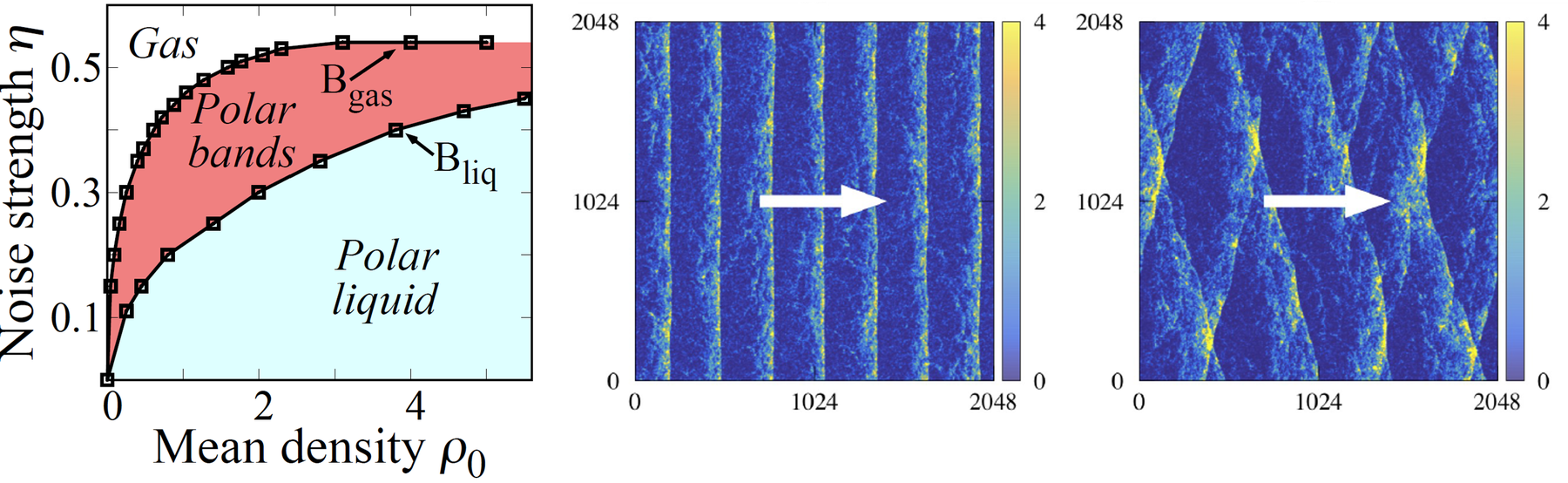}
    \caption{{\bf Phase diagram in the ($\rho_0, \eta$) plane of the Vicsek model}. Left - Phase diagram showing the three different phases of the Vicsek model. The $B_\text{gas}$ and $B_\text{liq}$ lines correspond to the binodal lines. Center - Snapshot of the traveling bands showing micro-phase separation (color encodes density) Right - Snapshot of the density field in the case of the cross-sea phase. Figures adapted from \cite{dadam_benoit}.}
    \label{phase_diagram_vm} 
\end{figure}

As we have already mentioned in the section dedicated to the DADAM classes, one of the main features of the VM model is the coupling between \textit{local order} and \textit{local density} induced by motion: indeed moving particles may gather in high density patches, increasing in turn the number of interacting neighbors. Locally high density has a positive feedback on the efficiency of the alignment interaction, so that high density patches may be able to order in a sea of disordered particles.\\
It has been shown \cite{bertin2006boltzmann, bertin2009hydrodynamic} that this feedback mechanism inevitably leads to a long wave-length instability near the onset of order, that destabilizes the homogeneous ordered phase and leads to phase separation (for a brief discussion see also Section \ref{pd_bgl_mu_dependence}). In the Vicsek class this (micro)-phase separation takes the form of many high density and ordered traveling bands which extend transversally w.r.t. the direction of motion: not too far from the gas binodal line $B_{gas}$, the bands are arranged in a smectic way forming a set of identical, equally-spaced, equal-speed objects, while close to the liquid binodal $B_{liq}$, the bands are more numerous and interact strongly with each other \cite{ihle_cross_sea}. The bands are portions of liquid whose number increases linearly with the dimension of the system along the polar order direction and/or the the global density $\rho_0$: one is thus in the presence of what is know as \textit{micro-phase} separation, where the liquid coexists with a disordered gas in the form of a micro-phase (in this case the bands), as shown by Fig. \ref{phase_diagram_vm} (right panels).

The two binodal lines defining the phase diagram correspond to non-critical transitions: crossing the liquid binodal $B_{liq}$ we enter the \textit{polar liquid} phase. 

\subsubsection{The Toner \& Tu phase} 
The homogeneous polar liquid phase of the Vicsek class is sometimes refereed to as \textit{Toner} \& \textit{Tu} phase, named after the authors of the pioneering papers of the hydrodynamics of flocking \cite{toner1995long, toner1998flocks, toner2012reanalysis}. \\
Numerical simulation of the VM model show that the system is capable of achieving long-range order (LRO) even in $d=2$. This is possible due to the non-equilibrium nature of the VM model, that does not have to obey the Mermin-Wagner-Hohenberg theorem \cite{mermin1966absence, hohenberg1967existence}, which would prevent LRO order for systems in $d=2$ at \textit{equilibrium} which break a continuous symmetry. A nice simplified argument, originally due to Toner, is reported in \cite{ginelli2016physics}. 
 
Since a continuous symmetry is spontaneously broken, we expect the entire ordered phase to be characterized by long range correlations between the fluctuations of the orientations. Moreover, the coupling between velocity (or orientation) and density, results into long range correlations also in the fluctuations of the density. 
Indeed, two-point correlation functions of density $C_\rho$ and velocity $C_v$ fluctuations display (a possibly anisotropic) algebraic decay
\begin{equation}\label{anisotropic_VM}
    C_\rho \sim C_v \sim C(\mathbf{r}) = |\mathbf{r}_\perp|^{2\chi} f(r_\parallel/|\mathbf{r}_\perp|^\xi)
\end{equation}
where $\parallel$ and $\perp$ pedices refer respectively to directions longitudinal and transverse the mean flocking direction. The roughness exponent $\chi$ and the anisotropy exponent $\xi$ other than the function $f$ are universal. Their exact values have been long debated and will be discussed in Sec.\ref{scaling_exp}. Anyhow, fluctuations in the Vicsek class are \textit{scale-free}. \\ 
This has a number of nontrivial consequences: the existence of giant number fluctuations (GNF) and in $d=2$ super-diffusive behavior, above all.\\
We have already seen in Sec.\ref{dadam_physical} that GNF are common to all three DADAM classes. Numerical simulations \cite{chate2008collective} of the VM show that the variance $\Delta n^2\sim\langle n \rangle^\phi$, where $\langle n \rangle = \rho_0 \ell^2$ is the mean number of particles in a box of size $\ell$ and $\phi \simeq 1.6$ both in $d = 2, 3$. A slow enough decay in space of local density fluctuations $\delta\rho(\mathbf{r},t)$ correlations, corresponds to an algebraic divergence at small frequencies $\mathbf{q}$ in Fourier space
\begin{equation}
    S_\rho(\mathbf{q},t) \equiv \langle \delta\rho(\mathbf{q},t)\delta\rho(-\mathbf{q},t)\rangle\sim \frac{1}{q^\sigma} \; \; \text{for} \;\; q\to 0 
\end{equation}
where $q = |\mathbf{q}|$. In systems without slow decaying correlations, instead, one would find $\sigma = 0$ and no small wave-number divergence of the structure factor. \\
The small frequency divergence of the stationary density structure factor $S_\rho(\mathbf{q})$ gives indeed the fluctuations to mean ratio of the local particles number in the limit of large particle numbers 
\begin{equation}
S_\rho(\mathbf{q}\to 0) = \rho_0\left[\frac{\Delta n^2}{\langle n \rangle}\right]_{n\to\infty} 
\end{equation}
Now, since $\langle n \rangle = \rho_0 \ell^d$, where $\ell$ is the linear size of a box in which $n$ and $\Delta n$ are measured and $d$ is the dimensionality. Transforming back into real space one has $S_\rho(\mathbf{q}\to0)\sim1/q^\sigma\sim \ell^\sigma$, one obtains
\begin{equation}
    \Delta n^2\sim\langle n\rangle^{1+\sigma/d}
\end{equation}
and finally
\begin{equation}
    \phi = 1 +\frac{\sigma}{d}
\end{equation}
recovering the normal, central limit theorem result $\phi = 1$ for $\sigma = 0$ as one would expect. \\
While this argument is somehow simplified as it ignores the possibly anisotropic nature of correlations (see Eq.\eqref{anisotropic_VM}), a more refined calculation leads to the same result with $\sigma$ now characterizing the leading divergence in the structure factor \cite{ginelli2016physics}. 

\section{Beyond the dilute limit}\label{other_ways}
Beyond the Vicsek model, there are of course several other sets of microscopic dynamics that can lead to collective motion.\\
The inertial spin model (ISM) \cite{cavagna2015flocking} was originally introduced in order to explain information propagation in flocks of birds: its short time dynamics allows to correctly capture the behavior of real flocks of birds (e.g. collective turns). Interestingly, the VM can be recovered as the overdamped limit of the ISM, indicating that the asymptotic the long-time physics is akin to that of the VM. The VM and the ISM both rely on explicit alignment interactions between the velocities of neighboring particles. Although this is probably the simplest way to make a collection of agents flock (since it is directly related to what is done in condensed matter with magnetic spins), there are also other ways of flocking that do not rely on explicit velocity alignment. \\
Models with visual cone interactions \cite{barberis2016large}, where particles orient their velocities towards the position of the neighbors perceived in their visual cone, have been introduced as to model the behavior of animals that move using the information inside their visual cone. It can be shown that in this set-up interactions turn out to be \textit{non-reciprocal}: the torque $\Gamma_{ij}$ exerted by particle $i$ on particle $j$ is different from $\Gamma_{ji}$. \\
Flocking behavior was also found to appear \cite{das2024flocking} through interactions that turn agents away from each other, through a torque $\Gamma$: at odds with what is found for aligning interactions, the emergence of polar order from turn-away interactions requires particle repulsion $F_{ij} \propto\mathbf{\hat{r}}_{ij}$, where $\mathbf{\hat{r}}_{ij}$ is the unit vector pointing from particle $i$ to particle $j$. While the Vicsek alignment rule couple the orientations of two particles and results in torques $\Gamma_{ij}\propto \mathbf{n}_i\times \mathbf{n}_j$, turn-away interactions couples the orientation of one particle to the relative position of particles $\mathbf{r}_{ij}$ with the result that the torque $\Gamma_{ij} \propto \mathbf{n}_j \times \mathbf{r}_{ij}$ is intrinsically non-reciprocal ($\Gamma_{ij}\neq-\Gamma_{ji}$). 

In the following we move beyond the dilute limit and focus on a model for flocking where volume exclusion interactions between particles are present and induce velocities misaligned with respect to self propulsion forces. We will argue that torques, that tends to align the latter with the former, may still induce global collective motion, and that also in this case it is possible to achieve a Toner \& Tu polar liquid phase that still belongs to the Vicsek universality class.

\subsubsection{The collisional Vicsek model}
Self-aligning particles \cite{baconnier2024self} are able to align their self-propulsion orientation towards their velocity. These models are important especially in the context of dense populations of active units (such as confluent epithelial tissues) and active solids.\\
An example of such a system is the collisional Vicsek model, first introduced in \cite{szabo2006phase, henkes2011active} and used to describe the flocking behavior of confluent epithelial tissues \cite{giavazzi2017giant}. In this agent-based model, $N$ soft disks with position $\mathbf{r}_i$ and interaction radius $\sigma_i$ (which might be poly-dispersed according to some distribution, e.g. Gaussian or uniform), move with self-propelled speed $v_0$ along their polarization $\mathbf{n}(\theta_i) = (\cos\theta_i, \sin\theta_i)$ (assuming $d=2$ for simplicity) and interact via short ranged harmonic repulsive forces, according to the following overdamped dynamics
\begin{equation}\label{velocity_cvm}
    \Dot{\mathbf{r}}_i = v_0 \mathbf{n}(\theta_i) + \mu \sum_{j = 1}^N \mathbf{F}_{ij}
\end{equation}
with $\mu$ controlling the intensity of the harmonic repulsion.\\
The repulsive force may be written as
\begin{equation}\label{CVM}
    \mathbf{F}_{ij} =
    \begin{cases}
    \quad 0 & \text{if } r_{ij} \equiv |\mathbf{r}_i - \mathbf{r}_j| \geq (\sigma_i + \sigma_j) \\
[r_{ij} - (\sigma_i + \sigma_j)] \hat{\mathbf{r}}_{ij} & \text{if } r_{ij} < (\sigma_i + \sigma_j)
\end{cases}
\end{equation}
with $\mathbf{\hat{r}}_{ij}$ being the unit vector from particle $i$ to particle $j$ and $r_{ij}$ their distance. 

\begin{figure}[hbt!]
    \centering
    \includegraphics[width=0.75\linewidth]{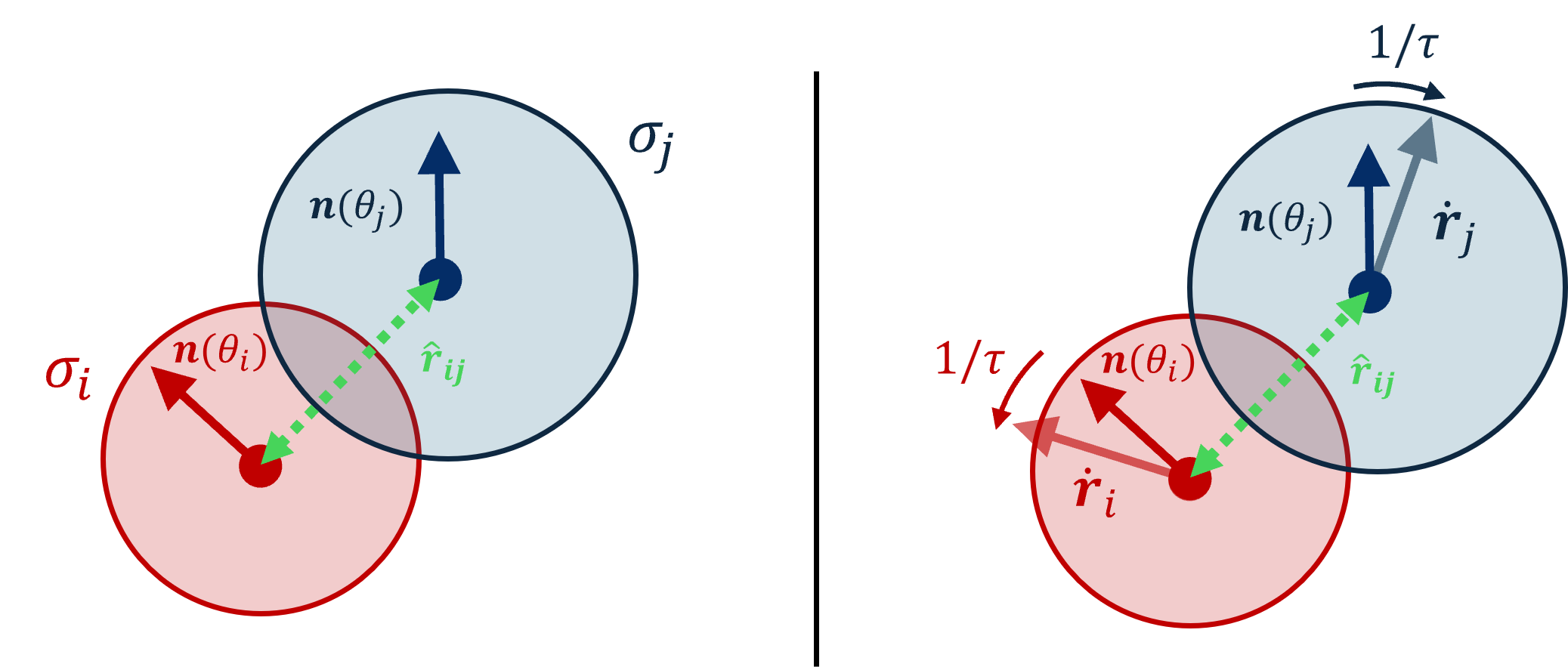}
    \caption{{\bf Schematic representation of the interaction rule for two particles in the CVM model} The two particles interact whenever the inter-particle distance $r_{ij} < \sigma_i + \sigma_j$: the repulsion, tuned by the parameter $\mu$, modifies the velocity as described in Eq.\eqref{velocity_cvm}. One can then include the tendency of particles to align their polarization $\mathbf{n}(\theta_i)$ to their velocity $\mathbf{\dot{r}_i}$ as described in Eq.\eqref{tendency}.}
    \label{schematic_cvm}
\end{figure}

It is possible to obtain a transition towards collective motion by including the tendency of the cell polarization $\mathbf{n}(\theta_i)$ to align towards its actual velocity $\mathbf{v}_i \equiv \dot{\mathbf{r}}_i = v_i (\cos \psi_i, \sin \psi_i)$, that is
\begin{equation}\label{tendency}
    \dot{\theta}_i = \frac{1}{\tau} (\theta_i - \psi_i) + \xi_i
\end{equation}
where $\tau$ is the realignment timescale \cite{szabo2006phase}.\\
Once the self-propelled velocity $v_0$ is fixed, the control parameters are the intensity of repulsion $\mu$, the realignment timescale $\tau$ and the packing fraction $\phi = \frac{1}{L^2}\sum_i^N\pi\sigma_i^2$, where $L$ is the linear system size.\\
In the $\tau \to \infty$ limit, the CVM is a model for non-ordering active Brownian particles interacting via repulsive forces and is characterized by a phase diagram showing a liquid phase with non-equilibrium clustering, motility induced phase separation (see Section \ref{sec_aim}) and a completely jammed state \cite{fily2014freezing}. For finite $\tau$ (and not too large noise values) the CVM is known to display a transition from a disordered gas-like state to a flocking state \cite{szabo2006phase} which shows a Toner \& Tu polar liquid phase that still belongs to the Vicsek universality class. Interestingly, this polar liquid phase can be also realized for confluent or hyper-confluent packing fractions $\phi$, that are rather far from the diluted limit \cite{giavazzi2017giant}.
\newpage

\section{Toner \& Tu hydrodynamic theory}\label{TT_theory}
In a hydrodynamic description one usually identifies the \textit{slow fields}, meaning those fields associated to symmetries and conservation laws that are expected to be slowly varying in time and space and that fully determine the dynamics of our system on sufficiently large timescales. In our context, this allows one to only consider the density $\rho$ (associated to particle number conservation) and the velocity $\bf{v}$ (associated to polar symmetry breaking) or some other orientational field such as the nematic tensor $\bf{Q}$  (associated with symmetry breaking towards a nematic state, that is, a state invariant under rotation by a $\pi$ angle).\\ 
The easiest way to obtain hydrodynamic equations is to just write them in the spirit of a gradient expansion in the slow fields, only retaining the terms compatible with the system symmetries and conservation laws. Since one is interested in the large scale and long time behavior, one can then only retain the lowest terms in the expansion. 
 
The Toner \& Tu theory for flocking \cite{toner1995long, toner1998flocks, toner2012reanalysis} presents a hydrodynamic description for the polar/Vicsek class, describing the flock through two continuous, coarse grained, slow fields: the number density $\rho(\mathbf{r},t)$ and velocity $\mathbf{v}(\mathbf{r},t)$. \\
As discussed above, the hydrodynamic equations of motion are written solely on symmetry grounds (they are invariant under arbitrary rotations) and, in the long-wavelength limit, read
\begin{equation}\label{TT_continuity}
    \partial_t\rho +\nabla\cdot(\rho\mathbf{v}) = 0
\end{equation}
\begin{equation}\label{TT_velocity}
\begin{split}
    \partial_t\mathbf{v} &+ \lambda_1(\mathbf{v}\cdot\nabla)\mathbf{v} + \lambda_2(\nabla\cdot\mathbf{v}) \mathbf{v} + \lambda_3\nabla(|\mathbf{v}|^2) \\
                         &= \mu\mathbf{v} - \kappa |\mathbf{v}|^2\mathbf{v} -\nabla P_1 - (\mathbf{v}\cdot\nabla P_2)\mathbf{v}\\
                         &+ D_1\nabla(\nabla\cdot\mathbf{v}) + D_v\nabla^2\mathbf{v} + D_2(\mathbf{v}\cdot\nabla)^2\mathbf{v} + \mathbf{f}\\
\end{split}
\end{equation}
The first equation simply describes conservation of particles number in an advective system: particles are immortal and the total density $\rho_0$ is conserved. A variation of this constraint, allowing for birth and death of particles leads to a different universality class and has been introduced elsewhere by Toner himself \cite{toner_malthusian}.\\
The second equation is akin to a Navier-Stokes equation where Galilean invariance is broken and a Ginzburg-Landau potential promotes symmetry breaking. The $\lambda_{1,2,3}$ terms are the convective terms: here $\lambda_1$ can be different from $1$ and also nonzero $\lambda_2$ and $\lambda_3$ terms are admitted given the absence of Galilean invariance. The $\mu$ and $\kappa$ together make up a force term $F$ induced by a Mexican hat force, typical of the Ginzburg-Landau theory, $F \equiv (\mu - \kappa|\mathbf{v}|^2)\mathbf{v}$, which allows the local velocity $\mathbf{v}$ to have a non zero magnitude $p_0=\sqrt{\mu/\kappa}$ in the ordered phase, thus inducing a {\it spontaneous symmetry breaking} of the continuous  rotational symmetry. The \textit{isotropic pressure} $P_1$ and \textit{anisotropic pressure} $P_2$ are, in general, functions of the density $\rho$ and the magnitude of the local velocity $|\mathbf{v}|$. It is convenient to expand them in a power series of the density fluctuations $\delta \rho({\bf r},t) \equiv \rho({\bf r},t) -\rho_0$  
\begin{equation}\label{pression_expansion}
    P_1 = \sum_{n=1}^\infty \sigma_n(|\mathbf{v}|)(\delta\rho)^n, \quad P_2 = \sum_{n=1}^\infty \sigma_n^A(|\mathbf{v}|)(\delta\rho)^n.
\end{equation}
$D_{1,2,v}$ are the diffusion constants reflecting the tendency of localized fluctuations to spread out because of the aligning interactions between the particles.\\
Moreover $\kappa, D_1, D_2$ and $D_v$ are all positive while $\mu < 0$ in the disordered phase and $\mu > 0$ in the ordered state. Note that, in principle, all the above coefficients could depend on $\delta \rho$ and $|{\bf v}|$. These dependencies are however irrelevant in what follows and we will ignore them, with the exception of the linear Ginzburg-Landau coefficient whose density dependence should not be overlooked, $\mu=\mu(\rho)$.\\
Finally the $\mathbf{f}$ term is a random force representing a Gaussian noise with correlations
\begin{equation}
    \langle f_i(\mathbf{r},t) f_j(\mathbf{r'}, t')\rangle = \Delta\delta_{ij}\delta^d(\mathbf{r}-\mathbf{r'})\delta(t-t')
\end{equation}
where $\Delta$ is the noise amplitude and $i,j$ denote Cartesian components.

\subsection{Linearized theory}\label{linearized_TT}
In the absence of fluctuations Eqs.\eqref{TT_continuity}-\eqref{TT_velocity} admit a homogeneous steady state solutions $\rho({\bf r}, t)=\rho_0$, ${\bf v}({\bf r},t)=p_{\rm 0} \hat{\mathbf{e}}_{\parallel}$. Here and in the following the subscripts $\parallel$ and $\perp$ denote, respectively, the longitudinal direction (that of flocking) and the direction(s) transversal to it. The value of $p_0$ is determined by the condition
\begin{equation}\label{condition_p0}
    \mu p_0 - \kappa p_0^3 = 0.
\end{equation}
For $\mu <0$, the only possible homogeneous solution that satisfies Eq.\eqref{condition_p0} is $p_0 = 0$, which is stable as long as $\mu < 0$ (see Ref.\cite{bertin2006boltzmann, bertin2009hydrodynamic}). When $\mu>0$, an ordered nonzero solution $p_0 = \sqrt{\mu/\kappa}$ appears, with the unit direction $\hat{\mathbf{e}}_{\parallel}$ randomly selected by the spontaneous symmetry breaking mechanism.\\
We now focus our attention deep in the ordered phase -- the Toner and Tu phase -- where $\mu$ is positive and finite and the ordered solution is linearly stable to perturbations. This is the regime relevant to the results obtained in Chapters 3 and 4. The stability of the ordered phase at the onset of collective motion, $0<\mu \ll 1$, will be briefly discussed in Sec. \ref{scaling_exp}.\\
In the following, we briefly review the Toner \& Tu results of Refs.\cite{toner1998flocks, toner2005hydrodynamics, toner2012reanalysis} and present more recent results \cite{chate_solon}.

To deal with fluctuating hydrodynamics we proceed to linearize Eqs. \eqref{TT_continuity}-\eqref{TT_velocity} around the homogeneous solution,
\begin{equation} \label{fluct_TT}
    \begin{split}
         \rho(\textbf{r},t)=\rho_0 &+ \delta\rho(\textbf{r}, t)\\
        \textbf{v}(\textbf{r},t) = [{p}_{\rm 0} + \delta v_{\parallel}&(\textbf{r},t)] \hat{\mathbf{e}}_{\parallel} + \textbf{v}_{\perp}(\textbf{r},t)
    \end{split}
\end{equation}
where $\mathbf{v}_{\perp}$ measures velocity fluctuations transversal to $\hat{\mathbf{e}}_{\parallel}$. \\
As we know, due to the spontaneous symmetry breaking of a continuous theory, the longitudinal velocity fluctuations $\delta v_\parallel$ are damped by the Mexican hat potential. As a consequence, only the transversal component ${\bf v}_\perp$ is a slow mode, while $\delta v_\parallel$ turns out to be fast mode enslaved to the slow fields of the unperturbed theory, $\textbf{v}_{\perp}$ and the density fluctuations $\delta \rho$. The enslaving of $\delta v_\parallel$ can be readily obtained and reads, to zeroth order in the derivatives
\begin{equation}
\delta v_\parallel \approx \frac{\mu'(\rho_0)}{2\beta p_0} -\frac{| {\bf v}_\perp|^2}{2p_0}
\label{enslave}
\end{equation} 
where $\mu'(\rho) = d\mu(\rho)/d\rho$ and thus the longitudinal velocity fluctuations can be eliminated from Eqs. \eqref{TT_continuity}-\eqref{TT_velocity}, leaving one with two linear equations for the slow fields $ \delta \rho $ and ${\bf
v}_\perp$ that can be readily solved. They read
\begin{equation}\label{rho_lin_TT}
\begin{split}
        \partial_t \delta\rho &+\rho_0 \nabla_{\perp}\cdot \textbf{v}_{\perp} +v_2 \partial_{\parallel}\delta\rho -\rho_0 \mu_2\partial_t \partial_{\parallel}\delta\rho \\
        & =D_{\rho_{\parallel}} \partial_{\parallel}^2 \delta\rho + D_{\rho_{\perp}} \nabla_{\perp}^2 \delta\rho +D_{\rho v}\partial_{\parallel} (\nabla_{\perp} \cdot \textbf{v}_{\perp})  \\ 
\end{split}
\end{equation}
and
\begin{equation}  \label{v_perp_lin_TT}
    \begin{split}
        \partial_t \textbf{v}_{\perp} &= - \gamma \partial_{\parallel}\textbf{v}_{\perp}-\frac{c_0^2}{\rho_0}\nabla_{\perp} \delta\rho +D_B \nabla_{\perp}(\nabla_{\perp}\cdot \textbf{v}_{\perp}) \\
        &+D_v \nabla_{\perp}^2 \textbf{v}_{\perp} +D_{\parallel}\partial_{\parallel}^2\textbf{v}_{\perp} 
        +g_t \partial_t \nabla_{\perp}\delta\rho + g_{\parallel} \partial_{\parallel} \nabla_{\perp}\delta\rho +\textbf{f}_{\perp}
    \end{split}\,,
\end{equation}
where all the various constants introduced above can be expressed as a function of the phenomenological constants appearing in Eqs.\eqref{TT_continuity}-\eqref{TT_velocity}. Their precise expressions are not important in what follows, but they can be found, together with all the details of the linearization, in Ref. \cite{kyriakopoulos2016leading}\footnote{In this thesis we use a slightly different notation where the diffusion constant $D_3$ is replaced with $D_v$. Also we indicate the magnitude of the mesoscopic velocity $|\mathbf{v}|$ with $p_0$ and not $v_0$ to distinguish it from the self-propelled velocity used in microscopic simulations of the Vicsek model.}.

The linearized equations \eqref{rho_lin_TT} and \eqref{v_perp_lin_TT} can be solved in Fourier space, where
\begin{equation}\label{fourier_campi_TT}
    \begin{split}
        &\hat{\bf v}_\perp(\textbf{q},\omega)\sim \int d\textbf{r}dt \,e^{-i(\textbf{q}\cdot \textbf{x}-\omega t)}\textbf{v}_\perp(\textbf{r},t)\,,\\ 
        & \delta\hat{\rho}(\textbf{q},\omega) \sim \int d\textbf{r}dt\, e^{-i(\textbf{q}\cdot \textbf{x}-\omega t)}\delta \rho(\textbf{r}, t)\,.
    \end{split}
\end{equation}
Indeed, from the analysis carried out in \cite{toner2012reanalysis} one finds that the system admits a pair of \textit{longitudinal} eigenmodes involving just the longitudinal velocity $v_L \equiv \mathbf{v}_\perp \cdot\mathbf{q}_\perp/q_\perp$ and $\rho$, and an additional $d-2$ \textit{transverse} modes associated with the transverse velocity $v_T \equiv \mathbf{v_\perp} - v_L \mathbf{q}_\perp/q_\perp$. The longitudinal modes are closely analogous to ordinary sound waves in a simple fluid, while the transverse modes are the analog of the diffusive shear modes in such a fluid.\\
In the hydrodynamic limit (i.e., when wave number $q \to 0$), the longitudinal eigenfrequencies are a pair of under-damped, propagating modes with complex eigenfrequencies
\begin{equation}
    \omega_{\pm}(\mathbf{q}) = c_{\pm}(\theta_q)q - i \Tilde{\epsilon}_{\pm}(\theta_q)q^2,
\end{equation}
where $q=|{\bf q}|$ and $\theta_q$ is the angle between $\textbf{q}$ and $\hat{\mathbf{e}}_\parallel$. We have also introduced the direction-dependent sound speeds $c_{\pm}(\theta_q)$ and the field dependent dampings $\epsilon_\pm ({\bf q})$ of Ref. \cite{toner2012reanalysis} are expressed as $ \epsilon_\pm ({\bf q}) = \Tilde{\epsilon}_\pm (\theta_q)q^2$  with the geometric factor $\Tilde{\epsilon}_\pm(\theta_q)$ depending only on the angle $\theta_q$. Their expression is given in App.\ref{linearized_SF} for compactness. Here it is sufficient to note that they only depend on the direction $\theta_q$.\\
The transverse modes have the far simpler character of anisotropic diffusion, with purely imaginary eigenfrequencies
\begin{equation}
    \omega_T(\mathbf{q}) = -i \widetilde{\Gamma}_T(\theta_q)q^2
\end{equation}
with the damping $\widetilde{\Gamma}_T(\theta_q) = D_v \cos(\theta_q)^2 + (D_v + D_2 v_0^2) \sin(\theta_q)^2 $.\\
With the eigenmodes at hand one can obtain the linearized density $S_\rho$ and velocity $S_v$ structure factors that read
\begin{equation}
\label{l_rho_TT}
S_\rho({\bf q}, {\bm \mu}) = A(\theta_q, {\bm \mu})\, q^{-2}
\end{equation}
\begin{equation}
\label{l_v_TT}
S_v({\bf q}, {\bm \mu}) = B(\theta_q, {\bm \mu}) \,q^{-2}\;,
\end{equation}
$A(\theta_q, {\bm \mu})$ and $B(\theta_q, {\bm \mu})$ depend on the TT equation coefficients, here collectively denoted as ${\bm \mu}$ and are given by
\begin{equation}\label{autocorr_eqt_TT}
    A(\theta_q, {\bm \mu}) = \frac{\Delta}{2}\left(\frac{\rho_0 \,\sin(\theta_q)}{2c_2(\theta_q)}\right)^2\times\bigg( \frac{1}{\Tilde{\epsilon}_+(\theta_q)}+ \frac{1}{\Tilde{\epsilon}_-(\theta_q)} \bigg),
\end{equation}
\begin{equation}     \label{autocorr_eqt_v_TT}
    B(\theta_q, {\bm \mu})\!\! =\!\! \frac{\Delta}{2}\!\left(\!\frac{[c_+(\theta_q)\! -\! v_2\cos(\theta_q)]^2}{4\Tilde{\epsilon}_+(\theta_q)[c_2(\theta_q)]^2} \!+\! \frac{[c_-(\theta_q) \!-\! v_2 \cos(\theta_q)]^2}{4\Tilde{\epsilon}_-(\theta_q)[c_2(\theta_q)]^2} \!+\! \frac{d - 2}{\widetilde{\Gamma}_T(\theta_q)}\!\right)\!\!.
\end{equation}

 It should be noted that the amplitude $A$ is singular in the longitudinal direction, $A(0, {\bm \mu})=0$, so that the linear behavior of $S_\rho$ for $q_\perp \to 0$ cannot be determined at this order in TT theory.

The equal time correlations are relevant, since they determine the size of density and velocity fluctuations. In particular the size of velocity fluctuations determines whether or not long-ranged order can exist in such a system, while the size of density fluctuations determines the presence or absence of giant number fluctuations, which we have discussed previously.

The most relevant result of the linearized theory is the prediction that both the density and velocity equal time auto-correlation functions scale as 
\begin{equation}
S_\rho({\bf q}) \sim S_v({\bf q}) \sim \,q^{-2}\;.
\end{equation}
This means, in particular, that long-ranged order is not possible according to the linearized theory since the real space fluctuations
\begin{equation}\label{real_space}
\langle|\mathbf{v}_\perp(\mathbf{r},t)|^2\rangle = \int \frac{d^dq}{(2\pi)^d}S_v(\mathbf{q})
\end{equation}
diverge in the infrared $(q\to 0$ or $L\to \infty)$ limit in all spatial dimensions $d\leq 2$ as in equilibrium systems such as the XY model, where this results is known as the Mermin-Wagner-Hohenberg theorem \cite{mermin1966absence, hohenberg1967existence}. This indicates that in order to stabilize long-ranged order in $d=2$ we must turn on the nonlinear terms. They are (see \cite{toner2012reanalysis} for the definition of the coefficients $w_{2,3}$ and $g_{1,2,3}$)
\begin{equation}\label{relevant_non_linearities}
\begin{split}
\nabla_\perp\!\cdot\!(\delta\rho \delta\mathbf{v}_\perp)&, \,
\lambda_1 (\delta \mathbf{v}_\perp \!\cdot\! \nabla_\perp) \delta \mathbf{v}_\perp, \,
w_2 \partial_\parallel \delta \rho^2, \,
w_3 \partial_\parallel |\delta \mathbf{v}_\perp|^2, \,\\
&g_1 \delta \rho \partial_\parallel \delta \mathbf{v}_\perp, \,
g_2 \delta \mathbf{v}_\perp \partial_\parallel \delta \rho,
g_3 \nabla_\perp \delta \rho^2. \,
\end{split}
\end{equation}

\subsection{Non-linear theory}\label{non_linear_and_DRG}
The dynamical renormalization group (DRG) approach can be used to deal with nonlinearities. Observing the system at different coarse-graining scales, it takes advantage of the scaling invariance of the Toner \& Tu phase to access relevant observables such as the correlation functions. Coarse-graining essentially consists of two steps. In the first one, the nonlinear equations of motion are averaged over the short-wavelength fluctuations: i.e., we average over the slow fields Fourier modes with wave-vector lying in the (hyper-cylindrical) shell of Fourier space $b^{-1} \Lambda \!\le q_{_\perp}  \le \Lambda$. Here $\Lambda$ is an ultra-violet cutoff (essentially dictated by the inverse of the microscopic interaction range), and $b>1$ is an arbitrary rescaling factor.\\
In the second step, in order to restore the ultraviolet cut-off to $\Lambda$, one rescales transversal distances $r_\perp=|{\bf r}_\perp |$ and wave numbers as 
\begin{equation}\label{rescale_perp}
r_\perp = b r_\perp'\;\;\;\;,\;\;\; q_\perp = b^{-1}q_\perp'\,.
\end{equation}
Also time, parallel distances and the fields rescale according to \footnote{It can be shown \cite{toner1998flocks} that transversal velocity and density fluctuations have the same scaling}
\begin{equation}\label{rescale_fields}
r_\parallel = b^{\xi} r_\parallel'\;,\;\;\; q_\parallel = b^{-\xi} q_\parallel'\;,\;\;\;
t= b^{z}t'\;,\;\;\;\delta \rho = b^{\chi} \delta \rho'\,.
\end{equation}
This procedure leads to a new \textit{renormalized} set of equations of motion with the same form w.r.t. to the original ones but with renormalized parameter values. If we denote collectively the initial full parameter set of the nonlinear equations of motion as $\{\bm\mu^{(1)}\}$, we can represent their evolution by the above DRG flow as $\{\bm\mu^{(1)}\}\to \{\bm\mu^{(b)}\}$. The scaling exponents  $\xi$, $z$, and $\chi$, known respectively as the anisotropy, dynamical and roughness exponents, are in principle arbitrary. For a suitable choice of their value, however, a renormalization group fixed point
$\{\bm\mu^*\}$, that is, a situation in which the renormalized parameters do not change under this renormalization group process, is obtained in the $b \to \infty$ limit. Analyzing the DRG flow near this fixed point one can thus deduce the system scaling properties. 

In the absence of nonlinearities, the small scale degrees of freedom can be integrated out harmlessly, and one is able to determine the scaling exponent for the linear fixed point,
\begin{equation}\label{exp_lin}
    \chi_L=1-d/2, \quad \xi_L = 1, \quad z_L=1.
\end{equation}
The stability of this linear fixed point can be analyzed by naively rescaling the nonlinear terms of Eq.\eqref{relevant_non_linearities} by Eqs.\eqref{rescale_perp}-\eqref{rescale_fields}. It can be shown that the linear fixed point (see \cite{toner2012reanalysis} for the details) is stable for $d>4$, while all non-linearities turn out to be relevant (i.e. they grow under this naive rescaling) for $d\leq d_c = 4$, where $d_c$ is the upper critical dimension. This means that the linearized theory is correct at long length and time scales only for $d>4$. For $d<4$, however, all of the non linearities grow, and the linear theory breaks down at sufficiently long length and time scales.

For $d\leq d_c$, when nonlinearities are relevant, the averaging step is non-trivial and it has to be performed perturbatively in the equations' of motion nonlinearities, and generally produces nonlinear corrections (the so-called {\it graphical} corrections) in the DRG flow equations, that makes arduous an exact treatment of the nonlinear DRG fixed point. This issue seem to be particularly severe in our present case due to the large number of different nonlinearities, see Eq.\eqref{relevant_non_linearities} and has long be prevented an exact determination of the nonlinear fixed point scaling exponents in $d<4$.

Being interested in the scaling properties of some physical quantity we can, nevertheless, deduced it at a qualitative level by a simple power counting argument. As an example consider the velocity structure factor $S_v(q_\perp, q
_\parallel, {\bm\mu}^{(1)})$. Its evolution under the DRG flow reads
\begin{equation}\label{S_v_drg_flow}
    S_v(q_\perp, q_\parallel, {\bm\mu}^{(1)}) = b^{d-1+\xi+2\chi} S_v(bq_\perp, b^\xi q_\parallel, {\bm\mu}^{(b)}),
\end{equation}
where the power of the rescaling factor $b$ on the r.h.s. can be deduced from the fact that the structure factor is the Fourier transform of the two-points equal time correlation function. The exponent $d-1$ comes from the scaling in the transversal directions, $\xi$ from the scaling of the longitudinal direction and $2\chi$ from the scaling of the two velocity fields. These exponent are now to be intended as the (still unknown) scaling exponents of the {\it nonlinear fixed point}.\\
Choosing $b = \Lambda/q_\perp$ and taking the small wavelength limit, we have ${\bm\mu}^{(b)} \simeq {\bm\mu}^*$. Therefore Eq. \eqref{S_v_drg_flow} becomes
\begin{equation}\label{S_v_drg_fp}
    S_v(q_\perp, q_\parallel, {\bm\mu}^{(1)}) = \left(\frac{q_\perp}{\Lambda}\right)^{-(d-1+\xi+2\chi)} S_v\left(\Lambda, \Lambda^\xi \frac{q_\parallel}{q_\perp^\xi}, {\bm\mu}^*\right).
\end{equation}
$\Lambda$ is an ultra-violet cutoff which is fixed by the inverse of some microscopic length-scale and it is thus a constant and Eq.\eqref{S_v_drg_fp} can be rewritten as
\begin{equation}\label{S_v_drg_final}
    S_v(q_\perp, q_\parallel, {\bm\mu}^{(1)}) = q_\perp^{-(d-1+\xi+2\chi)}f\left(\frac{q_\parallel}{q_\perp^\xi}\right).
\end{equation}
where $f$ is a scaling function that absorbs the dependence of the constants $\Lambda$ and ${\bm \mu}^*$.\\
It's easy to show that at the non-linear fixed point ${\bm \mu}^*$ the structure factors $S_\rho$ and $S_v$ share the same scaling in the transversal direction, that is for $q_\perp^\xi\gg q_\parallel$,
\begin{equation}\label{nlnSperp}
S_v(q_\perp, q_\parallel, {\bm \mu}^*)\sim S_\rho(q_\perp,
q_\parallel,  {\bm \mu}^*) \sim q_\perp^{-\zeta} \,.
\end{equation}
where $\zeta \equiv d-1 +\xi +2\chi$. Note that correlations may only
depend on the modulo of the wave-number in the symmetry unbroken transversal
directions.

In the longitudinal direction, for $q_\parallel \gg q_\perp^\xi$, Eq. \eqref{S_v_drg_final} implies that transversal velocity correlations scale as
\begin{equation}
S_v(q_\perp, q_\parallel, {\bm \mu}^*)\sim q_\parallel^{-\zeta/\xi} \,.
\label{nlnSpar}
\end{equation}
while the situation is less clear for the density structure factor. \\
TT theory predicts \cite{toner1998flocks}
\begin{equation}
S_\rho(q_\perp, q_\parallel, {\bm \mu}^*)\sim q_\perp^2 q_\parallel^{-2-\zeta/\xi} \,.
\label{nlnSpar2}
\end{equation}
while numerical simulations suggest a different and more complex
behavior \cite{Benoit2019}, which will be briefly discussed in Sec.~\ref{2C}.
This anomalous behavior, still poorly understood, is probably due to the
singular behavior of the linear density structure factor for
$q_\perp \to 0$ discussed above.

Finally we should note that this simple argument can only provide, by simple power counting, the expected scaling behavior of some physical quantities, leaving unspecified the exact value of such scaling exponents at the non-linear fixed point. 

\subsection{On the value of the scaling exponents}\label{scaling_exp}
The exact value of the nonlinear fixed point ${\bm \mu}^*$ scaling exponents for $d<d_c$ has long remained elusive due to the plethora of potentially relevant nonlinear terms in the TT equations. \\
All of the nonlinearities considered in the initial works of Toner \cite{toner1995long, toner1998flocks, toner2005hydrodynamics} could, at least in $d = 2$, be written as total $\perp$ derivatives. This implied that such nonlinearities could only renormalize terms which themselves involved $\perp$ derivatives; consequently, all of the terms that did not involve $\perp$ derivatives were argued to get no graphical corrections. Furthermore, it seemed possible to determine exactly the exponents in $d=2$ since the equations were invariant under a pseudo Galileian transformation $\mathbf{r}_\perp \to \mathbf{r}_\perp - \lambda_1 \mathbf{v}_G t$ and $\mathbf{v} \to \mathbf{v} + \mathbf{v}_G$, for arbitrary constant vector $\mathbf{v}_G \perp \hat{x}_\parallel$. This symmetry should hold through all scales and protects the involved nonlinear coefficients from graphic contributions. This, in turn, would make possible to determine the values of the $d=2$ scaling exponents
\begin{equation}
\chi = -1/5, \quad \xi = 3/5 \quad \text{and} \quad z = 6/5
\end{equation}
Unfortunately in \cite{toner2012reanalysis} additional non-linearities were shown to be present and it seemed \textit{impossible} to determine the exact values of such scaling exponents (beyond the fact that $\chi<0$ and $\xi<\leq1$ which, by Eqs. \eqref{real_space}-\eqref{nlnSperp}, implies long range order in $d=2$). It is interesting to note that these additional non-linearities that make it impossible to determine the exact exponents in $d=2$ involve density fluctuations $\delta\rho$. Therefore, whenever these fluctuations are absent, as in the case of Malthusian flocks \cite{toner_malthusian}, it is possible to determine the exact exponents.

Numerical estimates of the scaling exponents \cite{mahault2019quantitative} confirmed that the values obtained in $d=2$ were incompatible with large-scale numerical simulations of Vicsek flocks, highlighted in particular by the small or vanishing anisotropy 
\begin{equation}
\chi = -0.31(2), \quad \xi = 0.95(2) \quad \text{and} \quad z = 1.33(2).
\end{equation}
and by the apparent hyper-scaling relation $z=\zeta=d-1+\xi+2 \chi$. Interestingly, this latter relation can be exactly verified provided that the additive noise vertex does not acquire graphical corrections. Indeed, analytical developments \cite{chate_solon}, suggest that in the DRG procedure the noise vertex should not acquire graphical corrections due to the previously overlooked gradient structure of the symmetry broken theory. Furthermore it was argued that a generalized Galilean invariance of the theory preserves other nonlinear coefficients at least for $d=2$, making it possible to determine the scaling exponents exactly. In \cite{chate_solon} the authors write down hydrodynamic equations directly for the Goldstone mode emerging once the continuous rotational symmetry is broken, at odds with what is done in Toner \& Tu theory. Quite interestingly they find that the Goldstone mode obtained by eliminating the fast variable from the isotropic Toner-Tu equation is \textit{not} the direction of the velocity field, as one would naively expect.
Indeed, in $d=2$ one can parametrize the velocity field $\mathbf{v} = v(\mathbf{r},t)\mathbf{n}(\phi(\mathbf{r},t))$ highlighting the dependence on the phase $\phi(\mathbf{r},t)$. Once the modulo of the velocity is eliminated from the TT equations, since it is a fast mode, one obtains two coupled equations for $\phi$ and for the density fluctuations $\delta\rho$. While this holds also for the treatment reported in Eqs. \eqref{rho_lin_TT}-\eqref{v_perp_lin_TT}, it is argued that a suitable change of variable (the expression of coefficients $\alpha$ and $\beta$ can be found in \cite{chate_solon})
\begin{equation}\label{goldstone_mode}
    \Tilde{\phi} = \phi - \alpha\partial_\parallel\phi -\beta\delta\rho\partial_\parallel\phi
\end{equation}
leads to a dynamics with a well defined gradient structure that prevents the noise vertex to acquire graphical corrections. This indicates that the correct Goldstone mode is $\Tilde{\phi}$ and not $\phi$ as one would naively expect. \\
The values found, by this argument, for $d=2$ Vicsek flocks are in agreement with the values measured in microscopic simulations of Ref.\cite{mahault2019quantitative} and read
\begin{equation}
\chi = -1/3, \quad \xi = 1 \quad \text{and} \quad z = 4/3.
\end{equation}

It should be noted that the difference between $\phi$ and $\Tilde{\phi}$ disappears when working with linearized equations, since the difference between the two arises from the $\lambda_1$ non-linearity, and/or when the symmetries of the system (perhaps longitudinal invariance as in the Casimir geometry, as it will be shown later) fix the direction of flocking to a given direction. This conditions apply to the systems investigated in Chapters \ref{directed} and \ref{confined}, and therefore, in the remaining of this thesis we will continue to identify transversal velocity perturbations $\mathbf{v}_\perp$ as a slow field.

\newpage

\section{Boltzmann-Ginzburg-Landau approach}\label{BGL}
A major problem with deriving hydrodynamic equations directly on symmetry principles is the loss of connection between the parameters of the corresponding microscopic models and the value of the transport coefficients of the field theory. This connection is of course desirable since it would allow to compare results coming from microscopic models or experimental systems directly to a hydrodynamic description.
Fortunately, there are direct coarse-graining methods that allow to build a hydrodynamic description starting from a given microscopic model. Most, if not all of these methods, consist in building first a kinetic-level description of the problem and then build the hydrodynamic equations from the kinetic level.

In the case of DADAM one usually writes down a kinetic equation that governs the one-body probability $f(\mathbf{r}, \mathbf{v}, t)$, that is the probability of finding a particle at position \textbf{r}, with velocity \textbf{v} at time $t$. To obtain such an equation it is necessary to factorize a multi-body function into a product of one-body functions, which is legitimate under the \textit{molecular chaos} hypothesis, that assumes decorrelation of particles' orientation from one interaction to the next. This is a particularly strong assumption in the case of aligning particles as we will discuss further.

Two approaches are among the most popular in the context of dry active matter: the Smoluchowski/Fokker-Planck approach and the Boltzmann-Ginzburg-Landau (BGL) approach, which is well suited for DADAM since it formally requires the assumption of diluteness \cite{bertin2006boltzmann, bertin2009hydrodynamic}.
Since the molecular chaos hypothesis is fairly strong in the case of aligning active matter, we should \textit{not} expect quantitative agreement from the BGL hydrodynamic description, while it is reasonable to find results which are qualitatively faithful to the microscopic model at hand.\\
In the following, as an example of the technique, we briefly review the BGL approach for the polar Vicsek class in two spatial dimensions and use it to discuss the long-wavelength linear instability leading to phase separation at the onset of order \cite{bertin2009hydrodynamic}.

\subsection{Construction of the Boltzmann equation}
In $d=2$ given that the VM velocity is constant and fixed to $v_0$, it is convenient to work with the angular variable $\theta$, where $\mathbf{v} = v_0 \hat{\mathbf{e}}(\theta)$.\\
Here we will recall the coarse-graining for a continuous-time model in which particles align their velocities in a Vicsek-style way and undergo tumbles at a finite rate $\lambda$.
Following \cite{peshkov2014boltzmann}, they can be described by the following Boltzmann equation governing $f(\mathbf{r}, \theta, t)$ (here $f(\theta)$ for shorthand notation)
\begin{equation}
\partial_t f(\theta) + v_0\hat{\mathbf{e}}(\theta)\cdot\nabla f(\theta) = I_\text{sd}[f] + I_\text{col}[f]
\end{equation}\label{master_eq}
which is nothing but a master equation for $f(\mathbf{r}, \theta, t)$.\\
The advection term on the left hand side corresponds to the general form of the free motion dynamics: particles are moving at a mean speed $\bar{v}$ along their orientation. The two terms on the right hand side $I_\text{sd}$ and $I_\text{col}$ are the integrals accounting for angular self-diffusion via tumbling events (action of the noise on an isolated particle) and inter-particle interaction events. In the dilute limit they are modeled as {\it collisions}.\\
In order to account for the discrete time dynamics of the VM, the self-diffusion integral considers tumbling events occurring at a rate $\lambda$
\begin{equation}
    I_\text{sd}[f] = -\lambda f(\theta) +\lambda \int_0^{2\pi}d\theta' f(\theta') P_\eta(\theta-\theta')
\end{equation}\label{self_diffusion}
where $P_\eta$ is an angular noise Gaussian distribution of zero average and variance $\eta^2$.\\
The collisional integral is derived under the \textit{binary collision} assumption: in a dilute system collisions of more than two particles are rare and thus negligible, which is realized when the collisional mean free path $l_0\sim \/\sqrt{\rho_0}$ is much larger than the radius of interaction $R_0$.\\
As already mentioned, in order to have a closed equation for the single-particle distribution, we need to satisfy the \textit{molecular chaos} hypothesis\footnote{The molecular chaos hypothesis is satisfied whenever the typical flight distance is much larger than the interaction radius. However, since the interaction is aligning, it is quite a strong approximation to state that the velocities after collision will be independent from the one prior the collision: this is what renders the approach only qualitative.} which ensures that the states of two particles (labelled by 1 and 2) are uncorrelated prior to the collision
\begin{equation}
    f^{(2)}(\theta_1, \theta_2) = f(\theta_1) f(\theta_2)
\end{equation}
where we have indicated with $f^{(2)}(\theta_1, \theta_2)$ the probability of finding both particles 1 and 2 in the same (coarse-grained) position \textbf{r} at time $t$, with orientations $\theta_1$ and $\theta_2$ respectively.\\
The collisional integral gathers loss and gains due to collisions and reads
\begin{equation}
\begin{split}
     I_\text{col}[f] &= - 2R_0v_0\int_0^{2\pi} d\theta' K(\theta'-\theta)f(\theta) f(\theta')\\
     &+ 2R_0v_0\iint_0^{2\pi} d\theta_1d\theta_2 K(\theta_2-\theta_1)f(\theta_1) f(\theta_2) P_\eta[\theta-\Psi(\theta_1,\theta_2)]\\
\end{split}
\end{equation}
where $\Psi(\theta_1, \theta_2)$ is the post-collisional alignment rule and the noise is drawn from the same distribution $P_\eta$ as in Eq.\eqref{self_diffusion} (this condition can be relaxed).\\
For the Vicsek class the kernel of interaction $K(\theta_2-\theta_1)$ is simply given by \cite{bertin2009hydrodynamic}
\begin{equation}
    K(\theta_2-\theta_1) = |\hat{\mathbf{e}}(\theta_2) - \hat{\mathbf{e}}(\theta_1)| = 2 \left|\sin\left(\frac{\theta_2 -\theta_1}{2}\right)\right|
\end{equation}
The alignment rule $\Psi$, because of global rotational invariance, obeys $\Psi(\theta_1, \theta_2) = \theta_1 + H(\theta_2-\theta_1)$, meaning it can only depend on relative orientations.\\
In the case of polar alignment we get
\begin{equation*}
    H(\Delta) = \Delta/2 \quad \forall\Delta \in (-\pi;\pi], 
\end{equation*}
and is $2\pi$-periodic. 

\subsection{Angular Fourier modes expansion}
Since the variable $\theta$ is $2\pi$-periodic, it is natural to expand the distribution $f$ in terms of angular Fourier modes
\begin{equation}
    f(\mathbf{r}, \theta, t) = \frac{1}{2\pi} \sum_{k=-\infty}^\infty f_k(\mathbf{r}, t) e^{ik\theta}, 
\end{equation}
with $f_{-k} (\mathbf{r},t) = f_k^*(\mathbf{r},t)$, where the asterisk represents complex conjugation. \\
The zeroth mode correspond to the density field, while the next two modes correspond to the complex representations of the momentum $\mathbf{w} = \rho\mathbf{v}$ and nematic $\mathbf{S} = \rho\mathbf{Q}$ fields
\begin{equation}
f_0= \rho, \quad f_1 = w_x + i w_y, \quad f_2 = 2[S_{xx} + i S_{xy}]
\end{equation}
where the explicit dependence on \textbf{r} and $t$ has been dropped for compactness of notation.\\
Working near the onset of orientational order, where the order is low, it is safe to assume that the amplitude of higher order moments $f_k$ decays fast to zero as $k$ grows. Such a truncation is controlled only near the transition where the Ginzburg-Landau-like terms of the order parameter dominate the equation. Moreover, it can be shown (see \cite{dadam_benoit} for the details), that there is a unique convective \textit{scaling ansatz} that allows one to attribute an $\epsilon$ order to each term of the hierarchy
\begin{equation}
|f_k|\sim \epsilon^k \quad \text{and} \quad \partial_t\sim\nabla\sim\epsilon 
\end{equation}
Truncating to the first nontrivial order, which is found to be $\epsilon^3$, we can obtain closed hydrodynamic equations. In the case of Polar/Vicsek class the fast $f_2$ mode can be enslaved to $\rho$ and $f_1$ which can then be expressed in terms of the density $\rho$ and momentum $\mathbf{w}$. Setting $v_0 =1$, they read
\begin{equation}\label{continuity_bgl}
    \partial_t\rho +\nabla\cdot\mathbf{w} = 0
\end{equation}
\begin{equation}\label{velocity_bgl}
    \partial_t\mathbf{w} + \lambda_1(\mathbf{w}\cdot\nabla)\mathbf{w}+\lambda_2(\nabla\cdot\mathbf{w})\mathbf{w} +\lambda_3\nabla|\mathbf{w}|^2 = -\frac{1}{2}\nabla\rho +(\mu(\rho)-\kappa|\mathbf{w}|^2)\mathbf{w} + D_v\nabla^2\mathbf{w}
\end{equation}
that resemble quite closely Toner-Tu Eqs.\eqref{TT_continuity}-\eqref{TT_velocity}.\\
The expression of the transport coefficients are given by 
\begin{align*}
 \lambda_1 &= \left(P_1 - 2P_2 - \frac{16}{15}\right)\frac{2\rho_0}{\pi\psi}, &\quad \lambda_2 &= -\left(P_1 + 2 P_2 +\frac{4}{15}\right)\frac{2\rho_0}{\pi\psi},\\
    \mu(\rho) &= P_1 - 1 + \frac{4}{\pi} \left( P_1 - \frac{2}{3} \right) \rho_0 \rho, &\quad  \kappa &= -\frac{16 (5 P_1 - 2) (3 P_2 + 1) \rho_0^2}{15 \pi^2 \psi},\\
    D_v &= -\frac{1}{4 \psi}, &\quad \psi &= P_2 - 1 - \frac{8}{15 \pi} (7 + 5 P_2) \rho_0,
\end{align*}
where $P_k$ are the Fourier coefficients of the noise distribution. Here $\lambda_3 = -\lambda_2/2$ and are thus related in our microscopic model, unlike in TT equations, which should represent the entire universality class. Finally, note that these equations are also simpler and contain less terms than the TT equations: for example, only the $\mu(\rho)$ explicitly depends on the local density, the pressure term $\nabla\rho$ is as simple as possible and also diffusion $D_v$ is isotropic, as particles are point-wise.

\subsection{Phase diagram and $\rho$ dependence of $\mu$}\label{pd_bgl_mu_dependence}
\begin{figure}[hbt!]
    \centering
    \includegraphics[width=0.5\linewidth]{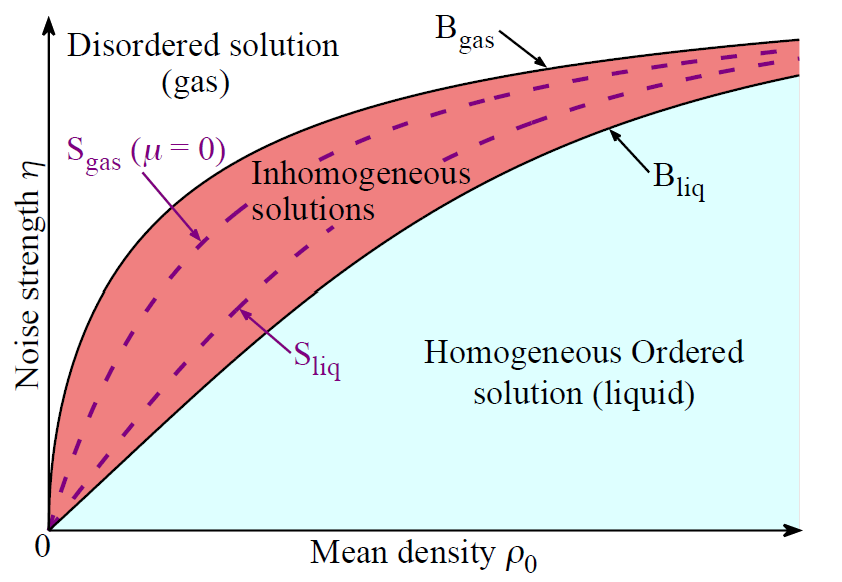}
    \caption{{\bf Phase diagram obtained from a linear stability analysis of the equations obtained with the BGL approach}. Figure adapted from Ref.\cite{dadam_benoit}.}
    \label{pd_bgl}
\end{figure}

Interestingly, one can also derive a qualitative phase diagram by studying at the linear level the stability of both the disordered and the homogeneous ordered phase.\\
In the phase diagram $(\rho_0,\eta)$, the line $\mu = 0$ ($S_{gas}$, see Fig.\ref{pd_bgl}) has the expected form, growing as a square root from the origin: above it the homogeneous disordered phase ($\mu <0$) is linearly stable, below ($\mu >0$) it is unstable and a homogeneous ordered phase solution exists: this solution is however linearly unstable to long-wavelength perturbations from the $\mu = 0$ line down to a similar line $S_{liq}$. These two lines are akin to the spinodal lines described for the AIM in Sec.\ref{liquid_gas_pt}.\\
Indeed, no homogeneous solution can be sustained between $S_{gas}$ and $S_{liq}$: nonlinear analysis \cite{solon2015pattern} shows that the emerging inhomogenoeus solutions resemble the traveling wave bands of the VM. Due to nonlinear effects, They can be sustained even outside the $S_\text{gas}$-$S_\text{liq}$ lines, up to $B_\text{gas}$ and $B_\text{liq}$ which correspond to the binodal lines of phase separation. These are the lines corresponding to those delimiting the coexistence region at the microscopic level.

We now describe the linear instability showing that it arises from the $\rho$-dependence of the linear coefficient $\mu$ and more precisely from the fact that it grows with $\rho$. The growth of $\mu$ with $\rho$ is the translation at the hydrodynamic level of the positive feedback mechanism between local density and local order, which is at the origin of phase separation. \\
Consider, for instance, a small perturbation around the homogeneous solution $\rho (\mathbf{r}, t) = \rho_0$ and $\mathbf{w}(\mathbf{r},t) = \mathbf{w}_0$ such that
\begin{equation}\label{linear_instability}
    \rho (\mathbf{r}, t) = \rho_0+ \delta\rho (\mathbf{r}, t), \quad  \mathbf{w}(\mathbf{r},t) = \mathbf{w}_0 +\delta\mathbf{w}(\mathbf{r},t)
\end{equation}
We can now plug Eq.\eqref{linear_instability} in Eqs.\eqref{continuity_bgl}-\eqref{velocity_bgl} and expand the resulting equations to first order in the perturbations, keeping into account the density dependence of $\mu$. This gives the following linearized equations
\begin{equation}\label{linear_inst_rho}
    \partial_t\delta \rho + \nabla \cdot \delta \mathbf{w} = 0,
\end{equation}
\begin{equation}\label{linear_inst_w}
\begin{split}
    \partial_t\delta\mathbf{w} = &- \lambda_1 (\mathbf{w}_0 \cdot \nabla) \delta \mathbf{w}  -\frac{1}{2} \nabla \delta \rho + 2\lambda_3 \nabla (\mathbf{w}_0 \cdot \delta \mathbf{w}) \\
    &+ (\mu'\delta\rho - 2\kappa\mathbf{w}_0 \cdot \delta \mathbf{w} - 2\lambda_3\nabla \cdot \delta \mathbf{w}) \mathbf{w}_0 \\
    &+ ( \mu - \kappa\mathbf{w}_0^2) \delta\mathbf{w} + D_v\nabla^2 \delta \mathbf{w},
\end{split}
\end{equation}
where $\mu'$ is a shorthand notation for $d\mu/d\rho$ (which is constant, being $\mu\propto\rho$). \\
Eqs.\eqref{linear_inst_rho}-\eqref{linear_inst_w} can be studied in Fourier space, transforming the fields as follows
\begin{equation}\label{linear_instability_Fourier}
    \delta\rho (\mathbf{r}, t) = \delta\rho_0 e^{st + i\mathbf{q}\cdot\mathbf{r}}, \quad  \delta\mathbf{w}(\mathbf{r},t) = \delta\mathbf{w}_0 e^{st + i\mathbf{q}\cdot\mathbf{r}}
\end{equation}
with the amplitudes $|\rho_0|$ and $|\mathbf{w}_0|$ assumed to be small.\\
Given a wave-vector $\mathbf{q}$, one looks for the dispersion relation $s(\mathbf{q})$: if the real part of $s(\mathbf{q}) > 0$, then the mode with wavenumber $\mathbf{q}$ is unstable. The strongest instability that arises from the analysis carried out in Ref. \cite{bertin2009hydrodynamic} is a longitudinal one, where $\mathbf{w}_0, \delta\mathbf{w}_0$ and $\mathbf{q}$ are all aligned. One then finds two dispersion relations for $s_{1,2}$ that have a rather complicated expression. Nevertheless, by expanding the solution with largest real part, $s_1$, for small $q = |\mathbf{q}|$ and near the onset of order ($\mu \ll 1$), one obtains
\begin{equation}\label{linear_instability_q2q4}
    Re(s_1) = \frac{\mu'}{8\kappa\mu^2}q^2 - \frac{5\mu'^4}{128\kappa^2\mu^6}q^4 + \mathcal{O}(q^6)  
\end{equation}
The positivity of the coefficient of the $q^2$ term confirms that, close to the transition line, long wavelength modes are unstable. This is actually consistent as long as $q\ll q_c$, with $q_c = \sqrt{\frac{16\kappa\mu^4}{5\mu'^3}}$.

Finally, due the truncation to order $\epsilon^3$ a \textit{spurious instability} arises: this is situated below the $S_{liq}$ spinodal line and it grows much faster than the band instability. This has no analogy with anything known to happen in the microscopic model and it disappears increasing the truncation order \cite{peshkov2014boltzmann}.\\
Deep in the fluid phase higher order gradients might play a role in stabilizing the homogeneous ordered solution.
For the Vicsek class it has been noted that introducing a small positional diffusion $D_\rho$ greatly reduces the size of this region. Vicsek-style models are thus singular since they do not incorporate positional diffusion: when simulated at discrete time some effective diffusion arises, preventing spurious instability deep in the fluid phase \cite{mahault2019quantitative}.
\fancyhead[LO]{{\it Perturbations in dry active matter}}
\fancyhead[RE]{{\it Perturbations in dry active matter}}
\chapter{Perturbations in dry active matter} 
So far we have considered bulk unperturbed active matter. In this chapter we review known results on the far reaching impact that boundary and bulk perturbations have on the properties of free dry active systems: bulk perturbations are widely intended as perturbations affecting the free system with inclusions (being them self-propelled or not), bulk disorder or even the prototypical homogeneous field (akin to what is done in equilibrium ferromagnetic systems) orienting the individual particles self-propulsion directions. Boundary perturbations are instead coming from the interaction of particles with boundaries that can be either homogeneous or disordered.

On general grounds, these perturbations can be divided in two main categories: perturbations that do not destroy a given bulk phase but only affect certain aspects of it as, for example the response of a flock to a global external field \cite{Nikos} or perhaps the effect of reflecting boundaries in flocking active matter \cite{fava2024casimir} and perturbations that may dramatically destroy the unperturbed bulk behavior, for instance with a small obstacle \cite{codina2022small}. In this chapter we will briefly review the results obtained for both cases of perturbations, while in our research from Ch. 3 we will focus on the first ones.

Here we will first discuss the mechanical pressure in dry active matter system, considering explicitly the torque free case and the case where torques are exerted at the boundaries: at odds with what happens at equilibrium, the mechanical pressure exerted on a boundary depends on the kind confining potential, leading, for example, to the spontaneous compression of an asymmetric wall. The dependence of the mechanical pressure on microscopic details of the interactions between active particles and boundaries underlines the impossibility to generally define a thermodynamic pressure as a state function in active systems. These generic considerations will play a role in Ch. \ref{confined} in our treatment of the pressure exerted on a reflecting boundary by a polar fluid. \\
The second part of this chapter will focus on the effect of perturbations in the context of dry scalar active matter: interesting results were obtained for MIPS, for example showing that it is unstable to boundary disorder \cite{ben2022disordered}.\\
The third and last section will instead focus on results obtained in the context of flocking active matter: recent results have shown that bulk perturbations can indeed destroy the ordered phase of polar flocks \cite{codina2022small, benvegnen2023metastability} and will be discussed explicitly here. On the other hand, very little is known in the case of confined vectorial active matter that will be the main topic of Ch. \ref{directed} and \ref{confined}.

\newpage
\section{Mechanical pressure on confining boundaries}\label{mechanical_pressure}
\begin{figure}[hbt!]
    \centering
    \includegraphics[width=0.75\linewidth]{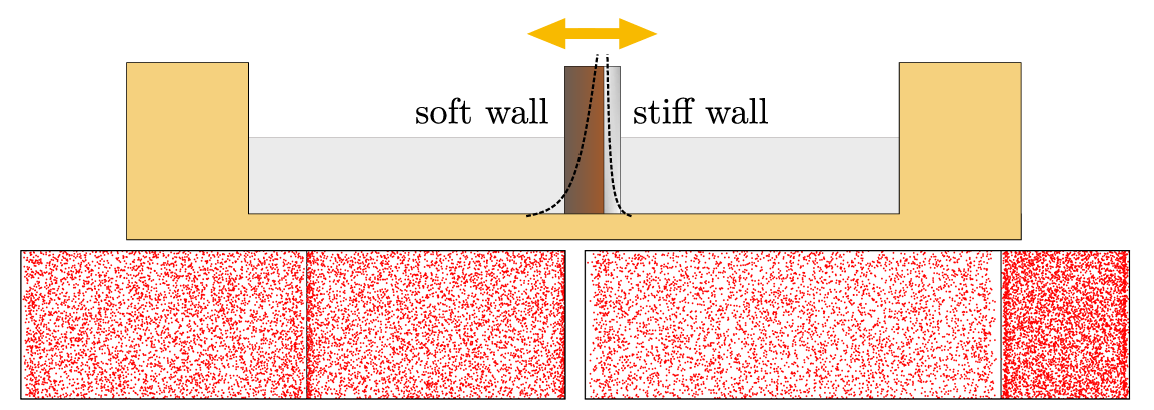}
    \caption{{\bf Spontaneous compression of a mobile asymmetric partition} Top: The system is divided by a mobile partition with a stiffer side into two compartments with the same density of particles. Each side of the partition feels a distinct force from the active particles since the pressure is dependent on the stiffness of the wall. Thus, until the forces on both sides is equal, the mobile partition moves. Bottom: Numerical simulations with circular (left panel) and elliptical (right panel) ABPs, that match the configuration shown in the top panel, show the spontaneous compression of one half of the system in the case of elliptical particles, indicating the absence of an equation of state (EOS). Figure adapted from \cite{solon2015pressure}.}
    \label{mobile_partition}
\end{figure}

Active systems interacting with boundaries display a range of unusual behaviors that defy our conventional understanding of equilibrium. Perhaps the simplest way to demonstrate this is by placing an asymmetric mobile partition in a cavity filled with a homogeneous gas of self-propelled ellipses. At odds with the second rule of thermodynamics, one witnesses a spontaneous compression of one side of the system, as shown in Fig.\ref{mobile_partition}. In contrast, perfectly symmetric active Brownian particles (ABPs) behave like their equilibrium counterparts and do not exhibit such spontaneous compression. In this section we will give an explanation of why this happens in the context of active matter following Ref.\cite{granek2023inclusions} and Refs. therein, highlighting the far reaching impact of boundaries.

Consider, for instance, a gas of non-interacting active particles in $d=2$ confined by a vertical flat wall localized at $x = R$. The wall can be modeled by a repulsive potential $V(x)$ that vanishes for $x < R$ and diverges at larger values of $x$. The pressure exerted by the gas on the wall is then simply given by 
\begin{equation}\label{pressure_flat}
    P = \int_{\bar{x}}^\infty dx \rho(x, \bar{y})\partial_x V(x, \bar{y})
\end{equation}
where $(\bar{x}, \bar{y}) \equiv \mathbf{\bar{r}}$ is a point deep in the bulk of the active fluid and $\rho(\mathbf{r}, t) \equiv \langle \sum_i\delta(\mathbf{r-r}_i)\rangle$ is the average density. The dynamics of particle $i$ obeys
\begin{equation}\label{dynamics_ABPs}
    \dot{\mathbf{r}}_i = \mu\mathbf{f}_a^i -\mu\nabla V(\mathbf{r}_i) + \sqrt{2D_T}{\bm \eta}_i
\end{equation}
where $\mathbf{f}_a^i\equiv f_a\mathbf{n}(\theta_i)$ is the particle's self-propulsion force, $\mu$ its motility, $D_T$ a translational diffusivity and ${\bm \eta}_i$ a centered Gaussian white noise of unit variance. \\
Particle number conservation implies that $\rho$ satisfies a conservation law  $\partial_t\rho + \nabla\cdot\mathbf{J} = 0$, with the current \textbf{J} given by
\begin{equation}\label{current_J}
    \mathbf{J} = \mu \mathbf{F}_a(\mathbf{r}) - \mu\rho(\mathbf{r})\nabla V(\mathbf{r}) - D_T\nabla\rho(\mathbf{r})
\end{equation} 
and $\mathbf{F}_a\equiv \langle \sum_i\mathbf{f}_a^i\delta(\mathbf{r-r_i})\rangle$ is the \textit{active force} density.\\
In the steady state, confinement by a wall and the translational symmetry along the wall imply $\mathbf{J} = 0$. Inserting Eq.\eqref{current_J}, for $\mathbf{J} = 0$, in Eq.\eqref{pressure_flat}, the mechanical pressure $P$ can then be written as
\begin{equation}\label{pressure_active_1}
    P = \frac{D_T}{\mu}\rho(\mathbf{\bar{r}}) + \int_{\bar{x}}^\infty dx \mathbf{F}_a (x,\bar{y}).
\end{equation}
since we expect the density $\rho$ to vanish deep inside the confining wall.\\
The system's pressure on the wall is therefore equal to the sum of the ideal-gas pressure (passive contribution) and the active force density (which correctly vanishes in the limit of $f_a\to 0 $) which is usually present also in close proximity to confining walls contributing to the total pressure. In the following two subsections, we will compute this active force density for active particles without and with torques acting on the self-propulsion orientation.

\subsection{The torque free case}
Consider the simple case of ABPs, whose angular dynamics reads
\begin{equation}\label{angular_iso}
    \dot{\theta}_i = \sqrt{2D_R}\eta_i^R
\end{equation}
with $D_R$ a rotational diffusivity, and $\eta_i^R$ a centered Gaussian white noise with unit variance. \\
Following \cite{granek2023inclusions} we can introduce the \textit{active impulse} $\Delta\mathbf{p}_i^a$, which measures the average momentum that the particle will receive in its future from the substrate. For a circular ABP with orientation $\mathbf{n}[\theta_i(t)]$ the active impulse reads
\begin{equation}
    \Delta \mathbf{p}_i^a \equiv\int_t^\infty ds f_a \overline{\mathbf{n}[\theta_i(s)]} = \frac{f_a}{D_R}\mathbf{n}[\theta_i(t)]
\end{equation}
where the overline denotes an average over future histories for $s\geq t$ and we have used that $\overline{\mathbf{n}[\theta_i(s)]} = \mathbf{n}[\theta_i(t)]\exp[-D_R(s-t)]$. The active impulse at time $t$ is non-zero due to persistence, even though the particle is undergoing a random walk, which results in an average active force of zero. \\
The time evolution of the active impulse field $\Delta\mathbf{p}^a(\mathbf{r}) \equiv \langle \sum_i\Delta\mathbf{p}_i^a\delta(\mathbf{r-r_i})\rangle$ can then be expressed as
\begin{equation}\label{active_implulse}
    \partial_t\Delta\mathbf{p}^a(\mathbf{r}) = -\mathbf{F}_a(\mathbf{r}) + \nabla\cdot\sigma_a(\mathbf{r})
\end{equation}
where 
\begin{equation}\label{active_stress}
    \sigma_a\equiv-\langle\sum_i\dot{\mathbf{r}}_i\otimes\Delta\mathbf{p}_i^a\delta(\mathbf{r-r_i})\rangle + D_T\nabla\otimes\Delta\mathbf{p}^a(\mathbf{r})
\end{equation} 
is a tensor that measures the flux of active impulse. \\
In the steady state Eq.\eqref{active_implulse} shows that the density of active force is the divergence of the active stress tensor ($\mathbf{F}_a = \nabla\cdot\sigma_a$), indicating that the active impulse acts as a momentum reservoir for the particle. The pressure $P$ of Eq.\eqref{pressure_active_1} then takes the ideal gas law form 
\begin{equation}\label{pressure_Teff}
    P = \rho(\mathbf{\bar{r}}) \frac{D_T}{\mu} - \sigma_a(\mathbf{\bar{r}}) = \rho(\mathbf{\bar{r}}) T_\text{eff},
\end{equation}
where $T_\text{eff}\equiv D_\text{eff}/\mu$ is an effective temperature and $D_\text{eff} = D_T + (\mu f_a)^2/2D_R$ is the large scale diffusivity of the particle, showing that in this case we obtain an equilibrium-like expression where activity simply makes the system's temperature hotter. The pressure can thus be written as the sum of a passive and an active stress, which is remarkable due to the absence of momentum conservation. Note that in this case the mechanical pressure does not depend on the microscopic details of the confining potential, which is remarkable since $\mathbf{F}_a$ depends on the choice of the confining potential, through Eq.\eqref{current_J}.This is due to the fact that the dynamics of particle orientation are independent of all other degrees of freedom, hence the total active impulse that the particle can transfer to the wall is independent of the wall potential. 

\subsection{The case in the presence of torques}
If the particles are not isotropic (e.g. in the case of ellipsoidal particles) they will experience a torque $\Gamma$ when interacting with the confining wall. The angular dynamics of \eqref{angular_iso} will be then modified as 
\begin{equation}\label{angular_aniso}
    \dot\theta_i = \Gamma(\mathbf{r}_i, \theta_i) + \sqrt{2D_R}\eta_i^R
\end{equation}
In this case one finds that in the steady state the active-force density $\mathbf{F}_a$ can be written as
\begin{equation}\label{full_active}
\mathbf{F}_a(\mathbf{r}) = \nabla\cdot \sigma_a^\text{tf} + \langle\sum_i \frac{f_a}{D_R}\Gamma_i(\mathbf{r}_i,\theta_i)\mathbf{n}^\perp(\theta_i)\delta(\mathbf{r-r}_i)\rangle
\end{equation}
where $\mathbf{n}^\perp(\theta) = \partial_\theta\mathbf{n}(\theta)$ and $\sigma_a^\text{tf}$ is the torque free part of the flux of active impulse. Equation \eqref{full_active} splits the contribution to the density of active forces between conserved and non-conserved parts, showing that wall-induced torques can be seen as sources or sinks of active impulse.

The pressure can be conveniently rewritten as
\begin{equation}
    P = \rho(\mathbf{\bar{r})}T_\text{eff} + \Delta P_W
\end{equation}
where $\Delta P_W$ is a wall-dependent contribution given by
\begin{equation}
    \Delta P_W = \frac{\mu f_a}{D_R}\int d\theta\int_{\bar{x}}^\infty dx \psi(x, \bar{y}, \theta) \Gamma(x, \bar{y}, \theta) \sin\theta.
\end{equation}
where $\psi(\mathbf{r}, \theta) \equiv \langle\sum_i\delta(\mathbf{r-r}_i)\delta(\theta-\theta_i)\rangle$. The pressure is no longer independent of the wall's microscopic details, resulting in the spontaneous compression of the asymmetric piston. The piston stalls when the pressures on both sides (left/right, indicated respectively with $L/R$) are equal, implying $\Delta P_W^L (\rho_L) = \Delta P_W^R(\rho_R)$. To achieve equality between the left and right walls, densities on both sides must differ. This example shows how torques lead to a lack of equation of state. 

Finally the lack of an equation of state for dry active systems is not be exclusively associated with torques generated by confining walls, but it is rather general, with the previous symmetric ABPs case being the exception: alignment interactions and motility control, for example, have also been demonstrated to preclude the existence of an equation of state \cite{solon2015pressure}. 
\newpage

\section{Perturbations in scalar active matter}
In this section we briefly review the effect of bulk and boundary perturbations in scalar active matter system.
\subsection{Bulk perturbations}
\begin{figure}[hbt!]
    \centering
    \includegraphics[width=0.75\linewidth]{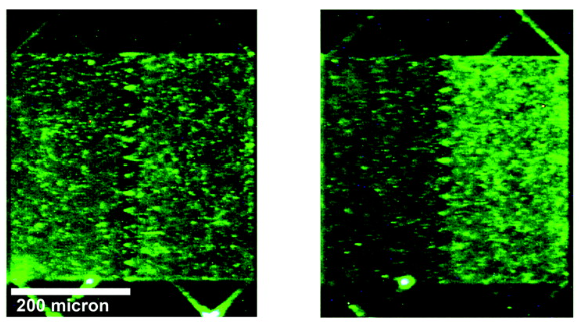}
    \caption{{\bf Bacteria distribution in a sample with a funnel wall made of several V-shaped obstacles} Left - Uniform distribution immediately after injection. Right - Steady-state distribution after 80 min. Figure adapted from \cite{galajda2007wall}}
    \label{galajda}
\end{figure}

The first evidence of the effect of bulk perturbation in scalar active matter was probably the experiment with asymmetric obstacles by Galajda \cite{galajda2007wall}, who showed that bacteria accumulate on one side of an array of V-shaped obstacles when the obstacles are placed in a bacterial bath, as shown by Fig.\ref{galajda}.

Several results have since then been obtained in this context: a single obstacle immersed in an active fluid was shown to induce a large-scale ratchet current (ultimately possible due to the lack of time-reversal invariance) that mass conservation turns into long-range density and current modulations. Since isolated obstacles create long-range density modulations, it is natural to expect that several obstacles immersed in the same active fluid will experience long-range mediated interactions, which turn out to be non-reciprocal \cite{baek2018generic}. \\
Passive tracers (particles with no activity, immersed in active fluid) have also attracted a lot of attention lately. When several of them are embedded in an active bath, their long-range interactions lead to interesting dynamical effects: it has been shown that the non-reciprocal interactions between the tracers lead to non-reciprocal phase transitions \cite{fruchart2021non} and the interaction between a tracer and its boundary-induced image leads to a localization transition \cite{dor2022passive}.

We have seen that localized obstacles might have a far-reaching impact on the properties of active fluids. When organized coherently, they can, for instance, act as pumps and generate large-scale flows or density gradients. Naturally, this raises the question as to whether disordered assemblies of obstacles can also impact the properties of active fluids. In the following we will focus our attention on dry active systems that phase separate and briefly recall how disorder impacts their properties.   

\begin{figure}[hbt!]
     \centering
     \includegraphics[width=0.75\linewidth]{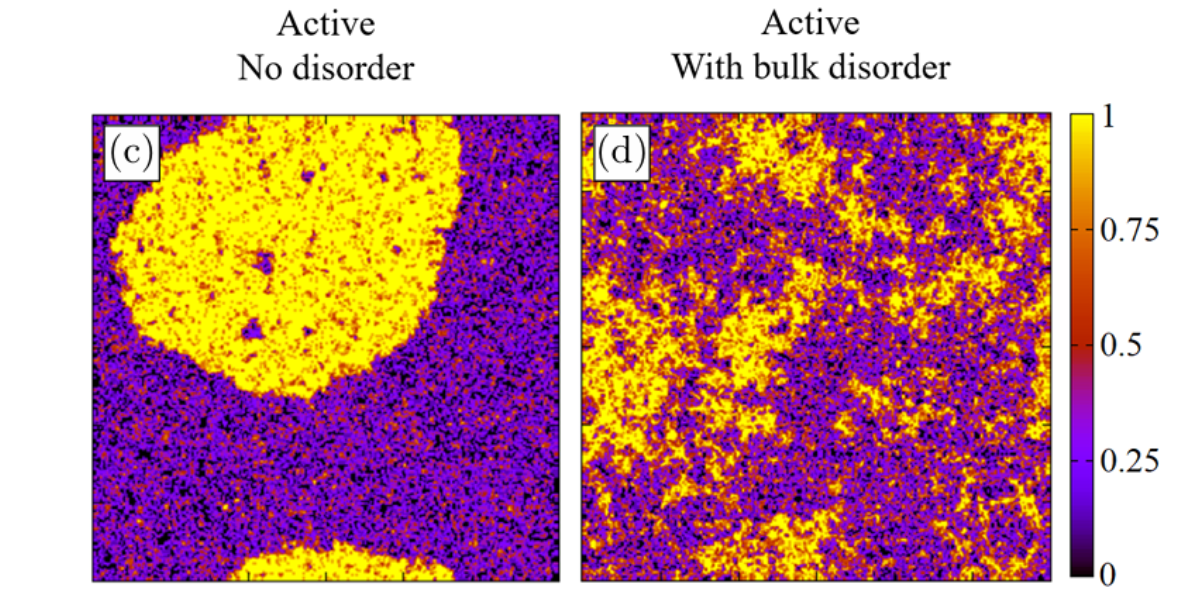}
     \caption{{\bf The effect of quenched disorder in active systems} Active lattice gas with repulsive interactions undergoing bulk phase separation. When a random potential is added (right panel), a scale-free distribution of particles replaces the bulk phase separation (left panel). Color encodes density. Figure adapted from \cite{ro2021disorder}.}
     \label{passive_bulk}
 \end{figure}

\textbf{Bulk disorder}\\ 
To characterize the impact of quenched disorder (i.e. a fixed realization of a random potential that affects the local particles velocities) on a scalar system, it is natural to consider density-density correlations, which are encoded in the density structure factor. For a system defined on a lattice of $N$ sites, in the presence of disorder, it may be written as
 \begin{equation}
     \overline{S_\rho(\mathbf{q})}\equiv \frac{1}{N}\sum_\mathbf{r}\overline{\langle\delta\rho(\mathbf{r})\delta\rho(0)\rangle}e^{i\mathbf{q\cdot r}},
 \end{equation}
where $\delta\rho(\mathbf{r}) \equiv \rho(\mathbf{r})-\rho_0$ measures density fluctuations w.r.t a reference global density $\rho_0$, brackets stand for steady-state averages in a particular realization of disorder, and the overline for an average over disorder realizations.\\ 
In the homogeneous phase, the disorder leads to a scale-free steady state with a structure factor that it may be shown to diverge in the low $q$ regime as (see Ref.\cite{granek2023inclusions} for details)
\begin{equation}
     \overline{S_\rho(\mathbf{q})} \underset{q\to 0}{\propto} \frac{1}{\mathbf{q}^2}
\end{equation}
Fluctuations at small $q$ are therefore scale-free: the correlation length is infinite and density-density correlations decay as $\overline{\langle\delta\rho(\mathbf{r})\delta\rho(0)\rangle}\propto r^{2-d}$ for $d > 2$, and logarithmically in $d = 2$. \\
The long-ranged correlations induced by disorder have a strong impact on the existence of a long-range ordered phase. For a system undergoing MIPS, disorder prevents phase separation in both $d=2, 3$, and the lower critical dimension is $d_c = 4$. As shown in Fig.\ref{passive_bulk}, the phase-separated state is replaced by a frozen scale-free distribution with large-amplitude density fluctuations. Technical details, including a simple argument and a more refined field theoretical description can be found in \cite{ro2021disorder, granek2023inclusions}. 

 \newpage
 
\subsection{Boundary perturbations}\label{boundary_scalar}
At equilibrium, boundaries and boundary conditions generically do not alter bulk phase behaviors, due to the sub-extensive nature of their contributions to the free energy. In active matter, due to the presence of non-equilibrium currents, the situation may be different.

\begin{figure} [hbt!]
    \centering
    \includegraphics[width=0.75\linewidth]{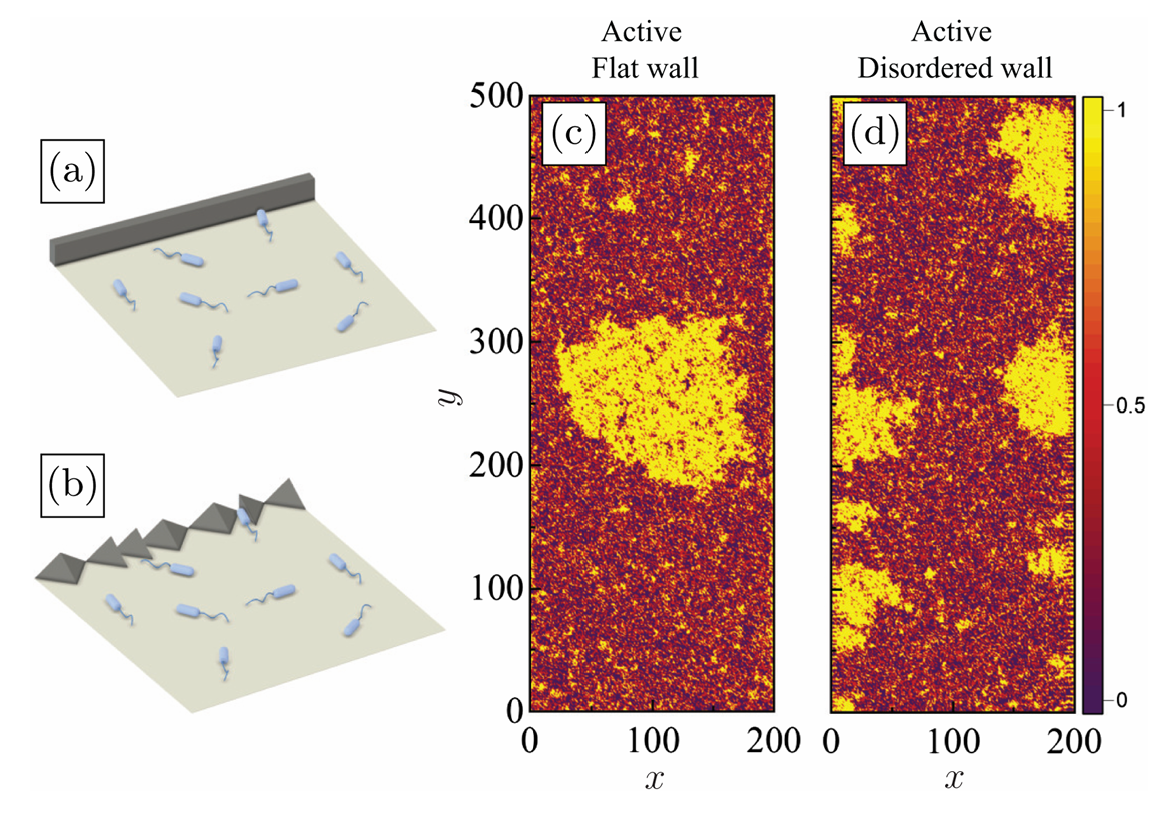}
    \caption{{\bf The impact of flat and disordered walls on phase separation in active systems}. Disordered boundaries destroy the phase separation observed in the presence of a flat wall, leading to a scale-free distribution of particles. Color encodes density. Figure edited from \cite{ben2022disordered}.}
    \label{boundaries_scalar}
\end{figure}

\textbf{Boundary disorder}\\
Disordered boundary conditions can be implemented through a wall potential $V(x,\mathbf{r}_\parallel)$, where $x$ is the coordinate normal to the wall and $\mathbf{r}_\parallel$ is a $(d-1)$-dimensional vector parallel to the wall. For example, $V(x,\mathbf{r}_\parallel)$ can be modeled by setting $V(x<0,\mathbf{r}_\parallel) = \infty$ and placing wedge-shaped asymmetric obstacles along the wall whose orientations are chosen randomly (see Fig.\ref{boundaries_scalar}(b) for a schematic representation). The obstacles then have a finite extent $x_W$ in the positive $\hat{\mathbf{x}}$ direction. 

In scalar active matter, these rough \textit{disordered} walls destroy bulk phase separation, leading to a nontrivial density distribution. This was shown \cite{ben2022disordered} to hold even in the thermodynamic limit where the boundaries are sent to infinity. Furthermore the phase-separated state is indeed replaced by scale-free density modulations in $d=2$. This result highlights again how boundaries can play a much more important role in active systems, w.r.t. equilibrium ones.

\textbf{Active particles in Casimir geometry}\\
The so-called \textit{Casimir} geometry \cite{casimir:originale}, consisting of two flat and parallel walls separated by a distance $R$, provides an excellent framework to study the effect of confinement in active matter. In the context of scalar active matter, an early numerical study of an ensemble of RTPs \cite{ray2014casimir} in $d = 2$, confined by parallel lines of finite length, revealed the emergence of an attractive force, the magnitude of which increases upon increasing the run length of the particles, but decreases exponentially upon increasing the distance $R$ between the lines. This attraction may be turned into a repulsion via varying $R$ and the line length. These systems lack genuine long-range correlations and the forces arise from geometrical constraints rather than from a genuine collective behavior.

Here we rather discuss explicitly the effect that confinement has on a collection of non-interacting active Ornstein-Uhlenbeck particles (AOUPs) of \cite{Caprini2018}. In a system with a simple geometry, such as infinite parallel plates (which can be realized by assuming periodic boundary conditions in the longitudinal direction w.r.t. the plates, as it will be shown later) the coordinates parallel to the walls play just a minor role. A one-dimensional description can therefore be used in order to account for the effect of confinement and provides a good approximation. In particular, AOUPs are well suited for such a calculation since they can be easily implemented in one-dimension.

Consider, for instance a $1d$ system of $N$ non-interacting active particles suspended in a fluid, driven by an active force $\gamma f_a$, where $\gamma$ is a Stokes friction constant, confined by an external potential $U(x)$ and subject to a random force representing the effect of the collisions with the molecules of the fluid. The self-propulsion force fluctuates both in intensity and direction and is modeled by a colored noise term, $f_a(t)$, evolving according to an Ornstein-Uhlenbeck process of correlation time $\tau$. The equations of motion for an AOUP thus read
\begin{equation}\label{x_dot}
    \dot{x} = f_a - \frac{1}{\gamma}\partial_xU(x) + \sqrt{2D_T}\eta
\end{equation}
\begin{equation}\label{f_a_dot}
    \tau\dot{f_a} = -f_a +\sqrt{2D_A}\xi
\end{equation}
where $\eta$ and $\xi$ are two independent white noises with unitary variance and zero average. The confining potential $U(x)$ is taken to be a piece-wise harmonic potential, namely
\begin{equation}
    U(x) = \theta(-x)\frac{kx^2}{2} + \theta(x - R) \frac{k(x-R)^2}{2}
\end{equation}
with $\theta$ being the Heaviside function acting outside of the interval $[0, R]$. The central region is thus a free-free region. The harmonic force is proportional to the stiffness $k$, allowing particles to explore the regions for $x\leq 0$ and $x\geq R$. The limit of impenetrable walls can then be retrieved when $k \to \infty$.

It possible to determine the steady state distribution of particles by solving the Fokker-Planck equation associated to \eqref{x_dot}-\eqref{f_a_dot}. Under some reasonable closure approximation of the single particle distribution momenta, the density distribution can be determined in all of the domain. In particular, the steady-state density $\rho(x)$ obeys the following equation in the region $[0, R]$
\begin{equation}\label{eq:density_caprini}
     \rho'''(x) - \frac{1}{\lambda^2} \rho'(x) = 0
\end{equation}
where $\lambda^2 = \tau D_T [(D_T/(D_A + D_T)]$. This equation, when complemented with the other ones outside of the region $[0, R]$, correctly predicts accumulation of particles at the walls. Nevertheless, the boundary region where particles accumulate is of order $\lambda$ as, which is an intrinsic length scale. Indeed, close to either boundaries, the density distribution can be shown to simply be
\begin{equation}
    \rho(x) = (\rho_W - \rho_B) e^{-x/\lambda} + \rho_B
\end{equation}
with $\rho_W$ being the density at the wall and $\rho_B$ being the density deep in the bulk of the system. This means that the extension of the accumulation is a finite region close to the boundaries and this effect, in the thermodynamic limit ($R\to \infty$) will then be limited to a negligible boundary layer of order $\lambda$ near the wall. \\
Moreover, the fluctuation-induced force acting between the boundaries that the authors also report is, yet again, a consequence of the geometrical constraints, rather than arising from truly long-range correlations in the system which is in general argued to be at the origin of both quantum and classical Casimir forces. We will show in Chapter \ref{confined} that in confined polar flocks, due to the presence of long-range correlations of the fluctuations of velocity (and density, being the two coupled) Eq.\eqref{eq:density_caprini} needs to be modified to include a contribution from long-ranged correlations. This will result in {\it extensive} boundary layers and in a fluctuation-induced force, acting between the confining walls and akin to critical Casimir forces in equilibrium systems \cite{gambassi2024critical}.
\newpage

\section{Perturbations in flocking active matter}
In this section we will briefly review some relevant results obtained in the context of flocking active matter. We will only treat bulk perturbation since very little has been done in the context of boundary perturbations in vectorial active matter.

\subsection{A small global external field}
We begin this section by recalling probably the easiest way to perturb a system made of flying spins: a small static external field of amplitude $h$ which aligns the individual particles self-propulsion directions.\\
At equilibrium, the response of systems breaking a continuous symmetry to a small external field is a classic problem of statistical field theory. Also in our non-equilibrium context this is a problem of great interest since in many experimental realizations the system is not truly isotropically homogeneous and, collective motion is actually directed by some external clue and not spontaneous. The effect of dynamically changing external field (applied at the boundaries) has been investigated in \cite{cavagna2013boundary}, where it was shown that the response to such a fluctuating field may give rise to a dynamical information flow from boundary to bulk giving rise to strong correlations, which decays algebraically with a near zero exponent and are akin to the ones observed in real flocks.

In \cite{Nikos}, on the other hand, it was shown how a small static and \textit{global} field of amplitude $h$, inserted directly at the dynamical level, affects the asymptotic \textit{longitudinal response} 
\begin{equation*}
    \chi_\parallel \equiv \frac{\delta\Phi(h)}{h},
\end{equation*}
where $\delta\Phi(h) = \Phi(h) - \Phi(0)$ is the change in the magnitude of the time-averaged scalar order parameter (the mean velocity), due to the applied field $\mathbf{h}$. \\
The longitudinal response can be computed considering a Toner \& Tu theory perturbed by a small homogeneous field and is found to scale as follows
\[
\chi_\parallel = h^{-\nu} f\left(Lh^{\frac{1}{z}}\right) \propto \begin{cases}
h^{-\nu}, & L \gg L_c(h), \\
L^\gamma, & L \ll L_c(h),
\end{cases}
\]
with the universal exponents $\nu = 4-d/(d+1)$ and $\gamma = 2(4-d)/5$ depending only on dimensionality. Also, $z$ is the dynamical exponent of TT theory and $L_c(h) \propto h^{-1/z}$. This shows a diverging longitudinal response (diverging susceptibility) as $h\to0$ in the thermodynamic limit ($L\to \infty$). For $d \geq d_c = 4$, where $d_c$ is the upper critical dimension, they instead predict $\delta \Phi \propto h$.

In Chapter \ref{directed} we will build on this approach to discuss how an external field suppresses the low $q$ divergence of polar flocks correlations.
\newpage

\subsection{Dissenter in a polar flock}
In \cite{yllanes2017many} it was shown that a very tiny fraction of dissenters, that is, particles with do not align with neighbors but do influence other particles orientation (note that this is a rather drastic form of non-reciprocal interactions), is sufficient to disrupt a well formed flock. Consider, for instance, a system composed of $N$ active particles with the same radius in a two dimensional box with periodic boundary conditions. Particle $i$ is characterized by its position $\mathbf{r}_i$ and an angle $\theta_i$ that defines the direction of self-propulsion. The dynamics for most particles is aligning, akin to that presented in Sec.\ref{other_ways} for the collisional Vicsek model, is then defined by the coupled Langevin equations
\begin{equation}\label{aligners}
    \dot{\mathbf{r}}_i = v_0 \hat{\mathbf{n}}_i(t) + \beta \sum_j \mathbf{F}_{ij}(t), \quad
\dot{\theta}_i = \frac{1}{\tau} \left[\psi_i(t) - \theta_i(t)\right] + \eta_i(t).
\end{equation}
A fraction $p$ of the particles instead are \textit{dissenters}: their dynamics follows Eq.\eqref{aligners}, without any explicit torque, thus their EOM boils down to that of repelling ABPs  
\begin{equation}\label{dissenters}
    \dot{\mathbf{r}}^{(d)}_i = v_0 \hat{\mathbf{n}}_i(t) + \beta \sum_j \mathbf{F}_{ij}(t), \quad
\dot{\theta}^{(d)}_i = \eta_i(t).
\end{equation}
Quite surprisingly a small number of dissenters can break up a well-formed flock: a few percent of dissenters are able completely destroy the polar ordered phase. Moreover, the fraction of dissenters needed in order to do so is much lower than the number of static obstacles (quenched disorder) required for an equally disruptive effect, when deep in the ordered phase.\\
Previous work with static obstacles \cite{chepizhko2013optimal} found, indeed, that tuning particle properties, such as their noise, can have a non-monotonic effect on their order and that, therefore, there are optimal values that maximize flocking in a disordered environment. In the case of moving dissenters, changing the intensity of rotational noise (but using the same value for aligners and dissenters) has almost no effect on the critical fraction of dissenters.

\subsection{Are polar flocks stable after all?}\label{polar_flock_stable}
\begin{figure} [hbt!]
    \centering
    \includegraphics[width=0.25\linewidth]{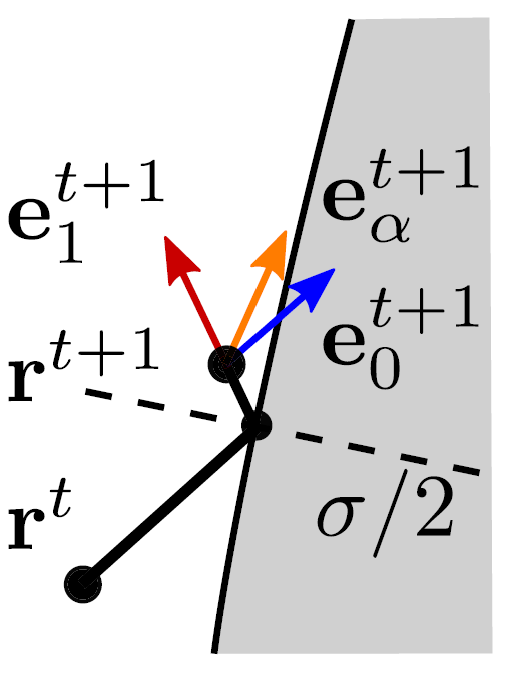}
    \caption{{\bf Sketch of collision rules} The postcollision orientation $\mathbf{e}_\alpha^{t+1}$ interpolates between the precollision one ($\alpha = 0$, blue arrow) and that of the reflected trajectory ($\alpha = 1$, red arrow). Figure adapted from \cite{codina2022small}.}
    \label{collision rule}
\end{figure}

In the last couple of years there have been a few relevant papers trying to clarify if polar flocks where stable to bulk perturbations. \\
For instance, even a single small obstacle in a large polar flock \cite{codina2022small} was shown to be capable of triggering counter-propagating dense bands that might destroy global order, near the onset of collective motion for the free system. In \cite{codina2022small} the authors consider a small disk, with radius typically of the order of 10 (with unit particles interaction radius), immersed in a square system of side $L = 1024$ or bigger. The model is a simple Vicsek flock, in $d = 2$, with vectorial noise. Whenever the position $\mathbf{r}_i^t$ of one particle is retrieved below the disk boundary one has a collision with the obstacle and the postcollisional position $\mathbf{r}_i^{t+1}$ is given by a simple reflection off the obstacle surface. The postcollisional orientation $\mathbf{e}_i^{t+1}$ is taken to vary, through a parameter $\alpha$, between the reflected trajectory ($\alpha = 1$) and the precollisional orientation ($\alpha = 0$) as sketched in Fig.\ref{collision rule}.\\
It is important to stress that the disruptive effect of such an obstacle on polar order has been reported for values of $\alpha < 1/2$: for $\alpha > 1/2$ the postcollisional particles will be oriented and will tend to move away from the obstacle and this results in a very different scenario, in particular one does not expect such a disruptive effect but rather accumulation of particles and enhancement of local density, consistent to the case of confining parallel walls \cite{Benoit2019, fava2024casimir}. Consistently with this picture, the authors find that at $\alpha = 1/2$ the size of radius needed to disrupt the order actually diverges.\\

The origin of such instability can be rationalized as follows: for sufficiently small $\alpha$, the polarity of particles colliding with the obstacle persists and particles glide along the obstacle. A circular obstacle immersed in a polar flock will thus tend to split the active fluid flow, with the resulting boundary flows following the obstacle contour and meeting on the opposite side (w.r.t. the flow direction). At this point, they align and accumulate, possibly forming a dense-enough blob that can then glide back on one side of the obstacle and eventually detach; when it detaches it can then evolve to a large scale perturbation as a counterpropating band. For smaller system sizes these bands are able to wind across boundaries and lead to an ordered phase with reversed polarity. At very large system sizes ($L > 1024$) the dynamics looks quite complicated and bands tend to cross each other and nucleate more bands when crossing the obstacle: this leads to a never-ending chaotic dynamics. It is concluded that asymptotically a single obstacle is able to destroy the ordered phase. It is however important to stress that, such an effect, is present in a finite fraction of the phase diagram, close to the disordered phase and it is \textit{not} generic in the entire flocking phase.

\begin{figure} [hbt!]
    \centering
    \includegraphics[width=0.75\linewidth]{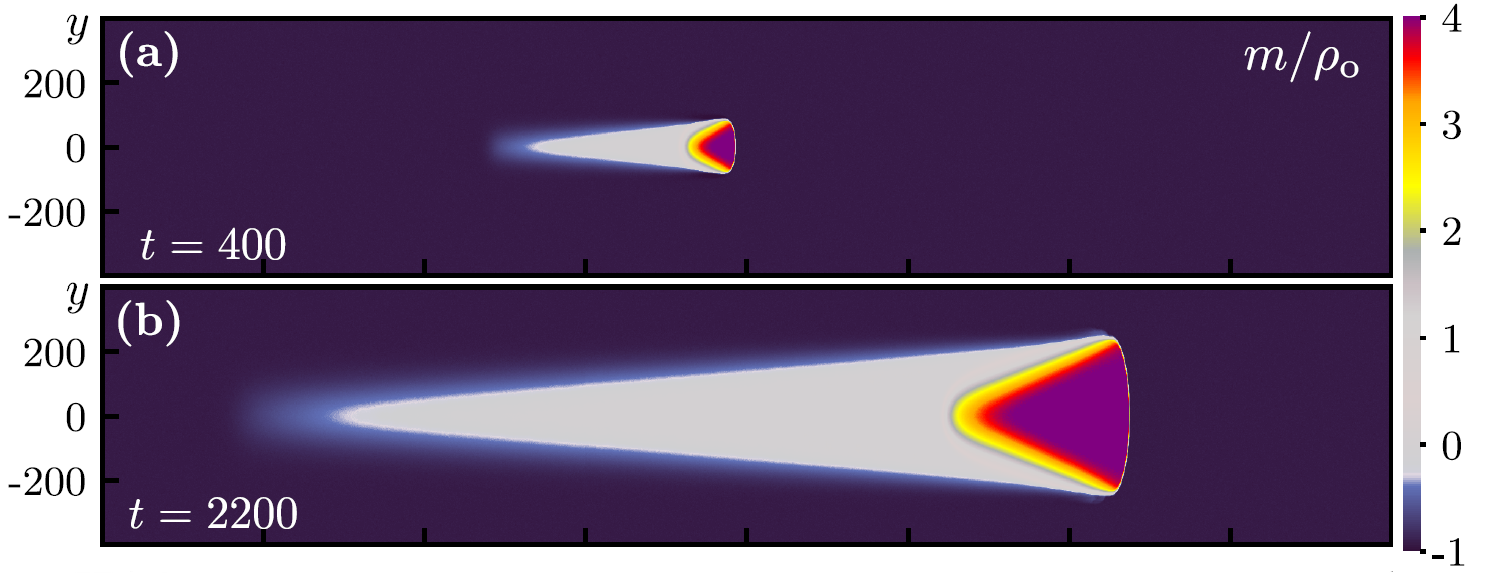}
    \caption{{\bf Snapshots of the magnetization field} The introduction in the ordered phase of a counter-propagating droplet (top) eventually disrupting polar order. Figure adapted from \cite{benvegnen2023metastability}}
    \label{aim_instability}
\end{figure}

Also the active Ising model \cite{benvegnen2023metastability} is found to be susceptible to droplets of particles moving in the direction opposite to that of the ordered phase: these nucleate and grow ballistically in all directions. Quite interestingly, since these droplets can occur/nucleate spontaneously, discrete-symmetry flocks are meta-stable in all dimensions $d\geq2$. As shown in Fig.\ref{aim_instability} a circular droplet of a sufficiently large radius $r$ and density $\rho_d > \rho_0$, with $\rho_0$ being the global density, of particles oriented opposite to the ordered phase grows at the expense of the ordered phase, leaving a behind a dilute, disordered comet-like structure.

An important difference with \cite{codina2022small} is that the critical radius at which this effect becomes relevant (when the probability of observing a reversal is $1/2$) weakly depends on the temperature $T$, meaning that more ordered flocks are just as likely to experience this effect w.r.t. flocks where fluctuations are bigger. Interestingly, by choosing a suitable region of parameter space, nucleations droplets within the ordered phase are spontaneous. Such nucleation events happen at a rate comparable to local fluctuations which would imply that the whole flocking phase is meta-stable for systems with discrete rotational symmetry in the thermodynamic limit.
\fancyhead[LO]{{\it Directed flocking}} 
\fancyhead[RE]{{\it Directed flocking}} 
\chapter{Directed flocking - Explicit symmetry breaking in the bulk or at the boundary}\label{directed}
As we have seen in Chapter \ref{dry_am}, when polar order emerges in an isotropic environment, it does so by \textit{spontaneously} breaking the underlying continuous rotational symmetry, a fact that largely determines the large scale physics of flocking systems \cite{Ginelli2016}. While the non-equilibrium nature of active matter systems allows flocks to escape the constraints of the Mermin-Wagner-Hohenberg theorem \cite{MW,H} and achieve long-range order even in  $d =2$ \cite{TT2}, the symmetry-broken phase is nevertheless characterized by massless modes (the {\it Nambu-Goldstone modes}) and consequently by long-range correlations.

Flocking, however, may also take place in anisotropic environments, where the rotational symmetry is explicitly broken and a (perhaps only slightly) favoured direction, determines the mean flocking direction. Cell motility, for instance, is known to be sensitive to a wide range of external gradients of chemical (chemotaxis), mechanical (durotaxis), and electrical (electrotaxis) origin which often direct cellular migrations. One relevant question regards the signatures of this directed motion. That is, how can an experimental observer detect a small anisotropy from the observation of the flocking system without an {\it a priori} knowledge of the lack of perfect environmental isotropy.

The above example regards an anisotropic perturbation that, in principle, should be felt by every active particle in the system and breaks the rotational symmetry {\it globally}. A different {\it local} source of anisotropy may be due to specific boundary conditions, often unavoidable in experiments, such as the channel geometry employed in Ref. \cite{bartolo_prx}. Moreover, confinement along one spatial direction has also been employed in numerical investigations of collective motion models \cite{TT3, Benoit2019} as a way to control the diffusion of the mean flock orientation in finite systems. Confinement effects have been so far mostly investigated in scalar active matter, either in \textit{dry} systems \cite{Kudrolli2008, Deseigne2010, Ray, Marchetti2014a, Marchetti2014b, Fily2014, ABPs, Caprini2018, Das2018} or active suspensions \cite{Goldstein2012, Goldstein2013, Boffetta2022}. Interestingly, recent work in active nematics \cite{ConfinedAN} and confined active Brownian particles \cite{Tailleur2022a, Tailleur2022b} revealed non-trivial long-range effects on the bulk dynamics of the system, as discussed in Section \ref{boundaries_scalar}.\\
Reflecting channel boundaries not only break the translational invariance, but also introduce a local anisotropy between the longitudinal and perpendicular directions w.r.t. the boundary. In this latter case, one may wonder how this local symmetry breaking at the boundaries may affect the flocking bulk behavior in large but finite systems. This chapter presents original research which addresses these two closely related questions.

In the first section of this chapter we address global symmetry breaking and derive scaling relations for the fluctuations of the mean direction of motion and for the static density structure factor (encoding static density fluctuations) in the presence of a homogeneous, small external field $\bf h$. This allows us to formulate two different and complementary criteria capable of detecting instances of directed motion exclusively from easily measurable dynamical and static signatures of the collective dynamics,  without the need to detect correlations with environmental cues. \\
The {\it static} one is informative in large enough systems, while the {\it dynamical} one requires large observation times to be effective. We believe these criteria may prove useful to detect or confirm the directed nature of collective motion in {\it in vivo} experimental observations.

In the second part, we investigate flocking in the presence of an explicit symmetry breaking at the system boundaries in a relatively simple set-up consisting of two reflecting and infinite parallel walls (i.e., a channel geometry). In particular, we discuss how this boundary symmetry breaking affects the scaling properties of bulk correlation functions and the polar order parameter fluctuations. As we noted, these results are relevant for many experimental realizations with active colloids, where confinement by hard boundaries (e.g., within a ring \cite{Bartolo2013}) is practically unavoidable.

Interestingly, it will turn out that -- at least in the case of confinement between two parallel reflecting walls -- the behaviour of our confined system is formally equivalent to the one of a boundary free system perturbed (in the linear regime) by an homogeneous external field of amplitude $h$, provided that
\begin{equation*}
h \sim L_\perp^{-z}
\end{equation*}
where $L_\perp$ is the separation between the reflecting boundaries and $z$ is the dynamical scaling exponent of TT theory.
\newpage

\section{Homogeneous external field}
A natural question when observing flocking phenomena, such as cellular migration or the coordinated movement of animal groups, thus regards the nature of collective motion. Is it a spontaneous phenomenon, exclusively driven by the interactions between individual active units, or the observed group movement is being {\it directed} by some external factor? \\
While this distinction can be simply made {\it in vitro}, where experimental factors are easily controlled, {\it in vivo} observations are typically conducted in rather complex environments, making the task much more arduous \cite{VitrotoVivo}. Chemotactic guidance, for instance, is known to be involved in many instances of embryonic development \cite{embryogenesis}, but {\it in vivo} chemical or mechanical gradients have not systematically been observed for all cellular migration phenomena, leading authors to speculate about other types of spatial guidance clues which may be at play in certain instances \cite{voituriez2015}. 

It may thus be desirable to formulate simple criteria capable of discriminating between spontaneous collective motion, taking place in an isotropic environment and directed one, simply by observing the static and dynamical features of the active units involved, without the need to detect and establish correlations (or the lack of) with gradients or other environmental cues. \\
Intuitively, the simplest signature of such a difference should lie in the persistence of the mean direction of motion, which is expected to be lower for spontaneous flocking. However, as we will show in the following, this simple criteria may fail for short observation times and/or small environmental anisotropies.\\
An alternative approach we propose involves the observation of large wavelength {\it static} fluctuations in the active particles density, which are encoded by the density structure factor $S_\rho(q)$. We have seen that, spontaneous collective motion implies a diverging structure factor at small wave-numbers $q$ \cite{toner1995long, toner1998flocks}, as experimentally confirmed in {\it in vitro} experiments of cellular migration \cite{giavazzi2017giant}. Here, we argue that such a divergence is suppressed in directed collective motion and that this fact can be used to successfully detect directed motion even on short or instantaneous observation timescales.

In the following, we will discuss collective motion in the presence of  static and homogeneous small external field of amplitude $h$ (i.e. in a {\it linear response} regime), showing that directed motion can be inferred from the fluctuations of the mean group direction (our so-called {\it dynamical approach}) only for observation times $t > \tau_c \sim h^{-1}$.\\
On the contrary, the study of the system density fluctuations in large enough systems, may reveal directed motion also for much shorter observation times. In particular, we show that the free system structure factor's divergence is suppressed for wave-numbers $q<q_c \sim h^{1/z}$, where $z$ is the dynamical scaling exponent of the TT theory (see Sec. \ref{non_linear_and_DRG}), This defines a complementary {\it static} approach.   
\subsection{Dynamical approach}
We begin discussing the more intuitive dynamical approach: in the absence of a driving field or any other anisotropy, collective motion is achieved by spontaneous breaking of the continuous rotational symmetry. The mean direction of any finite flock, however, is not constant, but freely diffuses since, as we have seen, small transverse perturbations are not damped and free to propagate. 

On the other hand, when collective motion is driven by an external field breaking rotational isotropy, fluctuations of the mean direction are confined and do not lead to free diffusion. Thus, one may discriminate between spontaneous and driven collective motion of any finite flock by simply observing the mean direction dynamics for a sufficiently long timescale. In the following, we precisely quantify this idea by developing a closed stochastic equation for the mean flocking direction.

\subsubsection*{Mean field dynamics of the flock's orientation}
For simplicity, we work in two spatial dimensions and consider Vicsek dynamics \eqref{vicsek_theta}-\eqref{position_vicsek} in the presence of an homogeneous external field of amplitude $h$ and orientation $\theta_h$, ${\bf h}= h (\sin\theta_h,\cos\theta_h)$, which tends to align the particles self propulsion orientations, as first introduced in Ref. \cite{kyriakopoulos2016leading}. It reads
\begin{equation}\label{ev_r}
{\bf r}_i^{t+1} = {\bf r}_i^t + v_{\rm 0} {\bf n}_i^t 
\end{equation}
\begin{equation}\label{micro_evo}
    \theta_i^{t+1}= \arg\bigg[\bigg(\sum_{j \sim i} \textbf{n}_j^t +\textbf{h}\bigg) \bigg] + \eta_i^{t},
\end{equation}
where $v_{\rm 0}$ is the particles speed and $\arg({\bf v})$ gives the angle defining the orientation of ${\bf v}$. Moreover, $\eta_i^t$ is a microscopic zero-average white noise such that $\langle \eta_i^t \eta_j^{t'}\rangle = \Gamma \delta_{ij}\delta_{tt'}$ and the sum is intended over the $m_j^t$ neighbours of particle $i$ (including $i$ itself) at time $t$. The neighbouring criteria may be metric, such that $i$ and $j$ are neighbors if $\lvert {\bf r}_i^t - {\bf r}_j^t \rvert < R_0$, or topological \cite{ginelli2010relevance}. 

We consider the Vicsek model (VM) \eqref{ev_r}-\eqref{micro_evo} in the homogeneous and highly ordered regime, deep in the Toner and Tu (TT) phase discussed in section \ref{TT_theory}.\\
In the presence of an external field, the direction of motion $\theta_i^t$ of particle $i$ can be expressed in terms of deviations  $\delta \theta_i^t$ from the field direction,  $\theta_i^t = \theta_h +\delta \theta_i^t$. If we assume $\delta \theta_i^t \ll 1$, which is reasonable in the ordered phase, we can expand \eqref{micro_evo} in a {\it spin-wave} approximation \cite{nishimori2010elements}. \\
To the first order in $\delta\theta$ we get 
\begin{equation}\label{micro_evo_2}
        \theta_i^{t+1}  \approx \arg \bigg[\mathbf{\hat{e}}_{\parallel} \bigg(\sum_{j \sim i}1 + h \bigg) +\mathbf{\hat{e}}_{\perp}\sum_{j \sim i} \delta\theta_j^t \bigg] +\eta_i^t.
\end{equation}
where $\mathbf{\hat{e}}_\parallel = (\cos\theta_h, \sin\theta_h)$ is the unit vector identifying the direction of the field ${\bf h}$ and $\mathbf{\hat{e}}_\perp=(-\sin\theta_h,\cos\theta_h)$ its perpendicular unit vector. 
We note that in Eq. \eqref{micro_evo_2} the component along the perpendicular direction is small with respect to the longitudinal one. We can thus expand the arg function as $\arg(\textbf{v}+\delta \textbf{w} ) \approx \arg(\textbf{v}) + \frac{\textbf{v} \vee \delta \textbf{w}}{\lvert v \rvert^2}$ where $\delta {\bf w} \ll {\bf v}$ and $\vee$ denotes the \textit{skew product}\footnote{For vectors ${\bf a}, {\bf b}$ lying in a plane perpendicular to a unit vector $\hat{\bf e}_3$ one has $\textbf{a}\vee \textbf{b}=\hat{\bf e}_3 \cdot(\textbf{a}\times\textbf{b})$.} 
By the above first order expansion Eq. \eqref{micro_evo_2} becomes
\begin{equation}\label{arg_expansion}
    \theta_i^{t+1} \approx \theta_h + \frac{\sum_{j\sim i}\delta \theta_i^t}{m_i^t+h}+\eta_i^t
\end{equation}
where $m_i^t$ is the number of neighbours of particle $i$ (including particle $i$ itself) at time $t$.\\
Further expanding the denominator for $h \ll m_i^t$ (we assume a sufficiently high local density, as in many systems of interest such as confluent tissues \cite{Giavazzi2018}) one gets
 \begin{equation}\label{theta_h_tilde}
     \delta\theta_i^{t+1} \approx \frac{1}{m_i^t}\sum_{j \sim i } \Big( 1-\Tilde{h}_i^t \Big)\delta\theta_j^t+\eta_i^t
 \end{equation}
where we have defined $\Tilde{h}_i^t \equiv h/m_i^t$. \\
The flocking mean direction $\psi(t)$ can be similarly expanded around the field direction 
\begin{equation}   \label{psi_t}
    \psi(t) \equiv \arg(\Omega(t)) \approx \theta_h +\frac{1}{N}\sum_{i=1}^N\delta\theta_i^t
\end{equation}
where $\Omega(t) \equiv \frac{1}{N} \sum_{i=1}^{N} \textbf{s}_i$ is the flock's global order parameter.\\
Eq. \eqref{psi_t} implies
\begin{equation}\label{deltapsi_t}
    \delta\psi(t)\equiv\psi(t) - \theta_h \approx \frac{1}{N}\sum_{i=1}^N\delta\theta_i^t
\end{equation}

Feeding Eq. \eqref{theta_h_tilde} in \eqref{deltapsi_t} we thus obtain
\begin{equation}  \label{psi_tp1_t}
        \delta \psi(t+1) \approx \frac{1}{N} \sum_{i=1}^{N} \frac{1}{m_i^t}(1-\Tilde{h}_i^t)\sum_{j \sim i}\delta\theta_j^t +\xi(t),
\end{equation}
where we have defined $\xi(t) \equiv\frac{1}{N}\sum_{i=1}^N \eta_i^t$ and for the central limit theorem we have that
\begin{equation}
   \langle \xi(t) \rangle =0 \;,\;\;\;\;\langle \xi(t) \xi(t') \rangle = \frac{\Gamma}{N}\delta_{tt'}
\end{equation}
In order to obtain a closed equation for $\delta \psi^t$, we finally resort to a mean field approximation, $m_i^t \approx \bar{m} \equiv \langle m_i^t\rangle_{i,t}$ where the latter average is conducted over all particles $i$ and time $t$. Noting that in \eqref{psi_tp1_t} every fluctuation $\delta\theta_j$ appears roughly $\bar{m}$ times we get
\begin{equation}\label{OU_samp}
    \delta \psi(t+1) \approx  \delta\psi(t) (1-h')+ \xi(t)\,,
\end{equation}
where we have defined the reduced field amplitude $h'\equiv h/\bar{m}$. \\
Eq. \eqref{OU_samp} defines an auto-regressive process of order one, which is the time-discrete sampling of an {\it Ornstein-Uhlenbeck} process \cite{livi2017nonequilibrium}
\begin{equation}  \label{OU}
    d x(\tau) = - \gamma \, x(\tau) d\tau + \sqrt{2 D} \, d W_\tau 
\end{equation}
where $d W_\tau$ is a Wiener process, $h' = \gamma \Delta t$, $\Gamma/N = 2 D \Delta t$ and $\tau=t\,\Delta t$ with $\Delta t \ll 1$.  
The two processes share the same statistics, 
\begin{equation}  \label{FLutt_OU}
 \langle \delta \psi(t)^2 \rangle = \langle x(\tau)^2 \rangle =
 \frac{\Gamma}{2 N} \frac{1}{h'} \left(1-e^{-2 h' t}\right) \,,
\end{equation}
showing that the mean flocking direction behaves as a one dimensional Brownian particle in an harmonic potential with stiffness proportional to the external field amplitude $h$. For large times, $t \gg 1/(2 h')$, the process is stationary with
\begin{equation}  \label{FLutt_OUS}
 \langle \delta \psi(t)^2 \rangle =
 \frac{\Gamma}{2 N} \frac{1}{h'}  \,,
\end{equation}
while for $t \ll 1/(2 h')$ we recover diffusive behavior
\begin{equation}  \label{FLutt_OUD}
 \langle \delta \psi(t)^2 \rangle =
 \frac{\Gamma}{ N} \,t  \,.
\end{equation}
For $h\to 0$ therefore, our mean field approximation yields a free diffusive behavior. One may indeed repeat the above argument in the zero external field case with a spin wave expansion around the mean flocking direction, $\theta_i^t = \psi(t) + \delta \theta_i^t$. By the same mean field approximation, one finally obtains the discrete time diffusive dynamics
\begin{equation}
    \psi(t+1) = \psi(t) +  \xi(t) \,.
\end{equation}

Arguably, our mean field approximation is rather crude; in particular, by assuming a constant number of interacting neighbours, $m_i^t \approx \bar{m}$, it ignores the non-reciprocal part of Vicsek model (VM) interactions \cite{vitelli2021, chepizhko2021revisiting}. However, it should be noted that the comparison between flocking dynamics with reciprocal and non-reciprocal interactions carried on in Ref. \cite{chepizhko2021revisiting} mainly reveals significant differences at the onset of order and in confined geometries. Our theory, on the other hand, describes the behavior of the flock's mean direction (a global quantity) in the strongly ordered regime, where we expect our approximation to be harmless. In the next section, we will verify its correctness by direct numerical simulations.

\subsubsection{Numerical simulations}\label{numsim}
We simulate the microscopic Vicsek dynamics \eqref{ev_r} and \eqref{micro_evo} in a $2d$ system of linear size $L$ with periodic boundary conditions and with metric interactions (setting $R_0=1$). We take the white noise term $\eta_i^t$ to be uniformly distributed in the interval $[-\eta_{\rm 0} \pi, \eta_{\rm 0} \pi]$, so that its variance is
\begin{equation}\label{eq_noise}
    \langle \eta_i^t \eta_j^{t'} \rangle = \Gamma \delta_{ij}\delta_{tt'} = \frac{\eta_{\rm 0}^2 \pi^2}{3}\delta_{ij}\delta_{tt'} 
\end{equation}
In the following, we fix the global particle density $\rho_{\rm 0}=N/L^2=2.0$ and particle speed $v_{\rm 0}=0.5$. Noise amplitude $\eta_{\rm 0}$ is chosen so that our the zero field system lies in the homogeneous ordered phase \cite{solon2015phase}, comfortably far away from the ordered band regime \cite{gregoire2004onset, chate2008collective}.

Particles positions are initialized from a uniform distribution, while initial velocity are aligned in the external field direction (or in a given direction for $h=0$). A proper transient $T_0 \approx 10^4$ is then discarded from the dynamics to ensure convergence to the stationary ordered state. Squared fluctuations 
\begin{equation} \label{mean_fluct}
\langle \delta \psi(t)^2\rangle =\langle [\psi(t) - \psi(0)]^2 \rangle
\end{equation}
in the flock's mean direction \eqref{psi_t} are then typically evaluated averaging $10^2$ independent runs for $t\leq T=10^4$.

\begin{figure}[hbt!]	
\centering
 \includegraphics[width=0.8\textwidth]{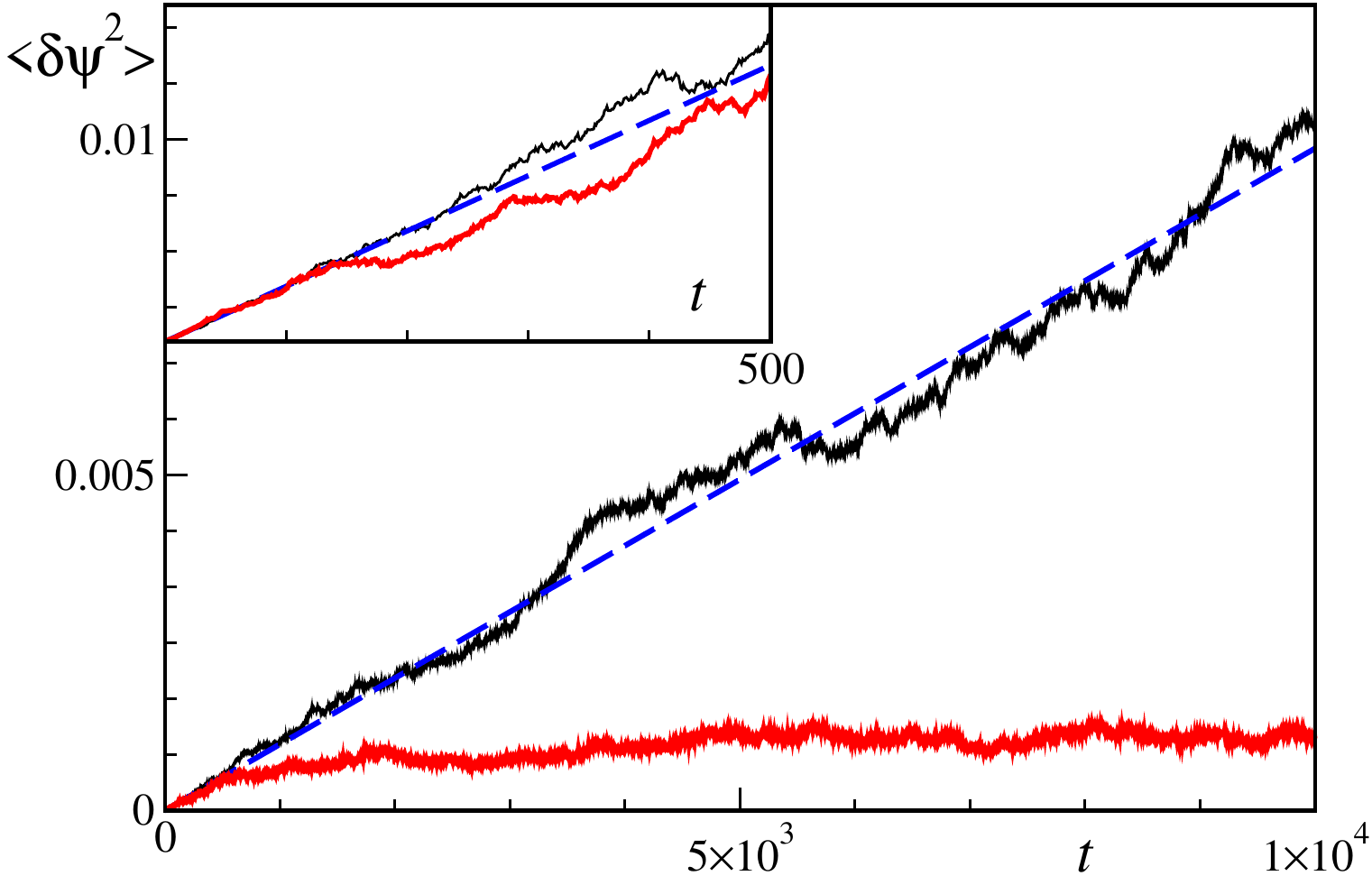}
	\caption{{\bf Spontaneous vs. directed mean direction squared fluctuations} Spontaneous (full black line) vs. directed ($h=0.01$, full red line) mean direction squared fluctuations as a function of time. The dashed blue line is the best fit of the spontaneous symmetry breaking data (see text). We have chosen noise amplitude $\eta_{\rm 0}=0.18$ and system size $L=256$, so that $N=\rho_{\rm 0} L^2 \approx 10^5$. In the inset: zoom of the first $500$ time-steps.}
	\label{diff_spont_vs_driv}
\end{figure}

As shown in Fig.~\ref{diff_spont_vs_driv}, the difference between spontaneous ($h=0$, black line) and a directed collective motion ($h=0.01$, red line) is readily evident for long enough observation times. However, if one is restricted to a shorter time interval (e.g.: $10^3$ timesteps as in the inset Fig.~\ref{diff_spont_vs_driv}), discrimination between the two cases becomes problematic (especially if one is not comparing a directed with a spontaneous case but has only access to a single set of data). 

\begin{figure}[hbt!]
\centering
\includegraphics[width=0.75\textwidth]{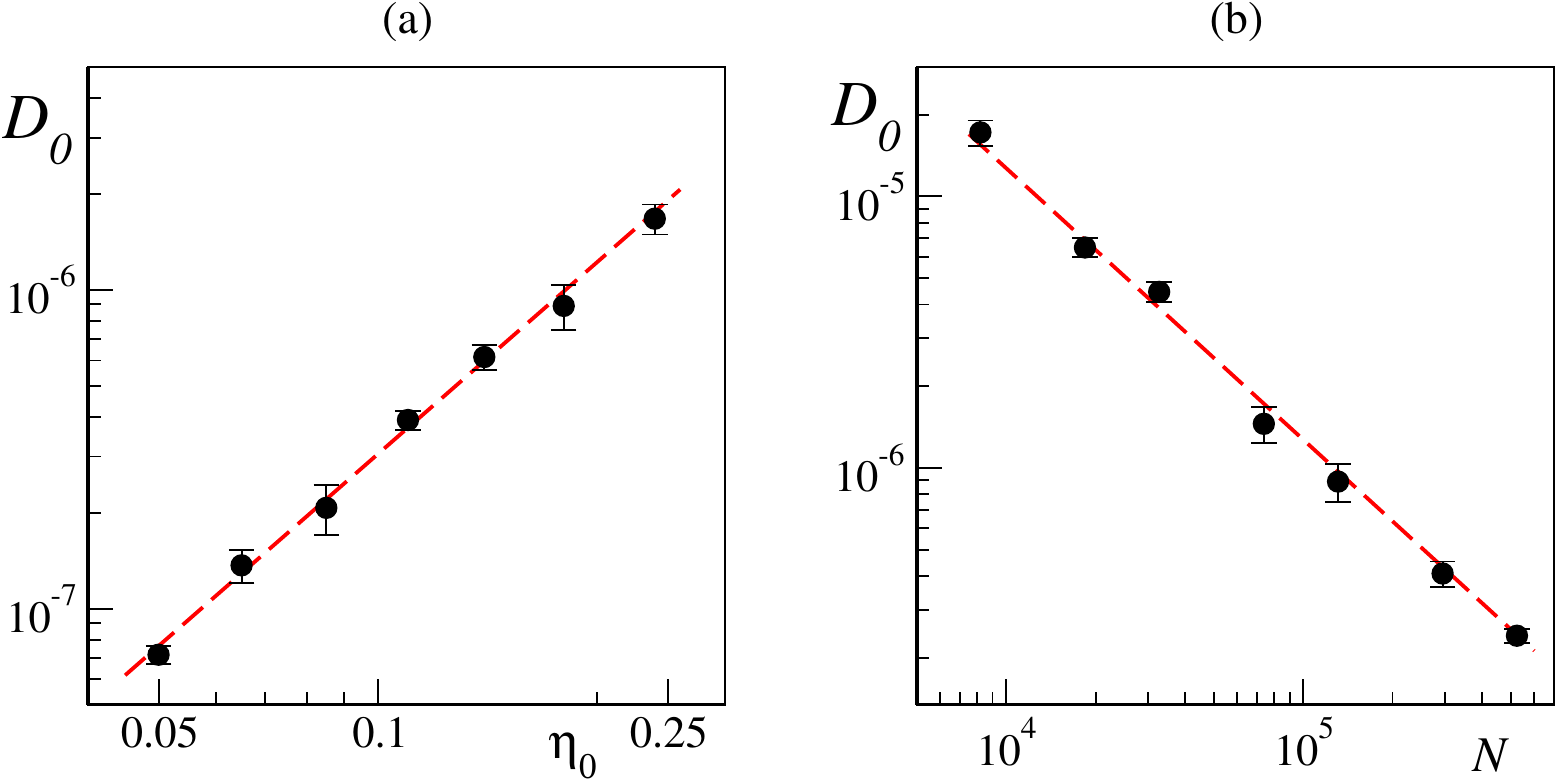}
\caption{{\bf Scaling of the diffusion constant in the zero field case} - (a) Diffusion vs. noise amplitude for system size $L=256$. The dashed red line marks quadratic growth, $D_{\rm 0}\sim \eta_{\rm 0}^2$. (b) Diffusion vs. total particle number for noise amplitude $\eta_{\rm 0}=0.18$ and constant density. The dashed red line marks $1/N$ decay.
Error bars are given by two standard errors and plot are in a double logarithmic scale.} 
\label{Diff_scal}
\end{figure}

A linear fit of the spontaneous case
\begin{equation}
    \langle \delta \psi^2 \rangle \approx 2D_{\rm 0}\,t
\end{equation}
returns a diffusion constant $D_{\rm 0} \approx 5.0(1) \cdot 10^{-7}$, to be compared with our mean-field prediction \eqref{FLutt_OUD}
\begin{equation}\label{mf1}
D_{\rm 0}=\frac{\Gamma}{2 N} = \frac{\eta_{\rm 0}^2 \pi^2}{6 N}
\end{equation}
where we have made use of Eq. \eqref{eq_noise} and, in $d=2$, $N=\rho_{\rm 0}L^2$. With our choice of parameters, Eq. \eqref{mf1} gives $D_{\rm 0} \approx 4.1\cdot 10^{-7}$. Despite being around $20\%$ off quantitatively, Fig. \ref{Diff_scal} shows that the qualitative scaling with noise amplitude and particles number predicted by mean field theory is nicely verified. 
 
\begin{figure}[hbt!]
\centering
 \includegraphics[width=0.425\textwidth]{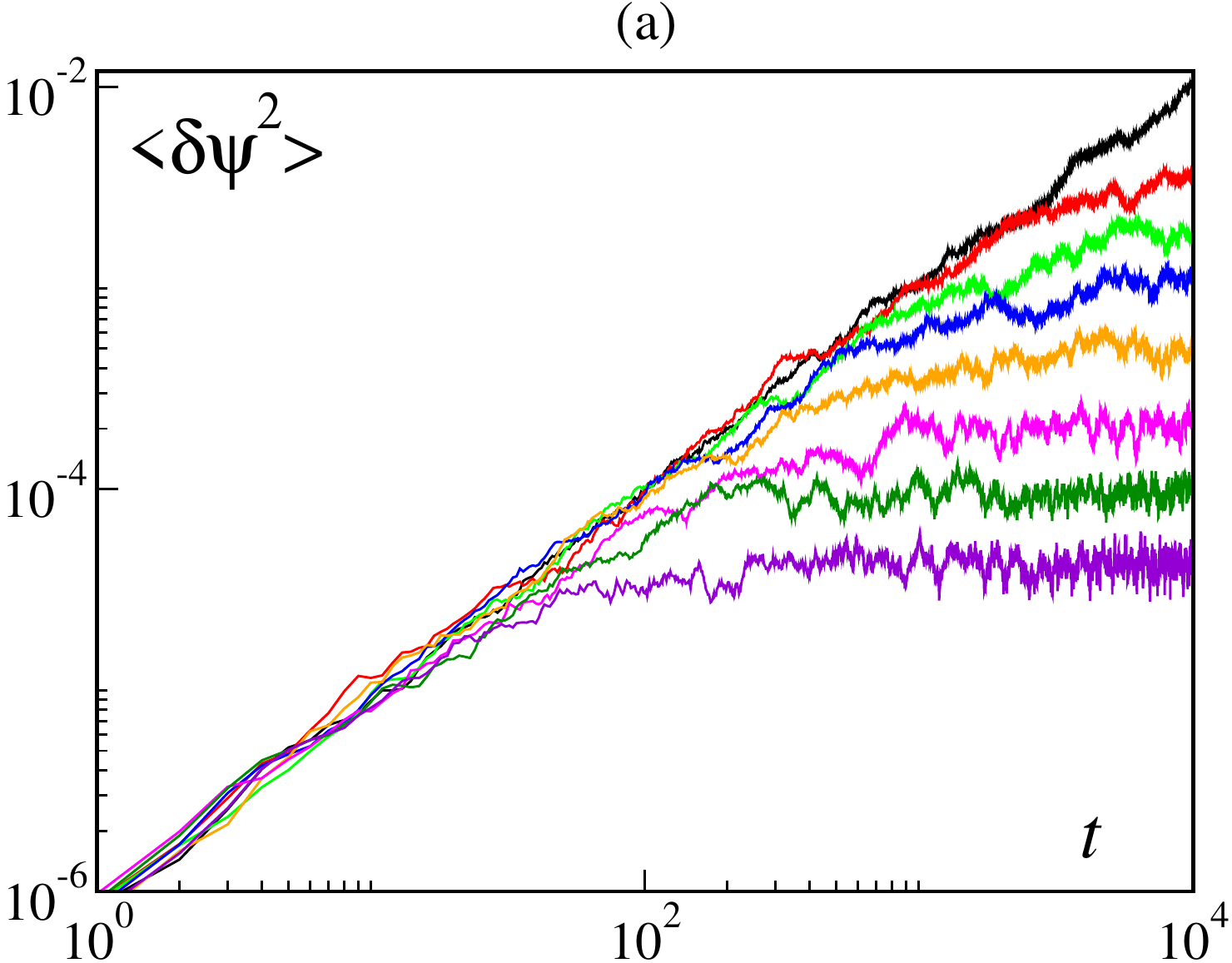}
    \includegraphics[width=0.425\textwidth]{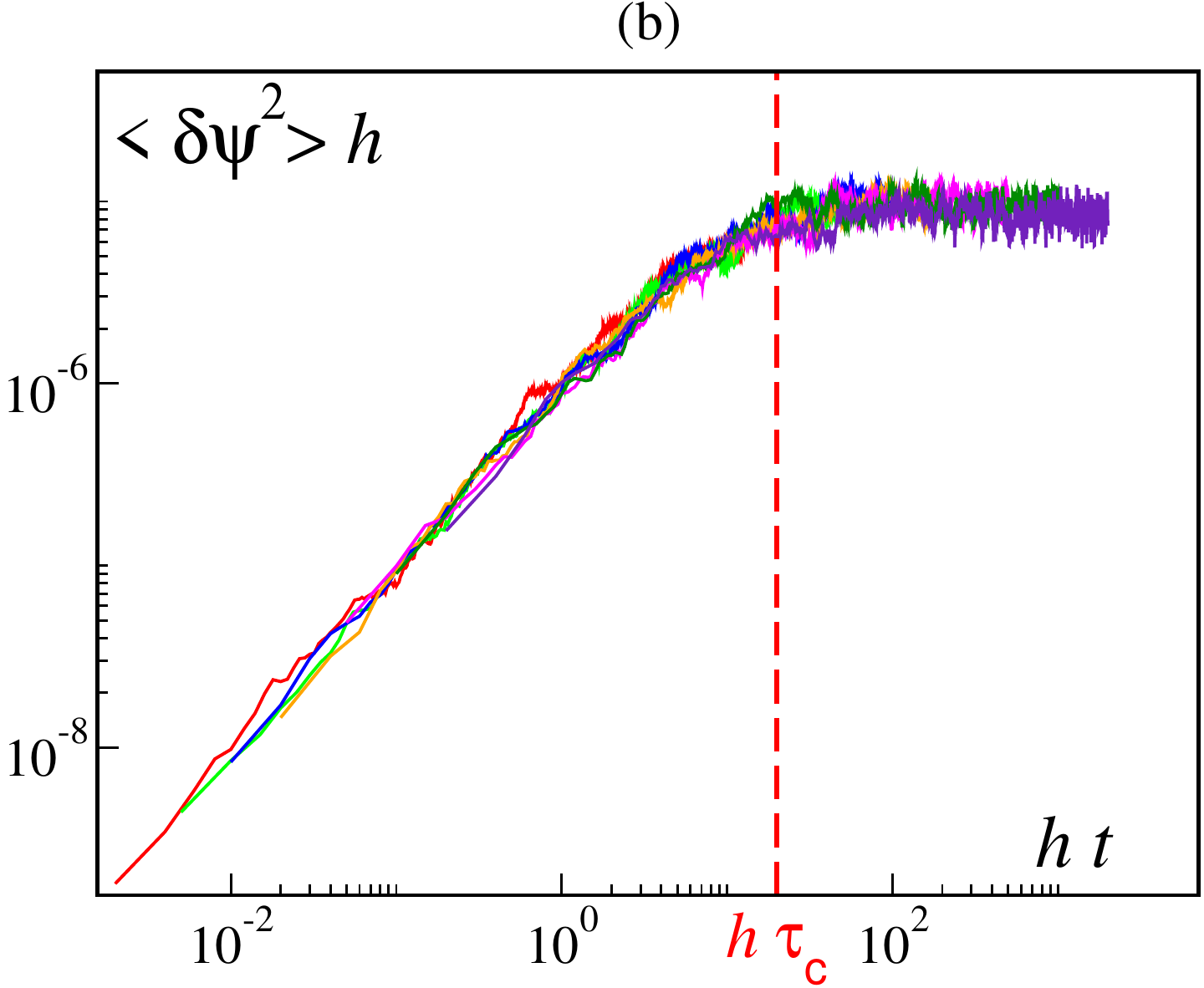}
    \caption{{\bf Rescaling of the mean squared fluctuations of the flocking average direction} (a) Mean squared fluctuations of the flocking average direction as a function of time $t$ for different external field values. From bottom to top $h=0.2, 0.1, 0.05, 0.02, 0.01, 0.005, 0.002, 0.001, 0$. Data has been obtained with noise amplitude $\eta_{\rm 0}=0.18$ and system size $L=256$. 
    (b) Same data with both axes rescaled by $h$ (the $h=0$ case has been removed). The dashed red line marks the rescaled crossover time $h \tau_c$. Both graphs are in a doubly logarithmic scale.}
    \label{all_collassi}
\end{figure}

We now turn to the directed case scaling, testing different values of the field intensity $h$. Our result are shown in Fig.~\ref{all_collassi}. According to Eqs. \eqref{FLutt_OUS}, \eqref{FLutt_OUD}, rescaling by $h$ both time and the mean squared fluctuations nicely collapses the curves obtained with different $h \in [0.001,0.2]$.  This confirms that the minimum observation time needed to discriminate spontaneous from directed motion scales as the inverse of the field amplitude, with a crossover time
\begin{equation}\label{tauc}
\tau_c(h) \sim h^{-1}\,.
\end{equation}
Note that the number of active particles $N$ controls the magnitude of mean direction fluctuations but not the scaling of the crossover time. 
 
\begin{figure}[hbt!]
\centering
\includegraphics[width=0.66\textwidth]{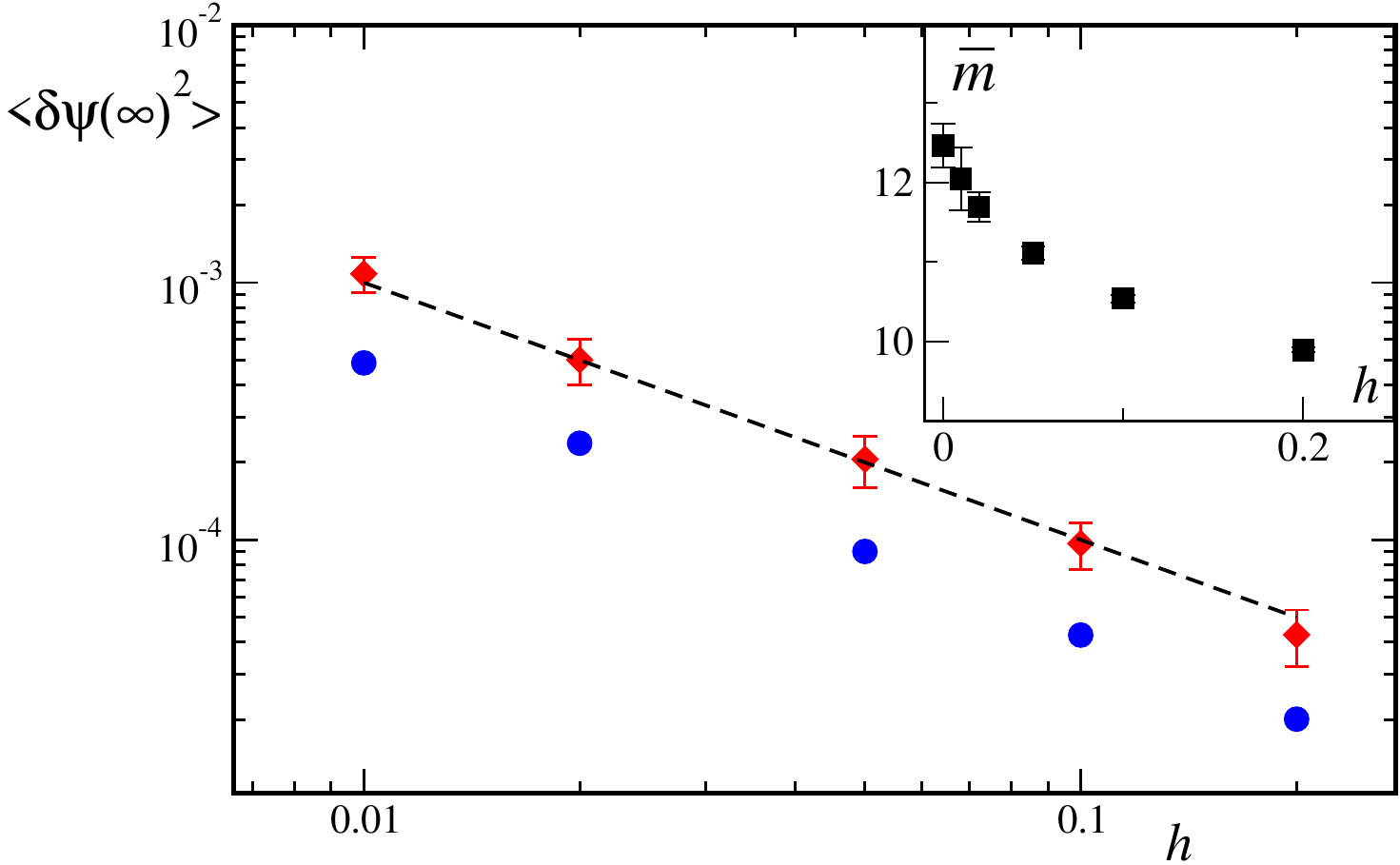}
  	\caption{{\bf Asymptotic value of $\langle\delta\psi^2\rangle$ as a function of $h$}. Red diamonds report numerical estimates, while blue dots are the mean field estimate of Eq. \eqref{FLutt_OUS2} (see text). The dashed black line marks the $\sim 1/h$ scaling. Here $\eta_{\rm 0}=0.18$ and $L=256$ and the plot is in double logarithmic scale. Inset: linear plot of the mean number of interacting active particles as a function of $h$. Error bars are given by two standard errors.}
   	\label{num_theo}
\end{figure}

Finally, we verify quantitatively the asymptotic expression for the flock's direction mean squared fluctuations \eqref{FLutt_OUS} that, by our mean field approximation reads
\begin{equation}  \label{FLutt_OUS2}
 \langle \delta \psi(\infty)^2 \rangle =
 \frac{\Gamma}{2 N} \frac{\bar{m}}{h} = \frac{\eta_{\rm 0}^2 \pi^2}{6 N} \frac{\bar{m}}{h}\,.
\end{equation}
We estimate $\bar{m}$ from direct numerical simulations and use it to compute the mean field prediction for $\langle \delta \psi(\infty)^2 \rangle$ from Eq. \eqref{FLutt_OUS2}. They are reported in Fig.~\ref{num_theo} as blue dots. The numerically measured values (red diamonds) turn out to be larger by a factor 2, possibly due to the inability of the mean field fluctuations to properly account for the anomalously large \cite{ginelli2016physics} local fluctuations in the number of interacting neighbours. Note also that numerical estimates of $\bar{m}$ (inset of Fig.~\ref{num_theo}) exhibit a weak dependence on $h$.  
 
In any case, we conclude that despite the quantitative mismatch, mean field theory faithfully captures the Ornstein-Uhlenbeck scaling with $h$ (the external field playing the role of a stiffness) of the mean squared fluctuations in the flocking direction. More generally, our analysis shows that one is able to discriminate between collective and directed motion by the analysis of the mean-squared fluctuations of the average flock's direction, but only provided the observation time is larger of a crossover threshold, scaling with the inverse of the external field amplitude. As we anticipated, this might not be possible in several scenarios where the observation time is limited and/or the magnitude of spatial anisotropy is too small.

\subsection{\label{static method}Static approach}
We now propose a second approach, based on the observation of spatial correlations. In particular, here we focus on the static density structure factor, which is easily accessible in experimental set ups. As we have seen, in the absence of any external field or other anisotropies, TT theory (as recalled in Sec.\ref{TT_theory}) predicts a long-ranged behavior for the slow fields spatial correlations. Equivalently, the static structure factors, which may be obtained by Fourier transforming spatial correlation functions, exhibits a diverging behavior at small wave numbers $q$.

In the presence of an external field, on the other hand, transversal fluctuations are damped (they acquire a \textit{mass} in the pseudo-particles language) and spatial correlations are cut-off exponentially at a finite characteristic length scale $L_c$. Correspondingly, the structure factor's divergence for $q \to 0$ is suppressed for $q < q_c \sim 1/L_c$. The cut-off length $L_c$ depends on the external field amplitude $h$ and diverges for $h \to 0$. This is well known at equilibrium since the pioneering studies of \cite{pata73, patashinskii1979fluctuation}, but as we show here it is relatively straightforward to derive the exact scaling of the cut-off length and of the static structure factor with $h$. 

Here we focus on the static density structure factor
\begin{equation}\label{eq:sf}
    S_\rho({\bf q})= \frac{1}{N}\left\langle \sum_{n, m} e^{i {\bf q}\cdot({\bf r}_n - {\bf r}_m)}\right\rangle
\end{equation}
where ${\bf r}_n$ with $n=1,\ldots,N$ are the position of the $N$ active particles, $i$ the imaginary unit and in principle the average $\langle \cdot \rangle$ should be taken over different experimental realizations or uncorrelated snapshots of the same experiment.
In particular, its average over all wavevector orientations 
\begin{equation} \label{eq:sf_iso}
    S_\rho(q)\equiv \langle S_\rho({\bf q})\rangle _{|{\bf q}|=q}
\end{equation}
can be measured relatively easily in experimental data \cite{giavazzi2017giant}. 

Introducing the number density $\rho({\bf r},t)=\sum_n \delta ({\bf r}-{\bf r}_n^t)$ and the density fluctuations $\delta \rho({\bf r},t)=\rho({\bf r},t)-\rho_0$, the density structure factor becomes the familiar
\begin{equation}\label{eq:sf2}
S_\rho({\bf q},h) = \frac{1}{N} \langle \lvert \delta \hat{\rho}(\textbf{q},t) \rvert^2 \rangle\,.
\end{equation}
For a zero external field, we have seen that Toner $\&$ Tu  theory predicts that the isotropically-averaged density structure factor of spontaneous collective motion diverges algebraically for small wave numbers $q$ as
\begin{equation}    \label{div_S}
    S_\rho(q) \sim q^{-\zeta}
\end{equation}
where $\zeta = d-1+\xi+2\chi>0$ is a universal exponent. In the following, we show that, in the presence of a small and static external field of modulo  $h$, this divergence is suppressed and
\begin{equation} \label{div_S_h}
    S_\rho(q,h) \sim \frac{1}{ q^\zeta+ C \,h}
\end{equation}
with $C$ being a phenomenological constant.

\subsubsection*{\label{theory static method}Linearized density structure factor}
We briefly recall the behaviour of Toner and Tu hydrodynamic equations \eqref{TT_continuity}-\eqref{TT_velocity} in the presence of a constant and homogeneous driving field ${\bf h}$ \cite{kyriakopoulos2016leading}. While the continuity equation for the density field evolution is left unchanged, long-wavelength dynamics of the conserved density $\rho({\bf r}, t)$ and velocity ${\bf v}({\bf r}, t)$ fields and consist in the continuity equation
\begin{equation}
\partial_t\rho +\nabla\cdot(\rho{\bf v})=0
\label{rho1}
\end{equation}
the velocity field dynamics isotropy is broken by the field ${\bf h}$
\begin{equation} \label{second_v}
    \begin{split}
        \partial_t \textbf{v} &+\lambda_1(\textbf{v} \cdot \nabla)\textbf{v} +\lambda_2 (\nabla \cdot \textbf{v})\textbf{v} + \lambda_3\nabla \lvert \textbf{v} \rvert^2 \\ =& (\mu - \kappa \lvert \textbf{v} \rvert^2 ) {\bf v} -\nabla P_1 -\textbf{v}(\textbf{v} \cdot \nabla P_2) + D_1 \nabla(\nabla \cdot \textbf{v})\\& +D_v \nabla^2 \textbf{v} +D_2(\textbf{v} \cdot \nabla)^2 \textbf{v} + \textbf{f} +\textbf{h}\,.
    \end{split}
\end{equation}
The coarse-grained constant field ${\bf h}=h \mathbf{\hat{e}}_{\parallel}$ is, by analyticity and rotational invariance of the free system, linearly proportional to the applied microscopic field, as long as it is sufficiently small.

In the absence of fluctuations Eqs.\eqref{rho1}-\eqref{second_v} admit a homogeneous steady state solution
$\rho({\bf r}, t)=\rho_0$,
${\bf v}({\bf r}, t)=p_{\rm 0}(h) \mathbf{\hat{e}}_{\parallel}$\,,
where $p_0(h)$ is determined by the condition
\begin{equation}
    \mu p_0 - \kappa p_0^3 + h = 0 \,.
\end{equation}
In the zero external field case (and for $\mu>0$), $p_0 = \sqrt{\mu/\kappa}$, with the direction $\mathbf{\hat{e}}_{\parallel}$ randomly selected by the spontaneous symmetry breaking mechanism. Assuming analyticity of the symmetry breaking coefficients, we have for small $h$
\begin{equation}
p_0(h)-p_0(0) \propto h
\end{equation}

To deal with fluctuating hydrodynamics, we follow \cite{toner1998flocks, toner2012reanalysis, kyriakopoulos2016leading} and proceed as usual to linearize Eqs. \eqref{rho1}-\eqref{second_v} around the homogeneous solution,
\begin{equation} \label{fluct}
\centering
    \begin{split}
        & \rho(\textbf{r},t)=\rho_0 +\delta \rho(h,\textbf{r}, t)\\
        & \textbf{v}(\textbf{r},t)=[{p}_{\rm 0}(h)+ \delta v_{\parallel}(h,\textbf{r},t)] \mathbf{\hat{e}}_{\parallel}+ \textbf{v}_{\perp}(h,\textbf{r},t)
    \end{split}
\end{equation}

where $\textbf{v}_{\perp}$ measures velocity fluctuations transversal to $\mathbf{\hat{e}}_{\parallel}$. \\
As we know, in the Toner and Tu phase, the longitudinal velocity fluctuations $\delta v_\parallel$ are a fast mode enslaved to the slow fields of the unperturbed theory, $\textbf{v}_{\perp}$ and the density fluctuations $\delta \rho$. Thus, they can be eliminated from Eqs. \eqref{rho1}-\eqref{second_v} to yield the linearized hydrodynamics
\begin{equation}\label{rho_lin}
\begin{split}
        \partial_t \delta\rho &+\rho_0 \nabla_{\perp}\cdot \textbf{v}_{\perp} +v_2 \partial_{\parallel}\delta\rho -\rho_0 \mu_2\partial_t \partial_{\parallel}\delta\rho \\
        & =D_{\rho_{\parallel}} \partial_{\parallel}^2 \delta\rho + D_{\rho_{\perp}} \nabla_{\perp}^2 \delta\rho +D_{\rho v}\partial_{\parallel} (\nabla_{\perp} \cdot \textbf{v}_{\perp})  \\ 
\end{split}
\end{equation}
and
\begin{equation}  \label{v_perp_lin}
    \begin{split}
        \partial_t \textbf{v}_{\perp} &= - \gamma \partial_{\parallel} \textbf{v}_{\perp} -\frac{c_0^2}{\rho_0}\nabla_{\perp} \delta\rho +D_B \nabla_{\perp}(\nabla_{\perp}\cdot \textbf{v}_{\perp}) +D_v \nabla_{\perp}^2 \textbf{v}_{\perp} \\ & +D_{\parallel}\partial_{\parallel}^2\textbf{v}_{\perp} 
        +g_t \partial_t \nabla_{\perp}\delta\rho + g_{\parallel} \partial_{\parallel} \nabla_{\perp}\delta\rho +\textbf{f}_{\perp}-h_v \textbf{v}_{\perp} 
    \end{split}\,,
\end{equation}
where we have introduced the reduced field
\begin{equation}\label{reduced-f}
    h_v \equiv \frac{h}{p_{\rm 0}(0)}
\end{equation}
and all the various constants introduced above can be expressed as a function of the phenomenological constants appearing in Eqs. \eqref{rho1}-\eqref{second_v}. Their precise expressions are not important in what follows, but they can be found, together with all the details of the linearization, in Ref. \cite{kyriakopoulos2016leading}.

The linearized density structure factor can straightforwardly computed from \eqref{rho_lin} and \eqref{v_perp_lin} in Fourier space. This calculation closely resembles the one for the zero field case, seen in Sec. \ref{TT_theory}, and for compactness we report its details in the App. \ref{linearized_SF}. To leading order in wave vector ${\bf q}$ the linearized structure factor is
\begin{equation} \label{autocorr_eqt}
\begin{split}
        S_\rho({\bf q},h)\equiv \langle \lvert \delta \hat{\rho}(\textbf{q},t) \rvert^2 \rangle=&\frac{1}{2} \bigg( \frac{\Delta\, \rho_0^2 \,\sin(\theta_q)^2}{[c_+(\theta_q)-c_-(\theta_q)]^2} \bigg) \\
        &\times \bigg( \frac{1}{\epsilon_+(h,{\bf q})}+ \frac{1}{\epsilon_-(h,{\bf q})} \bigg)\,,
        \end{split}
\end{equation}
where  $q=|{\bf q}|$ and $\theta_q$ is the angle between $\textbf{q}$ and $\mathbf{\hat{e}}_\parallel$.\\
The field dependent dampings do not simply scale with $q^2$ as in the zero field case and read
\begin{equation}\label{epsilon_pm}
\epsilon_\pm (h, {\bf q}) = \epsilon_\pm (0, {\bf q})+ a_\pm(\theta_q) h \equiv \Tilde{\epsilon}_\pm(\theta_q) q^2 + a_\pm(\theta_q) h\,,
\end{equation}
where
\begin{equation}\label{eq:apm}
a_\pm(\theta_q)=\frac{(c_{\pm}(\theta_q)-v_2\cos(\theta_q)^2)}{p_{\rm 0}(0)[2c_{\pm}(\theta_q)-(v_2+\gamma)\cos(\theta_q)]}\,.
\end{equation}
The precise form of $c_\pm(\theta_q)$ and $\Tilde{\epsilon}_\pm(\theta_q)$ is not relevant for what follows and its reported in the appendix \ref{linearized_SF} for compactness. Here it suffice to note that they are only a function of the angle $\theta_q$. 

We conclude that a small static and homogeneous external field only affects the damping terms, which acquire a correction linear in the field amplitude but independent of the wavevector magnitude $q$. This suppresses the small wavelength divergence of the free theory, since
\begin{equation} \label{autocorr_q0}
       \lim_{q \to 0} S_\rho({\bf q},h) \sim h^{-1}
\end{equation}

\subsubsection{Nonlinear corrections}
As in the isotropic, field-less case, the structure factor \eqref{autocorr_eqt} is only valid in linear approximation. We have seen that nonlinear terms are relevant in $d<d_c=4$ and can be accounted for by a dynamical renormalization group (DRG) analysis (see Sec.\ref{non_linear_and_DRG}). In particular, the driven case has been first analyzed in \cite{kyriakopoulos2016leading} in order to compute linear response. Here we follow the same approach to compute the scaling behavior of the density structure factor.

Including an external field in the DRG analysis is, as usual, fairly simple, and we actually may discuss it without the need to specify the exact form of the nonlinear equations of motion (they may be checked in \cite{kyriakopoulos2016leading}). It suffice to know that they have to be rotationally invariant with the only exception of the rescaled field $h_v$, the only term explicitly breaking the rotational symmetry. As a consequence, it may not gain any graphical correction in the short wavelength averaging step. Moreover, trivial dimensional analysis of Eq.\eqref{v_perp_lin} shows that $h_v$ scales as the inverse of time, yielding the linear recursion
\begin{equation}
    h_v \to h_v'=b^{z} h_v
\end{equation}
We are now able to write down the recursive equation for the density structure factor in the presence of an external field,
\begin{equation}\label{scaling1}
S_\rho({\bf q}_\perp, q_\parallel, \{\bm \mu^{(1)}\}, h_v) = b^{\zeta} S_\rho(b {\bf q}_\perp, b^{\xi} q_\parallel, \{\bm \mu^{(b)}\}, b^{z} h_v)\,,
\end{equation}
where we recall that $\zeta\equiv(d-1) + \xi + 2 \chi$ and the scaling of the structure factor has been determined considering that it is given by the Fourier transform of the equal time, real space density correlation function. Thus, it involves two powers of the density fluctuations and one volume element. 

The scaling of the structure factor with the field can be now deduced by fixing a reference value $h_v^*$ for the reduced field \eqref{reduced-f} and choosing the rescaling factor $b$ such that $b^{z} h_v=h_v^*$, which
implies 
\begin{equation}
    b=\left(\frac{h_v}{h_v^*}\right)^{-1/z}\,.
\end{equation}
In practice, one adapts the DRG magnifying glass to the value of the external field \cite{nishimori2010elements}. Furthermore, if $h\ll 1$ and thus also $h_v$, this choice implies $b\gg 1$, so that all other parameters will flow to their fixed point value, $\{\bm \mu^{(b)}\} \to \{\bm \mu^*\} $.\\
Therefore, from Eq. \eqref{scaling1} we have 
\begin{equation}\label{scaling2}
S_\rho({\bf q}_\perp, q_\parallel, \{\bm\mu^{(1)}\}, h_v) = h^{-\zeta/z} g(h^{-1/z} {\bf q}_\perp, h^{-\xi/z} q_\parallel)\,,
\end{equation}
where we have introduced the universal scaling function
\begin{equation}\label{scaling-g}
g({\bf x},y)=h_0^{\zeta/z} S_\rho(h_{\rm 0}^{1/z} {\bf x}, h_{\rm 0}^{\xi/z} y , \{\bm\mu^*\}, h_v^*) \,,
\end{equation}
and $h_{\rm 0}=h_v^* p_{\rm 0}(0)$. This expression further simplifies by the hyper-scaling relation $\zeta=z$ suggested by numerical simulations \cite{mahault2019quantitative} and recently put forward by the arguments of Ref. \cite{chate_solon}. However, we prefer to develop our arguments in a more general setting and to invoke $\zeta=z$ only for the final result.

We now briefly discuss the role of the anisotropy exponent $\xi$. As we have seen, numerical simulations \cite{mahault2019quantitative} suggest $\xi \approx 1$ in $d=2, 3$, as confirmed more recently (at least for $d=2$) by the analytical arguments of Ref.\cite{chate_solon}. For $\xi=1$ Eq.\eqref{scaling2} implies strictly isotropic correlations. However, the density structure factor is nevertheless known to present an anomalous scaling in the longitudinal direction \cite{mahault2019quantitative} (see Section \ref{scaling_exp}), so that in the following we find convenient to consider the isotropically-averaged density structure factor 
\begin{equation}
    S_\rho(q,h) = \langle S_\rho({\bf q},h)\rangle_{|{\bf q}|=q}
\end{equation}
which is more easily accessible in experimental measures \cite{giavazzi2017giant}. In any case, we develop our scaling theory without assuming isotropic correlations, using the fact that the short wavelength structure factor is dominated by contributions in the $q \sim q_\perp \sim q_\parallel$ direction \cite{ginelli2016physics} so that
\begin{equation}\label{s_free}
    S_\rho(q,0) \sim q^{-\zeta}\,.
\end{equation}
Along the $q_\perp \sim q_\parallel$ line and for small $h$ one has $h^{-1/z} q \sim h^{-1/z} q_\perp \gg h^{-\xi/z} q_\parallel$, so that the structure factor is dominated by transverse wavevectors ${\bf q}_\perp$ and we can replace $g$ with the universal scaling function $w(x)=g(x,0)$ (if $\xi=1$ one otherwise chooses $w(x)=g(x,x)$). This finally yields the scaling for the isotropic structure factor
\begin{equation}\label{scaling3}
S_\rho(q, h) = h^{-\zeta/z} w(h^{-1/z} q)\,.
\end{equation}

The behavior of the universal scaling function $w$ can be inferred by the request that, for $h\to0$, the structure factor's scaling coincides with the one of Eq. \eqref{s_free}, so that
\begin{equation}
w(x) \sim \left\{ \begin{array}{cc}
   x^{-\zeta}&\;\;\;\mbox{for}  \;\;x\gg1    \\
 \mbox{const.} & \;\;\;\mbox{for}\;\; x \ll 1
\end{array}\right.\,, 
\end{equation}
and the isotropically averaged structure factor takes the form
\begin{equation} \label{scaling5}
    S_\rho(q,h)\sim  \frac{h^{-\zeta/z}}{h^{-\zeta/z} q^\zeta +C} = \frac{1}{q^\zeta+C\, h^{\zeta/z}}\,.
\end{equation}
where $C$ is a phenomenological parameter. We can finally make us of the hyperscaling relation 
\begin{equation}\label{hyper}
z=d-1+\xi+2\chi=\zeta
\end{equation}
conjectured in \cite{toner1995long, toner1998flocks, toner2012reanalysis} and recently proven by the arguments of Ref.\cite{chate_solon}. As a consequence, Eq. \eqref{scaling5} simplifies to finally yield Eq. \eqref{div_S_h}.

\subsection{\label{simulations static method}Numerical scaling }
We tested our static method on two dimensional synthetic data generated from the Vicsek model \eqref{ev_r}-\eqref{micro_evo} with metric interactions  and $R_0=1$.\\
We consider systems in the stationary TT phase with periodic boundary conditions, with or without an external field. We compute the structure factor from Eq. \eqref{eq:sf2}, starting from real space density field, coarse-grained in boxes of size one. The resulting structure factor $S_\rho({\bf q})$ is further averaged over all the ${\bf q}$ orientations to obtain its isotropic average, $S_\rho(q)$ (see Eq. \eqref{eq:sf_iso}; the isotropic average is then binned over channels of width $2 \pi/L$). Invoking the ergodicity of the stochastic process, the average over different realizations may be replaced by time-averages. 

\begin{figure}[hbt!]
	\centering
	\includegraphics[width=0.75\textwidth]{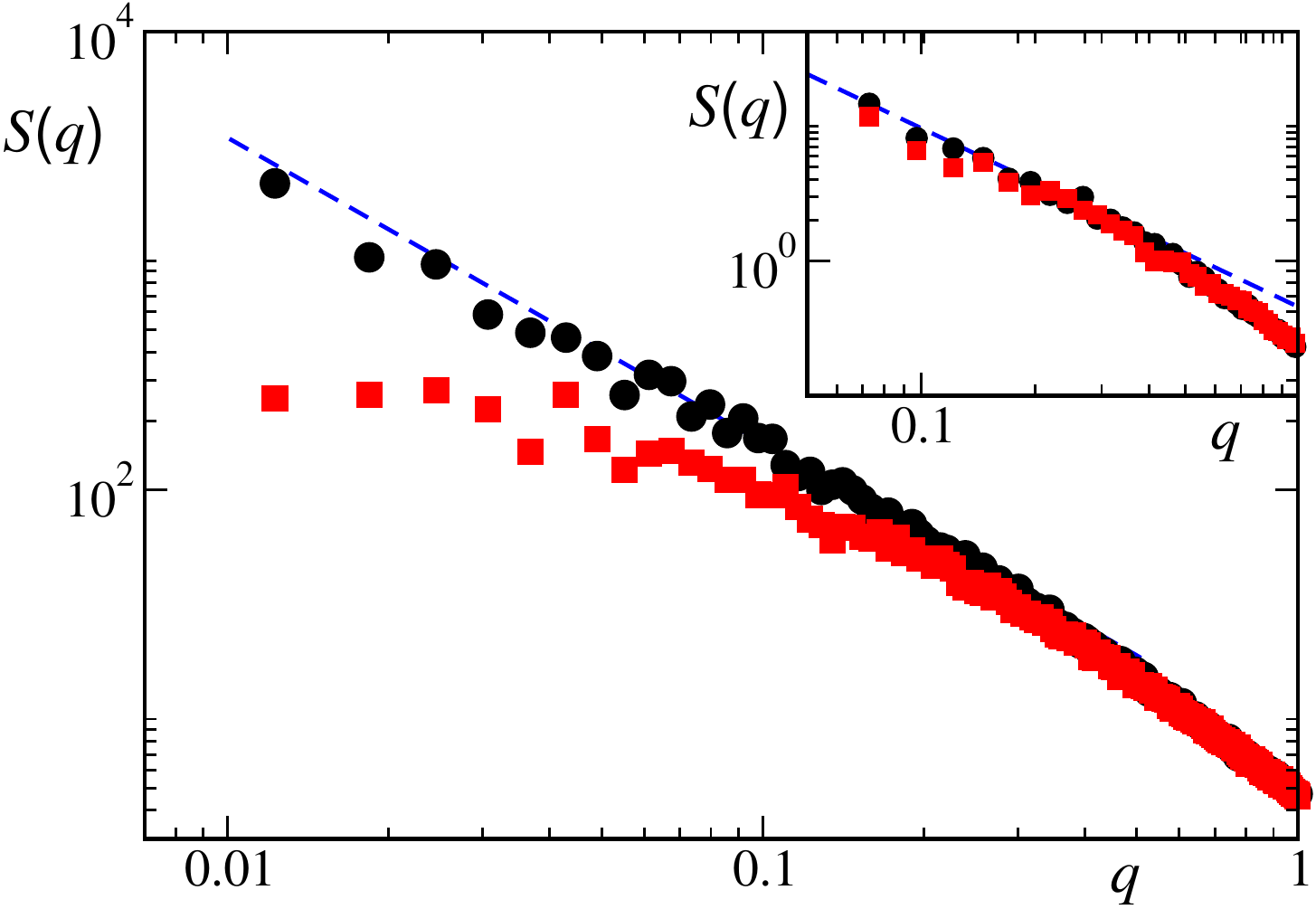}
	\caption{{\bf Isotropically averaged density structure factor $S_\rho(q,h)$ for the spontaneous and directed case} Black dots indicated the spontaneous and red squares the directed ($h=0.01$) isotropically averaged density structure factor $S_\rho(q,h)$ for systems of size $L=1024$.  Data has been obtained averaging the structure factor of ten different configurations sampled every 50 time-steps (see text). The dashed blue line marks the power law divergence $\sim q^{-\zeta}$, with $z=1.33$. Other parameters are $\rho = 2.0$, $v_0=0.5$, $\eta=0.18$. Inset: same parameters as in the main panel, but with system size $L=256$. All axes are in a log-log scale.}
	\label{2048_final_sq}
\end{figure}

We first consider an external field of magnitude $h=0.01$. In Section \ref{numsim}, we have seen that its presence may be inferred from the dynamic of the flock's mean direction only when observation times are larger than $\tau_c \approx 500$ (see Fig. \ref{diff_spont_vs_driv}). In order to test the static approach in a regime where the dynamic one fails, we have restricted the time averages of the structure factor over a time-window of $T=500$ time-steps. Due to temporal correlations between subsequent spatial configurations, this may yield a low statistics, especially for the lowest $q$ modes, so we have cut-off frequencies $q\leq \pi/L$. \\
In Fig.~\ref{2048_final_sq} we consider a relatively large ($L=1024$) system and compare results for the directed ($h=0.01$) collective dynamics with the ones for spontaneous collective motion ($h=0$). The latter case shows a behavior compatible with the free scaling $S_\rho(q) \sim q^{-\zeta}$ with $\zeta=1.33$ \cite{mahault2019quantitative}, while in the former case, the suppression of the low $q$ divergence is rather evident for $q<q_c$, where $q_c$ is a crossover wave-number.  This shows the viability of our static approach for large enough systems.

\begin{figure}[hbt!]
\centering
    \includegraphics[width=0.75\textwidth]{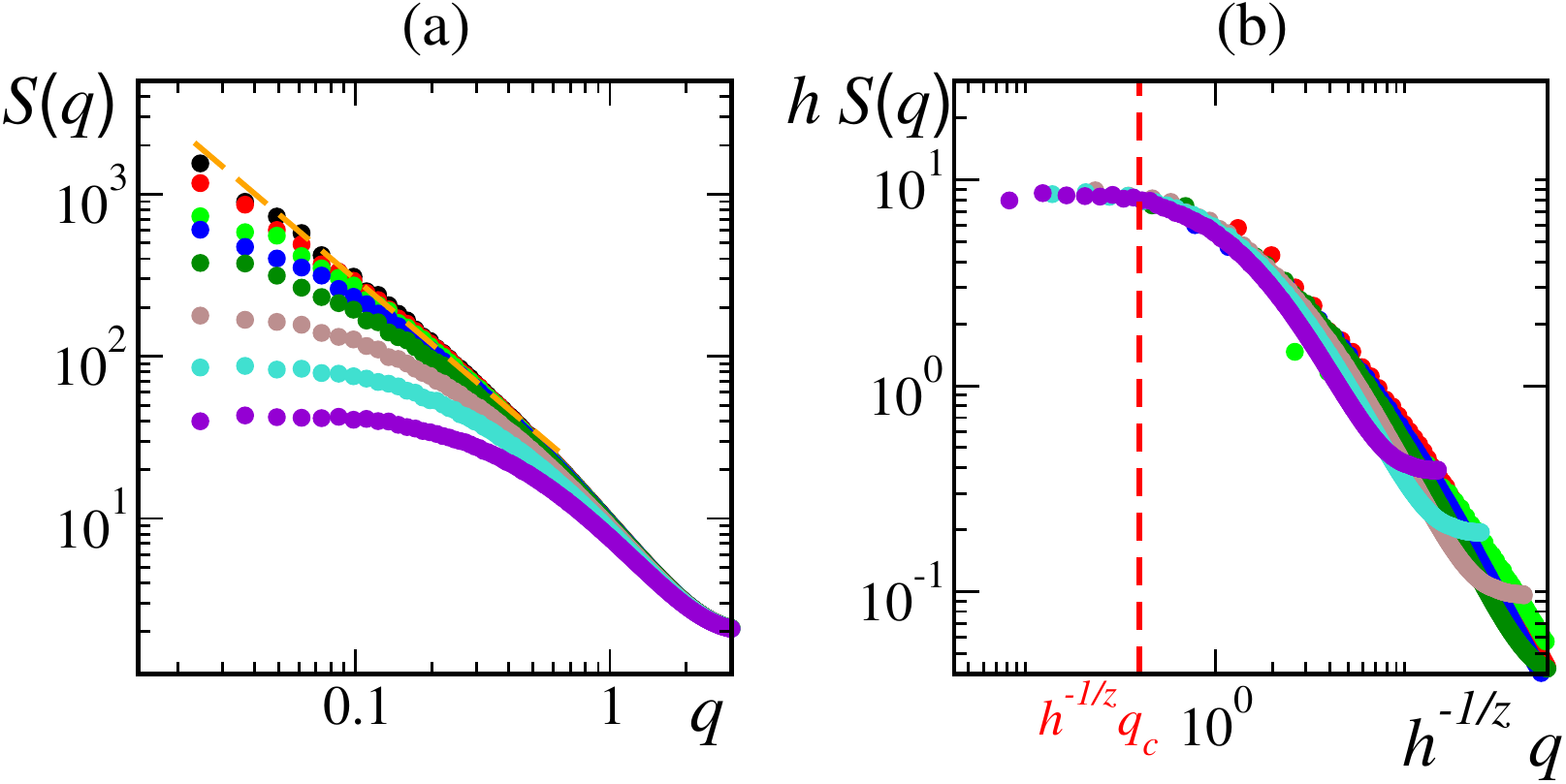}
    \caption{{\bf Rescaling of the isotropically averaged structure factor $S_\rho(q,h)$ for different values of $h$} (a) Isotropically averaged structure factor $S_\rho(q,h)$ for different external field amplitudes $h$. From top to bottom $h=0, 0.002, 0.005, 0.01, 0.02, 0.05, 0.1, 0.2$. The dashed The dashed orange line marks the power law divergence $\sim q^{-\zeta}$, with $z=1.33$. 
    (b) same as (a), but with the data rescaled by $h^{-1/z}$ (horizontal axis) and $h$ (vertical axis) in order to collapse the structure factor curves. The $h=0$ data has been omitted.
    The vertical dashed red line marks the location of the (rescaled) crossover wavenumber (see text).
    All axes are in a log-log scale, and model parameters are $\rho = 2.0$, $v_0=0.5$, $\eta=0.18$ and $L=512$.}
	\label{final_sq}
\end{figure}

In smaller systems ($L=256$, inset of Fig.~\ref{2048_final_sq}), on the other hand, directed collective motion cannot be detected by the observation of the low $q$ behavior since the divergence suppression becomes effective at wave-numbers not accessible due to the limited system size.\\
To better probe the structure factor scaling, we run longer time-averages for systems of size $L=512$ and different external field amplitudes $h\in [0, 0.2]$. They are shown in Fig.~\ref{final_sq}a. According to the scaling law \eqref{scaling3} and the hyperscaling relation \eqref{hyper}, the field rescaled structure factor $h S_\rho(q)$ should be only a function of $h^{-1/z} q$, and this is verified nicely in Fig.~\ref{final_sq}b where we have once again used used the $2d$ numerical estimate $z=1.33$. This also implies the crossover wave-number scaling
\begin{equation}\label{crossover_size1}
    q_c(h) \sim h^{1/z}\,.
\end{equation}
Equation \eqref{crossover_size1} immediately implies that observations need to be carried on in large enough systems, with a threshold linear system size
\begin{equation}\label{crossover_size2}
    L_c(h) \sim h^{-1/z}\,.
\end{equation}
Note that the crossover temporal and spatial scales are indeed related through the dynamical exponent,
\begin{equation}\label{crossover_size3}
    \tau_c \sim L_c^z \,.
\end{equation}

\subsubsection{Discussion} 
To summarize, we have discussed the behavior and scaling properties of the mean flocking direction and static density correlations in the presence of a small homogeneous external field. For large enough times, fluctuations in the mean direction depart from the diffusive behavior exhibited by finite flocks in the zero field case, following the statistics of an Ornstein-Uhlenbeck process. For the first time, we have also discussed explicitly the scaling of the diffusive process exhibited by the mean direction of motion of finite flocks in the spontaneous, zero-field case.\\
In large enough systems, a complementary signature of directed motion can be found in the small $q$ behavior of the static density structure factor, which saturates in the presence of an external field rather than showing the divergent behavior typical of spontaneous symmetry breaking. These facts can be used to detect directed collective motion, that is flocking behavior guided by external cues such as concentration gradients or other global anisotropies.

In Ref \cite{giomi}, it has been shown that the effect of an external static field only affecting a finite fraction of the flocking particles is equivalent to that of a rescaled (by the fraction of affected particles) homogeneous field. The equivalence holds provided the affected particles are randomly distributed in the flock. On the contrary, we do not expect our considerations to be directly applicable to localized or time-varying perturbations \cite{cavagna2013boundary}.

It is also fair to stress that while our method can detect the presence of external fields/environmental cues, it cannot exclude their presence: negative results may simply mean that the external field is so small that the accessible spatial or temporal observational scales are too small to detect it. On the other hand, strong external fields or anisotropies beyond the linear regime should result in a complete suppression of the static structure factor low $q$ divergence and of the diffusive, short time behavior of the mean flock direction. In this regime, however, our scaling relations are no longer valid.\\
We have focused on the explicit symmetry breaking of the continuous rotational symmetry by a {\it vectorial} external field, which we believe to be the most biologically relevant situation. However, at least in principle, one may also conceive spatial anisotropies inducing more complex discrete symmetries. \\
A prominent example is the AIM (seen in Section \ref{sec_aim}), where the discrete $\mathbb{Z}_2$ symmetry completely suppresses the structure factor low $q$ divergence, and strongly pins the mean flock direction either in the upward or downward direction, similar to what is expected in the presence of a strong external field.\\
Perhaps more intriguing, is the generalization to the so-called active clock model recently carried on in \cite{Solon2022}. Here, each particle orientation ${\bf n}_i$ can be in $Q$ different states equally distributed around the unit circle (the AIM being recovered for $Q=2$), and the hopping bias along the lattice is proportional to the projection of ${\bf n}$ in the hopping direction. The clock model is characterized by a crossover scale which grows {\it exponentially} with $Q$. Below such a crossover scale, the system shows a wandering order parameter and a diverging structure factor, while above it the order parameter is pinned to a discrete clock direction and the divergence is suppressed at low wave-numbers.\\
This analysis is clearly complementary to ours: we have developed scaling relations for continuos particle orientations slightly biased by a small vectorial field, while Ref. \cite{Solon2022} strongly constrains orientations along a discrete set of $Q$ directions. We derived scaling w.r.t. the field intensity, while Ref. \cite{Solon2022} studies scaling w.r.t. the number of strongly constrained directions.

In any case, its is fair to note that the approaches put forward cannot discriminate, from the analysis of a single given collective motion instance,  between collective motion in the presence of a small external field and collective motion that somehow follows the discrete symmetry of, say, a large $Q$ active clock model. Or to disentangle the role of the two effects, when both are present.\\
However, while our analysis cannot technically exclude the presence of more than one preferred direction of motion, we believe that this latter possibility should not be relevant in many situations of practical interest, and thus that our approach conveys useful information on the nature of collective motion. 
\newpage

\section{Boundary symmetry breaking}\label{Leo}
We now turn to the case of boundaries induced anisotropy. More specifically, we consider a $d$ dimensional flocking system confined in an infinite channel of width $L_\perp$ by reflecting boundaries conditions in $d-1$ {\it transversal} spatial directions. Such a confinement explicitly breaks rotational invariance at the boundaries, selecting a preferred direction along the free direction (the {\it longitudinal} one) and effectively forcing collective motion along the channel. In finite systems, one can assume periodic boundary conditions along the free directions, obtaining a ring geometry.\\
This setup has been employed in numerical investigations of flocking models \cite{TT3, Benoit2019} as a way to suppress the diffusion of the mean flock orientation in finite systems. The underlying assumption of these numerical studies is that fluctuations correlations measured in the bulk (that is, sufficiently far away from  the reflecting boundaries) are left unperturbed by the boundary symmetry breaking.

In this work we verify explicitly these assumptions. Our finite size scaling analysis and numerical simulations \cite{Lenzini_2024} show that confinement induces an effective mass term ${M_c} \sim L_\perp^{-\zeta}$ (with positive $\zeta$ being equal to the dynamical scaling exponent of the free Toner \& Tu theory) suppressing scale free correlations at small wave-numbers. However, due to the finite system size in the transversal direction, this effect can only be detected for large enough longitudinal system sizes (i.e. narrow ring geometries), effectively validating the assumption of unperturbed
bulk correlations. 
Furthermore, in the longitudinal direction, density correlations are characterized by an anomalous effective mass term. The effective mass term also enhances the global scalar order parameter and suppresses fluctuations of the mean flocking direction.
\subsection{Scaling of transversally confined flocks}\label{sec1}
We first discuss the effect of confinement on the correlation functions of the relevant slow hydrodynamic fields. In particular, we focus on the equal time Fourier space fluctuations correlations or structure factors. Here and in the following the subscripts $\parallel$ and $\perp$ denote, respectively, the longitudinal and transversal vector components w.r.t. the reflecting boundaries.\\
As argued in Sec.\ref{scaling_exp}, recent results show that the anisotropy exponent $\xi$ should be equal to $1$ and this implies, at least in $d=2$, no spatial anisotropy between the longitudinal and transversal directions in a boundary free system. In the following we will anyhow derive our result for the general case $\xi \leq 1$ and we will see how transversal confinement affects differently correlations in the longitudinal and transversal directions even when $\xi=1$.

In the presence of transversal confinement by reflecting or partially
reflecting walls, the average direction of collective motion aligns
along the channel. The scaling behavior of the structure factors can
then be deduced by a finite size scaling analysis and the request that
for a diverging confinement length $L_\perp \to \infty$ one should
recover the scaling results of bulk Toner \& Tu theory.

Taking into account the dependence on $L_\perp$, the structure factors
$S_\rho$ and $S_v$, here collectively denoted
as $S$, obey the following finite size DRG scaling law
\begin{equation}
S(q_\perp, q_\parallel, {\bm \mu}^{(1)},
L_\perp^{-1})=b^\zeta S(b q_\perp, b^\xi q_\parallel,
{\bm \mu}^{(b)}, bL_\perp^{-1})
\end{equation}
where we recall that the scaling of the structure factors has been determined
considering that they are given by the Fourier transform of the equal
time, real space density correlation function. Thus, it involves two
powers of the density fluctuations and one volume element, leading to
the scaling exponent $d-1+\xi+2\chi = \zeta$. \\
In order to extract the final size scaling, we choose $bL_\perp^{-1}=1$ which implies $b=L_\perp$. For a
sufficiently large separation, $L_\perp\gg1$, we have ${\bm
  \mu}^{(b)}\simeq {\bm \mu}^*$ and we obtain
\begin{equation}
S(q_\perp,q_\parallel,{\bm \mu}^{(1)},
L_\perp^{-1})=L_\perp^{\zeta}S(L_\perp q_\perp,L_\perp^\xi
q_\parallel,{\bm \mu}^*,1)
\end{equation}
where $S(x,y,{\bm \mu}^{*},1) \equiv g(x,y)$ is a universal scaling function.

We analyze two different regimes, depending on whether the behavior of $S$ is dominated by the longitudinal or transverse wave numbers. \\
When $q_\perp^\xi\gg q_\parallel$, in the long wavelength limit we
consider the one parameter universal scaling function $w_\perp(x)
\equiv g(x,0) = S(x,0,{\bm \mu}^{*},1)$ which yields the scaling
\begin{equation}
S({\bf q},L_\perp^{-1})=L_\perp^{\zeta}w_\perp(L_\perp q_\perp)
\end{equation}
The behavior of the universal scaling function $w_\perp$ can be
inferred by the request that, for $L_\perp \to \infty$ , the structure factor
scaling coincides with the one of Eq. (\ref{nlnSperp}) and that a tight confinement $L_\perp \to 0$ suppresses the scale free
divergence at small wave-numbers preserving a finite variance, 
\begin{equation}
w_\perp(x)\sim
\begin{cases}
x^{-\zeta} \,\, & x\gg1 \\
\text{constant} \,\,\, & x\ll1
\end{cases}
\end{equation}
The simplest expression for the
scaling function is thus
\begin{equation}
w_\perp(L_\perp q_\perp)\sim\dfrac{1}{(L_\perp q_\perp)^\zeta+G_\perp},
\end{equation}
where $G_\perp$ is a phenomenological parameter depending (among other
things) on microscopic boundary conditions.\\
Then the structure factors $S_\rho$ and $S_v$ take the form
\begin{eqnarray}
\label{Sperp}
S_\rho({\bf q},L_\perp^{-1})&\sim&\dfrac{L_\perp^\zeta}{(L_\perp
                              q_\perp)^\zeta+G^{(\rho)}_\perp}=\dfrac{1}{q_\perp^\zeta+G^{(\rho)}_\perp
                              L_\perp^{-\zeta}}\nonumber\\\\
S_v({\bf q},L_\perp^{-1})&\sim&\dfrac{L_\perp^\zeta}{(L_\perp q_\perp)^\zeta+G^{(v)}_\perp}=\dfrac{1}{q_\perp^\zeta+G^{(v)}_\perp L_\perp^{-\zeta}}\nonumber
\end{eqnarray}
for $q_\perp^\xi\gg q_\parallel$ and
with possibly different parameters $G^{(\rho)}_\perp $ and $G^{(v)}_\perp$.

An analogous derivation for the transversal velocity correlation in the
longitudinal case, $q_\perp^\xi\ll q_\parallel$, introduces
the longitudinal scaling function  $w_\parallel(y)
\equiv g(0,y) = S_v(0,y,{\bm \mu}^{*},1)$, whose behavior is also
determined by matching with Eq.  (\ref{nlnSpar}) for $L_\perp \to \infty$,
\begin{equation}
w_\parallel(L_\perp q_\parallel)\sim\dfrac{1}{(L_\perp^\xi q_\parallel)^{\zeta/\xi}+G_\parallel},
\end{equation}
with the phenomenological parameter $G_\parallel$ also depending on
boundary conditions. It follows that when $ q_\parallel \gg q_\perp^\xi$ the structure factor for the transversal velocity takes the form
\begin{equation}
\label{Spar}
S_v({\bf q},L_\perp^{-1})\sim\dfrac{L_\perp^\zeta}{(L_\perp^\xi q_\parallel)^{\zeta/\xi}+G_\parallel}=\dfrac{1}{q_\parallel^{\zeta/\xi}+G_\parallel L_\perp^{-\zeta}}
\end{equation}

Eqs. (\ref{Sperp}) and (\ref{Spar}) express the scaling of the density and
transversal velocity static correlations in Fourier space for
transversally confined flocking. They show how confinement induces an
effective mass term $M_c \sim L_\perp^{-\zeta}$ nominally suppressing the
bulk fluctuations divergence at small wave-numbers.

Transversal confinement, however, implies that $q_\perp$ is a positive
integer multiple of $\pi/L_\perp$. The mass corrections is only
relevant for transversal wave-numbers such that $(L_\perp q_\perp)^\zeta \lesssim
G_\perp$, which implies 
\begin{equation}
G_\perp \gtrsim \pi^\zeta\,.
\label{Cperp}
\end{equation}
We conclude that for $G_\perp$ of order one or smaller the suppression
of the divergence may well not be detectable along the confined
directions. 

This is not the case in the longitudinal direction. For an infinite system, $q_\parallel$ is unbounded from zero and a plateau in the
transversal velocity structure factor should appear for $q_\parallel \lesssim
G_\parallel^{\xi/\zeta} L_\perp^{-\xi}$. In numerical or experimental
systems one typically deals with periodic boundary
conditions in the longitudinal direction (i.e. a ring geometry) and a
longitudinal size $L_\parallel$ with a corresponding smallest longitudinal
wavenumber  $\pi/L_\parallel$. Boundary effects thus generate a
detectable suppression of the small wavelength divergence for
\begin{equation}
G_\parallel \gtrsim
\left(\frac{L_\perp^\xi}{L_\parallel}\pi\right)^{\zeta/\xi}\,.
\label{Cpar}
\end{equation}

As we will discuss in Sec. \ref{2C}, numerical simulations suggest a poorly understood anomalous behavior of density correlations in the longitudinal direction. The lack of a clear analytical understanding of the free theory behavior in this case prevents us from formulating a precise scaling form for confined systems, nevertheless we still expect that explicit symmetry breaking will result in 
\begin{equation}
S_\rho(0, q_\parallel, L_\perp^{-1}) \xrightarrow{q_\parallel \to 0}
\frac{1}{\Sigma(L_\perp)}
\end{equation} 
with $\Sigma(L_\perp)$ an anomalous mass term scaling with a negative power of $L_\perp$.

We have derived the scaling of velocity correlations in the transversal and longitudinal
directions, even if, as previously discussed, recent results predict no
scaling anisotropy ($\xi=1$) in Vicsek flocks. However, this scaling
isotropy does not imply that also the prefactors have to be equal,
leading to the general scaling form
\begin{equation}\label{Siso}
S_v({\bf q}, L_\perp^{-1})\sim\dfrac{1}{q^{\zeta}+G(\theta_{\bf q}) L_\perp^{-\zeta}}
\end{equation}
where $G(\theta_{\bf q})$ is a phenomenological parameter with $G(0)=G_\parallel$ and $G(\pi/2)=G^{(v)}_\perp$.

Note finally that these results have been derived without any explicit
modelling of the interaction between the active particles and the
confining walls. The only requirement is that the rotational
invariance of the self-propulsion orientation is explicitly broken at
the boundaries. This clearly applies to reflecting or partially
reflecting boundaries, or more generally to any systems in which the
proximity with the wall induces -- directly or indirectly - torques on
the self-propulsion orientation, avoiding trapping of active particles at the wall.

\subsubsection{Numerical evidence}\label{2C}
\begin{figure}[hbt!]
    \centering
    \includegraphics[width=0.5\linewidth]{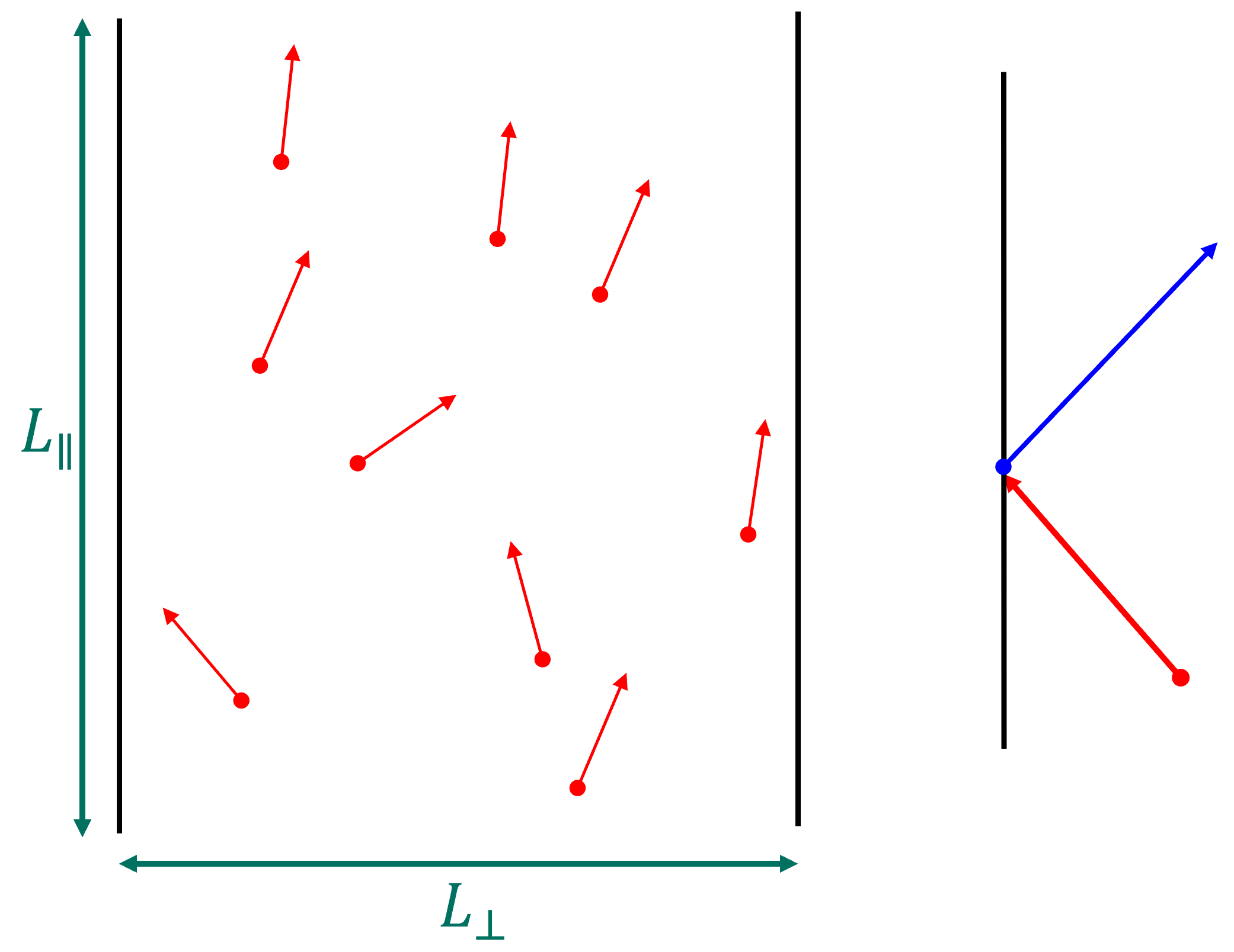}
    \caption{{\bf Schematic representation of the system and its boundary conditions}: Left - The system is confined between two parallel plates at distance $L_\perp$ (where reflecting boundary conditions are assumed), while periodic boundary conditions are assumed in the longitudinal direction (that of flocking) of large size $L_\parallel$. Right - Schematic representation of the reflecting boundary conditions. The incoming particle (in red) gets reflected by the following collision rule $n_\perp \to -n_\perp$ and the position $r_\perp \to 2B - r_\perp$, where $B = 0, R$. }
    \label{bc_confined_leo}
\end{figure}

These predictions can be verified considering the Vicsek model (VM) \eqref{vicsek_theta}-\eqref{position_vicsek}, transversally confined by reflecting boundary conditions. We consider periodic boundary conditions in the longitudinal direction of size $L_\parallel$ and implement the transversal reflecting boundaries by the following collision rule
\begin{equation}
n_\perp \!\!\to\! - n_\perp\;\;\;,\;\;\; r_\perp
\!\!\to\! 2 B \!- r_\perp
\end{equation} 
with either $B\!=\!0$ (left boundary) or $B\!=\!L_\perp$ (right boundary), which is applied whenever the Vicsek dynamics would result in a particle position with transversal component $r_\perp$, outside the region $r_\perp\!\in\! [0,L_\perp]$, see Fig.\ref{bc_confined_leo}.

In the following, we place ourselves in the polar liquid phase of the
non-confined model \cite{Chate2, VicsekPD} by fixing $v_0=0.5$, $\eta=0.2$ and
$\rho_0=N/(L_\parallel L_\perp)=2$ and we consider more than two
orders of magnitude range of values for the transverse separation
distance, from $L_\perp=32$ to  $L_\perp=8192$, keeping the
longitudinal size constant, $L_\parallel=2048$. For numerical
convenience, we mainly consider uniform
initial positions with alignment parallel to the confining walls, but
we have also verified that different initial conditions always lead, after a
transient, to a polar state with the mean flocking direction in the longitudinal direction.

In order to measure the density and transversal velocities structure
factors, we first obtain fluctuating fields for the density
$\delta\rho({\bf r})$ and the
perpendicular velocity $v_\perp ({\bf r})$ by coarse-graining the microscopic
particles number and transversal velocities over boxes of unit linear
length. In order to avoid densities inhomogeneities near the confining
walls \cite{fava2024casimir} we only evaluate the fields in a central channel
of extension $0.7 L_\perp$, thus excluding two regions of size $0.15 L_\perp$
adjacent to the walls. The resulting Fourier space static correlations 
\begin{equation}
S_\rho({\bf q}) \equiv \langle |\delta \hat{\rho} ({\bf
  q})|^2\rangle\;\;\;,\;\;\; S_v({\bf q}) \equiv \langle
| \hat{v}_\perp ({\bf q})|^2\rangle
\end{equation}
are also averaged in time (after a proper transient has been
discarded) over typically $10^6$ timesteps. 

\begin{figure}[hbt!]
\centering
\includegraphics[width=0.8\textwidth]{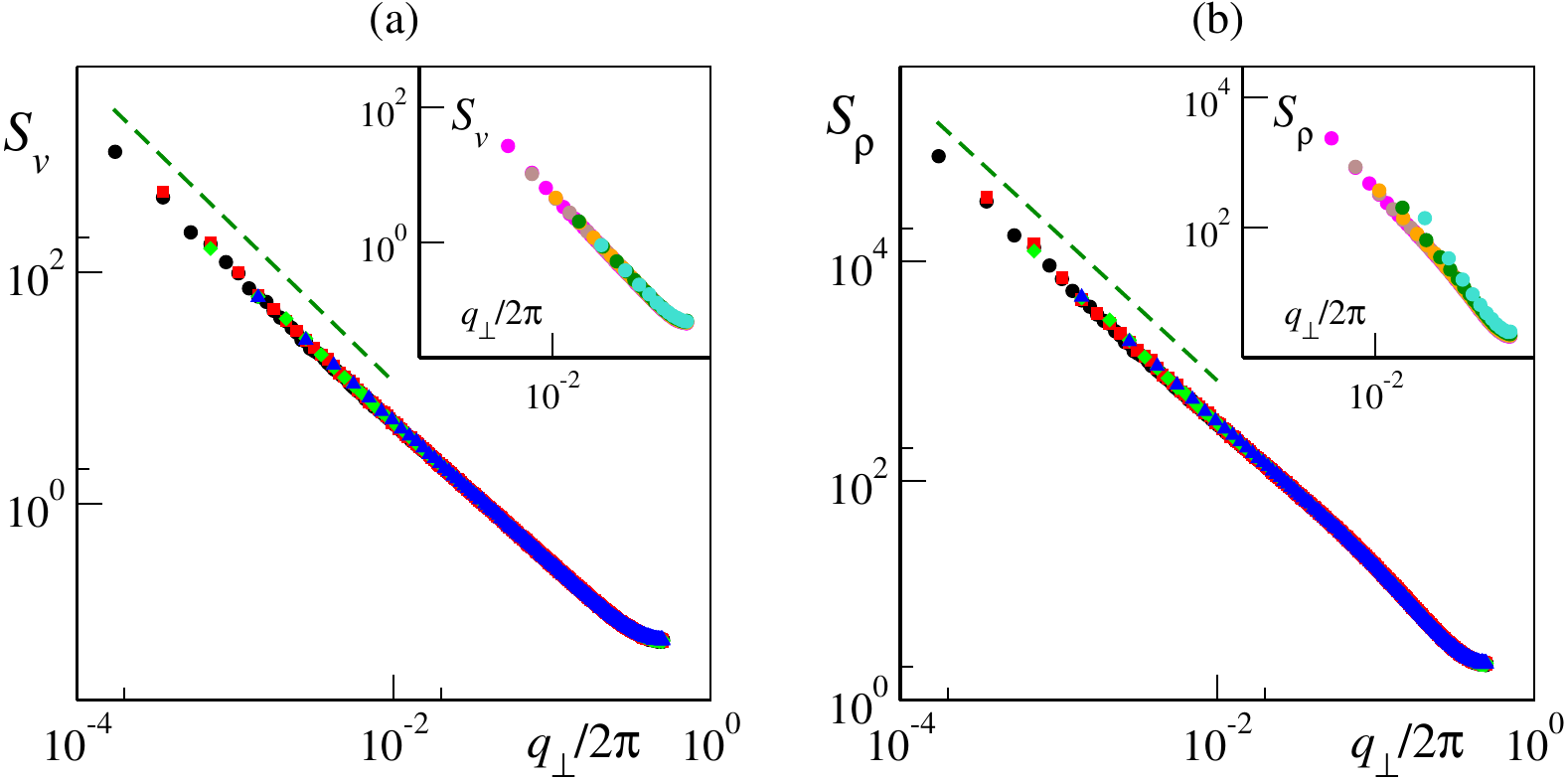}
\caption{{\bf Equal time Fourier correlations in the transverse
  direction.} (a) Transversal velocity correlations and (b) Density
  fluctuation correlations. Transversal system sizes are
  $L_\perp=8192$ (black circles), $L_\perp=4096$ (red squares),
  $L_\perp=2048$ (green diamonds), $L_\perp=1024$ (blue
  traingles). The dashed green line marks a power law
  divergence with exponent $\zeta=1+\xi+2\chi=4/3$ \cite{Solon,
    Benoit2019}. Correlations for smaller transversal separations,
$L_\perp=32, 64, 128,256,512$, are shown in the two insets.}
\label{Fig1}
\end{figure}

We first fix $q_\parallel=0$ in order to probe correlations in the transversal direction. Their behavior, reported in Fig.~\ref{Fig1}, does not show any sign of suppression of the small $q_\perp$ divergence which, for sufficiently large separations $L_\perp$, shows an
excellent agreement with the predicted $d=2$ bulk exponent $\zeta=1+\xi+2\chi=4/3$ \cite{Ikeda, Solon}. This implies that, at least for typical Vicsek dynamics transversally confined by reflecting walls, condition (\ref{Cperp}) on the phenomenological parameters
$G_\perp^{(\rho)}$ and $G_\perp^{(v)}$ is not met.

\begin{figure}[hbt!]
\centering
\includegraphics[width=0.8\textwidth]{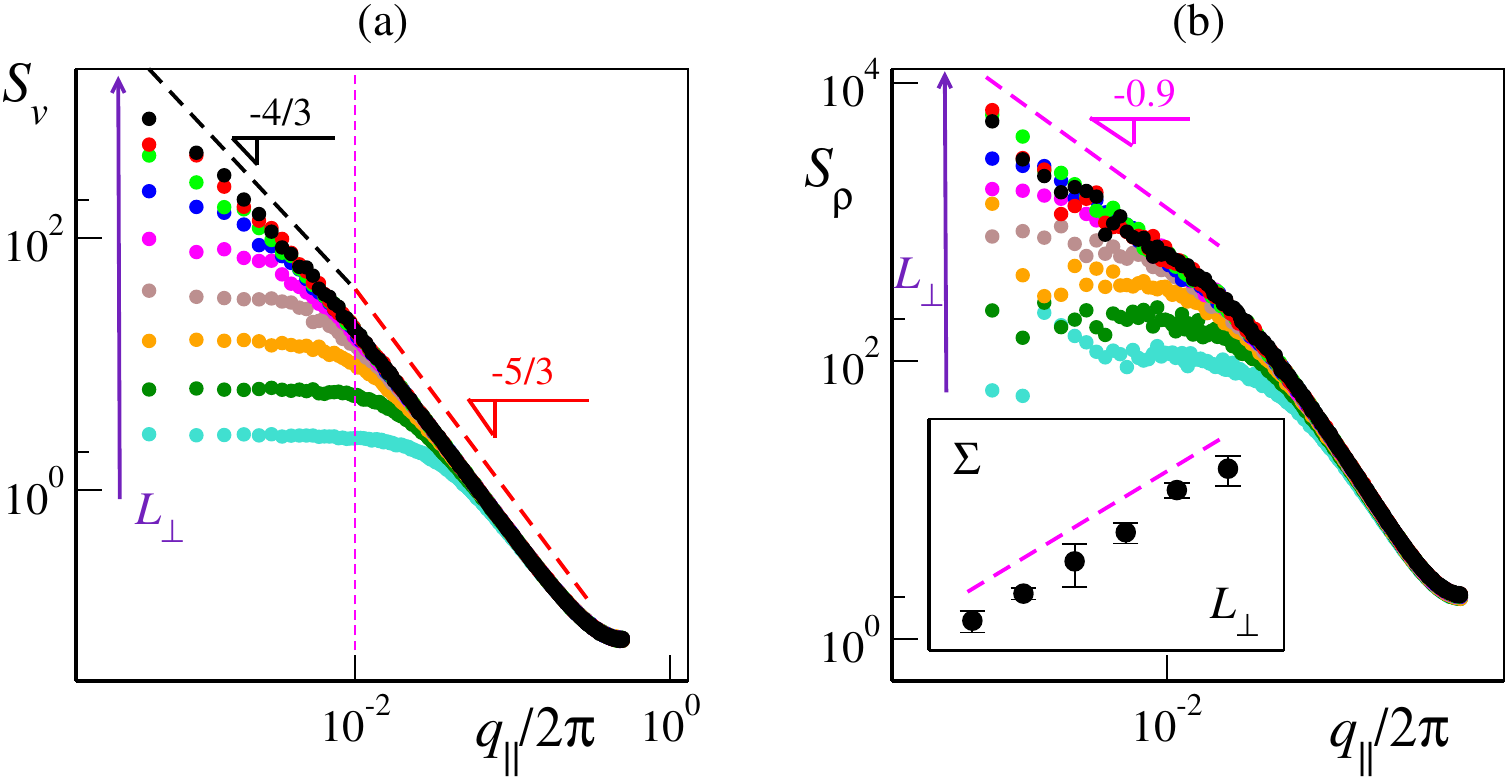}
\caption{{\bf Equal time Fourier correlations in the longitudinal
  direction.}  (a) Transversal velocity correlations. The vertical dashed line marks the crossover scale
  $q_c/(2\pi) = 10^{-2}$ separating the asymptotic regime
  (dashed black line with scaling exponent $4/3$) from the finite size 
  one  (dashed red line with scaling exponent $\approx 5/3$) (see text).
(b) Density
  fluctuation correlations. In both panels transversal separations are, from bottom
  to top, $L_\perp = 32, 64, 128, 256, 512, 1024, 2048, 4096,
  8192$. The dashed magenta line marks an
  algebraic divergence with an exponent $0.9$.
Inset: Scaling of the effective mass term $\Sigma$ as a
  function of transversal separation. The dashed magenta line marks an
  algebraic growth with an exponent $0.9$.}
\label{Fig2}
\end{figure}

The behavior of velocity correlations in the longitudinal direction (see Fig.~\ref{Fig2}a), obtained by setting $q_\perp=0$, is different, with a clear suppression of the low $q_\parallel$ divergence for small transversal separations $L_\perp\lesssim L_\parallel$, where condition (\ref{Cpar}) is met. Only when $L_\perp\gtrsim L_\parallel$ one cannot clearly identify a plateau for small $q_\parallel$, and the structure factor approaches the free theory
algebraic divergence (\ref{Spar}).

We have also tested the density structure factor in the longitudinal direction fixing ($q_\perp=0$), results shown in Fig.~\ref{Fig2}b. Also in this case, a plateau for $q_\parallel\ll 1$ is evident for $L_\perp \lesssim L_\parallel$. At larger longitudinal wave-number, an anomalous slow scaling behavior $\sim q_\perp^{-\gamma}$, first reported in \cite{Benoit2019}, becomes apparent\footnote{In Ref. \cite{Benoit2019} it was also
  reported an intermediate scaling regime, in qualitative agreement with Eq. (\ref{nlnSpar2}) which, however, should be unobservable for $q_\perp \to 0$.}, although our numerical estimates suggest a slightly larger exponent, $\gamma \approx 0.9$. Applying the scaling arguments developed in Sec.~\ref{sec1} to this empirical scaling behavior immediately gives the scaling
\begin{equation}
\label{Spar2}
S_\rho(0,q_\parallel,L_\perp^{-1})\sim\dfrac{1}{q_\parallel^{\gamma}+\Sigma(L_\perp)}
\end{equation}
with the anomalous effective mass term scaling
\begin{equation}
\Sigma(L_\perp) \sim L_\perp^{-\gamma}\,.
\label{anomalous}
\end{equation}
When $L_\perp \lesssim L_\parallel$ is possible to measure $\Sigma$
by evaluating $S_\rho(0,q_\parallel, L_\perp^{-1})$ in the
limit $q_\parallel \to 0$. Our results, shown in the inset of
Fig.~\ref{Fig2}b confirm the scaling (\ref{anomalous}) with $\gamma
\approx 0.9$.

The scaling behavior predicted by
Eq. (\ref{Spar}) also implies that the velocity structure factor $S_v$
measured for different
transversal separation $L_\perp$ should collapse to a universal curve $f(x)$
when properly rescaled,
\begin{equation}
L_\perp^{-\zeta} S_v(0, L_\perp^{\xi} q_\parallel)\equiv f(x)\,.
\label{collapse}
\end{equation}
When testing this
result by data-collapse, however, one should be aware that the most accurate
numerical simulations of longitudinal structure factors (see
Ref. \cite{Benoit2019}) revealed large finite size effects. In
particular in $d=2$
the anisotropy exponent $\xi$ shows a crossover behavior from a
finite-size (for $q>q_c$) value $\xi \approx 0.8$ to the asymptotic
(for $q<q_c$) one $\xi =1$ as also shown in Fig.~\ref{Fig2}a. For Vicsek dynamics with scalar noise it was
found $q_c/(2 \pi) \approx 10^{-2}$, as marked in Fig.~\ref{Fig2} by the
vertical dashed lines. Transversal correlations, on the other hand, do
not show such crossover and one can confidently use $\zeta=4/3$ over a
wider range of scales. 

\begin{figure}[hbt!]
\centering
\includegraphics[width=0.8\textwidth]{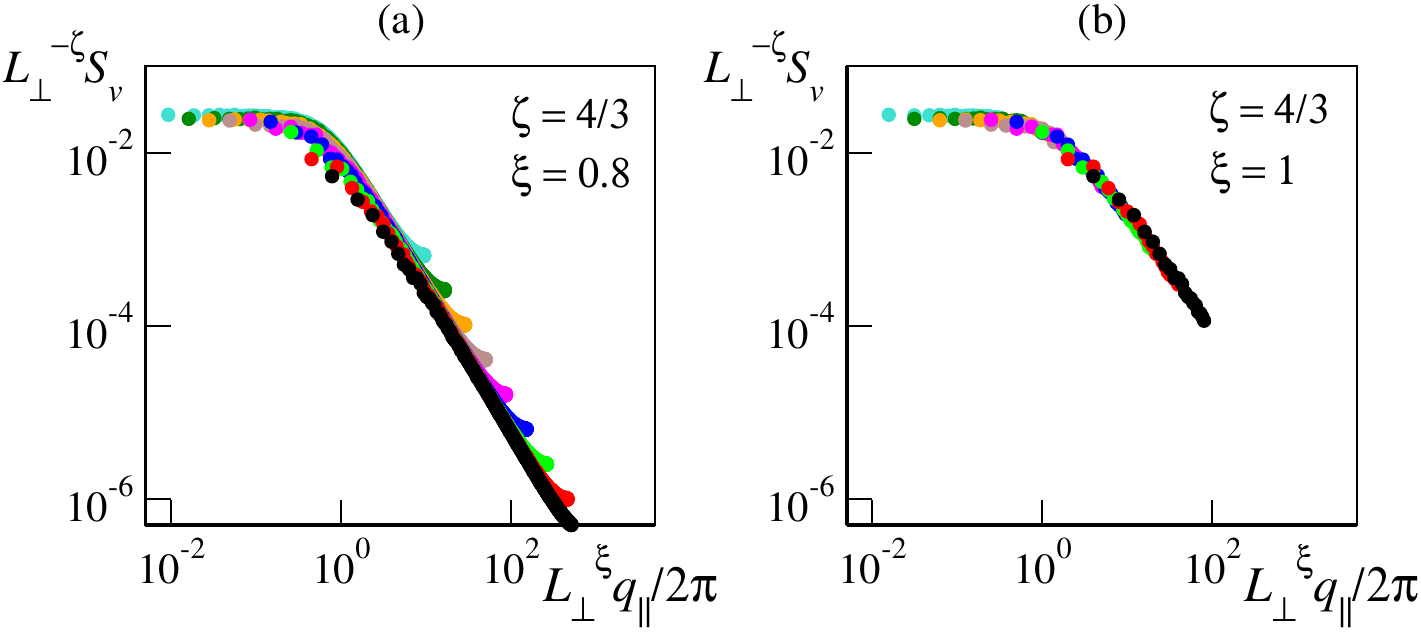}
\caption{{\bf Velocity structure factor data collapses in the
    longitudinal direction.}
  (a) Structure factor data from Fig.~\ref{Fig2} rescaled according to
  Eq. (\ref{collapse}) with scaling exponents $\zeta=4/3$ and
  $\xi=0.8$. (b) Same data, but only for $q_\parallel < q_c$ (see
  text) rescaled with scaling exponents $\zeta=4/3$ and
  $\xi=1$. Color coding for different $L_\perp$ as in Fig. ~\ref{Fig2}a.}
\label{Fig3}
\end{figure}

In Fig.~\ref{Fig3}a we first attempt to rescale the data
of Fig.~\ref{Fig2}a by using $\zeta=4/3$ and the pre-crossover scaling exponent
$\xi=0.8$. While one may observe a reasonable
collapse at large enough $q_\parallel$ values, a
closer inspection reveals a less satisfactory collapse at smaller
wave-numbers. The small $q_\parallel$ collapse may be improved
rescaling the data with the post crossover value $\xi=1$, as shown in
Fig.~\ref{Fig3}b where only the data for $q_\parallel < q_c$ is considered.
 
To summarise, numerical simulations of the VM in the confined
polar liquid phase confirm that bulk
correlations in the transversal direction are not affected by the reflecting
boundaries. In the longitudinal direction, our numerics support the scaling form (\ref{Spar}) for the
transversal velocity structure factor and suggest the empirical
anomalous effective mass term (\ref{anomalous}) for density correlations.
\newpage

\subsection{Order parameter behavior}\label{sec2}
We now turn our attention to the behavior of the order parameter in
the presence of transversal confinement. \\
Microscopically, the instantaneous global order parameter is defined as
\begin{equation}
\Omega(t)=\frac{1}{N} \sum_{i=1}^N {\bf n}_i^t \,.
\label{SOP}
\end{equation}
The scalar order parameter is given by $\Phi = \langle | \Omega(t) |
\rangle_t$, where $\langle \cdot \rangle$ denotes temporal
averages. Here we are interested in the static longitudinal response, that is, the difference in the scalar order parameter between the confined and the bulk system,
\begin{equation}
\delta \Phi(L_\perp) = \Phi(L_\perp)-\Phi(\infty)
\label{response}
\end{equation}
At the hydrodynamic level, the scalar order parameter can be obtained
by the spatial and temporal average of Eq. (\ref{enslave}). Since
linear terms in $\delta \rho$ and ${\bf v}_\perp$ all vanish under
such averages, one is left with 
\begin{equation}
\Phi = p_0 + \langle \delta v_\parallel \rangle \approx p_0 -
\frac{\langle|{\bf v}_\perp|^2\rangle}{2 p_0}\,,
\label{modulo}
\end{equation}
the analogous of the principle of conservation of the modulus that links longitudinal and transversal fluctuations in equilibrium ferromagnets \cite{Pata}.\\
From Eq. (\ref{response}) it follows that the order parameter response is given by 
\begin{eqnarray}
\delta \Phi(L_\perp) &=& \frac{1}{2 p_0}\left[ \langle |{\bf v}_\perp (\infty)|^2
  \rangle  -\langle  |{\bf
  v}_\perp (L_\perp)|^2\rangle \right] \nonumber\\
& =& \frac{1 }{2 p_0} C_v (0, L_\perp^{-1},{\bm \mu}^{(1)}) 
\end{eqnarray}
where we have recognized the real space correlation function
\begin{equation}
C_v (0, L_\perp^{-1},{\bm \mu}^{(1)}) = \langle|{\bf v}_\perp (\infty)|^2
\rangle - \langle |{\bf
  v}_\perp (L_\perp)|^2\rangle 
\end{equation}
and assumed $L_\parallel \to \infty$.\\
We can deduce the scaling behavior of the response  $\delta \Phi$ by repeating in real space essentially the same finite size scaling analysis performed in section \ref{sec1}. Performing a DRG step with a rescaling factor $b$ one obtains
\begin{equation}
C_v (0, L_\perp^{-1},{\bm \mu}^{(1)}) = b^{2\chi} C_v (0, b L_\perp^{-1},{\bm \mu}^{(b)}) 
\end{equation}
where the scaling of the correlation function is determined by the
fact that it just involves a squared field. Choosing as before $b=L_\perp$ we finally obtain, for a sufficiently large spatial separation $L_\perp \gg 1$,
\begin{equation}
C_v (0, L_\perp^{-1},{\bm \mu}^{(1)}) = L_\perp^{2\chi} C_v (0, 1,{\bm \mu}^{*}) 
\label{rFP}
\end{equation}
with the r.h.s. parameters now evaluated at the nonlinear fixed point.\\
Eq.(\ref{rFP}) immediately implies the response scaling
\begin{equation}
\delta \Phi(L_\perp) \sim L_\perp^{2\chi} \,.
\label{OPscale}
\end{equation}

\begin{figure}[hbt!]
\centering
\includegraphics[width=0.85\textwidth]{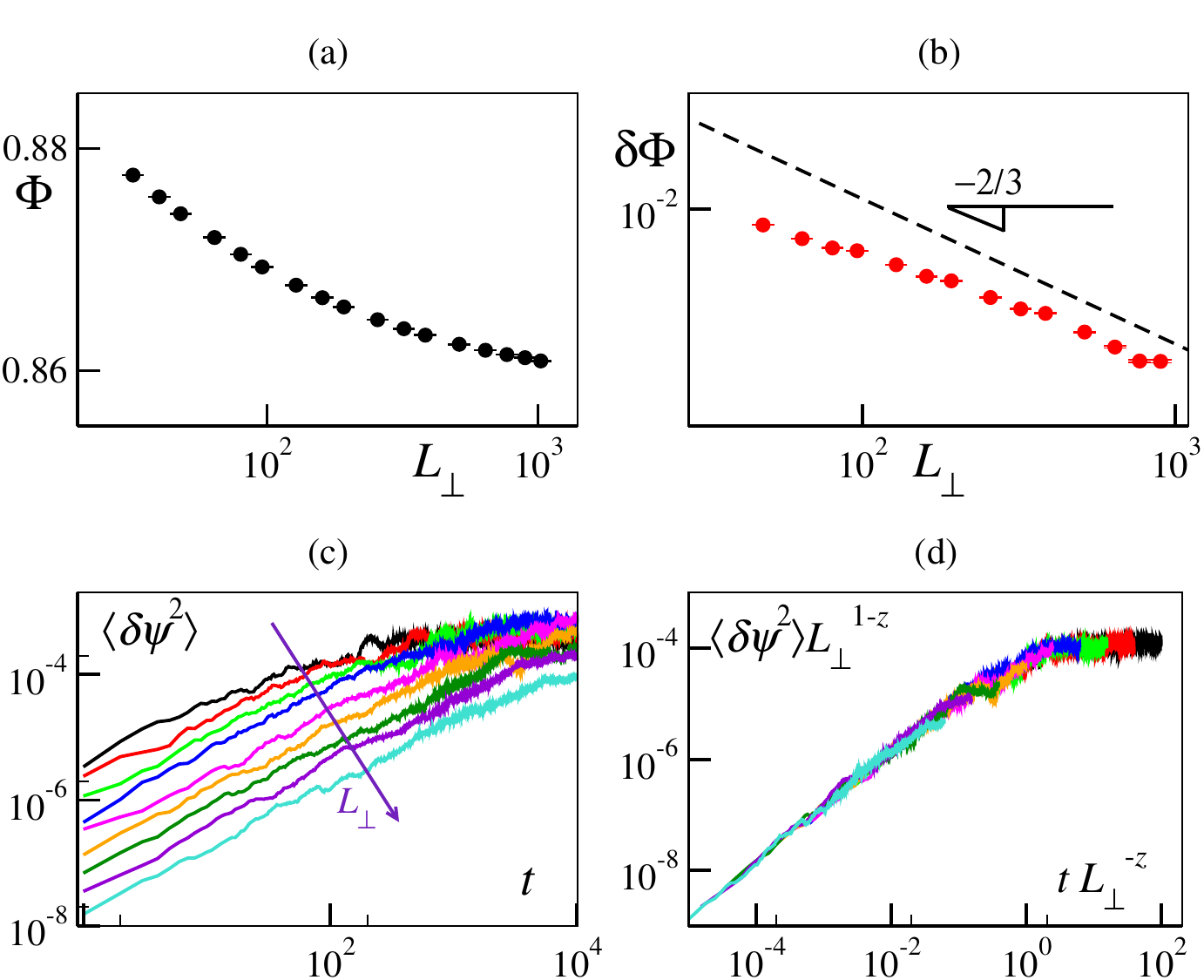} 
\caption{{\bf Order parameter scaling.} (a) Scalar order parameter
  $\Phi$ as a function of the transversal separation size. 
(b) Order parameter static response 
$\delta \Phi$, as a function of the transversal separation size,
evaluated from panel (a) by Eq. (\ref{centered}).
System parameters are $L_\parallel=512$,
$\rho_0=2$, $v_0=0.5$ and $\eta=0.21$.
Data has ben averaged over around $10^5$ {\it independent} datapoints.
Error bars mark two standard errors. 
The dashed line marks the expected asymptotic power law decay with an exponent
$2\chi=-2/3$.
(c) Mean squared fluctuations in the flocking direction for
increasing transversal separation sizes, from top to bottom
$L_\perp=32,64,128,256,512,1024,2048,4096, 8192$. (d) Data collapse of the
data in panel (c) according to Eq. (\ref{eq:01}) with dynamical
scaling exponent $z=4/3$. 
Data in panels (c)-(d) has been averaged over $10^2$
different realizations with parameters $L_\parallel=2048$,
$\rho_0=2$, $v_0=0.5$ and $\eta=0.2$.
}
\label{Fig4}
\end{figure}

This asymptotic scaling can be verified by microscopic simulations of the Vicsek dynamics. Fig.~\ref{Fig4}a shows the convergence of the scalar order parameter to its asymptotic value fixed by $p_0$.\\
In order to estimate the convergence to this asymptotic value we consider the centered finite difference $\partial_c \Phi(L_\perp)$ of the average order parameter, which approximates the first derivative of $\Phi(L_\perp)$ up to corrections of third order in the derivatives. We can then estimate the response as
\begin{equation}
\delta \Phi \sim L_\perp \partial_c \Phi(L_\perp)\,.
\label{centered}
\end{equation}

From Fig.~\ref{Fig4}b we can see that its asymptotic behavior well
matches our prediction, while a crossover is observed for smaller transversal
separation sizes. This should not come as a surprise: in  $d=2$ the anisotropy exponent $\xi$ shows a finite size crossover from $\approx 0.8$ to its asymptotic value $\xi=1$ \cite{mahault2019quantitative}. Since $\zeta=d-1+\xi+2\chi=4/3$ remains constant over a wider range of scales, this implies that also the roughness exponent should cross-over from the finite size value $\chi \approx -0.2$ to its asymptotic value $\chi = -1/3$. 

\subsection{Equivalence between homogeneously driven and confined systems}
It is instructive to compare these results with the behavior of
flocking systems in the presence of a homogeneous external field ${\bf
  h}$ of
amplitude $h=|{\bf h}|$ driving each active particle orientation. \\
Ref. \cite{Nikos} it was shown that in the linear regime (valid for
small field amplitudes) and large system sizes, the response scales as 
\begin{equation}
\delta \Phi(h) \equiv \Phi(h) - \Phi(0) \sim h^{-2\chi/z}\,.
\label{Phih}
\end{equation}
Similarly, it is known from Ref. \cite{BFG} that an external homogeneous field breaks
explicitely the rotational symmetry and thus induces an
effective mass term $\sim h^{\zeta/z}$ or, thanks to the hyperscaling
relation $\zeta=z$, see Sec.\ref{scaling_exp}, more simply $\sim h$. The
structure factors thus read
\begin{equation}
\label{Hperp}
S({\bf q},h)\sim\dfrac{1}{q_\perp^{\zeta}+D_\perp h^{\zeta/z}}\,,
\end{equation}
in the direction transversal w.r.t. to the external field and
\begin{equation}
\label{Hpar}
S({\bf q},h)\sim\dfrac{1}{q_\parallel^{\zeta/\xi}+D_\parallel h^{\zeta/z}}
\end{equation}
in the direction parallel to ${\bf h}$ (these two scalings
coincide for $\xi=1$).

Comparison between the response scalings (\ref{OPscale})-(\ref{Phih})
and the structure factors (\ref{Sperp})-(\ref{Hperp}) and
(\ref{Spar})-(\ref{Hpar}) immediately suggests the equivalence between
transversal confinement by reflecting boundaries, a local form of explicit
symmetry breaking, and the presence of a homogeneous external driving field
(which explicitly breaks the rotational symmetry at bulk level) of
amplitude \footnote{Note that this relation holds on dimensional grounds since the field
amplitude scales as the inverse of a time \cite{Nikos}.}
\begin{equation}
h \sim L_\perp^{-z}\,.
\label{equivalence}
\end{equation}

We test this equivalence on the fluctuations of the fluctuations of
the mean flocking direction $\psi(t) = \arg[\Omega (t)]$. \\
In Ref. \cite{BFG}, it has been shown by mean-field like approximations of the microscopic VM dynamics, that $\psi(t)$ obeys an Ornstein-Uhlenbeck process. In particular, we are interested in the
fluctuations of the mean flocking direction, $\delta \psi(t) =
\psi(t)-\psi(0)$ with the initial condition $\psi(0)$ aligned in the
external field direction.\\
The mean squared fluctuations are thus given by
\begin{equation}
\langle \delta \psi(t)^2 \rangle \approx 
\frac{D_\psi}{h'}\left(1-e^{-2 h' t} \right) 
\label{eq99}
\end{equation}
where $h' = h/\bar{m}$, with $\bar{m}$ being the average number of interacting
particles in the Vicsek dynamics \eqref{vicsek_theta}-\eqref{position_vicsek}, plays the role of the confining potential stiffness \footnote{Note that this potential confines the mean flocking orientation, not the position of the active particles.} and 
\begin{equation}
D_\psi=\frac{\eta^2 \pi^2}{6 N}
\end{equation}
is (twice) the effective
diffusion acting on the mean flocking orientation. \\
Eq. (\ref{eq99}) implies that $\langle \delta \psi^2 \rangle$ will grow
linearly as $2D_\psi t$ on short timescales, $t \ll 1/(2h')$, eventually saturating to a constant
value $D_\psi/h'$ for $t \gg 1/(2h')$. In practice, the short time
dynamics of $\psi(t)$ is diffusive and cannot be distinguished from
the zero field behavior, while at large enough time the external field
suppresses orientation diffusion and acts as a confining potential on
the mean flocking direction.

We now consider fluctuations of the mean flocking directions in the
absence of an external field but for a transversally confined flock. Carrying on the correspondence expressed by Eq. (\ref{equivalence})
and using $N=\rho_0 L_\parallel L_\perp$ one gets by direct
substitution in Eq. (\ref{eq99}) 

\begin{equation}
\langle\delta\psi(t)^2\rangle=\dfrac{\eta^2 \pi^2}{6\rho_0
  L_\perp
  L_\parallel}\dfrac{\bar{m}}{L_\perp^{-z}}\left[1-\mbox{exp}\left(-\frac{2
      L_\perp^{-z} } {\bar{m}} t \right)\right] 
\label{eq:00}
\end{equation}
or, introducing the scaling function
\begin{equation}
f(x)\equiv\dfrac{\eta^2 \pi^2 \bar{m}}{6\rho_0 
  L_\parallel}\left[1-\mbox{exp}\left(-\frac{2}{\bar{m}} x \right)\right] \,,
\end{equation}
\begin{equation}
\langle\delta\psi(t)^2\rangle=L_\perp^{z-1}f( L_\perp^{-z} \,t)\,.
\label{eq:01}
\end{equation}
Transversal confinement between parallel reflecting walls thus
suppresses diffusion of the mean flocking directions at large enough
times, with
\begin{equation}
\langle\delta\psi(\infty)^2\rangle \sim L_\perp^{z-1}
\label{eq:02}
\end{equation}
while at short times ($t \ll L_\perp^z \bar{m}/2$) one has a diffusive behavior with
\begin{equation}
\langle\delta\psi(t)^2\rangle \approx \dfrac{\eta^2 \pi^2}{3\rho_0
  L_\parallel L_\perp}\,t\,.
\end{equation}
\label{eq:03}
Microscopic numerical simulations, reported in Fig.~\ref{Fig4}c,
clearly show these two regimes. In particular, data collapse (see Fig.~\ref{Fig4}d) confirms the scaling form (\ref{eq:01}), thus
supporting the equivalence conjectured in (\ref{equivalence}).

\subsubsection{Discussion}
\label{sec4}
We have shown by analytical and numerical arguments that transversal confinement between parallel reflecting walls separated by a distance $L_\perp$ introduces an
effective mass term ${M_c} \sim L_\perp^{-\zeta}$ damping the free theory
Nambu-Goldstone mode. The effect of confinement on bulk connected correlations can be deduced by standard finite size scaling analysis under the rather generic assumption that the transversal boundaries break the rotational symmetry of the active particles self propulsion orientation. In principle, the mass term should suppress the small wavenumber divergence of fluctuations correlations in Fourier space (or, equivalently, introduces an exponential cut-off in the real space connected correlations), introducing a crossover towards a finite value
\begin{equation*}
S({\bf q} \to 0) \sim M_c^{-1}\,.
\end{equation*}
This crossover is however controlled by phenomenological constants which in principle depends on microscopic parameters and on the details of the interaction with the confining boundaries. In the transversal direction, where wave-numbers are bounded from below due to the finiteness of the system, this implies that the saturation regime does not appear for a phenomenological constants $G_\perp$ of order one or smaller. We have verified numerically that this is for instance the case for the standard VM model confined between parallel reflecting
walls.\\
In the longitudinal direction, on the other hand, such a crossover can be observed in a narrow enough ring configuration, that is for $L_\perp\lesssim L_\parallel$. Density correlation in the longitudinal direction, however, seem to be characterized by an anomalous effective mass term $\Sigma \sim L_\perp^{-\gamma}$, with $\gamma \approx 0.9$ an empirical scaling
factor of unknown origin.

Confinement also increases the scalar order parameter values, which shows an algebraic decay to its asymptotic value with increasing separation sizes,
\begin{equation}
\Phi(L_\perp) - \Phi(L_\perp \to \infty) \sim L_\perp^{2\chi}\,.
\label{FSOP}
\end{equation} 
Since this result has been obtained only by finite scaling analysis,
it does not depend in any way on the nature of boundary conditions,
and should also apply to the standard numerical setup of finite
systems with periodic boundary conditions (PBC), as it has been
already realized in \cite{Duan}. Indeed, numerical
simulation of the VM in two dimensional tori of linear size $L$ show a power
law decay of the scalar order parameter to its asymptotic value, $\Phi(L) - \Phi(L \to \infty) \sim L^{\alpha}$, with
an exponent $\alpha \approx -0.64$ \cite{Chate2020} compatible with
the theoretical prediction $\alpha=2\chi=-2/3$. Interestingly, this
result suggest an alternative and more robust way to estimate
numerically the theory scaling exponent from the finite size scaling
of global observables rather than from the measure of the small
wavelength behavior of correlation functions.

Altogether, these results suggest an equivalence between the effects
of an explicit symmetry breaking due to boundary conditions and the
one induced by an homogeneous (and small) external driving of
amplitude $h$. Both induce an effective mass term, with the
equivalence $M_c \sim h \sim L_\perp^{-z}$. 
This mass term, finally, constrains the mean flocking
direction $\psi(t)$ fluctuations acting as an effective harmonic potential
stiffness which, at large times, prevents the diffusion of $\psi(t)$
that characterize free finite flocks. 

We believe our results could be of experimental relevance and can be tested by confining flocking systems such as active colloids \cite{Bartolo2013} or flocking epithelial tissues \cite{Giavazzi}. 
\fancyhead[LO]{{\it Confined flocking}} 
\fancyhead[RE]{{\it Confined flocking}} 
\chapter{Confined flocking - Extensive boundary layers and Casimir-like forces}\label{confined}
While in the previous chapter we have shown how reflecting boundaries affect the bulk properties of confined flocks, here we will focus on the boundary effects induced by confinement. We will show that a confined vectorial active fluid is indeed characterized by {\it extensive boundary layers} where particles accumulate and which are a consequence of the long-range correlations of fluctuations, present in all of the ordered flocking phase. We will also show how, as a further consequence, these also induce a {\it Casimir-like force} on the confining walls. Remarkably both these results hold generically in all the homogeneous polar liquid phase, even beyond the strict dilute limit, showing a certain degree of universality.

In the first section we will show how the boundaries (either perfectly or partially reflecting) have a far-reaching impact, concluding that their influence extends well beyond a finite boundary layer typically observed in the context of \textit{scalar} active matter \cite{Caprini2018}, as discussed in Sec. \ref{boundaries_scalar}. As a consequence of the long ranged correlations present in the bulk system, the ordered flocking phase of a confined active \textit{vectorial} fluid is instead characterized by extensive boundary layers. Interestingly, we are also able to predict, by means of a mesoscopic description, both the density and velocity fluctuations profiles, that match to a very good degree the results coming from microscopic simulations.\\
In the second section we will show how non-equilibrium fluctuations, present deep in the ordered phase, induce an unusually strong attractive Casimir-like force \cite{gambassi2024critical}, a fluctuation-induced force characterized by an algebraic decay with the boundary separation $L_\perp$. Interestingly, the scaling with the wall separation of the force is found to be remarkably slow and should thus be detectable with relative easiness also in experimental set-ups.\\
The third and final section is an extension of our results beyond the dilute limit: 
our numerical simulations of the CVM (see Sec. \ref{other_ways}) indicate that our results are robust, even in the presence of repulsive interactions between the particles and in the hyper-confluent regime, a condition typical of experimental set-ups \cite{giavazzi2017giant}. This former result paves the way towards experimental tests of our findings. 
\newpage

\section{Extensive boundary layers}
\begin{figure}[hbt!]
    \centering
    \includegraphics[width=0.5\linewidth]{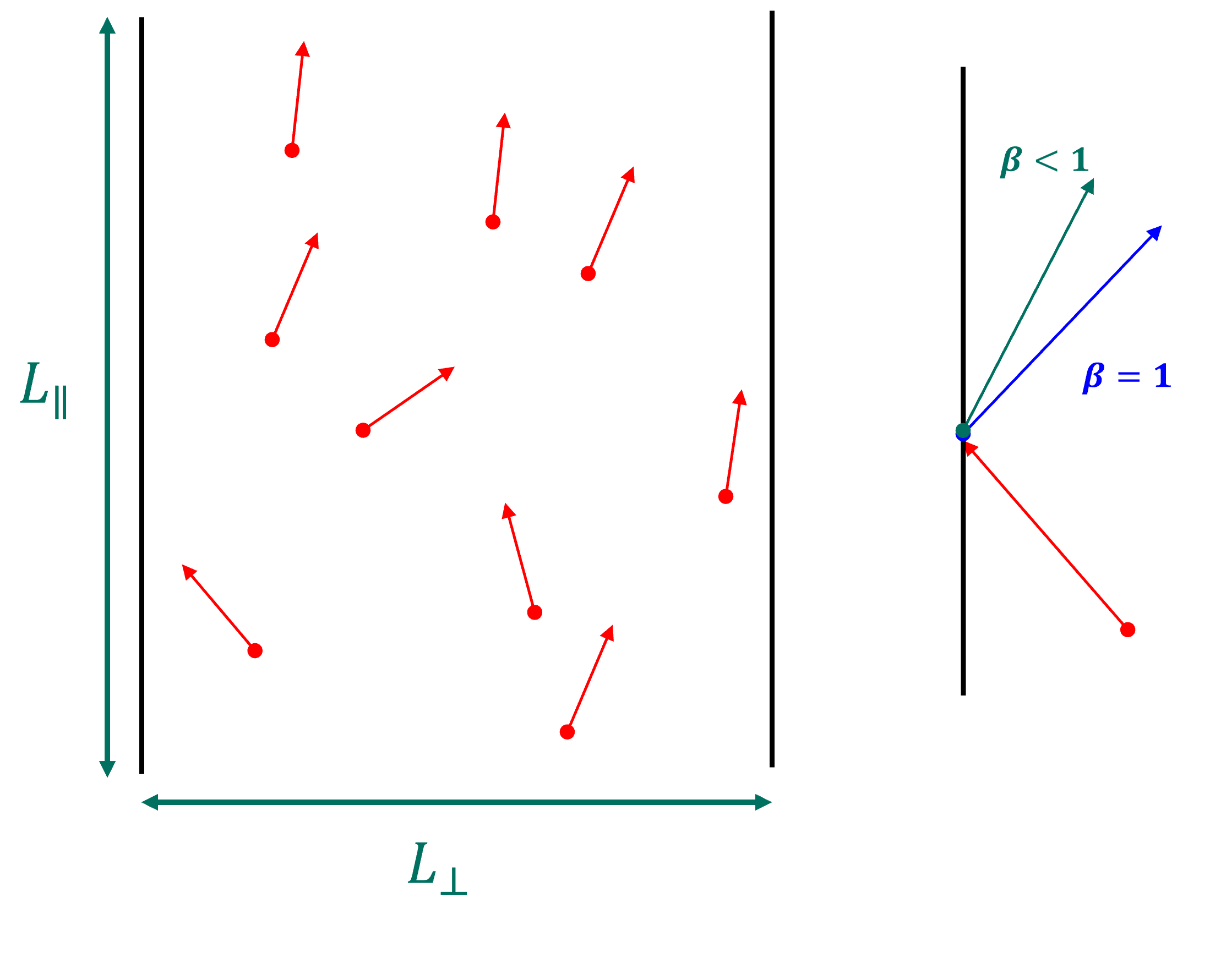}
    \caption{{\bf Schematic representation of the system and its boundary conditions}: Left - The system is confined between two parallel plates at distance $L_\perp$ (where reflecting boundary conditions are assumed), while periodic boundary conditions are assumed in the longitudinal direction (that of flocking) of large size $L_\|$. Right - Schematic representation of the reflecting boundary conditions. The incoming particle (in red) gets reflected: its postcollisional orientation is varied between $\beta = 0$ (aligned with the wall) to $\beta = 1$ (simple elastic reflection).}    
    \label{post_coll_cas}
\end{figure}

We consider a slight generalization of the perfectly reflecting boundaries introduced in the previous section, with a flocking system of $N$ particles in $d = 2$ spatial dimensions confined between two parallel hard walls at distance $L_\perp$. Here, instead, we assume partially reflecting boundary conditions at the walls (see Fig.\ref{post_coll_cas}), so that, as we have seen, the mean velocity of the flock fluctuates around the direction parallel to them, which we refer to as longitudinal ($\|$). Periodic boundary conditions are assumed in this direction of length $L_\|$.

The overdamped, time-discrete particle dynamics is provided by the VM \eqref{vicsek_theta}-\eqref{position_vicsek}. The partially reflecting boundaries reverse the perpendicular component (denoted by the subscript $\perp$) of the particle orientation with respect to the walls. Algorithmically, whenever the dynamics \eqref{vicsek_theta}-\eqref{position_vicsek} renders a particle position, with transversal component $r_\perp$, outside the region $r_\perp \!\in\! [0,L_\perp]$, we apply the collision rule 
\begin{equation}\label{eq:reflection_casimir}
    n_\perp \to -\beta n_\perp \quad \text{and} \quad r_\perp \to 2 B - \beta r_\perp
\end{equation}
with either $B\!=\!0$ (left boundary) or $B\!=\!L_\perp$ (right boundary), and $0<\beta \leq 1$ a collision parameter which slightly generalizes the fully reflecting set-up of the previous chapter (which is recovered for $\beta=1$). Note that since $\beta>0$ this rule still models the tendency of particles to move away from the wall after a collision (see Fig.\ref{post_coll_cas} for a schematic representation).\\
Finally, recall that the VM describes the overdamped dynamics of self-propelled particles which locally inject and dissipate energy. In this overdamped approximation, fast speed fluctuations are ignored and particles self-propel with a fixed speed and $\hat{\bf n}=1$. Therefore, in our collision rule the unit norm of the orientation after a collision is instantaneously restored by the corresponding increase of its longitudinal component, that is $n_\parallel \to$ sgn$(n_\parallel)[1-\beta(1-n_\parallel^2)]^{1/2}$, (which leaves $n_\parallel$ unchanged for $\beta=1$).

\subsection{Microscopic density and velocity profiles}\label{micro_rho_vel}
\begin{figure}[hbt!]
    \centering
    \includegraphics[width=1\linewidth]{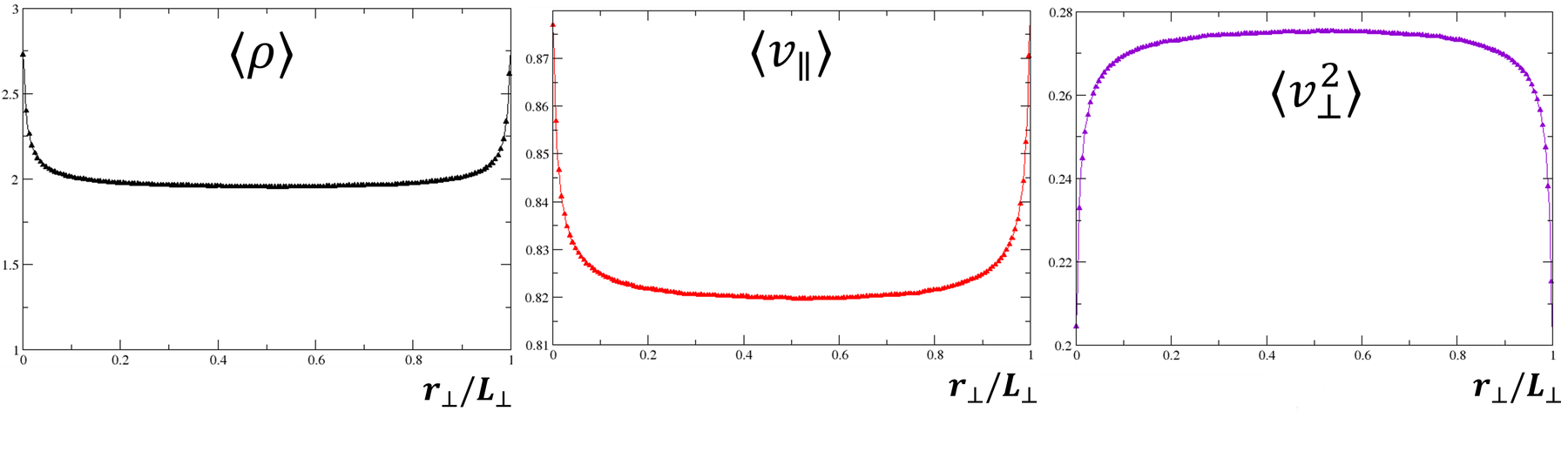}
    \caption{{\bf Profiles of average density $\rho$, longitudinal velocity $v_\parallel$ and transverse velocity fluctuations $v_\perp^2$} Average density $\langle\rho\rangle$ (black), longitudinal velocity $\langle v_\parallel\rangle$ (red) and transverse velocity fluctuations $\langle v_\perp^2\rangle$ (purple) as a function of the rescaled transverse separation $r_\perp/L_\perp$, showing how the suppression of velocity fluctuations induces accumulation of particles and increase in the longitudinal velocity at the boundaries. Here $v_0 = 0.5$, $\rho_0 = 2.0$, $\eta = 0.21$, $\beta = 1$, $L_\| = 512$ and $L_\perp = 512$.}
    \label{micro_profiles}
\end{figure}

We first investigate the boundary behavior of the confined system with a series of microscopic simulations conducted in the Toner \& Tu phase at fixed (but rather arbitrary) parameter values (see caption of Fig.\ref{micro_profiles}). Particles in the flock tend to accumulate at the walls: while this is quite generic for active particles \cite{Lowen}, here we find that the accumulation of particles extends well beyond the characteristic particle-wall interaction length scale and has an extensive nature w.r.t. the channel extension $L_\perp$. .

We measure the time-averaged value of the density $\rho(\mathbf{r}, t)$, longitudinal velocity $v_\parallel(\mathbf{r}, t)$ and transverse velocity fluctuations $v_\perp^2(\mathbf{r}, t)$ in the stationary state, after discarding a proper transient $T_0\approx 10^5$ time-steps.
As in the previous Chapter, the instantaneous fields are obtained by coarse-graining the microscopic particles number and velocities over boxes of unit linear length. The Casimir geometry (parallel infinite walls) implies that the time-averages of such fields are invariant along the longitudinal direction: this enables us to further take averages over this invariant direction, which in turn greatly increases the statistics. For the remaining of this chapter $\langle \cdot \rangle$ indicates averages over time and the longitudinal direction.\\
Transverse confinement, on the other hand, breaks translational invariance in the perpendicular (w.r.t. the walls) direction, thus we expect the averages $\braket{\rho(r_\perp)}, \braket{v_\parallel(r_\perp)}$ and $ \braket{v_\perp^2(r_\perp)}$ to be inhomogeneous across the transverse direction. In fact, as it was already noted in Ref.~\cite{Benoit2019}, reflecting boundaries tend to suppress transversal velocity fluctuations in their vicinity, inducing an excess density and increasing longitudinal velocity, such as that shown in Fig.\ref{micro_profiles}.

For $\beta \!\!\neq\!\! 1$ (partial reflection), these profiles turn out to coincide with those for $\beta\!\!=\!\!1$ (complete reflection), with the sole exception of the first (non extensive) bin closest to the wall (we will further discuss this fact in the Appendix \ref{coarse_grained}), as shown in Fig.\ref{fig2}(a) for the density and transversal fluctuations profiles, showing a degree of universality in terms of boundary interactions.

As we have already mentioned, the accumulation of particles at the boundaries extends well below the particle-wall interaction: varying the wall separation $L_\perp$ we can study how deep inside the system this boundary layer extends. As opposed to what has been found in scalar systems (for example, in the case of confined AOUPs, reported in Section \ref{boundary_scalar}) where boundary layers are found to be intensive, here we report \textit{extensive boundary layers}: if we define $d_0$ as the typical distance from the wall at which density deviations $\braket{\delta\rho} \equiv\braket{\rho}-\rho_0$ change sign we find that, in the thermodynamic limit, $L_\perp\to \infty$
\begin{equation}
    d_0(L_\perp) \propto L_\perp
\end{equation}
meaning the boundary layer extends well beyond the microscopic particle-wall interaction range, as shown in the scaling analysis reported in Fig.\ref{fig2}(b).

\begin{figure}[hbt!]
\includegraphics[width=1\textwidth]{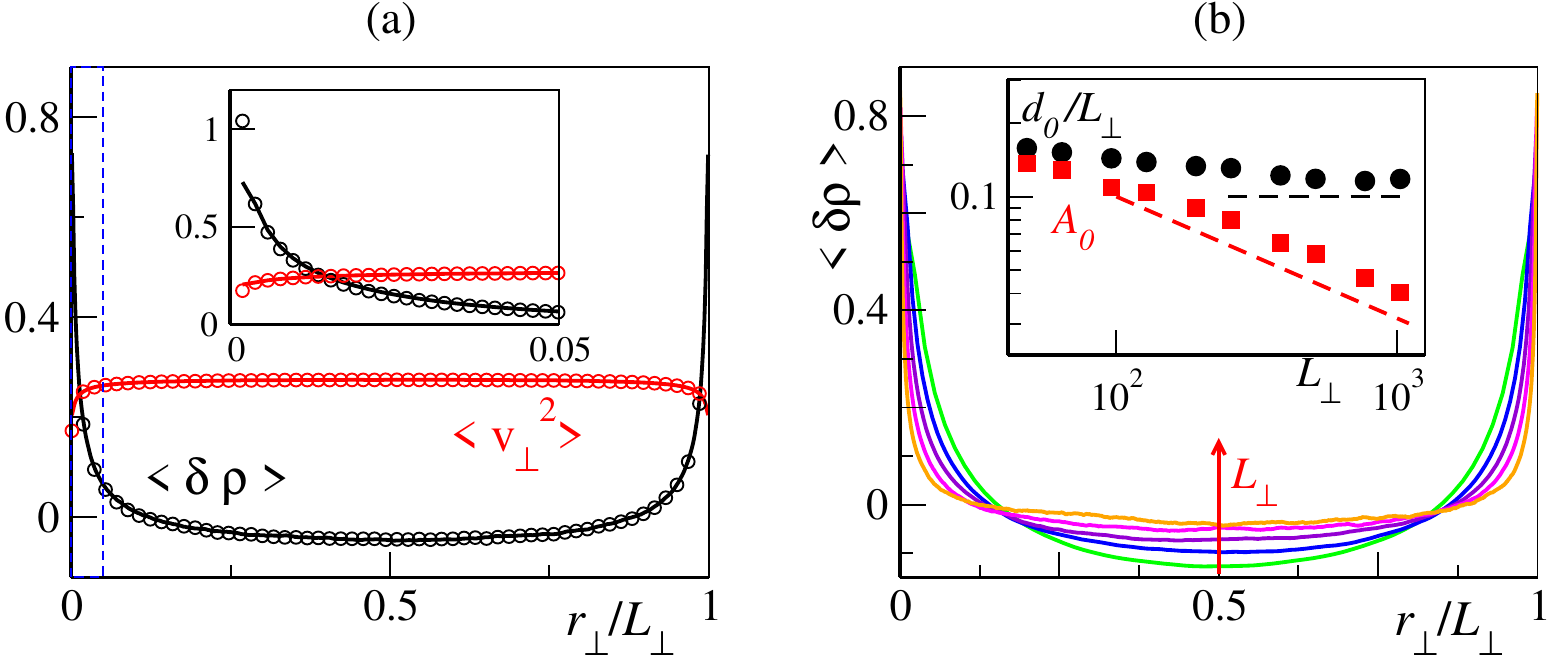}
\caption{{\bf Stationary profile of density and velocity fluctuation in a confined system} (a) Average density deviations (black) and transversal velocity fluctuations (red) for $L_\perp\!=\!512$ as a function of the rescaled transverse coordinate $r_\perp/L_\perp$. The case of fully reflecting (solid lines, $\beta=1$) and partially reflecting (open circles, $\beta\!=\!0.5$) boundaries are compared. Inset: zoom in the boundary layer on the left of the dashed vertical blue line in the main panel. (b) Density profiles as functions of the rescaled transverse coordinate $r_\perp/L_\perp$, for increasing separation $L_\perp\!=\!$ 64, 128, 256, 512, and 1024, with $\beta\!=\!1$. Inset: Magnitude of density deviations $A_0$ (red symbols) and rescaled zero-crossing $d_0/L_\perp$ of $\langle \delta \rho\rangle$ (see text) as functions of $L_\perp$. The dashed black line is a constant ($=0.1$) and shows that asymptotically $d_0\propto L_\perp$, while the red dashed line corresponds to the $1/\sqrt{R}$ expected behaviour of $A_0$. \\
Data has been obtained from microscopic numerical simulations with $\rho_0\!=\!2.0$, $\eta\!=\!0.21$, and $L_\|\!=\!1024$.} 
\label{fig2}
\end{figure} 

This result may rise fears that in a semi-infinite system, the boundary effect may be felt at arbitrary distances, so that the bulk behavior cannot be recovered even in the thermodynamic limit  far away from the boundary.  Note, however, that the amplitude of the density deviations decays asymptotically to zero, as shown in the inset of Fig.\ref{fig2}(b), $A_0 \!=\! \int_0^1 dx \left | \langle \delta \rho(x, t) \rangle\right | \!\sim\! 1/\sqrt{L_\perp}$ where $x\!=\!r_\perp/L_\perp$ is the rescaled wall separation. In a sense, density deviations inside the boundary layer become smaller with a growing $L_\perp$, so that the bulk TT hydrodynamic behavior is indeed recovered for $L_\perp \!\to\! \infty$, even if the boundary layer extension $d_0$ diverges with $L_\perp$.

Finally, it is important to stress that, while the exact value of $\beta \geq 0$ does not impact so much the large scale behavior of our system (as it will be shown in the next sections), we find that for $\beta < 0$ none of our results hold: this condition (which essentially corresponds to $\alpha < 1/2$ in the language of Section \ref{polar_flock_stable}) models persistent boids whose dynamics is typically that of getting stuck at boundaries and crawling/moving along them. Indeed this is what was found in \cite{codina2022small}, where persistency and a small perturbation lead asymptotically, in a finite fraction of the phase diagram, to disruption of the ordered phase. A negative $\beta$ (which means persistence at the boundaries) invalidates our treatment. Preliminary numerical simulations indicate that in this latter set-up active the particles simply accumulate at the walls in a very narrow region (of the order of a few units) leaving an empty bulk region. This regime has not be explored in detail and clearly cannot be captured by a hydrodynamic description at the level of Toner \& Tu equations, as done in the following section for $\beta>0$.

\subsection{Coarse grained description}\label{coarse_grained}
We now develop an hydrodynamic description of the average slow field profiles. As usual, the flocking fluid can be described in terms of suitably coarse-grained slow hydrodynamic fields, i.e., the density deviations  $\delta\rho \!\equiv\! \rho({\bf r}, t)-\rho_0$ from the mean density $\rho_0$ and the transverse velocity fluctuations $v_\perp^2 ({\bf r}, t)$  (the longitudinal component of the velocity is a fast mode enslaved to the slow ones, see Sec.\ref{TT_theory}).

In order to determine the average density at the confining walls, we consider the TT equations in a slightly simplified form, previously considered in the literature \cite{Bartolo2017}. The derivation from the full TT equations will be discussed in the appendix \ref{full_TT}. The simplified equations ease our discussion while retaining all the essential non-equilibrium terms describing the advection of $\rho ({\bf r}, t)$ and the velocity ${\bf v}({\bf r}, t)$ \cite{Ramaswamy, HydroReview, tailleur2022active}, i.e.,
\begin{align}\label{a: rho evolution}
\centering
\partial_t \rho + \nabla \cdot (\rho{\bf v})&=D_\rho \nabla^2 \rho, \\
\label{a: v evolution}
\partial_t{\bf v}+\lambda_1({\bf v}\cdot\nabla) {\bf v}=(\mu - \kappa {\bf v}^2) {\bf v}
&-\sigma_1 \nabla \rho +D_v\nabla^2\mathbf{v}+\mathbf{f},
\end{align}
where $\mathbf{f}$ is the usual delta-correlated white noise of amplitude $\Delta$.
In these equations we have also included a nonzero density diffusion term for $\rho$, proportional to $D_\rho$, which in the TT equations is generated by renormalization \cite{TTR}. The (positive) coefficients $D_v$, $\lambda_1$ and $\sigma_1$ are related, respectively, to diffusion, advection and pressure. The Ginzburg-Landau terms (i.e., the first two on the r.h.s. of Eq.~\eqref{a: v evolution}, with $\mu, \kappa >0$ in the ordered phase) determine as usual the modulus $p_0\equiv|{\bf p}_0|=\sqrt{\mu(\rho_0)/\kappa}$ of the average bulk velocity ${\bf p}_0$.

Due to confinement, ${\bf p}_0$ aligns with the unit vector ${\bf \hat{e}}_\parallel$ parallel to the walls. For simplicity, we consider constant coefficients, only retaining the local density dependence of $\mu=\mu(\rho)$ which, as we have seen in Section \ref{pd_bgl_mu_dependence}, plays a fundamental role in the onset of the phase separated regime.\\
We proceed by spin-wave approximation \cite{Nishimori}, decomposing the velocity ${\bf v}$ into its longitudinal and transversal components, ${\bf v} = (p_0 + \delta v_\parallel) {\bf\hat{e}}_\parallel + {\bf v}_\perp $, and expanding in small velocity deviations from $ {\bf  p}_0 =p_0 \hat{\bf e}_\parallel$. Taking averages and projecting Eqs.~\eqref{a: rho evolution} and \eqref{a: v evolution} along the transversal direction one gets
\begin{align}\label{rho1_cas}
\centering
\partial_\perp \langle {\bf v}_\perp \rangle &\approx \rho_0^{-1}  D_\rho \partial_\perp^2 \langle \delta\rho\rangle,\\
\label{vperp}
\frac{\lambda_1}{2} \partial_\perp \langle {\bf v}_\perp^2 \rangle =
&-\sigma_1 \partial_\perp \langle \delta \rho \rangle +
D_v \partial_\perp^2 \langle {\bf v}_\perp \rangle,
\end{align}
where we have retained up to quadratic terms in the transversal velocity\footnote{For a vectorial field ${\bf v}$ with a locally constrained modulo $w_0$, small fluctuations can be expressed in terms of an angle $\delta \theta \ll 1$, so that $| {\bf v}_\perp| = w_0 \sin \delta \theta \approx w_0 \delta \theta$ and $w_0+\delta v_\parallel = w_0 \cos \delta \theta \approx w_0 (1-\delta \theta^2/2)$. It follows  $\delta v_\parallel \sim | {\bf v}_\perp|^2\sim \delta \theta^2$} and used translational invariance with respect to time and the longitudinal direction, i.e., $\partial_t \langle \cdot \rangle = \partial_\parallel \langle \cdot\rangle=0$ (this last one being provided by the symmetry of the Casimir geometry). We also discarded a higher-order contribution $\langle {\bf v}_\perp \delta\rho\rangle$ from Eq.~\eqref{rho1_cas}. \\
Note that by projecting Eq.~\eqref{a: v evolution} along the longitudinal direction we simply obtain an equation which expresses the enslaving of $\langle \delta v_\parallel \rangle$ to $\langle\delta \rho \rangle$ and $\langle {\bf v}_\perp^2\rangle$ (see Section \ref{linearized_TT} and Ref. \cite{TTR}). To zeroth order in the derivatives and to first order in $\langle \delta \rho\rangle$ it reads
\begin{equation}
\langle \delta v_\parallel\rangle \approx \frac{\mu'(\rho_0)}{2\beta p_0} \langle\delta \rho\rangle - \frac{\langle \mathbf{v}_\perp^2 \rangle}{2 p_0}\,,
\end{equation}
where $\mu'(\rho) = d\mu(\rho)/d\rho$.\\
Substituting Eq.~(\ref{rho1_cas}) into Eq.~(\ref{vperp}) we obtain that $\langle\delta \rho\rangle$ satisfies the equation 
\begin{equation}\label{eq:profile}
2 \rho_0^{-1}D_v D_\rho\, \partial_\perp^3 \langle\delta \rho\rangle - 2 \sigma_1 \partial_\perp \langle\delta \rho\rangle= \lambda_1 \partial_\perp \langle {\bf v}_\perp^2 \rangle.
\end{equation}
Neglecting the nonlinear term $ \overline{w} \equiv \langle {\bf v}_\perp^2 \rangle$ in Eq.~\eqref{eq:profile} renders the (linear) equation for the density profile discussed in section 2.2.3 for confined {\it scalar} active matter, resulting in non-extensive boundary layers. In the present {\it vectorial} case, instead, we will see that bulk long-ranged correlations lead to an extensive boundary layer, due to having $\overline{w}\neq 0$ in Eq.~\eqref{eq:profile}. 

\subsubsection{Boundary conditions and slow fields expansion}
Because of the symmetry of the transverse confinement, the average profiles $\overline{\delta \rho} \equiv \langle \delta\rho\rangle$ and $\overline{w}=\langle {\bf v}_\perp^2\rangle$ are even functions of $r_\bot$ with respect to the mid-line of the slab at $r_\bot =L_\perp/2$. Moving to Fourier space, this implies that 
\begin{equation}
    \overline{\delta \rho}(r_\perp)=\sum_{q_\perp} \widehat{\delta\rho}(q_\perp) \cos (2 q_\perp r_\perp),\quad \overline{w}(r_\perp) = \sum_{q_\perp} \widehat{w}(q_\perp) \cos (2 q_\perp r_\perp),
\end{equation}
where $q_\perp=\pi n/L_\perp$, $n$ is a non-negative integer and the average field Fourier transform $\widehat{\delta\rho}(q_\perp)$, can be simply obtained from \eqref{eq:profile} and for any $q_\perp >0$ (note that $\hat{\delta \rho(0)}=0$ due to total density conservation) is given by
\begin{equation}
\label{F:exp}
\widehat{\delta
    \rho}(q_\perp)=\frac{-\lambda_1\,\widehat{w}(q_\perp)}{(2
  D_v D_\rho/\rho_0)q_\perp^2+2\sigma_1}.
\end{equation}
We can derive the Fourier modes $\widehat{w}(q_\perp)$ of the average nonlinear term $\overline{w}(r_\perp)$ from the two-point static bulk correlations in Fourier space: this will enable us to obtain an expression for $\widehat{\delta\rho}(q_\perp)$ and consequently obtain a prediction for the density profile.

\subsubsection{Velocity fluctuations}

\begin{figure}[hbt!]
    \centering
    \includegraphics[width=0.6\linewidth]{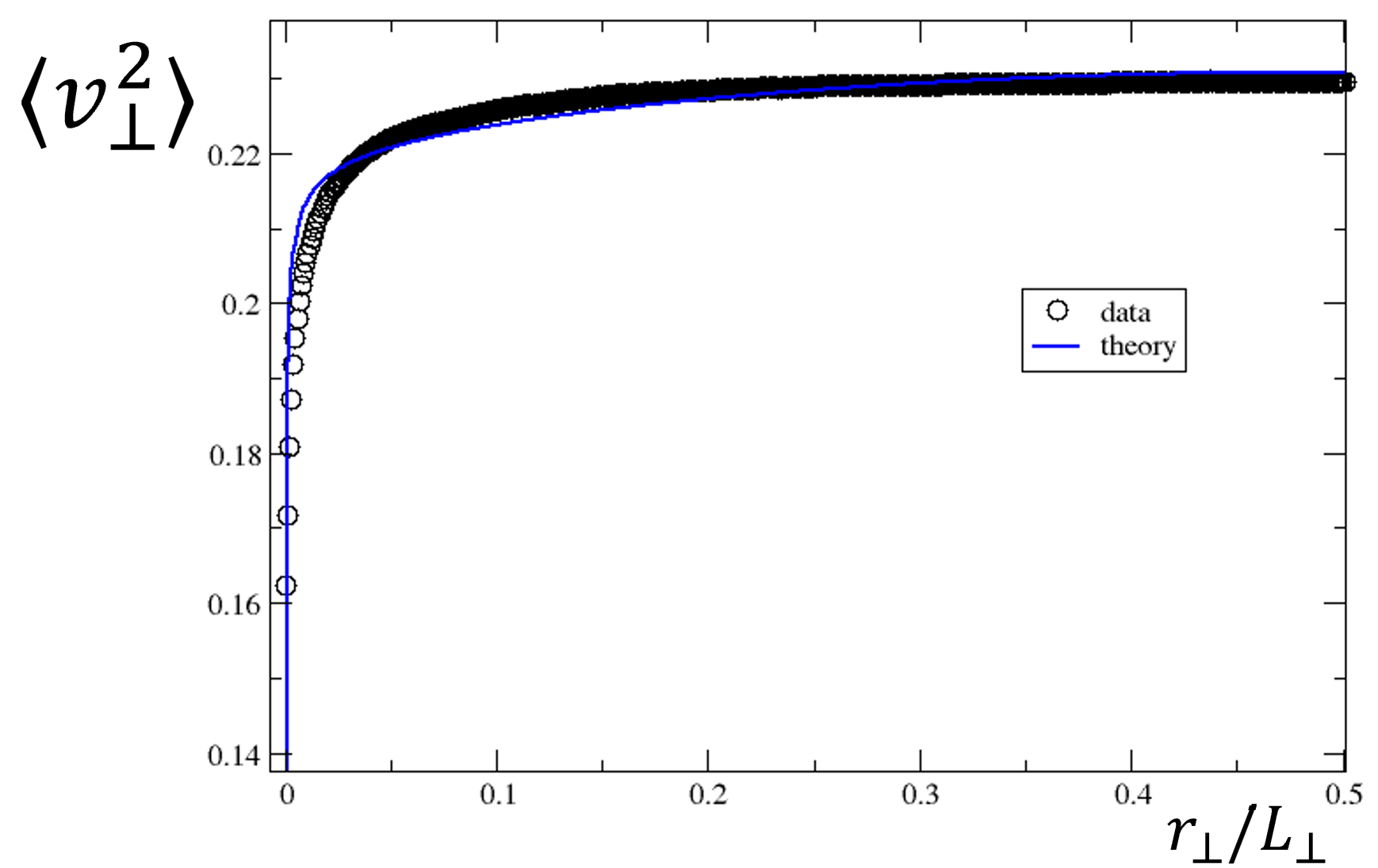}
    \caption{{\bf Profile of the transverse velocity fluctuations} - Comparison between the average profile of transverse velocity fluctuations $\langle v_\perp^2\rangle$ obtained from microscopic simulation of the VM (black open circles) and our theoretical prediction (blue line). Both profiles are plotted against the rescaled wall separation $r_\perp/L_\perp$.} 
    \label{numerical profiles fluctuations}
\end{figure}

In order to model reflecting walls, we adopt zero Dirichlet boundary
conditions for the instantaneous transversal velocity field $v_\perp ({\bf r}, t)$. In the limit of infinite longitudinal extension $L_\| \to \infty$, the instantaneous, fluctuating field $v_\perp ({\bf r}, t)$ can be expressed in terms of its Fourier transform $\hat{v}_\perp ({\bf q}, t)$ as
\begin{equation}
\label{eq:v1}
v_\perp ({\bf r}, t) = \int_{-\infty}^{\infty} dq_\parallel \sum_{q_\perp>0}
\hat{v}_\perp ({\bf q}, t) \sin\left( q_\perp
  r_\perp\right) e^{i q_\parallel r_\parallel},
\end{equation}
where ${\bf q}=(q_\parallel,q_\perp)$, with $q_\parallel \in \mathbb{L_\perp}$ while $q_\perp$ takes discrete values $q_\perp=q_\perp(n)=\pi n/L_\perp$, with $n$
a positive integer. The Fourier modes are given by
\begin{equation}
\label{eq:v1bis}
\hat{v}_\perp ({\bf q}, t) = \frac{1}{\pi L_\perp} \int_{-\infty}^{\infty} \!\!dr_\parallel \int_0^{L_\perp} \!\! dr_\perp 
v_\perp ({\bf r}, t) \sin\left( q_\perp
  r_\perp\right) e^{-i q_\parallel r_\parallel}.
\end{equation}
An analogous expansion of the delta-correlated noise term ${\bf f} ({\bf r}, t)$ appearing in the TT equations (see Eq.\eqref{a: v evolution}) leads to Fourier space correlations
\begin{equation}
\langle \hat{\bf f}_i({\bf q}, t) \hat{\bf f}_j({\bf q}',
t') \rangle = \frac{\Delta}{\pi L_\perp} \delta_{q_\perp, q_\perp'} \,\delta(q_\parallel
+q_\parallel')\delta(t-t')\delta_{i,j},
\end{equation}
where $i$ and $j$ indicate the coordinate components of ${\bf f}$ and which, in turn, constrain Fourier space correlations of the slow fields at the linear level \cite{TT1}. \\
In particular, for equal times,
\begin{equation}
\label{eq:v2}
\langle \hat{v}_\perp ({\bf q}) \hat{v}_\perp
({\bf q}')\rangle = \frac{S_v(\mathbf{q})}{\pi L_\perp}  \delta_{q_\perp, q_\perp'} \,\delta(q_\parallel
+q_\parallel').
\end{equation}
where $S_v(\mathbf{q}) \equiv \langle |\hat{v}_\perp({\bf q}) |^2\rangle $.\\
The bulk behavior of these correlations\footnote{Repeating the analysis done in Ref~\cite{TTR}, it is easy to show that the simplified form of the TT equations introduced in the main text have the same linearized structure and nonlinearities of the complete TT equations. Accordingly, their correlations functions are characterized by the same scaling behavior.}of the transversal velocity for small $|{\bf q}|$, as we have seen in Section \ref{sec1} is modified due to the presence of confining walls and is given by Eqs.\eqref{Sperp}-\eqref{Spar}
\begin{equation}\label{S_v_app}
S_v({\bf q},L_\perp^{-1})\sim\begin{cases}
			\dfrac{1}{q_\perp^\zeta+G^{(v)}_\perp L_\perp^{-\zeta}} &  q_\perp^\xi\gg q_\parallel\\
            \dfrac{1}{q_\parallel^{\zeta/\xi}+G^{(v)}_\parallel L_\perp^{-\zeta}} & q_\perp^\xi\ll q_\parallel
		 \end{cases}
\end{equation}
where $\zeta\!\equiv\!d-1+2\chi + \xi$ and $G^{(v)}_{\perp,\parallel}$ are phenomenological constants. Once again the subscripts $\|$ and $\perp$ refer to the directions which are, respectively, parallel and perpendicular to the polarization ${\bf p}_0$ of the flock in the bulk. We recall that in the confined system, ${\bf p}_0$ is naturally parallel to the confining walls. \\
We now consider the average fluctuations of the transversal velocity in real space $\langle v_\perp^2(r_\perp)\rangle$. Due to translational invariance along the longitudinal direction, from Eqs.~\eqref{eq:v1} and \eqref{eq:v2}, one finds
\begin{equation}
\label{eq:v3}
\langle v_\perp^2  (r_\perp)\rangle = \frac{1}{\pi L_\perp}   \sum_{q_\perp>0}  \sin^2\left( q_\perp
  r_\perp\right) \int
dq_\parallel S_v(\mathbf{q}, L_\perp^{-1}).
\end{equation}
Integration over the longitudinal modes $q_\parallel$ can be split in two parts
\begin{equation}\label{S_v_L_perp}
\int_{-\infty}^{\infty}\!\!\!\!\! dq_\parallel S_v({\bf q},L_\perp^{-1}) \simeq \!a_1 \!\int_0^{q_\perp^\xi} \!\!\!\!\!dq_\parallel \frac{1}{q_\perp^\zeta + G_\perp^{(v)}L_\perp^{-\zeta}} + a_2\!\int_{q_\perp^\xi}^\infty \!\!\!\!\! dq_\parallel \frac{1}{q_\parallel^{\zeta/\xi} + G_\parallel^{(v)}L_\perp^{-\zeta}}
\end{equation}
where we have used Eq.\eqref{S_v_app}. Note also that the two integrals on the r.h.s. of the first line of Eq.~\eqref{S_v_L_perp} are weighted by possibly different unknown multiplicative positive factors $a_1$ and $a_2$. The exact solution to the second integral on the r.h.s. is a rather complicated ${}_2F_1$ hyper-geometric function, but since here we are interested only in the first corrections induced by the confining boundaries, we can proceed by calculating the first correction to the behavior expected in the case of the free system.\\
In order to do so, we start by simply expanding the denominator of the integrand, which is legitimate since $q_\parallel\gg q_\perp^\xi$ and $L_\perp\sim q_\perp^{-1}$, and we get
\begin{equation}
    \frac{1}{q_\parallel^{\zeta/\xi} + G_\parallel^{(v)}L_\perp^{-\zeta}} \approx q_\parallel^{-\zeta/\xi}(1- G_\parallel^{(v)}L_\perp^{-\zeta}q_\parallel^{-\zeta/\xi} + \text(h.o.))
\end{equation}
The second integral in the r.h.s. of Eq.\eqref{S_v_L_perp} can then be readily solved
\begin{equation}
     \int_{q_\perp^\xi}^\infty \!\!\!\!\! dq_\parallel q_\parallel^{-\zeta/\xi}(1- G_\parallel^{(v)}L_\perp^{-\zeta}q_\parallel^{-\zeta/\xi}) = \left(-\frac{q_\perp^{-\zeta+\xi}}{1-\zeta/\xi} - G_\parallel^{(v)}L_\perp^{-\zeta}\frac{q_\perp^{-2\zeta +\xi}}{1-2\zeta/\xi}\right)
\end{equation}
since the contributions at infinity go to zero.\\
The first integral, once integration over $q_\parallel$ is carried out, can also be expanded to the same order
\begin{equation}
    \frac{a_1}{q_\perp^\zeta + G_\perp^{(v)}L_\perp^{-\zeta}}q_\perp^\xi \approx a_1 q_\perp^{\xi-\zeta} (1-G_\perp^{(v)}L_\perp^{-\zeta}q_\perp^{-\zeta} + h.o.)
\end{equation}
where we have expanded the first term for $q_\perp^\zeta\gg G_\perp^{(v)} L_\perp^{-\zeta}$.\\
Finally, collecting terms with the same power of $q_\perp$ and discarding higher order terms, the approximated expression reads
\begin{equation}\label{c_1_c_2}
    \int_{-\infty}^{\infty}\!\!\!\!\! dq_\parallel S_v({\bf q},L_\perp^{-1}) \approx 2\pi C q_\perp^{\xi-\zeta}\left[1-\frac{G}{C} (L_\perp q_\perp)^{-\zeta}\right]
\end{equation}
with 
\begin{equation}
    C\! \equiv\!  \frac{1}{2\pi}\left(a_1 + \frac{a_2}{\zeta/\xi -1}\right), \quad \text{and}\quad G\!\equiv\!\frac{1}{2\pi}\left(a_1 G_\perp^{(v)} - \frac{a_2 G_\parallel^{(v)}}{1-2\zeta/\xi}\right)
\end{equation}
As expected, the higher order correction originates from the corrections to bulk correlation functions (see Section \ref{sec1}), depending on the wall separation $L_\perp$, and goes asymptotically to zero when $L_\perp$ goes to infinity. The two positive constants $C$ and $G$ are unknown phenomenological parameters.\\
Substituting Eq.\eqref{c_1_c_2} into Eq.~\eqref{eq:v3}, we finally obtain
\begin{equation}
\langle v_\perp^2 (r_\perp)\rangle =\frac{C}{L_\perp} \sum_{q_\perp > 0}\left[1-\cos\left(2q_\perp  r_\perp\right) \right] q_\perp^{-\gamma}\left[1-\frac{G}{C} (L_\perp q_\perp)^{-\zeta}\right].
\label{eq:2}
\end{equation}
where $\gamma \equiv \zeta-\xi= d-1+2\chi$. As shown in Sec.\ref{scaling_exp}, numerical estimates and analytical predictions in $d = 2$ give $\chi-1/3$ and therefore $\gamma=1/3$.\\
Eq.\eqref{eq:2} can be compared with the numerical results for transversal velocity fluctuations. Numerically performing the sum for different $q_\perp = \pi n/L_\perp$, up to the ultraviolet cut-off $N = R\Lambda/\pi$, we obtain a theoretical prediction for the velocity fluctuations, with $C$ and $G$ as positive free parameters that can be varied in order to fit the numerical profiles. The result, shown in Fig.\ref{numerical profiles fluctuations}, demonstrates that our theoretical prediction is in good agreement with the profile obtain from microscopic simulations of the VM. 

\subsubsection{Density deviations profile}
From Eq.~\eqref{eq:2} we readily identify the coefficients of the Fourier series of $\langle v_\perp^2 (r_\perp) \rangle$, taking into account that because of symmetry, it is an even function for spatial reflection about the mid-line of the slab, located at  $r_\perp = L_\perp/2$:
\begin{equation}
\label{eq:v4}
\langle v_\perp^2 (r_\perp) \rangle = \hat{w}_0+\sum_{q_\perp>0} \hat{w}(q_\perp) \cos\left( 2 q_\perp
  r_\perp\right). 
\end{equation}
Accordingly, the coefficients are given by
\begin{equation}
\label{eq:w0}
\hat{w}_0 = \frac{1}{L_\perp} \int_0^{L_\perp} dr_\perp \langle v_\perp^2
(r_\perp)\rangle = \frac{C}{ L_\perp}\sum_{q_\perp > 0} q_\perp^{-\gamma}\left[1-\frac{G(L_\perp q_\perp)^{-\zeta}}{C} \right],
\end{equation}
and
\begin{equation}
\begin{split}
\label{eq:wq}
\hat{w}(q_\perp) &= \frac{2}{L_\perp} \int_0^{L_\perp} dr_\perp \langle v_\perp^2
(r_\perp)\rangle \cos\left(2 q_\perp r_\perp\right)\\
                &= -\frac{C}{L_\perp} q_\perp^{-\gamma}\left[1-\frac{G(L_\perp q_\perp)^{-\zeta}}{C}\right],
\end{split}
\end{equation}
for $q_\perp>0$. Note that the formal divergence of $\hat w_0$ (whose actual value is unimportant in the following) is actually regularized by a microscopic short wavelength cutoff, implicitly present in TT theory \cite{TT1}.

The expression of $\hat{w}(q_\perp)$ in Eq.~\eqref{eq:wq} can now be used in order to calculate the full density profile $\overline{\delta \rho}(r_\perp)=\sum_{q_\perp} \widehat{\delta\rho}(q_\perp) \cos (2 q_\perp r_\perp)$. Here we follow \cite{fava2024casimir} and present our calculations ignoring the higher order terms coming from boundary corrections to the bulk structure factor, thus setting $G=0$ in Eq.\eqref{eq:wq}. We find that this further correction, while useful in the calculation of the velocity fluctuations, becomes negligible in the case of the density. Inserting Eq.\eqref{eq:wq} (for $G=0$) in Eq.\eqref{F:exp} we obtain the following expression for the density profile
\begin{equation}\label{rho_sum}
    \overline{\delta \rho}(r_\perp)=\frac{1}{L_\perp}\sum_{q_\perp>0}\frac{C\Omega}{(1+\Gamma^2q_\perp^2)q_\perp^\gamma}  \cos (2 q_\perp r_\perp)
\end{equation}
where $\gamma = 1 +2\chi$ in $d=2$, $\Omega = \lambda_1/2\sigma_1$ and $\Gamma = \sqrt{\frac{D_v D_\rho}{\sigma_1\rho_0}}$ acts as a regularizator in the sum, keeping into account the effect of higher order derivatives. Note that the sum runs for positive wave-numbers $q_\perp$ only, since the zeroth mode vanishes due to total density conservation. One can now sum numerically perform this sum for different $q_\perp = \pi n/L_\perp$, in analogy with what we have done for the the velocity fluctuations. Note that in this case we keep $C$, that controls the total magnitude of density deviations, and $\Gamma$, that tunes higher order corrections, as free parameters.

\begin{figure}[hbt!]
    \centering
    \includegraphics[width=1\linewidth]{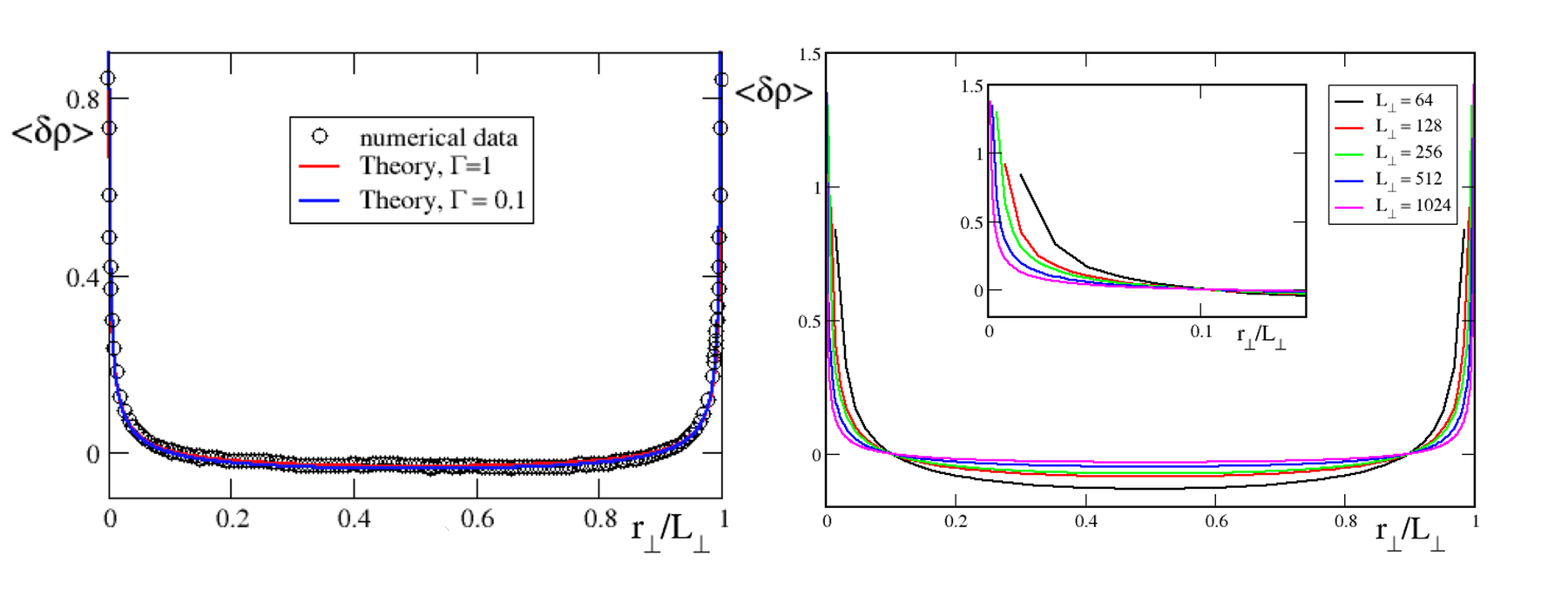}
    \caption{{\bf Profile of the density obtained from Eq.\eqref{rho_sum}} Left - Comparison between the average profile of density deviations $\langle \delta \rho\rangle$ obtained from microscopic simulation of the VM (black open circles) and from our theoretical prediction (red line for $\Gamma =1$ and blue for $\Gamma = 0.1$). Right - Scaling of the density deviations profile with wall separation plotted against the rescaled wall separation $r_\perp/L_\perp$ shows that the boundary layer $d_0$ is indeed a finite fraction of the system asymptotically $d_0 \sim L_\perp$.}
    \label{numerical profiles}
\end{figure}

As shown in Fig.\ref{numerical profiles} (left panel) the agreement between data coming from microscopic simulations and the one predicted from Eq.\eqref{rho_sum} is excellent and does not seem to depend on the exact value of $\Gamma$ indicating that the effect of higher order gradients is indeed small. From the data shown in Fig\ref{numerical profiles} (right panel) we also find that the prediction of the zero crossing $d_0\sim L_\perp$ is compatible with what we get from microscopic simulations, indicating that also from a theoretical approach we can predict extensive boundary layers. 
\newpage

\section{Casimir effect in flocking active matter}
Casimir forces are long-range forces arising from the confinement of a fluctuating correlated field \cite{KB-94,Gambassi, GambassiLast}. They may be of {\it quantum} origin, as noted by Casimir \cite{Casimir}, who considered the quantum vacuum fluctuations of the electromagnetic field confined between two parallel conducting plates, but also of thermal origin. This is the case, e.g., when one confines a classical fluctuating field characterized by a diverging correlation length, such as in a binary liquid mixture approaching its demixing critical point \cite{deGennes}.\\
In both cases, the (free) energy of the confined medium depends on the separation $L_\perp$ between the boundaries (as the set of allowed eigenmodes of the fluctuating fields changes) and a weak, but measurable force emerges on the boundary walls, which Typically decays algebraically upon increasing $L_\perp$.\\
Casimir-like forces of thermal origin may also be found, inter alia, far from a critical point, in equilibrium systems such as liquid crystals \cite{PelitiProst}, where long-ranged soft modes emerge from the spontaneous breaking of a continuous symmetry. 

Statistical systems far from equilibrium may also develop spatially correlated fluctuations, in some cases of non-thermal nature. These fluctuations, when spatially confined by boundaries or inclusions, may give rise to short-range effective forces \cite{AngelPRL11}, possibly long-range Casimir-like forces \cite{AKK-15,CBM-06,BettoloMarconi,PRS-14}, and generalizations thereof \cite{NG-04}. For instance, Ref.~\cite{BettoloMarconi} suggests that long-range forces may arise upon confining a reaction-diffusion system near an absorbing phase transition. 

In the following we show how Casimir-like forces also arise in the non-equilibrium context of confined flocking active matter as a consequence of the long-range fluctuation correlations characteristic of the broken symmetry Toner \& Tu phase.

\subsection{Analytical derivation}\label{sec:casimir_anal}
The mechanical pressure is defined as the force per unit length exerted by the confined active fluid on the walls. For generic active fluids we have seen that it is not a state function and its exact 
form depends on the details of the interaction of the active particles with the wall as shown in Section \ref{mechanical_pressure}. In particular, it depends on the mechanisms involved in the reorientation of the transverse self-propulsion direction, which may also involve momentum transfer to the surrounding viscous fluid and/or substrate.\\
These non-universal features are not specifically captured by the overdamped dynamics of the Vicsek model which, for instance, glosses over the microscopic details of how the torques reorienting particles self-proelling direction are dissipated. However, assuming that a local mechanical interaction prevents the active particle from crossing the boundaries, the mechanical pressure can simply be expressed as minus the force per unit length that the walls exert on the particles positional degree of freedom. 

Consider, for instance, a wall which confines the particles in the region with $r_\perp < L_\perp$. 
At the mesoscopic scale, assuming translational invariance along the longitudinal direction and 
a sharp potential $U(r_\perp)$ in the transversal direction such that
\begin{equation}
    U(r_\perp<L_\perp) =0 \quad \text{and} \quad \partial_\perp U(r_\perp) \approx \delta(r_\bot - L_\perp)
\end{equation}
one obtains the following average pressure \cite{TailleurP} in terms of the local particle density $\rho({\bf r},t)$:
\begin{equation}
\label{eq:pressure}
\overline{P}_b(L_\perp) = \left\langle \int_{\bar{x}}^\infty \partial_\perp
  U(r_\perp) \rho(r_\perp) dr_\perp \right\rangle \approx  \langle  \rho(L_\perp) \rangle,
\end{equation}
where $\bar{x}$ is an arbitrary bulk point in $(0,L_\perp)$ and $\langle \cdot\rangle$ denotes time (or, due to ergodicity, ensemble) average.

Therefore, we can estimate the mechanical pressure exerted by the active fluid on the boundaries evaluating the average value of the density profile at the walls. While by doing so we cannot quantitatively determine its actual value (which depends upon microscopic, non-universal features) with our procedure we will be able to determine the scaling of the mechanical pressure with the separation $L_\perp$ between the reflecting walls. By fixing $r_\perp=0$ or, equivalently, $r_\perp=L_\perp$ one has
\begin{equation}
\langle \delta \rho (0) \rangle = \frac{C}{L_\perp}\sum_{q_\perp>0}
\frac{\Omega}{1+\Gamma^2 q_\perp^2}  \frac{1}{q_\perp^{\gamma}} \equiv C
  \,\Omega \,S_{L_\perp},
\label{eq:rho1}
\end{equation}
where we have introduced the sum
\begin{equation}
S_{L_\perp} \equiv \frac{1}{L_\perp}\sum_{n=1}^\infty
\frac{1}{1+\Gamma^2 q^2_\perp(n)}\,\,\,\frac{1}{q_\perp^{\gamma}(n) },
\label{sm:eq-SR}
\end{equation}
where $q_\perp(n) = \pi n/L_\perp$ is given after Eq.~\eqref{eq:v1}.\\
The dependence of $S_{L_\perp}$ on the (large) distance $L_\perp$ can be evaluated as explained in the appendix \ref{calculation_SR} and turns out to be
\begin{equation}
S_{L_\perp} =S_\infty + \frac{{\bm\zeta}(\gamma) }{\pi^\gamma}\frac{1}{L_\perp^{1-\gamma}}+{\cal O}\left(\frac{1}{L_\perp^{3-\gamma}}\right),
\label{eq:cas}
\end{equation}
where ${\bm\zeta}(\gamma)$ is the Riemann zeta function. \\
Note that $0<\gamma<1$ in the case we are interested in here. Accordingly, ${\bm\zeta}(\gamma)<0$ and therefore $S_{L_\perp}$ approaches $S_\infty$ from below. In the previous expression 
\begin{equation}
S_\infty =  \left[2\Gamma^{1-\gamma}\cos\left(\frac{\pi \gamma}{2}\right)\right]^{-1},
\label{eq:inf}
\end{equation}
corresponds to the contribution of the integral which emerges in the limit $L_\perp\to\infty$ from interpreting $S_{L_\perp}$ as a Riemann sum.\\
Putting everything together we can obtain the value of the density deviations at the boundaries $B = 0, L_\perp$ and thus, by Eq.\eqref{eq:pressure} the mechanical pressure's scaling with the separations between the reflecting walls. Up to higher-order corrections in $1/L_\perp$, it turns out to be given by
\begin{equation}
\overline{P}_b(L_\perp) \sim\overline{\delta \rho}(B)=\sum_{q_\perp>0} \widehat{\delta
    \rho}(q_\perp) = \Omega\, C\left[S_\infty + \frac{{\bm\zeta} (1+2\chi)}{\pi^{1+2\chi}}
 L_\perp^{2 \chi}\right].
 \label{eq:scal-drho}
\end{equation}
For $L_\perp \to \infty$ the finite-size contribution vanish and we retrieve the bulk pressure $\overline{P}_b(\infty) \equiv \Omega C S_\infty$, which is correctly found to be positive. For finite $L_\perp$ one instead must also consider the second term, which is negative\footnote{The argument of the Riemann zeta function ${\bm \zeta}$ is positive and bounded between $0<1+2\chi<1$ since, as we have seen in Section \ref{scaling_exp}, the value of the roughness scaling exponent $\chi=-1/3$ in $d=2$. In the interval $(0,1)$ the Riemann ${\bm \zeta}$ function is negative.} since ${\bm\zeta}(1+2\chi)<0$, and scales algebraically with the wall separation as $L_\perp^{2\chi}$. This means that the pressure can be expressed as
\begin{equation}\label{eq:pow_th}
\overline{P}_b(L_\perp) =\overline{P}_b(\infty) - A L_\perp^{2\chi},
\end{equation}
with $A\equiv \Omega C|{\bm\zeta}(1+2\chi)|/\pi^{1+2\chi} > 0$.\\
In the typical Casimir setup in which two infinite reflecting walls, separated by a distance $L_\perp$, are immersed in a much larger flocking active fluid of transverse size $L \!\gg\! L_\perp$, the bulk contributions exerted on the two faces of the walls cancel out and one is left with the net pressure
\begin{equation}
\Delta \overline{P}_b(L_\perp) \equiv
\overline{P}_b(\infty)\!-\!\overline{P}_b(L_\perp) = A L_\perp^{-\alpha},
\label{eq:Casimir}
\end{equation}
due to the interaction between the confining walls.\\
This shows that two infinite and parallel reflecting walls, separated by a distance $L_\perp$ and immersed in an active polar fluid in the flocking phase experience an \emph{attractive}, long-range Casimir-like pressure which decays algebraically with the boundary separation. 

\subsection{Numerical results in the dilute limit}\label{sec:casimir_num}
We now verify our theoretical prediction by numerical simulations of the dilute limit confined VM deep in the bulk homogeneous phase.\\
We start from the pressure definition of Eq.\eqref{eq:pressure}: microscopically, this amounts at defining the pressure as the average number of collisions per unit length
\begin{equation}
    \overline{P}_b(t)\equiv\langle M(t)\rangle/L_\|
\end{equation}
where $M(t)$ is the instantaneous number of collisions on the confining wall. The time-series of $M(t)$, once a transient of about $10^5$ time-steps is discarded in order to erase memory of the initial conditions and obtain a flocking stationary state along the walls, can be averaged to compute the average mechanical pressure $\overline{P}_b(L_\perp)$. We have verified that the autocorrelation time $\tau$ of the instantaneous pressure $\overline{P}_b(t)$ is roughly equal to $L_\perp$, scaling linearly with the transversal separation. Accordingly, varying $L_\perp$ we considered time-series of increasing length, also resorting to averages over different realizations to increase the statistics. Overall, our averages are performed over around $n = 10^5$ independent data-points. Error bars are then evaluated as one standard error $\sigma_S = \sigma/\sqrt{n}$, with $\sigma$ being the standard deviation of the pressure time-series.\\
In the following, we will adopt this definition of the pressure, but in Section \ref{universality_dilute} we will also verify that an alternative definition in terms of the nominal momentum exchanged in the collisions yields the same results (up to a constant scaling factor).

\begin{figure}[hbt!]
\includegraphics[width=1\textwidth]{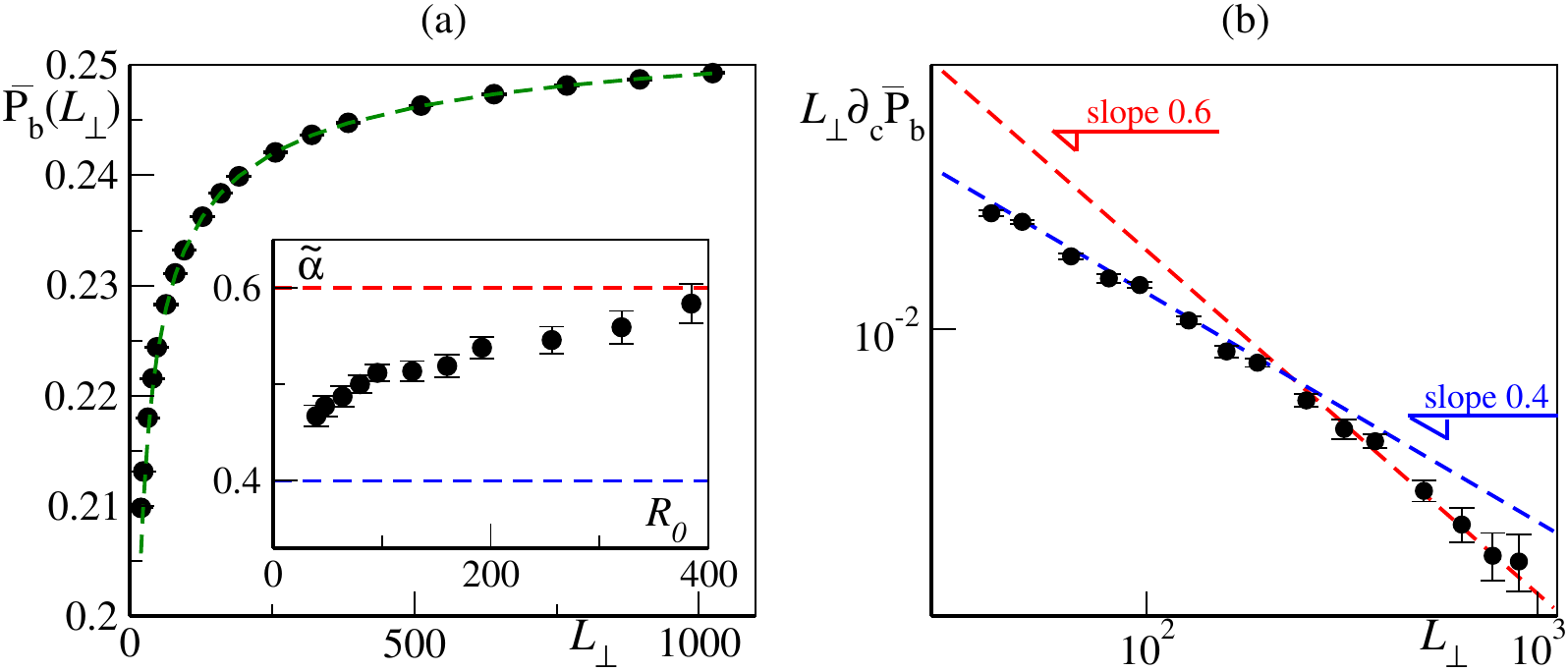}
\caption{{\bf Scaling of the Casimir-like pressure exerted on the confining walls} (a) Time-averaged mechanical pressure exerted by the polar fluid on the confining walls as a function of their separation $L_\perp$ for $\rho_0 = 2.0$ and $\eta\!=\!0.21$. The dashed green line is the best fit by Eq.~(\ref{eq:pow}) for $L_\perp\!\geq\!R_0=80$. Inset: Estimates of the effective exponent $\tilde{\alpha}(R_0)$ as a function of the lower extreme $R_0$ of the interval of values of $L_\perp$ of the data which are included in the least-square fit of the exponent. (b): Rescaled centered finite difference of $\overline{P}_b$ as a function of $L_\perp$. The dashed blue and red lines indicate respectively an effective exponent 0.4 and 0.6. (see text). Error bars represent one standard deviation.} 
\label{fig1}
\end{figure}

Fixing the longitudinal size $L_\|\!=\!512$, we have measured the average mechanical pressure $\overline{P}_b(L_\perp)$ in systems of various transverse sizes $L_\perp$. We first consider perfectly reflecting walls, i.e., $\beta = 1$.\\
Our results, reported in Fig.~\ref{fig1}(a), show that the pressure increases upon increasing $L_\perp$ and that indeed it approaches a bulk value $\overline{P}_b(\infty)$ as $L_\perp\!\to\!\infty$. As we have seen, this value characterizes the mechanical pressure exerted on a reflecting boundary by a semi-infinite system. In equilibrium, it typically originates from the bulk contribution to the free energy.

Our data, presented in Fig.~\ref{fig1}(a), can be indeed fitted by the three-parameter function
\begin{equation}
\overline{P}_b(L_\perp) =\overline{P}_b(\infty) - A L_\perp^{-\alpha},
\label{eq:pow}
\end{equation}
with $\overline{P}_b(\infty)$ and $A$ positive and $\alpha \approx 0.5$, which seems to be somehow off from the expected scaling given by Eq.\eqref{eq:scal-drho}, $\alpha=-2\chi$. We indeed recall that, as reported in Sec.\ref{scaling_exp}, the best large-scale numerical estimates of the TT exponents available in $d\!=\!2$ yield asymptotically $\chi\!=\!-0.31(2)$ and also recent theoretical argument suggests that $\chi = -1/3$. In the following we investigate more carefully the value of the exponent $\alpha$.

The three parameter fit of Eq.\eqref{eq:pow}, shown in Fig.\ref{fig1}(a), does not allow for an accurate estimate of the {\it Casimir exponent} $\alpha$. This can be estimated more accurately by considering the centered finite difference 
$\partial_c f(x)$ of the average pressure, which approximates the first derivative of $f(x)$ up to corrections of order $f'''(x)$. Assuming that Eq.~(\ref{eq:pow}) holds, we have $L_\perp \partial_c \overline{P}_b(L_\perp) \approx\alpha A L_\perp^{-\alpha}$, which, for instance, renders the estimate $\alpha \!=\! 0.50(1)$ by fitting the data in the region $L_\perp \ge R_0 = 80$. A closer inspection, however, reveals a systematic dependence of the effective exponent $\tilde{\alpha}(R_0)$ on the choice of the lower bound $R_0$ of the fitting region. This is shown in the inset of Fig.~\ref{fig1}(a) where the least square fits are limited to the regions $L_\perp\geq R_0$. This behavior suggests a finite-size crossover within the range $0.4 \lesssim \tilde{\alpha}(R_0)\lesssim 0.6$ as illustrated in Fig.~\ref{fig1}(b).\\
Note, indeed, that the analytical argument of Ref.~\cite{TTR} would give $\alpha=2/5$ under the assumption that certain nonlinear bulk contributions to TT equations are irrelevant. While numerics clearly shows this is not the case for large $L_\perp$, Ref.~\cite{Benoit2019} also reports a crossover which makes the effective value of $|\chi|$ increase upon increasing the system size, possibly due to the late onset of relevant nonlinear contributions. Overall, these considerations are compatible with the finite-size crossover of the Casimir exponent $\alpha$ from the small size behavior $\alpha\approx0.4$ to the asymptotic value $\alpha\approx0.6$ as reported  in Fig.~\ref{fig1}(b) for large enough $L_\perp$.

To summarize, our numerical simulations confirm the attractive character of Casimir-like forces as predicted by Eq.\eqref{eq:Casimir}, with a Casimir scaling exponent $\alpha$ which controls the algebraic decay with the separation between the boundaries and whose asymptotic value is compatible with the theoretical prediction $\alpha=-2\chi$.

\subsection{Universality of the Casimir force scaling in the dilute limit}\label{universality_dilute}
In this section we present additional numerical evidence of the robustness and of the universality of the results just presented.
\subsubsection{Alternative definition of the mechanical pressure}
The exact expression of the exchanged momentum at the reflecting boundaries depends on the details of the collisions between the particles and the confining wall. So far, we considered the pressure $\overline{P}_b(L_\perp)$ defined only on the basis of the number of collisions per unit length and unit time. \\
Here we consider, instead, an alternative definition of the microscopic mechanical pressure $\overline{P}^{\rm MOM}_b(L_\perp)$ as the average momentum $\Delta p$ exchanged per unit length and unit time. We assume that $\Delta p$ is entirely exchanged with the boundary (and not with the surrounding viscous fluid/substrate), so that
\begin{equation}
    \Delta p\! =\! 2 \beta v_0 n_\perp
\end{equation}
whenever the particle collides with one of the boundaries. Here $\beta$ is the reflection parameter defined earlier.

\begin{figure}[hbt!]
\centering
\includegraphics[width=1\textwidth]{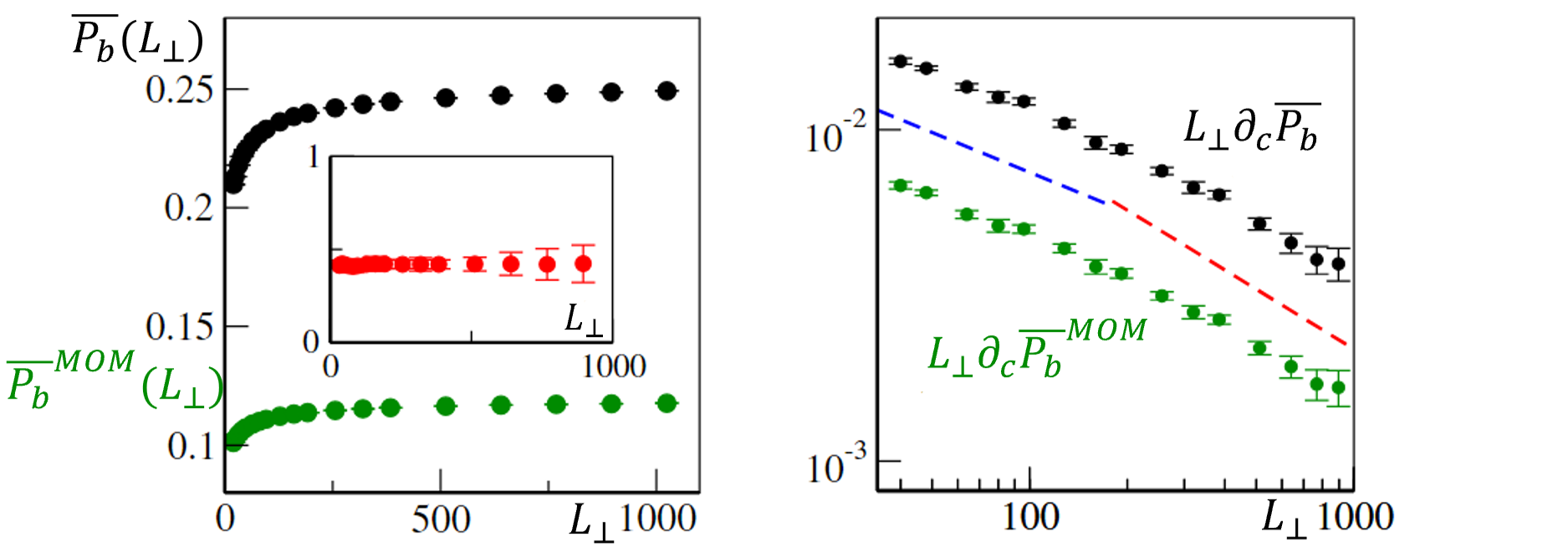}
\caption{{\bf Universality of the pressure's scaling - Different definition} Left - Numerical estimate of $\overline{P}_b^{\rm MOM}(L_\perp)$ (green dots) and $\overline{P}_b(L_\perp)$ (black) for fully reflecting boundaries, i.e., $\beta=1$, with $\eta = 0.21$, $\rho_0 = 2.0$, $v_0 = 0.5$, and $L_\| = 512$. Right - Rescaled centered finite differences of the pressures presented in panel (a) as a function of $L_\perp$. Their ratio is shown in the inset of panel (a). The dashed lines mark the pre-crossover (with exponent $\alpha=0.4$, blue) and asymptotic algebraic decays ($\alpha = 0.6$, red). For clarity, only the asymptotic regime is shown.}
\label{fig1_sm_1}
\end{figure}

If, at a certain time $t$, there are $M(t)$ such collisions with momentum exchange $\Delta p_n$, for $n\!=\!1,\ldots, M(t)$, the instantaneous mechanical pressure, considering both walls, can be defined as
\begin{equation}
P^{\rm MOM}_b (t) \equiv \frac{1}{2L_\|} \sum_{n=1}^{M(t)} |\Delta p_n|. 
\end{equation}
We then take a long-time average of this pressure, finally yielding $\overline{P}_b^{MOM} (L_\perp)= \langle P_b^{MOM}(t)\rangle_t$.\\
We performed numerical simulations with the same parameters (and averaging procedures) considered in Section \ref{sec:casimir_num}. Figure~\ref{fig1_sm_1} shows that $\overline{P}^{\rm MOM}_b(L_\perp)$ is actually characterized by the same scaling with the wall distance $L_\perp$ as the one exhibited by $\overline{P}_b(L_\perp)$ defined in Eq.\eqref{eq:pow}. 

Indeed, expressing the boundary pressure $\overline{P}_b$ as a function of the relevant hydrodynamics fields $\rho$, and ${\bf v}$ and expanding at the first order around $\rho_0$ and $p_0$ \cite{AKK-15} yields, after averaging
\begin{eqnarray}   
\label{expansion}
    \overline{P}_b (\rho, {\bf v}) &\approx& P_b(\rho_0, p_0) + \\ &+&\partial_\rho P_b (\rho_0, p_0) \langle\delta\rho\rangle + \partial_{v_\perp} P_b (\rho_0, p_0) \langle v_\perp\rangle\,.\nonumber
\end{eqnarray}
The third term on the r.h.s vanishes at the boundary due to the presence of reflecting walls, which we model by zero Dirichlet boundary condition for the transversal velocity $v_\perp$. The resulting expression for the pressure depends only on the density deviations and thus should exhibit the same scaling as the average mechanical pressure given by Eq.\eqref{eq:pow}.

\subsubsection{Partially reflecting boundaries}
\begin{figure}[hbt!]
\centering
\includegraphics[width=1\textwidth]{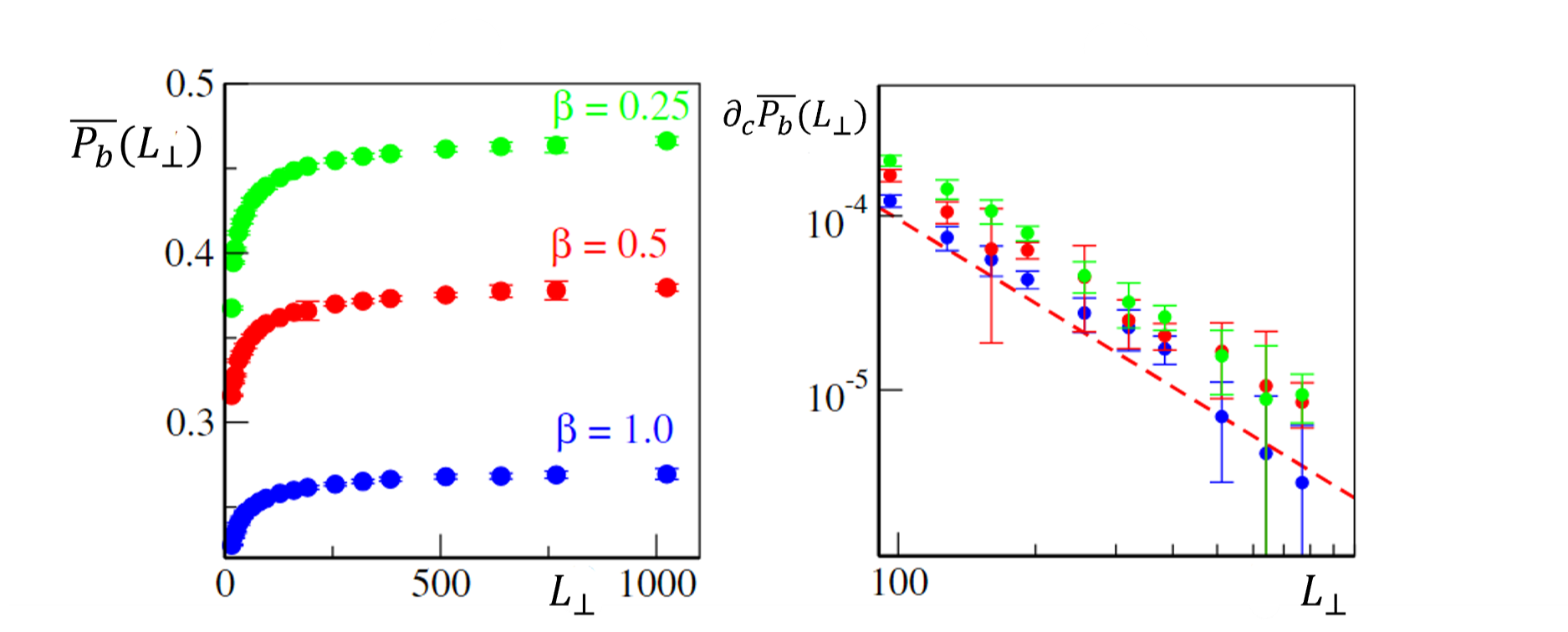}
\caption{{\bf Universality of the pressure's scaling - Boundary interaction} Left - Numerical estimate of $\overline{P}_b(L_\perp)$ for various values of the reflection parameter $\beta = 1.0$ (blue), $0.5$ (red), $0.25$ (green). The remaining parameters are $\eta = 0.24$, $v_0 = 0.5$, and $L_\| =512$. Data have been averaged over $10^6 \sim 10^7$ data-points. Right - Corresponding centered finite differences compared to the algebraic decay $\sim L_\perp^{-\alpha-1}$ with $\alpha = 0.6$ (red dashed line). For clarity, only the asymptotic regime is shown.}
\label{fig1_sm_2}
\end{figure}

We also consider the case in which the active particles have partially reflecting collisions with the boundaries.\\
Simulations with $\beta = 0.5$ and $0.25$, reported in Fig.~\ref{fig1_sm_2}, show that the exponent $\alpha$ of the algebraic approach $\propto L_\perp^{-\alpha}$ of the mechanical pressure $\overline{P}_b(L_\perp)$ to the bulk value $\overline{P}_b(L_\perp\to\infty)$ upon increasing $L_\perp$ is not affected by the value of $\beta$ and stays compatible with our analytical estimate $\alpha= - 2\chi$. This confirms a certain degree of universality for repelling confinements, with post-collision particles moving away from the wall. 
\\

\begin{figure}[hbt!]
\centering
\includegraphics[width=1\textwidth]{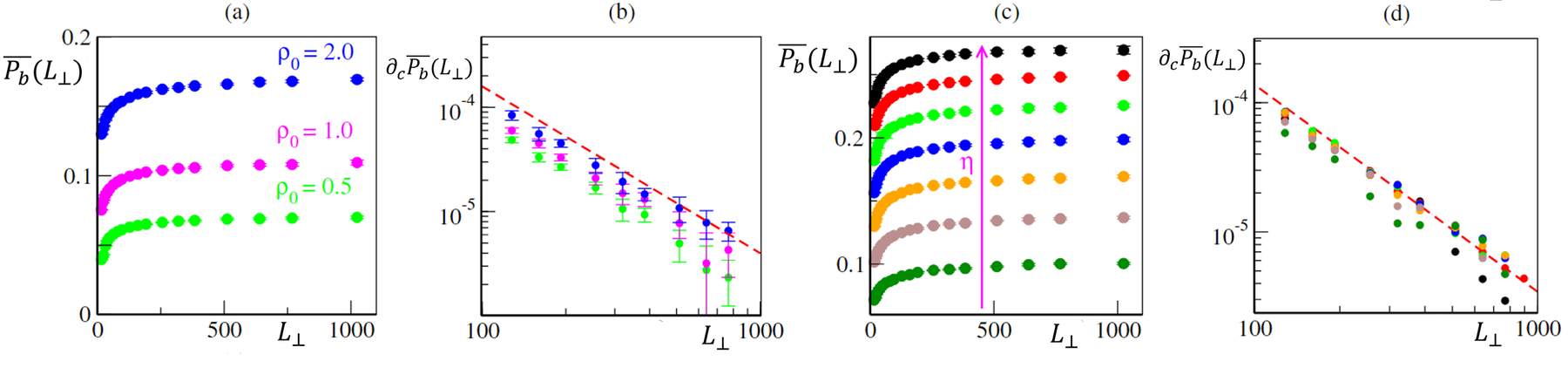}
\caption{{\bf Universality of the pressure's scaling in the ordered phase} (a) Numerical estimate of $\overline{P}_b(L_\perp)$ for various values of the density $\rho_0 = 2.0$ (blue), $\rho_0 = 1.0$ (magenta) and $\rho_0 = 0.6$ (green). The remaining parameters are $\beta = 1.0$, $\eta = 0.12$, $v_0 = 0.5$, and $L_\| = 512$. (b) Corresponding centered finite differences compared to the asymptotic algebraic decay $\propto L_\perp^{-\alpha-1}$ with $\alpha = 0.6$ (red dashed line). (c) Time-averaged mechanical pressure exerted by the polar fluid on the confining walls as a function of their separation $L_\perp$ for $\eta \in$ [0.06, 0.09, 0.12, 0.15, 0.18, 0.21, and 0.24]. (d) Centered finite difference of $\overline{P}_b(L_\perp)$ as a function of $L_\perp$. The dashed red line mark the asymptotic algebraic decay $\propto L_\perp^{-\alpha-1}$ with $\alpha = 0.6$. For clarity, only the asymptotic regime is shown in (c)-(d). Data have been averaged over $10^6 \sim 10^7$ data points. Error bars are omitted for the sake of clarity.}
\label{fig2_sm}
\end{figure}

\textbf{Exploration of the flocking phase}\\
We finally investigate the dependence of the mechanical pressure $\overline{P}_b(L_\perp)$ on the two control parameters: the global density $\rho_0$ and the amplitude of the microscopic noise $\eta$. In Fig.~\ref{fig2_sm}(a) we show the mechanical pressure $\overline{P}_b(L_\perp)$ as obtained for three different values of the particles density, i.e., for $\rho_0 = 2.0$, $1.0$, and $0.5$, keeping the value of the microscopic noise $\eta$ fixed. We find that the bulk pressure $\overline{P}_b(L_\perp\to\infty)$ increases monotonously with the global density $\rho_0$, while the value of the {\it Casimir} exponent $\alpha \approx 0.6$ does not depend on $\rho_0$ (see Fig.~\ref{fig2_sm}(b)), as long as the active fluid is within the flocking phase \cite{VicsekPD}. 

Furthermore, we also inestigated that the dependence of the mechanical pressure $\overline{P}_b(L_\perp)$ on the value of the microscopic noise $\eta$, now keeping the value of the global density $\rho_0$ fixed. The results of these simulations are reported in Fig.~\ref{fig2_sm}(c)-(d).      
The bulk pressure $\overline{P}_b(L_\perp\to\infty)$ is found to increase monotonously also with the amplitude of the noise $\eta$ while no significant deviation from the {\it Casimir} exponent $\alpha \approx 0.6$ is observed (see Fig..~\ref{fig2_sm}(d)). Overall, this confirms the validity of our result for a broad set of parameters within the polar liquid phase.

\subsection{Pressure in the disordered phase} 
The long-range nature of the Casimir-like force between the two (partially) reflecting confining walls is connected to the long-range correlations which characterize the ordered phase \cite{TT1}. As a counterexample, we have also performed numerical simulations in the disordered gas-like phase where, in the absence of spontaneous symmetry breaking, correlations are short-ranged over a length scale, the correlation length scale $\ell_c = \ell_c(\eta)$, which is finite in the disordered phase and increases upon lowering the noise $\eta$.\\
In particular, we have considered the case $\rho_0 = 2.0$ and noise strength $\eta = 0.60$. These values ensure that the system is in the disordered gas phase \cite{VicsekPD}, avoiding the region of parameter in which the system is micro-phase separated, which we do not discuss here. 

\begin{figure}[hbt!]
\centering
\includegraphics[width=0.9\textwidth]{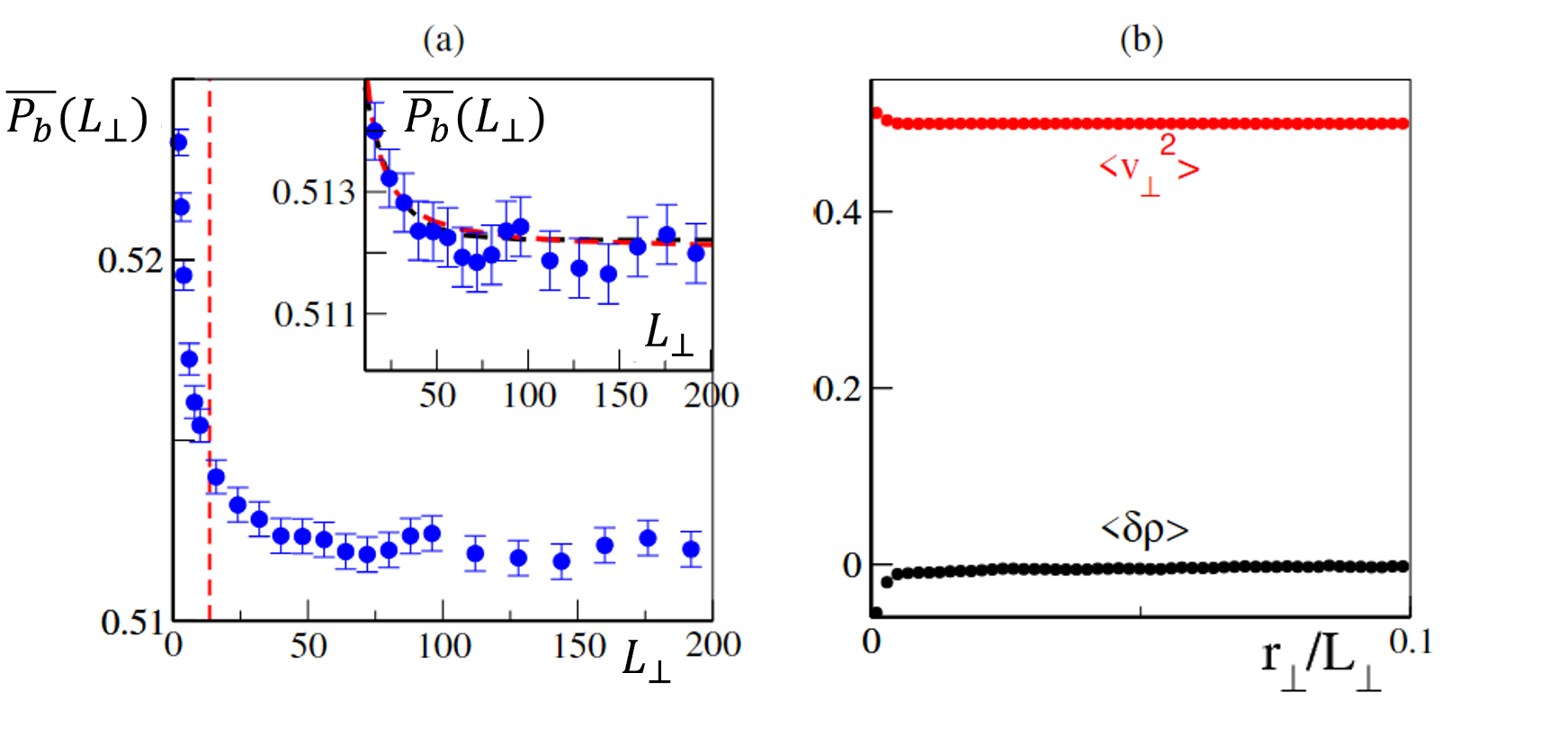}
\caption{{\bf Scaling of the pressure and field profiles in the disordered phase} Pressure and fields behavior in the confined disordered phase,  with parameters $\beta=1$, $\eta = 0.60$, $\rho_0 = 2$, $v_0=0.5$, and $L_\|=512$. (a) Mechanical pressure $\overline{P}_b(L_\perp)$ as a function of the transverse separation $L_\perp$. The vertical dashed red line corresponds to the correlation length $\ell_c$  in the disordered phase (see text). Inset: exponential (black dashed line, $\overline{P}_b(L_\perp) \approx 0.5+0.005\,e^{-0.07 L_\perp}$) and algebraic ($\overline{P}_b(L_\perp) \approx 0.5 + 0.06\,L_\perp^{-1.3}$, red dashed line) fits of the function $\overline{P}_b(L_\perp)$ for $L_\perp$ larger than the correlation length $\xi$ in the disordered phase. (b) Average density deviations (black) and transverse velocity fluctuations (red) as functions of the rescaled transverse coordinate $r_\perp/L_\perp$ (zoom of the near-wall behavior). Data have been averaged over $10^6$ data points.}
\label{fig3}
\end{figure}

The behavior of the confined disordered system, presented in Fig.~\ref{fig3}, is qualitatively different from that obtained in the ordered phase discussed above.\\
In particular, Fig.~\ref{fig3}(a) features a change of the sign of the finite-size contribution to the total boundary pressure, resulting in a net \emph{repulsive} force between the two boundaries. Moreover, this force has a significantly shorter spatial range, as shown by the fast decay of the pressure to its asymptotic value upon increasing $L_\perp$.

As we have noted, the disordered phase is characterized by a \emph{finite} correlation length $\ell_c$, which we have estimated numerically from the equal-time correlations of the transverse velocity and which is indicated in Fig.~\ref{fig3}(a) as a dashed red vertical line. Once the transverse separation $L_\perp$ exceeds this finite bulk correlation length $\ell_c$, the decay of the mechanical pressure to its bulk value $\overline{P}_b(\infty)$ becomes compatible with a fast exponential decay. 

Finally, we also note that the dependence of the density and of the transverse velocity fields on the distance from a wall, shown in Figure~\ref{fig3}(b), is much less pronounced than in the flocking phase, with a comparatively small boundary layer, characterized by a moderate decrease of the density and increase of velocity fluctuations near the boundaries. This behavior is the opposite of what is observed in the flocking phase; at the moment we cannot offer a theoretical explanation of these preliminary numerical results.
\newpage

\section{Universality beyond the dilute limit}
The Vicsek model is of course a simplistic model which takes a somehow abstract view of how active particles interact with the boundaries and glosses over steric interaction between particles. One might believe that introducing volume exclusion forces between particles affects the overall steady state distribution of particles in the confined geometry, suppressing density variations (especially in the hyper-confluent regime) and ultimately invalidating our dilute limit results. \\
Here we show that this is not the case and that both the extensitivity of boundary layers and the Casimir-like pressure are present even in the collisional Vicsek model (CVM) at high packing fractions.

\begin{figure}[hbt!]
    \centering
    \includegraphics[width=0.5\linewidth]{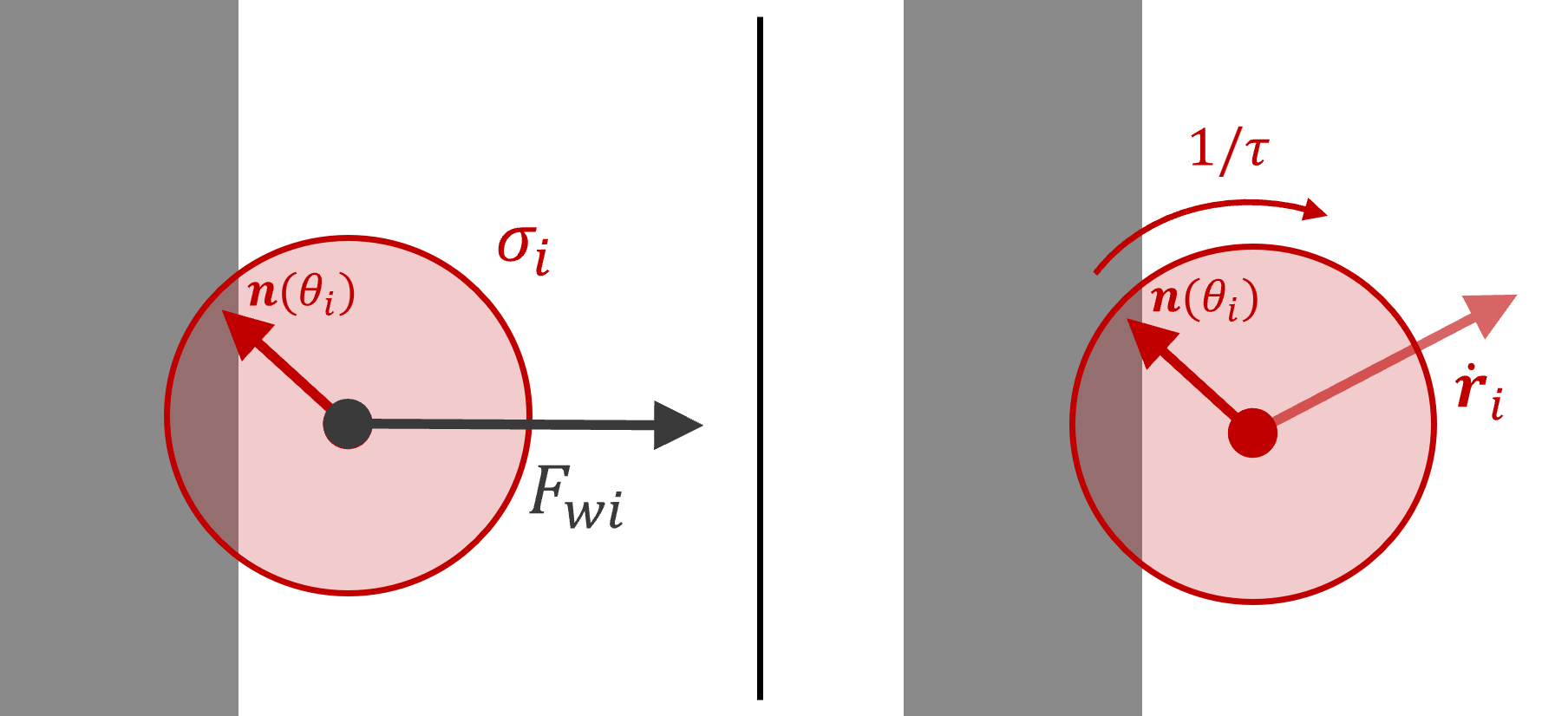}
    \caption{{\bf Schematic representation of the interaction between a particle and the wall} The force $F_{wi}$ acts on particle $i$ whenever the particle is sufficiently close to the wall ($r_{iw}<\sigma_i$). The wall harmonically repels the particle, which in turn will move  away from it} 
    \label{interaction_particles_walls}
\end{figure}

The CVM, presented in Section \ref{other_ways}, can be used to test the generality of our results, in particular the effect of volume exclusion between particles since in the CVM particles interact with each other through soft harmonic repulsion. Moreover, the alignment interaction that particles feel relies on the local force exerted by other particles, not on the neighbors velocity explicitly. This allows us to specify a precise and {\it natural} definition for the force exerted by the boundaries on the active particles, leading to an unambiguous definition of the mechanical pressure exerted by the active fluid on the walls.\\
The interaction with boundaries is naturally modeled as a harmonic repulsion with stiffness $k$ so that Eq.\eqref{velocity_cvm} is modified as follows
\begin{equation}\label{dynamics_cvm}
    \Dot{\mathbf{r}}_i = v_0 \hat{\mathbf{n}}(\theta_i) + \mu \sum_{j = 1}^N \mathbf{F}_{ij} + k\mathbf{F}_{wi}
\end{equation}
with $\mu$ controlling the intensity of the harmonic repulsion and $\mathbf{F}_{wi}$ being the interaction between particle $i$ and the wall which may be written as
\begin{equation}
    \mathbf{F}_{wi} =
\begin{cases}
    0 & \text{if }  r_{iw} > \sigma_i \\
[r_{iw} - \sigma_i] \hat{\mathbf{x}} & \text{if } r_{iw} < \sigma_i
\end{cases}
\end{equation}
where $\sigma_i$ is the (polydisperse) particle interaction radius and $r_{iw}$ is the distance from the closest boundary. The geometry is always the same confining geometry used in previous simulation with the VM, see Fig.\ref{interaction_particles_walls} for a schematic representation.\\
As in the free model, self propulsion orientations are subjected to (overdamped) torques aligning it with the velocity orientation $\psi_i$ of particle $i$
\begin{equation}\label{cvm_theta_ripe}
    \dot{\theta}_i = \frac{1}{\tau}(\theta_i -\psi_i) +\xi_i
\end{equation}
Note that the force exerted by the boundaries on the particles only involves their center of mass and do not explicitly involve the reflection or the partial reflection of the transversal component of the self-propulsion direction. Active particles, however, change their self-propulsion direction as a consequence of the torques that tend to align it with the particle velocity $\dot{\bf r}_i$, see Eq.\eqref{cvm_theta_ripe}. As in the case of bulk inter-particle interactions, these torques are assumed to be dissipated on some unspecified substrate.\\
It is therefore natural to define the mechanical pressure exerted on the boundaries as minus the total force per unit length exerted by the walls on the active particles. The pressure exerted on the wall is then simply given by 
\begin{equation}
    P_W = \frac{1}{L_\|}\sum_i |\mathbf{F}_{wi}|
\end{equation}
Note that this rather natural definition of the mechanical pressure at the boundaries does not involve the torques applied on the self propulsion direction, and directly maps into the mesoscopic definition of the pressure given in Eq.\eqref{eq:pressure}, further justifying our results obtained at the hydrodynamic level.

Numerical simulations carried over in the hyper-confluent Toner \& Tu phase \cite{giavazzi2017giant} confirm the we retrieve both of our main results obtained in the case of the VM: extensive boundary layers and a finite size scaling of the pressure.\\
Simulations are performed in the ordered phase, using the same parameter values that have been used to reproduce the flocking TT behavior of confluent epithelial tissues \cite{giavazzi2017giant}: self-propelling velocity $v_0 = 0.5$, turning time-scale $\tau = 2$, stiffness of particle particle interaction $\mu = 1$, noise $\eta  = 0.42$, packing fraction $\phi = 1.2$ (that is the area occupied by particles over the whole system area) and finally polydispersity $p = 0.4$ (the particles' radii is uniformly distributed between $[1-p/2, 1+p/2]$).\\
The system is initialized with random positions, orientation and radii extracted from a uniform distribution and are evolved for a total time of $T=10^6$ time units, after having discarded a transient time of $T_0 \approx 10^5$ time units. System size in the longitudinal (flocking) direction is fixed at $L_\| = 512$, while various values of the walls separation $L_\perp$ are considered, in order to verify as usual the correct finite size scaling. Finally the value of the stiffness at the wall is chosen to be $k = 4$. The value of $k$ can, in principle, be taken even larger: in the limit of $k\to \infty$ this results in very strong forces applied on the particle's velocities that will in turn result in strong torques applied on the orientation. On the other hand if $k$ is too small, say comparable to $\mu$, the confining force is not strong enough to keep the particles in a confined domain. In this spirit the value of $k = 4$ is a good compromise between these two effects, but we expect the results presented here hold for a wide range of values of $k$. 

\begin{figure}[hbt!]
    \centering
    \includegraphics[width=1\linewidth]{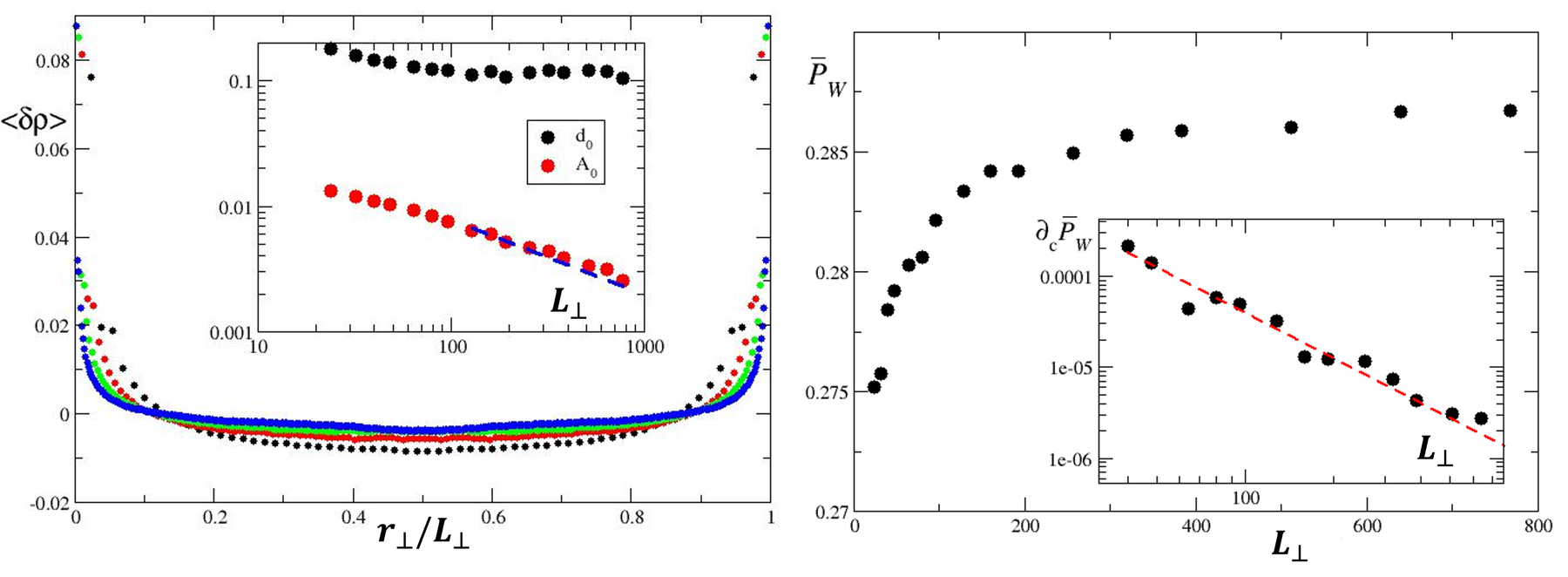}
    \caption{{\bf Average density fluctuations and pressure scaling in the CVM}: Left - Average density deviations $\langle\delta\rho\rangle$ profiles, for different $L_\perp = 64$ (black), $128$ (red), $256$ (green) and $512$ (blue), as a function of the rescaled wall separation $r_\perp/L_\perp$ show accumulation of particles at the boundaries. Inset - Scaling of $d_0$ (extension of the boundary layer) and $A_0$ (integrated fluctuations of the density) showing the same scaling $d_0\propto L_\perp$ and $A_0\sim1/L_\perp^{-0.6}$ (blue dashed line) as for Vicsek flocks, indicating the presence of extensive boundary layers. Right - Scaling of the mechanical pressure $P_W$ with the wall separation $L_\perp$. Inset - Centered finite differences of the pressure show that the pressure $P_W$ scales algebraically with an exponent $\alpha \approx 0.6$ (red dashed line).}
    \label{density_pressure_cvm}
\end{figure}

As it is shown in Fig.\ref{density_pressure_cvm}, particle accumulate at the boundaries: as in the case of the VM a finite-size scaling analysis (inset of left panel) reveals that boundary layer $d_0$ is extensive ($d_0\propto L_\perp)$. Note however that the amplitude of this effect (density enhancement) is much smaller than in the case of Vicsek flocks, possibly due to the repulsion between particles and their packing fraction $\phi > 1$ which might damp the density fluctuations. The amplitude of density fluctuations $A_0$, go to zero in the thermodynamic limit $A_0 \sim 1/L_\perp^{-0.6}$, showing that also in this case it is possible to recover the physics of the free unperturbed system.\\
Furthermore, the scaling of the pressure $P_W$ is also algebraic and compatible with a value of $\alpha \approx 0.6$, (right panel of Fig.\ref{density_pressure_cvm}), which we found to be the post-crossover value of the Casimir exponent for Vicsek flocks. These numerical results are therefore analogous to the one obtained via the VM in the dilute limit. 

While in the hyper-confluent regime the magnitude of the density variations and of the Casimir force is largely diminished by volume exclusion effects which reduce the compressibility of the active fluid, it is remarkable that they share the same extensive behavior and scaling of the dilute case. Moreover, the CVM set-up let us define rather unambiguously a mechanical boundary pressure that can be directly mapped in our mesoscopic definition, Eq.\eqref{eq:pressure}.\\ 
Overall, these results confirm the generality and universality of our results for an active polar fluid confined between two parallel reflecting boundary walls, which in particular extend well beyond the dilute limit to hyper-confluent flocking fluids.
\newpage

\section{Discussion}
We have shown that, in the flocking phase, an active polar fluid confined between two infinite and parallel reflecting walls separated by a distance $L_\perp$ is characterized  by an excess number density at the boundaries. This creates {\it extensive} boundary layers, as opposed to the finite ones usually observed in confined scalar active matter, discussed in Sec.\ref{boundaries_scalar}. This behavior is ultimately induced by the diverging bulk correlation length and by the coupling between the slow fields of density and velocity, which characterizes vectorial active matter. In turn, this leads to the emergence of an \emph{attractive} long-range force $\!\propto\! L_\perp^{2\chi}$ on the two walls. After a finite size crossover, our numerical results in $d=2$ are compatible with the numerical and theoretical estimate $\chi=-1/3$.\\
While the exact value of this force depends on the
microscopic details of the interactions with the wall, 
the power of its algebraic dependence on the transversal separation is
universal and ultimately controlled by bulk fluctuations.\\
We based our analytical discussion on a slightly
simplified version of TT equations, but the 
same conclusion can be drawn from the complete TT theory (see appendix \ref{full_TT}). 
In this latter case too, the Casimir force turns out to be attractive
provided one considers the specific values of the transport
coefficients computed from the direct coarse-graining of microscopic
Vicsek-like models \cite{Bertin, Ihle}.
Our result also implies that in principle the bulk scaling exponent $\chi$ can be alternatively measured by the finite-size scaling of the mechanical pressure exerted on the boundaries.

The non-equilibrium Casimir-like force investigated here is analogous to the critical
Casimir force $\!\propto\! T L_\perp^{-d}$ at equilibrium in $d$ spatial dimensions, which arises due to long-ranged correlated fluctuations, but with important differences. In the latter case, in fact, the algebraic dependence on $L_\perp$ is fixed by the finite-size scaling behavior of the free energy from which the force is derived (see, e.g., Ref.~\cite{KB-94}).
In the present case, instead, the long-ranged force is directly caused by a
fluctuation-induced accumulation of density at the boundaries and is controlled by the field scaling exponent $\chi$. \\
Note finally that, since the Vicsek fluid is not incompressible, the force discussed here bears no relation with the hydrodynamic interactions
arising in molecular fluids due to incompressibility \cite{kundu2015}.

A detailed study of the effects of confinement in other regions of the phase diagram, e.g., in the micro-phase separated band phase \cite{Chate2,VicsekPD} has yet to be performed. However, preliminary results obtained in the disordered phase indicate the emergence of a \emph{repulsive} force between the reflecting
walls with a much shorter range, possibly controlled by the
finite correlation length of fluctuations in the disordered phase. 
Correspondingly, the density profile is depleted near the reflecting walls. This may seem at odds with what mentioned in Section \ref{boundaries_scalar} (see Ref.~\cite{Ray}) for confined active Brownian particles (ABPs). Note, however, that the boundaries considered there are not reflecting and do not change the persistent directions of the ABPs.

Finally we have tested the generality of our results by exploring the effect of confinement in the collisional Vicsek model, where soft harmonic repulsion are present and we are far from the dilute limit. Our numerical results indicate that, even at high packing fractions, one can retrieve the extensitivity of boundary layers and the emergence of a Casimir-like pressure. This last test also provides an unambiguous definition of the pressure, which further legitimizes the results found for the VM, where the exact definition of the pressure is left unspecified. This confirms the universality of the depicted scenario for generic confined polar fluids.

\fancyhead[LO]{{\it Virtual confinement}} 
\fancyhead[RE]{{\it Virtual confinement}} 
\chapter{Nematic active matter under virtual confinement}\label{ch_nematic}
\begin{figure}[hbt!]
    \centering
    \includegraphics[width=0.75\linewidth]{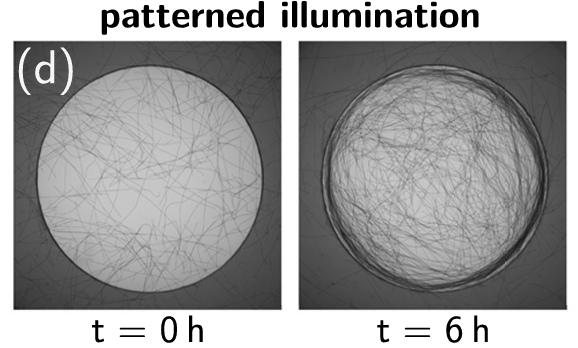}
    \caption{{\bf Accumulation of filamentous cyanobacteria at light boundaries} Left - Snapshot of the initial configuration where bacteria are homogeneously distributed both in and out of the illuminated region. Right - Steady state of the filamentous bacteria that have been exposed to a circular light shape for about 6 hours. Figure adapted from \cite{kurjahn2024collective}.}
    \label{fig:pattern_illumination}
\end{figure}

As we have have shown in Chapters 3, 4 the environment is crucial in determining the behavior of a collection of flocking particles, since it has a far-reaching impact also deep in the bulk. So far, we dealt with the mechanical confinement, by hard (VM) of soft walls (CVM), that generates aligning torques at the boundaries with particles re-orienting in order to move away from the wall after the collision.\\
Here, motivated by a recent work on filamentous cyanobacteria \cite{kurjahn2024collective}, we consider a more subtle confinement mechanism. It is known that these bacteria are able to respond to light gradients by reverting their direction of motion: more specifically, they can reverse their direction of motion either when they experience an increase (photobocic) or decrease (scoto-phobic) in illuminance \cite{wilde2017light}. By shaping light in order to produce compact light patterns (e.g. a circle, square ecc. which are strongly illuminated and surrounded by darker regions), bacteria with scotophobic response are able to accumulate and align their direction of motion with this {\it virtual} boundary, as shown in Fig.\ref{fig:pattern_illumination}.

In the first section of this chapter we will show how it is possible to obtain the same phenomenology in a minimal model for nematic polar rods (cyanobacteria align nematically and can be described by a nematic polar rod model), first introduced in \cite{patelli2019understanding}. Interestingly, accumulation and alignment of particles at the virtual boundaries happens in a very wide region of the parameter space, irrespectively of the phase (disordered, homogeneous nematic, nematic chaotic) of the corresponding boundary free system. Moreover, the alignment does not seem to depend on the curvature of the confining shape.\\
These suggests that some insights about photo-tactic confinement can be obtained studying a simple case such as the slab geometry. Our numerical analysis will confirm alignment with and accumulation at the boundaries. Quite surprisingly we also realize that accumulation and alignment can be obtained even in the absence of repulsion if we introduce an explicit escape rate at the boundary. This fact seems to suggest that the main effect of inter-particle repulsion could be to induce an effective escape rate at such boundaries. We will finally take advantage of this analogy to write down a simple mesoscopic level description of confined polar rods.
\newpage

\section{Microscopic model}
\begin{figure}[hbt!]
\centering
\includegraphics[width=0.5\linewidth]{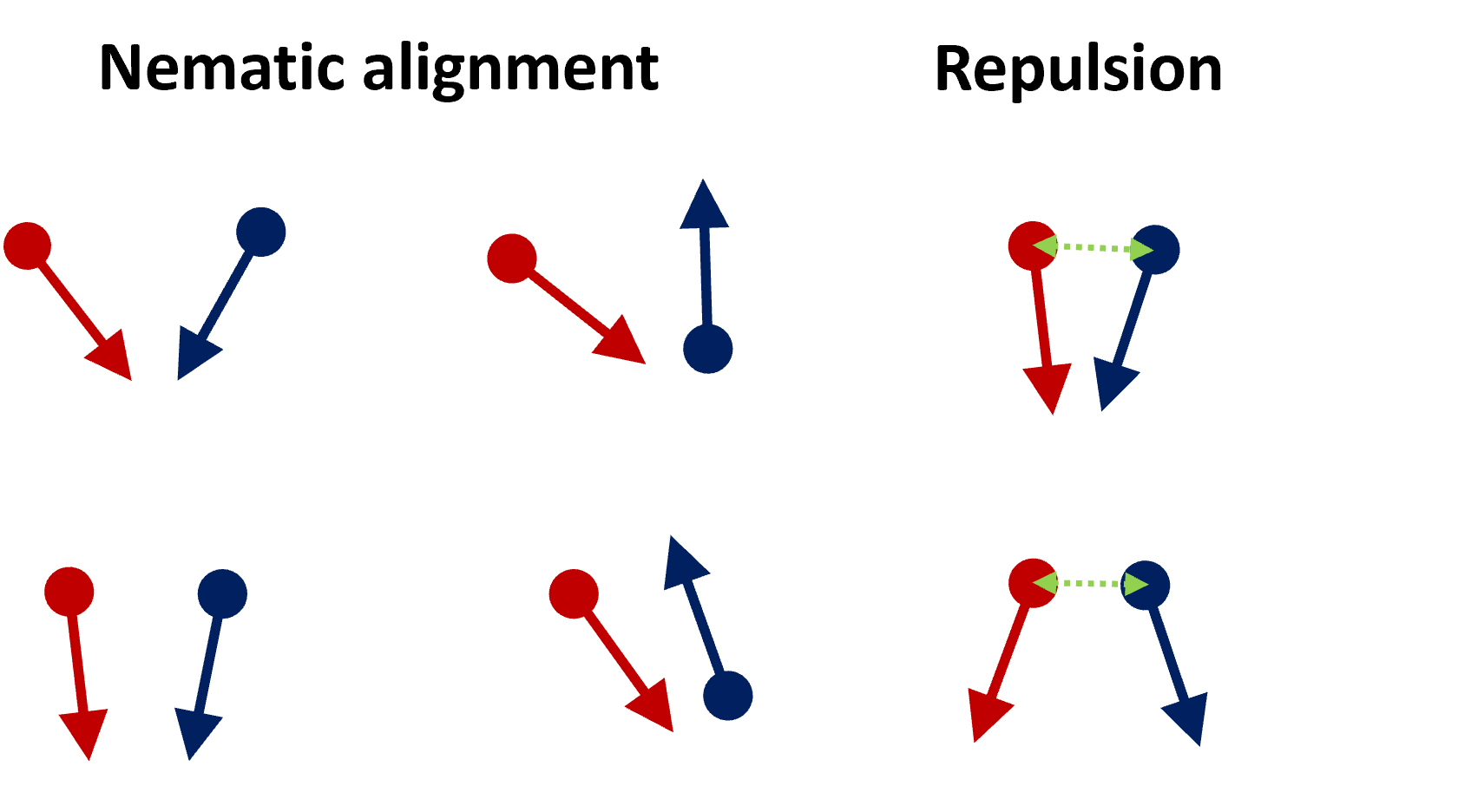}
 \caption{{\bf Sketch of the interactions between particles} Particles align nematically (left) and get repelled by other particles through a torque, which turns a particle self-propelling direction away from a neighboring particle. Top row shows the positions and orientations before the interaction, while the bottom row shows the updated positions and orientations.}
 \label{sketch_interactions}
\end{figure}

In \cite{kurjahn2024collective} it was argued that three ingredients are essential in order to get accumulation and alignment of cyanobacteria at virtual boundaries. Of course, the presence of such virtual boundaries where particle revert their direction of motion is perhaps the most obvious. Moreover, the elongated filaments interact by steric effect, resulting both in the tendency to align nematically and in a certain degree of repulsion. Finally, being in a quasi-2D geometry, filaments should also be able to cross each other: if this is not admitted, then boundary accumulation and alignment will be inhibited by the formation of traveling clusters.\\
Point-wise particles, as those considered in Vicsek-style models seen in Section \ref{dadam}, naturally allow for the agents to cross each other. Moreover, alignment interactions in the case of these models are explicit and can be either ferromagnetic or nematic: here we will consider nematic alignment interactions to mimic the interaction of filaments. Finally particles will also experience some repulsive interaction from their neighbors through a torque, which turns a particle self-propelling direction away from a neighboring particle, as schematically reported in Fig.\ref{sketch_interactions}. This choice models repulsive forces but still allows for some crossing between particles and therefore differs from repulsive forces directly applied on the positional degree of freedom, such as for the CVM introduced in Section \ref{other_ways}.

The microscopic model that we consider is a Vicsek-style model, first introduced in \cite{patelli2019understanding}, that does not strictly belong to the DADAM classes since particle interact also through repulsion. The overdamped dynamics of the $N$ point-wise particles, in $d=2$, can be as usual expressed in terms of the particles position $\mathbf{r}_i^t$ and orientation $\theta_i$. The time discrete evolution then reads
\begin{equation}\label{eq:r_i_patelli}
\mathbf{r}_i^{t+1} = \mathbf{r}_i^t + v_0 \mathbf{n}_i^{t+1},
\end{equation}
\begin{equation}\label{eq:theta_i_patelli}
\theta_i^{t+1} = \arg[\langle \mathcal{S} \mathbf{n}_j^t\rangle_{j\sim i} +\beta\langle\mathbf{\hat{r}}_{ji}^t\rangle_{j \sim i}] + \eta\chi_i^t
\end{equation}
where $v_0$ is the self-propulsion speed and $\mathbf{n}_i^t = (\cos{\theta_i^t}, \sin{\theta_i^t})$ is the usual unit orientation vector.\\
The first term in Eq.\eqref{eq:theta_i_patelli} simply describes nematic alignment with neighbors within unit distance ($R_0 = 1$), with the prefactor $\mathcal{S}=sgn(\mathbf{n}_i^t \cdot \mathbf{n}_j^t)$ of ${\bf n}_j^t$ accounting for the relative orientation of two particles, being $+1$ if the self-propelling directions form an acute angle and $-1$ otherwise. Pairwise repulsion, tuned by the parameter $\beta$, is here modeled as an overdamped torque, with $\mathbf{\hat{r}}_{ji}^t = (\mathbf{r}_i^t-\mathbf{r}_j^t)/|\mathbf{r}_i^t-\mathbf{r}_j^t|$. Finally, $\chi_i^t \in [-\pi, \pi]$ is an angular white noise drawn from a uniform distribution with $\eta$ setting its strength. For $\beta=0$, that is, in the absence of repulsive torques, this model reduces to the self-propelled nematic rods briefly introduced in Section \ref{dadam}.

\subsection{The free case}
A systematic study of the phase diagram has been carried out in \cite{patelli2019understanding} where three phases characterize the phase diagram (in the absence of spontaneous reversal, $k = 0$ in the language of \cite{patelli2019understanding}, where the possibility for polar rods to spontaneously revert their direction of motion with a rate $k$ was also considered). For high noises amplitudes the system is unable to synchronize and particles are indeed in the disordered or {\it isotropic} phase; lowering the value of the noise we enter the {\it homogeneous nematic} ordered phase, disordered phase where noise wins over alignment, a homogeneous nematic ordered phase, characterized by a global nematic order $|{\bf Q}| \equiv |\sum_i^N e^{2i\theta_i}|/N$ of order $1$. Finally, deep in the ordered phase a {\it nematic chaos} phase appears: this is characterized by patches of local nematic order, but global disorder, that is induced by $\pm1/2$ defects. \\
Here, before moving to the effect of phototactic confinement, we study the relative importance of alignment interactions (whose efficiency grows, increasing the the global density $\rho_0$) and repulsion (tuned by the parameter $\beta$), at fixed value of the self-propelled velocity to $v_0 = 0.5$ and the amplitude of the noise $\eta = 0.1$ values which will remain fixed through this entire chapter.

\begin{figure}[hbt!]
\centering
\includegraphics[width=0.66\linewidth]{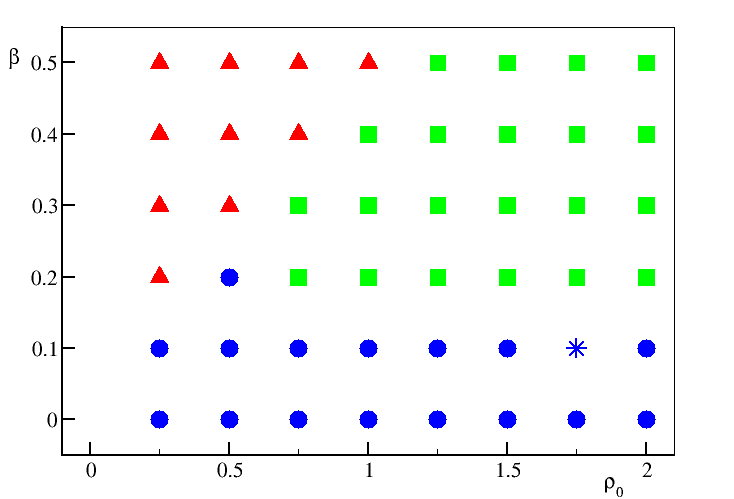}
 \caption{{\bf Phase diagram $(\rho_0, \beta)$ in the absence of confinement} We can identify three phases: a homogeneous nematic ordered phase (blue dots), where  $|\mathbf{Q}| > 0$; a disordered/isotropic phase (red triangles); and a nematic chaotic phase (green squares). We also observe a couple of polar traveling bands (blue star) which may be due to finite size effects but will not be discussed anymore here. Here the self-propelled velocity $v_0 = 0.5$ and $\eta = 0.1$. }
 \label{phase_diagram_free}
\end{figure}

As shown in Fig.\ref{phase_diagram_free} the $(\rho_0, \beta)$ phase diagram also shows three distinct phases: for low values of the repulsion $\beta$, particles are able to synchronize and displaying a homogeneous nematic ordered phase (blue circles). Increasing the intensity of repulsion we then obtain two distinct phases: for low enough densities, a strong enough repulsion is able to disrupt order and this results in a disordered, isotropic phase (red triangles). For higher values of the density we instead retrieve the \textit{nematic chaotic} phase (green squares), with no global nematic order but patches of local order and active $\pm 1/2$ topological defects, as in \cite{patelli2019understanding}.

In the next section we will show how, by introducing boundaries equipped with a simple reflection rule that mimics the effect of virtual confinement, accumulation at and alignment with such boundaries emerges: this is rather generic and irrespective of the phase of the system, possibly coexisting with the various phases as a background on the outside from the bounded region.

\subsection{The effect of virtual confinement}
\begin{figure}[hbt!]
    \centering
    \includegraphics[width=1\linewidth]{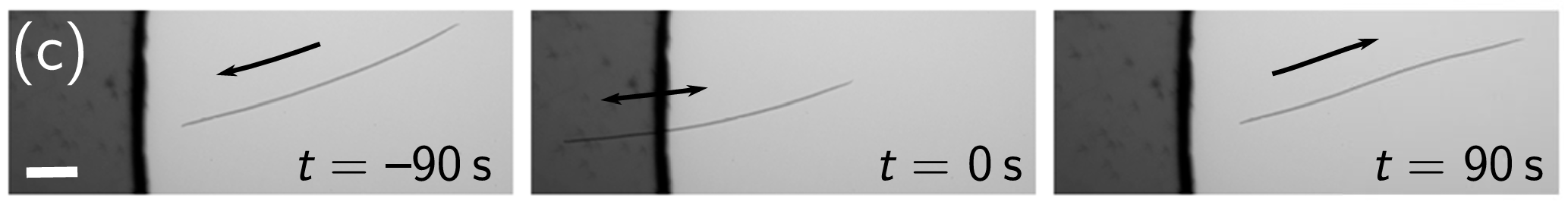}
    \caption{{\bf Example of the reversal mechanism for the filament} Figure adapted from \cite{kurjahn2024collective}.}
    \label{fig:reversal_filament}
\end{figure}

The effect of virtual confinement is that of reverting the particle's direction of motion whenever decreasing in illuminance (crossing from lighter to darker regions). In \cite{kurjahn2024collective} scoto-phobic filaments revert their direction of motion when a finite fraction of their body crosses the boundary separating the lighter from the darker region., as reported in Fig.\ref{fig:reversal_filament}. No direction reversal is experienced in the opposite case, crossing from darker to lighter region.

\begin{figure}[hbt!]
\centering
\includegraphics[width=0.45\linewidth]{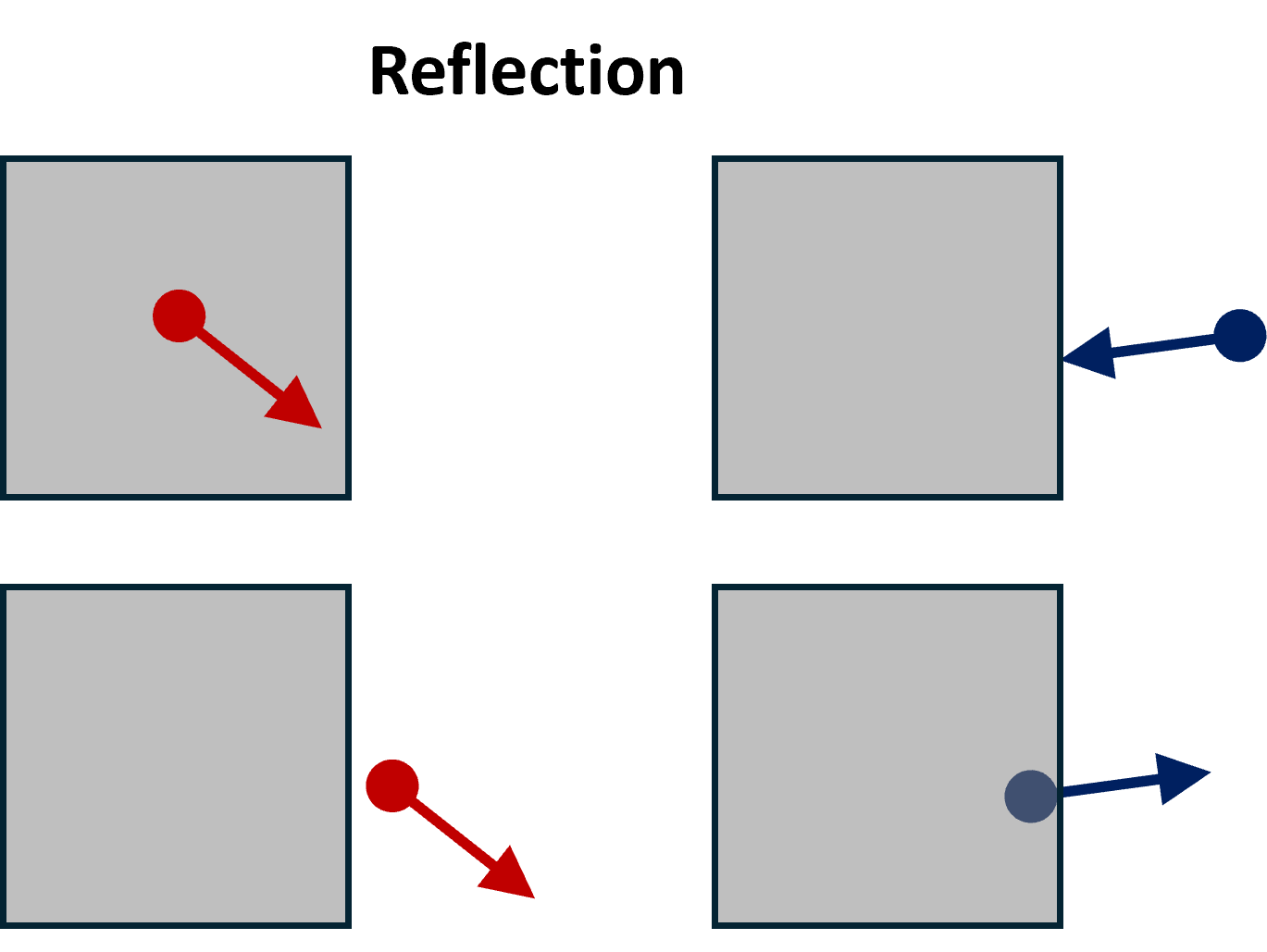}
 \caption{{\bf Example of the selective reversal mechanism} The particle coming from the darker region (red) does not change its self-propelling direction when crossing from a dark to an illuminated region, while the particle coming from the illuminated region (blue) reverses its direction of motion.}  
 \label{reflection}
\end{figure}

In microscopic simulations point-wise particles obviously do not have an extension: a particle $i$ will then revert its direction of motion
\begin{equation}\label{eq:micro_reflection}
    \mathbf{n}_i \to -\mathbf{n}_i \quad
\end{equation}
whenever it experiences a (sharp in our case) decrease in illuminance (crossing from bright to darker regions in a single time-step). Note that the reflection rule of Eq.\eqref{eq:micro_reflection}, mimicking the effect of virtual boundaries, does not affect the positional degrees of freedom of the particles allowing, at least in principle, for particles to also explore the darker regions if their immediate re-entrance in the lighter region is prevented by some other change in the direction of ${\bf n}_i$. This is a remarkable difference with Eq.\eqref{eq:reflection_casimir} where also the positions of the particles where affected in order to prevent, microscopically, the crossing of a mechanical boundary. \\
Finally, when particles are crossing in the opposite direction (feeling an increase in illuminance) they will proceed unaltered. A schematic representation of this selective mechanism is shown in Fig.\ref{reflection}.

This rules can be used to confine polar rods in illuminated regions of various shapes. We begin by presenting the effect of a perfectly scoto-phobic circular confinement in analogy with the work of \cite{kurjahn2024collective}: a circular region is illuminated and surrounded by a darker region. \\
We perform microscopic simulations based on Eqs.\eqref{eq:r_i_patelli}-\eqref{eq:theta_i_patelli} and the boundary rule \eqref{eq:micro_reflection} in square domains of size $L = 512$ with periodic boundary conditions assumed in both directions. The circle, of radius $R=100$, is placed at the center of the domain, so that we are actually simulating a system which is roughly 8 times bigger than the illuminated confining circle. Note that this particular choice of confinement preserves continuous rotational symmetry. \\
In order to quantify the degree of nematic order we measure the local coarse-grained nematic angle $\Phi(\mathbf{r},t)$ and the local norm of the nematic tensor $|{\bf Q}(\mathbf{r},t)|$. Knowing the self-propulsion direction $\theta_i$ of one particle one can readily define the two independent components of the nematic tensor, in $d=2$, simply as 
\begin{equation}
Q_{xx}(i) = \frac{1}{2}\cos(2\theta_i) \quad \text{and} \quad Q_{xy}(i) = \frac{1}{2}\sin(2\theta_i),
\end{equation}
and the nematic angle $\Phi_i$ and norm of the nematic tensor $|\mathbf{Q}_i|$ as
\begin{equation}\label{eq:nem_angle_norm}
    \Phi_i = \frac{1}{2}\arctan(Q_{xy}(i)/Q_{xx}(i)) \quad \text{and} \quad  |\mathbf{Q}_i|= 2\sqrt{Q_{xx}(i)^2+Q_{xy}(i)^2}
\end{equation}
where the $\arctan$ function will return an angle $\Phi_i \in [-\pi/2, \pi/2]$, while the norm of the nematic tensor $|{\bf Q}_i|\in [0,1]$. The coarse-grained fields are then obtained averaging the single particle $\Phi_i$ and $|{\bf Q}_i|$ in boxes of unit size centered around $\bf{r}$.

\begin{figure}[hbt!]
\centering
\includegraphics[width=0.75\linewidth]{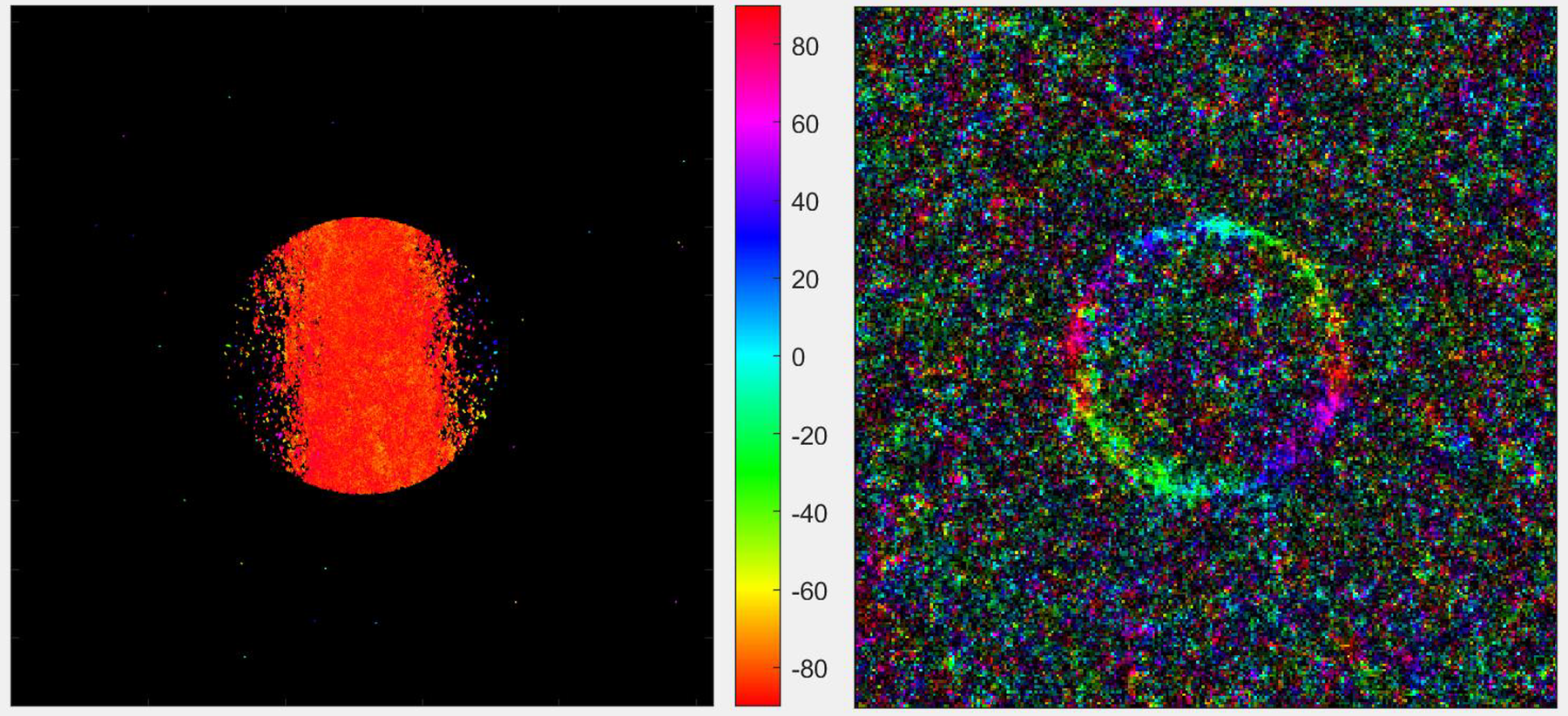}
 \caption{{\bf Snapshots of the coarse-grained system in presence of circular virtual confinement} - Left - In the absence of repulsion ($\beta = 0$) the particles form a nematically ordered band, whose direction $\Phi$ is spontaneously chosen $\Phi\in [-\pi/2, \pi/2]$. Right - When repulsion is turned on ($\beta = 0.3$) the system re-organizes shows a different pattern, akin to what has been observed in \cite{kurjahn2024collective}, showing accumulation and alignment with the virtual boundary. The color-bar refers to the nematic angle $\Phi$ obtained by coarse-graining in square boxes on unit size (see text). Black indicates the lack of particles. Here $\rho_0 = 0.5$.}
 \label{snapshots}
\end{figure}

We first consider a global density $\rho_0=0.5$. From the phase diagram of Fig.\ref{phase_diagram_free}, in the free case this results in an homogeneous nematic phase for zero or low repulsion and to a disordered phase for large repulsion ($\beta \ge 0.3$).
In the absence of repulsion $(\beta = 0)$, as shown in Fig.\ref{snapshots} (left panel) the particles accumulate almost exclusively inside the illuminated region and tend to form a nematically ordered band, with the direction of it spontaneously chosen in $[-\pi/2, \pi/2]$. The width of the band, namely the extension of the band in the direction perpendicular to that of global order, grows when increasing the global density $\rho_0$, as shown in Fig.\ref{phase_diagram_ring} below, possibly reaching the size of the confining circle. Note that active particles accumulate almost exclusively inside the illuminated circle, which therefore reaches rather high local densities. This is due to the one-way nature of the boundary conditions (preventing crossing from light to dark but not vice-versa) and the fact that in the absence of repulsion reversal of the self-propulsion direction immediately results in a reentrance in the illuminated region which is, therefore, almost fully trapping.

As we introduce repulsion ($\beta \geq 0.1$), the nematically ordered band is no longer stable and the system orders along the virtual boundaries forming a nematically aligned {\it ring structure} as it is shown in the right panel of Fig.\ref{snapshots} for $\beta=0.3$. Note that now the illuminated circle is not anymore strictly trapping, and many particles can be found outside of it. This is due to the effect of the repulsion, that may deflect the reverted orientation of particles that just crossed the light-to-dark boundary, thus preventing their immediate reentrance. For large enough $\beta$ (e.g., $\beta\ge0.3$ for $\rho_0=0.5$), the particles escaping the circle are found in a disordered state, consistently with the free case phase-diagram of Fig.\ref{phase_diagram_free}. Moreover, also the center of the ring is disordered. This is indeed strongly reminiscent of the results of Ref.\cite{kurjahn2024collective} (see also Fig.\ref{fig:pattern_illumination} here).

We now proceed to more carefully explore the ($\rho_0, \beta$) phase diagram in the presence of photo-tactic confinement, considering time-averaged coarse-grained fields. The time-averaged nematic fields for different densities and repulsion values (nematic order magnitude in the right panel and nematic orientation in the left one) are reported in Fig.\ref{phase_diagram_ring}. Our scan of the parameter space shows that the ring structure persists regardless of the background phase of the free system: correspondingly, the outside active fluid may be found both in the ordered or in the disordered regimes (isotropic or nematic chaos). For instance, with $\rho=0.5$ and $\beta=0.1$ the ring appears with an ordered outside background, for $\rho=0.5$ and $\beta=0.3$ we have an isotropic background and for $\rho=2$ and $\beta=0.1$ an outside background showing nematic chaos. Finally, note that the thickness of the nematically ordered ring depends on the parameter choice and, in particular, tends to increase with the global density.

\begin{figure}[hbt!]
\includegraphics[width=1\linewidth]{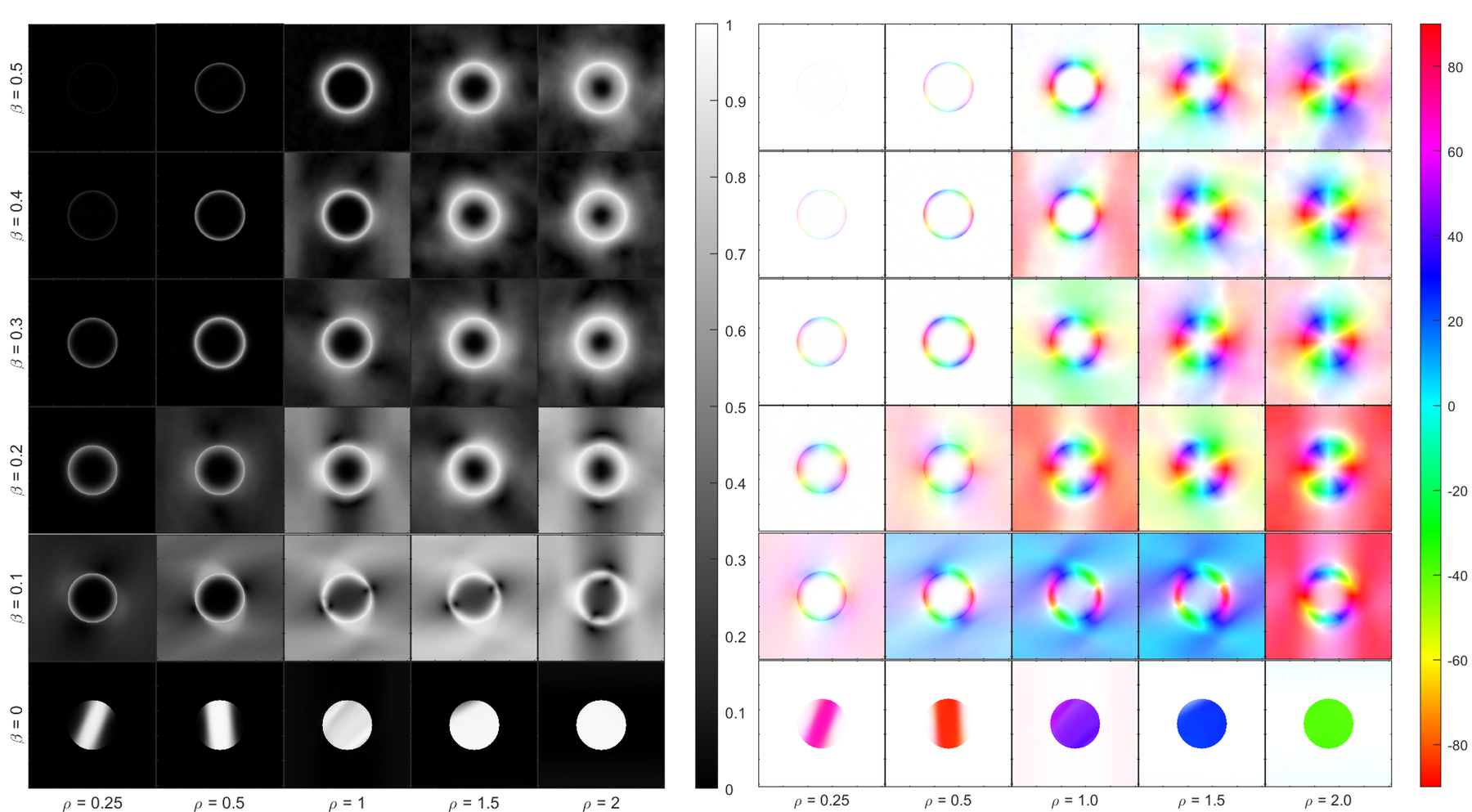}
 \caption{ {\bf Time averaged nematic order parameter $|Q|$ and nematic angle $\Phi$} Left - Phase diagram in the ($\rho_0, \beta$) for the nematic scalar order parameter $|\mathbf{Q}|$. Gray-scale coding is set by the amplitude of the nematic scalar order parameter $|\mathbf{Q}|\in [0, 1]$. Here white corresponds to nematic order and black to disorder. Right - Same as the left panel, showing the nematic angle $\Phi$. The color intensity are faded according to the value of the nematic scalar order parameter of the left panel (with white representing full disorder here), while color coding is set by the nematic angle $\Phi$. Time-averages are taken over $T = 5\times 10^5$ time steps, after discarding a proper transient $T_0$.}
 \label{phase_diagram_ring}
\end{figure}

\begin{figure}[hbt!]
\centering
\includegraphics[width=0.8\linewidth]{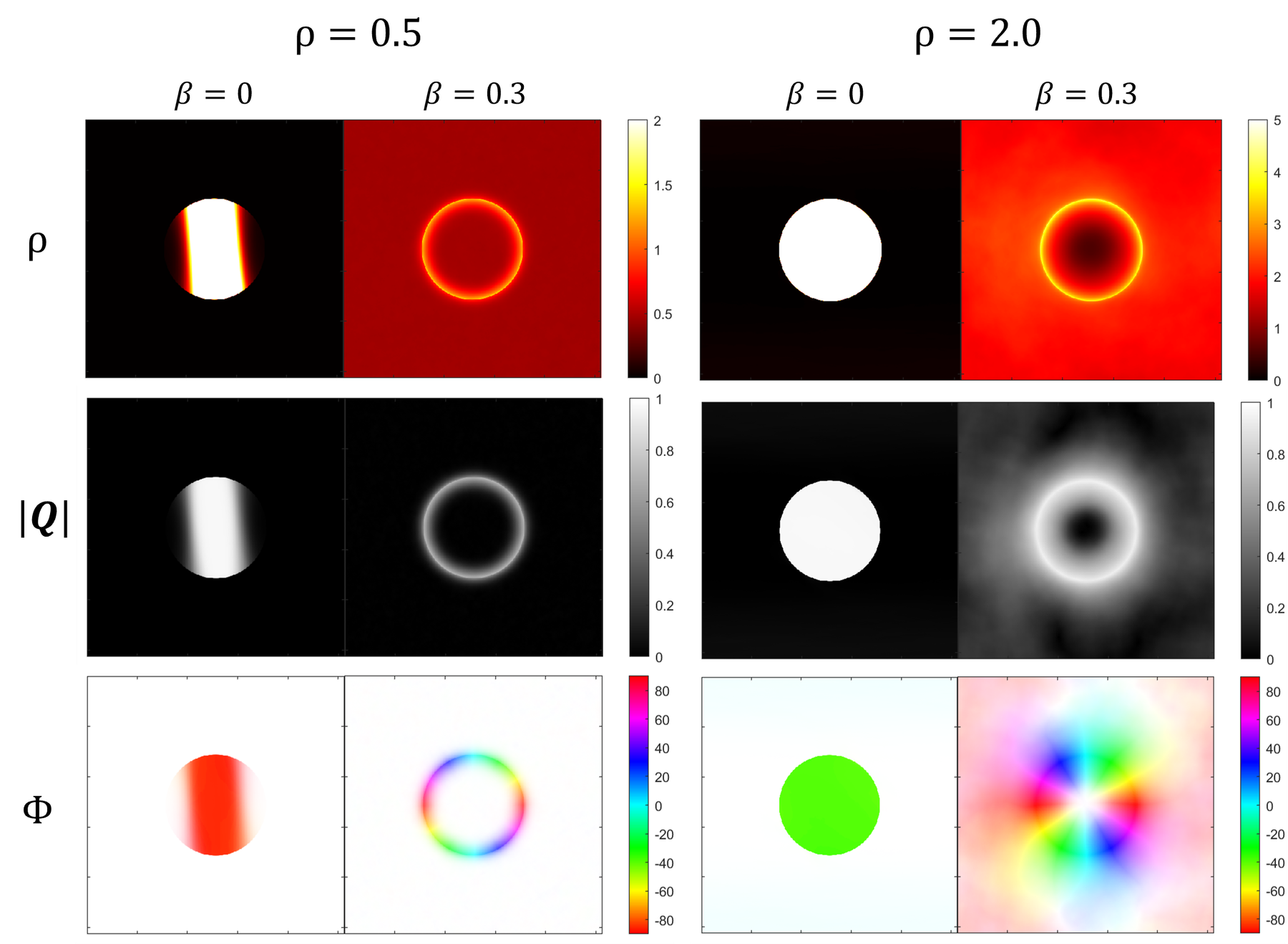}
\caption{{\bf Time averaged maps of density, nematic order parameter and nematic angle} Left - Detail of the effect of circular virtual confinement in the case of disordered background ($\rho_0 = 0.5$ and $\beta = 0$ for the first column while $\beta = 0.3$ for the second one) on the density (top), nematic order parameter (middle) and nematic angle (bottom). Right - Same as left, but for higher global density $\rho_0 = 2.0$ showing how the ring structure survives also in the presence of a chaotic background.}
\label{details}
\end{figure}

To better characterize these configurations we now focus on four different points in the parameter space, considering relatively low ($\rho_0=0.5$) and high densities ($\rho_0=2$) without ($\beta=0$) and with inter-particle repulsion ($\beta=0$). The corresponding density, nematic amplitude and nematic orientation are shown, top to bottom, in Fig.\ref{details}. \\
As we have seen for the polar case, see Section \ref{BGL}, also in the case of nematic alignment, the magnitude of the local nematic order seems to be strongly correlated with the local density. In the absence of repulsion and at sufficiently low densities ($\rho_0 = 0.5$) one observes a single confined band. The nematic field aligns in the band direction and thus meets the virtual boundary perpendicularly. 
At higher densities ($\rho=2.0$) the confined bands gives way to a confined high density and nematically ordered homogeneous state. Remember also that in the absence of repulsion all particles get trapped inside the ring, resulting in a higher (roughly eight times in this specific case) confined density.\\
In the presence of non-negligible inter-particle repulsion ($\beta \ge 0.1$), particles are no more strictly trapped inside the illuminated area and a ring structure emerges at the light/dark boundary. In this ring structure, particles are highly aligned with the virtual boundary and this leads to accumulation of the particles at the boundaries. Moreover, the central region inside the ring shows density depletion and no nematic order.

\begin{figure}[hbt!]
\centering
\includegraphics[width=1\linewidth]{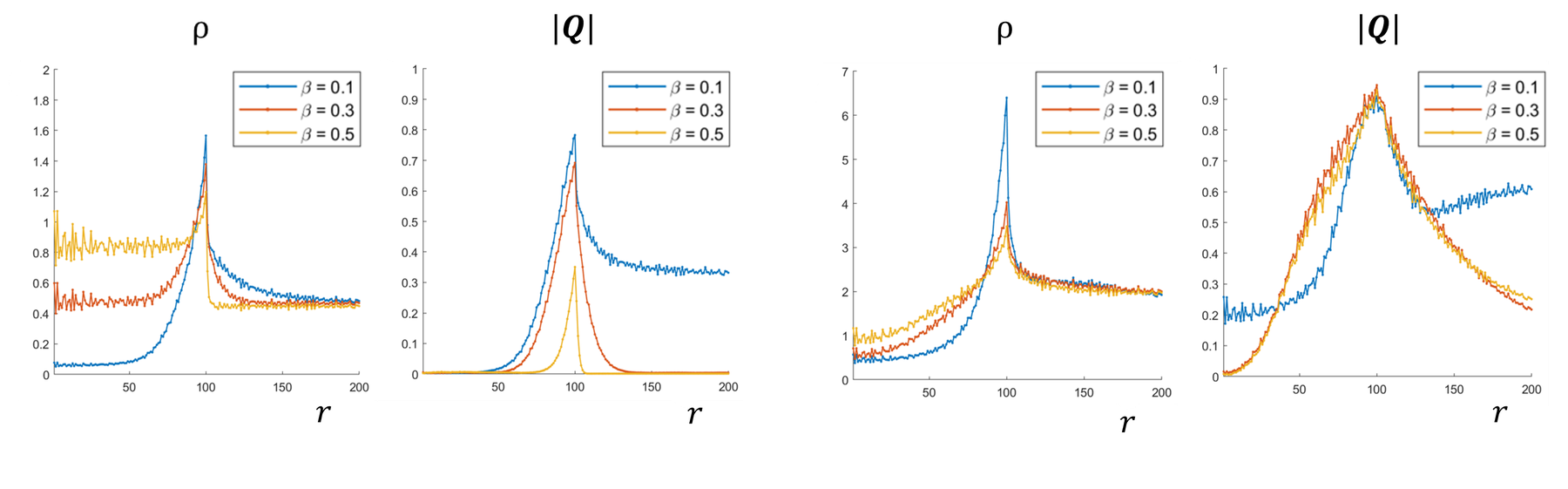}
\caption{{\bf Time averaged profiles of density and nematic order parameter} Left - Profiles of the density $\rho$ and nematic scalar order parameter $|\bf Q|$ averaged over the polar angle, against the radial distance $r$ from the center, for three different values of $\beta = 0.1, 0.3, 0.5$, that clearly show accumulation of density at the boundary and alignment with it. Right - Same as left, but for higher global density $\rho_0 = 2.0$ showing how the ring structure survives also in the presence of a chaotic background.}
\label{profiles}
\end{figure}

Since the system's stationary state is rotationally invariant, at least in the case when the ring forms, it is convenient to consider profiles of both density and nematic order parameter, averaged over the polar angle (see Fig.\ref{profiles}).\\
In particular, in our simplified model the density profiles are found to peak exactly at the boundary of the virtual confinement ($r = R = 100$), at odds with what has been observed in \cite{kurjahn2024collective} where the physical length of the filaments sets an offset between the virtual boundary and the density peak of the ring structure formed by the filaments. Outside the illuminated area we observe a non-zero density of particles (and either a zero or non-zero scalar order parameter). Inside the ring, one observes lower densities (but increasing with $\beta$) and virtually no order.

\begin{figure}[hbt!]
    \centering
    \includegraphics[width=0.5\linewidth]{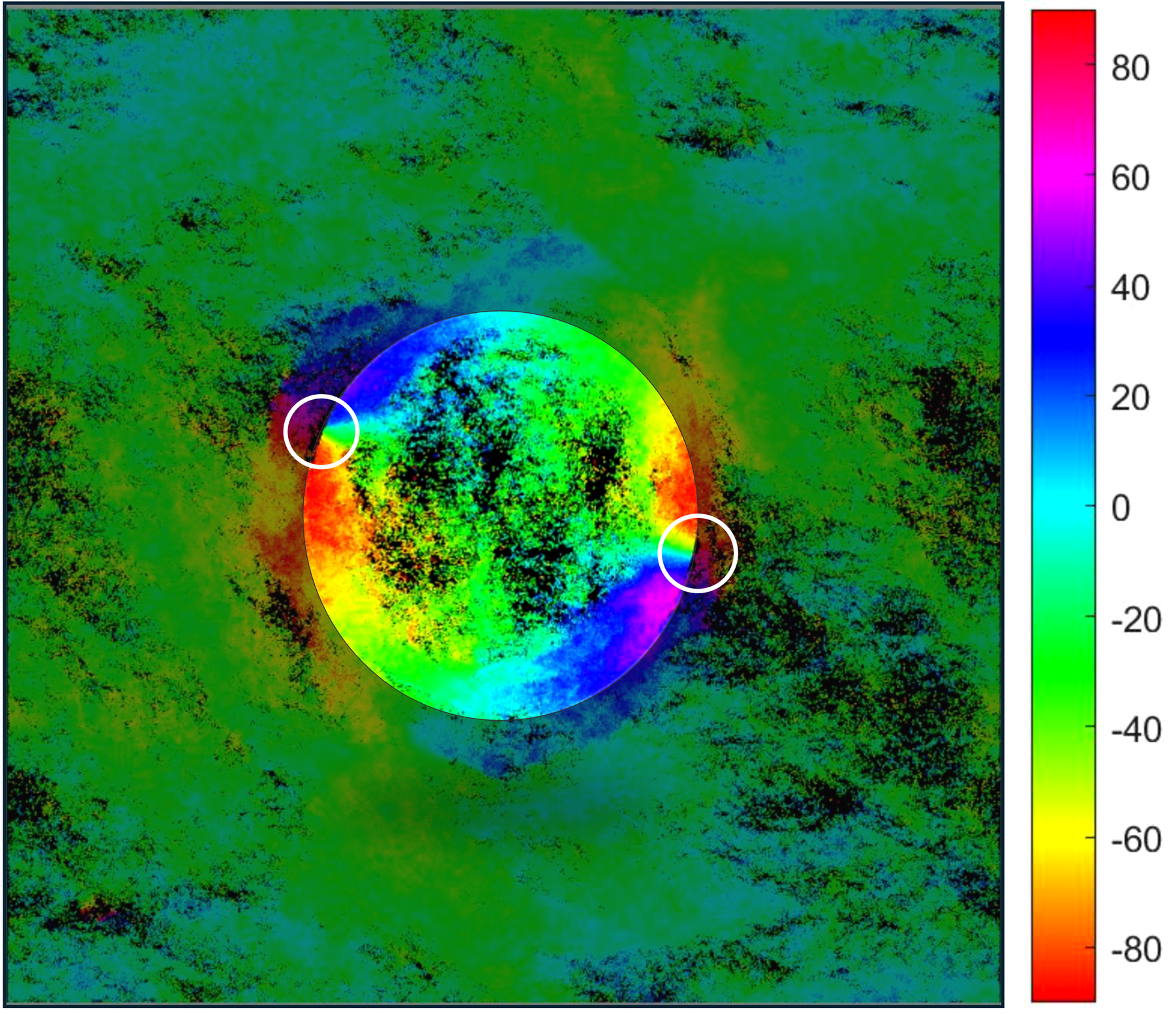}
    \caption{{\bf Snapshot of the coarse-grained ring coexisting with a homogeneous ordered background} For values of the repulsion $\beta$ close to the transition value $\beta_c(\rho)$ we observe the formation of a ring structure which perhaps interacts with the homogeneous background. In this specific case the homogeneous background (green) when interacting with the boundaries creates a couple of $+1/2$ defects inside the ring (white circles). Here $\beta = 0.1$ and $\rho = 2.0$. }
    \label{fig:defects}
\end{figure}

The case $\beta=0.1$ and $\rho_0=2$ is however an exception, showing some residual order inside the ring. Indeed, sharp ring structures do not appear for any arbitrarily small repulsion values, but seem to depend on a (density dependent) small threshold, $\beta > \beta_c (\rho)$. The transition between the $\beta=0$ band structure and the ring geometry may not be sharp, and one can observe the effect of the interaction between a homogeneous ordered background and the ring structure: this results in the presence of 2 $\pm 1/2$ nematic defects (see Fig. \ref{fig:defects}). The exact characterization of this transition is beyond the scope of this work.

So far we have shown that the accumulation and alignment at the virtual boundaries holds quite generically, regardless the background phase. In the next section we will show how this effect is also general regardless of the geometry and the curvature of the confining shape.

\subsection{Different closed geometries}
\begin{figure}[hbt!]
    \centering
    \includegraphics[width=1\linewidth]{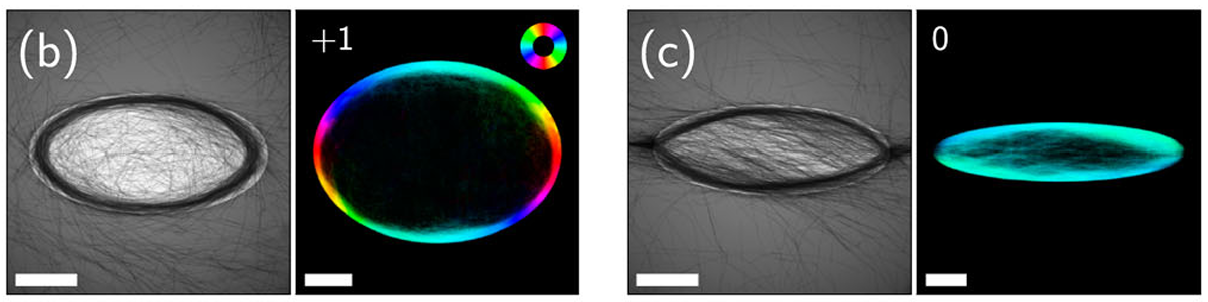}
    \caption{{\bf Accumulation patterns for the ellipse} Left - For high eccentricity the filaments align similarly to the circle, bending continuously from point to point. Right - When the eccentricity is high enough to produce very tight turns the filaments re-organize in a structure with discontinuous orientation. Figure adapted from \cite{kurjahn2024collective}.}
    \label{fig:bend_splay}
\end{figure}

In \cite{kurjahn2024collective} it was shown that the accumulation of particles at boundaries also appears in shapes with inhomogeneous boundary curvatures. The first and perhaps more natural shape to consider, in this direction, is the ellipse: interestingly they find that for low eccentricity the boundary accumulation is akin to that of the circle, with filaments aligning along the boundary and bending continuously from point to point (see Fig.\ref{fig:bend_splay}, left panels). For high eccentricity, on the other hand, Ref.\cite{kurjahn2024collective} reports a transition to a splay pattern, where filaments change discontinuously their orientation along the boundary (see Fig.\ref{fig:bend_splay}, right panels). 

\begin{figure}[hbt!]
    \centering
    \includegraphics[width=1\linewidth]{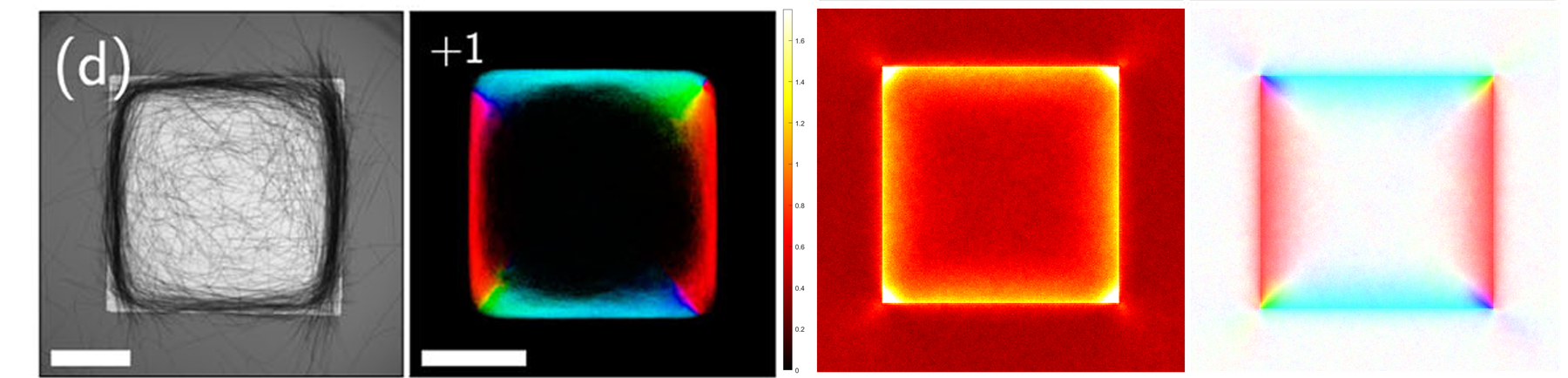}
    \caption{{\bf Comparison between filamentous cyanobateria and microscopic model in the square shape} Left - Snapshot of cyanobacteria confined in a square region and relative coarse-grained orientation (taken from \cite{kurjahn2024collective}). Right - Coarse-grained density and nematic angle fields obtained in our microscopic simulation. Here $\rho_0 = 0.5$ and $\beta = 0.3$.}
    \label{fig:square_comparison}
\end{figure}

In microscopic simulations, particles do not have a physical length so that they can bend continuously also in very acute angles, as we will show in the following. Here we will just show a few examples of different shapes where the analogous of the circular domain ring effect is observed: for consistency, in our analysis of different shapes we keep the area of the confining shape the same as the one of the circular case. The general behavior that we observe is analogous to that obtained for the circle, with band forming in the absence of repulsion and, for $\beta \geq 0.1$ the formation of a ring structure with different kind of backgrounds outside of the confining shape.\\
In the case of a square shape, where the boundary shows a strictly sharp bend, we indeed get the same results obtained in \cite{kurjahn2024collective} with particles accumulating and aligning with the boundaries and sharp bending at the four corners. There is anyhow a difference since in our case particles tend to accumulate in the corners of the square (see Fig.\ref{fig:square_comparison} right panels), while filaments tend to avoid the corners and in a sense they smoothen them, as shown in Fig.\ref{fig:square_comparison} (left panels). The alignment pattern is nevertheless found to be the same.

\begin{figure} [hbt!]
    \centering
    \includegraphics[width=0.66\linewidth]{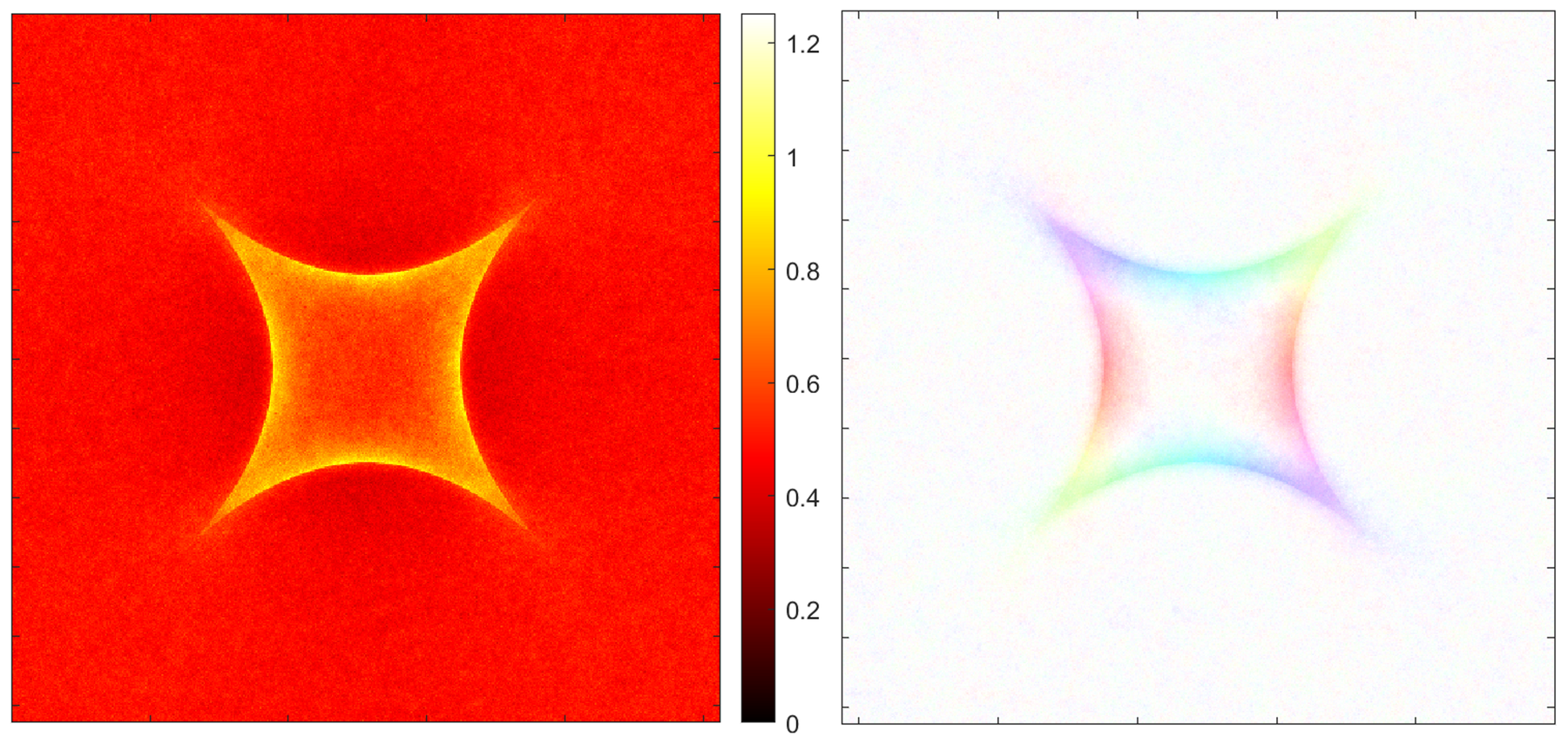}
    \caption{{\bf Microscopic simulation showing the effect of virtual confinement in a super-ellipse geometry} Map of the density (left) and nematic angle (right) fields obtained in microscopic simulation. Here $\rho_0 = 0.5$ and $\beta = 0.3$.}
    \label{fig:superellipse}
\end{figure}

Finally, in order to further investigate the effect of the curvature, we considered a concave confining shape with negative curvature along the boundaries: this can be realized considering a super-ellipse, as shown in Fig.\ref{fig:superellipse}. Also in this case we obtain accumulation and alignment with the boundaries.

In particular note that we do not observe a bend to splay transition, since also in this case, due to our peculiar geometry, particles accumulating in corners are able to follow the curvature of the boundary continuously. This is a significant difference with \cite{kurjahn2024collective}. \\
Nevertheless this tells us that it is possible to obtain accumulation of particles at the boundaries, regardless the shape and the curvature. This fact can be used at our advantage to study photo-tactic confinement in possibly the simplest geometry: a slab geometry, akin to the Casimir geometry of Ch.\ref{confined}.

\section{Slab geometry and the effective escape rate}

\begin{figure}[hbt!]
    \centering
    \includegraphics[width=1\linewidth]{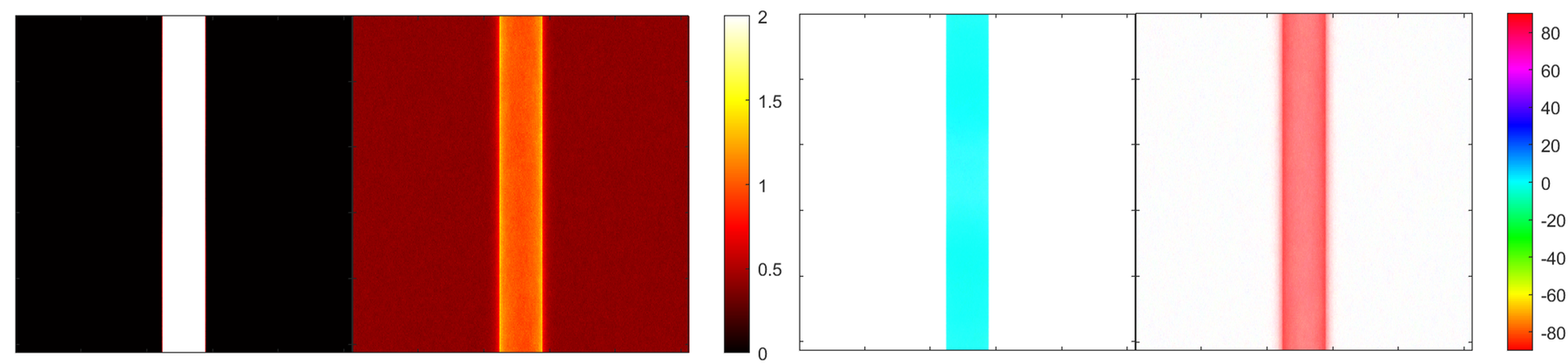}
    \caption{{\bf The effect of two virtually confining boundaries separated by a distance $R$} Left - Time averaged maps of the density in the absence (left) or presence of repulsion (right, for $\beta = 0.3$) in the presence of two virtually parallel confining boundaries separated by a distance $R = 64$. Color encodes local density. Right - Same as top for the nematic angle (faded according the value of $|\mathbf{Q}|$, white denotes disorder). Color encodes local orientation of the nematic angle $\Phi$ w.r.t. the transversal direction. Here $\rho_0 = 0.5$ and $L = 512$.}
    \label{fig:slab_repulsion}
\end{figure}

In this section we investigate a simpler slab geometry, where the confining illuminated shape is actually defined by two confining parallel and infinite walls separated by a distance $R$: in the direction along the walls periodic boundary conditions are assumed (we will refer to this direction as the longitudinal $(\|)$) while in the direction transverse to the walls $(\perp)$ particles will experience reversals given by Eq.\eqref{eq:micro_reflection} when leaving the internal illuminated region.\\
In the absence of repulsion particles fill homogeneously the region in the slab aligning transversally to the virtual walls. When repulsion is turned on, particles instead align along the boundaries displaying an inhomogeneous density distribution, as shown from the microscopic simulations (see Fig.\ref{fig:slab_repulsion}).

\begin{figure}[hbt!]
    \centering
    \includegraphics[width=0.9\linewidth]{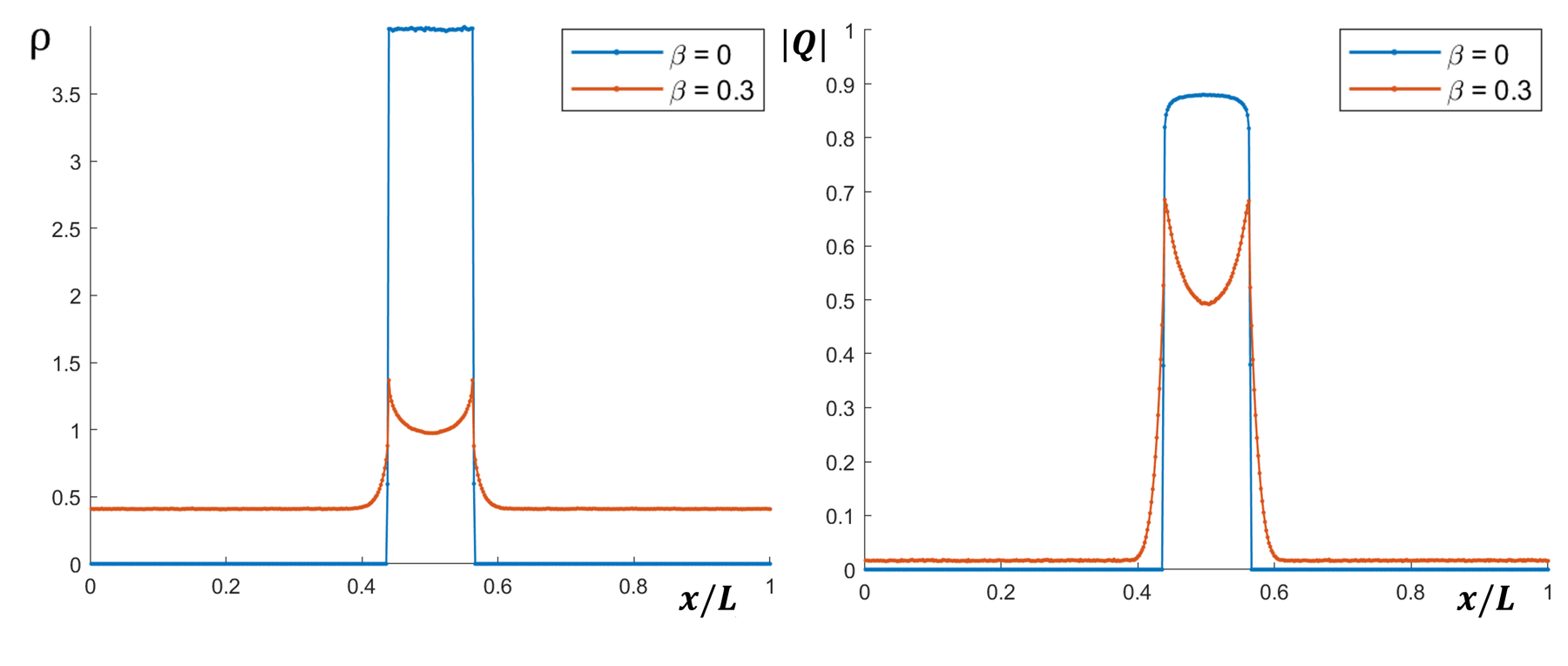}
    \caption{{\bf Longitudinally averaged profile of density and nematic order in the absence or presence of repulsion} Left - Longitudinally averaged profiles of the density $\rho$ for different values of $\beta = 0, 0.3$, showing accumulation in presence of repulsion. Right - Same as left, showing the nematic order parameter$|\mathbf{Q}|$. Data is plotted against the rescaled distance $x/L$, where $x$ is the coordinate in the direction perpendicular to the walls, $L$ is system size.} 
    \label{fig:profiles_slab_rep}
\end{figure}

Longitudinal invariance can be used in order to obtain the density and nematic order longitudinally averaged profiles, which are shown in Fig.\ref{fig:profiles_slab_rep}.
The nematic order profile closely follows that of density: in both cases the system is ordered inside and disordered outside. In the case of repulsion nematic order is anyhow enhanced by the presence of accumulation at virtual boundaries in a way similar to what happens for circular confinement (see Fig.\ref{phase_diagram_ring}).

\begin{figure}[hbt!]
    \centering
    \includegraphics[width=0.9\linewidth]{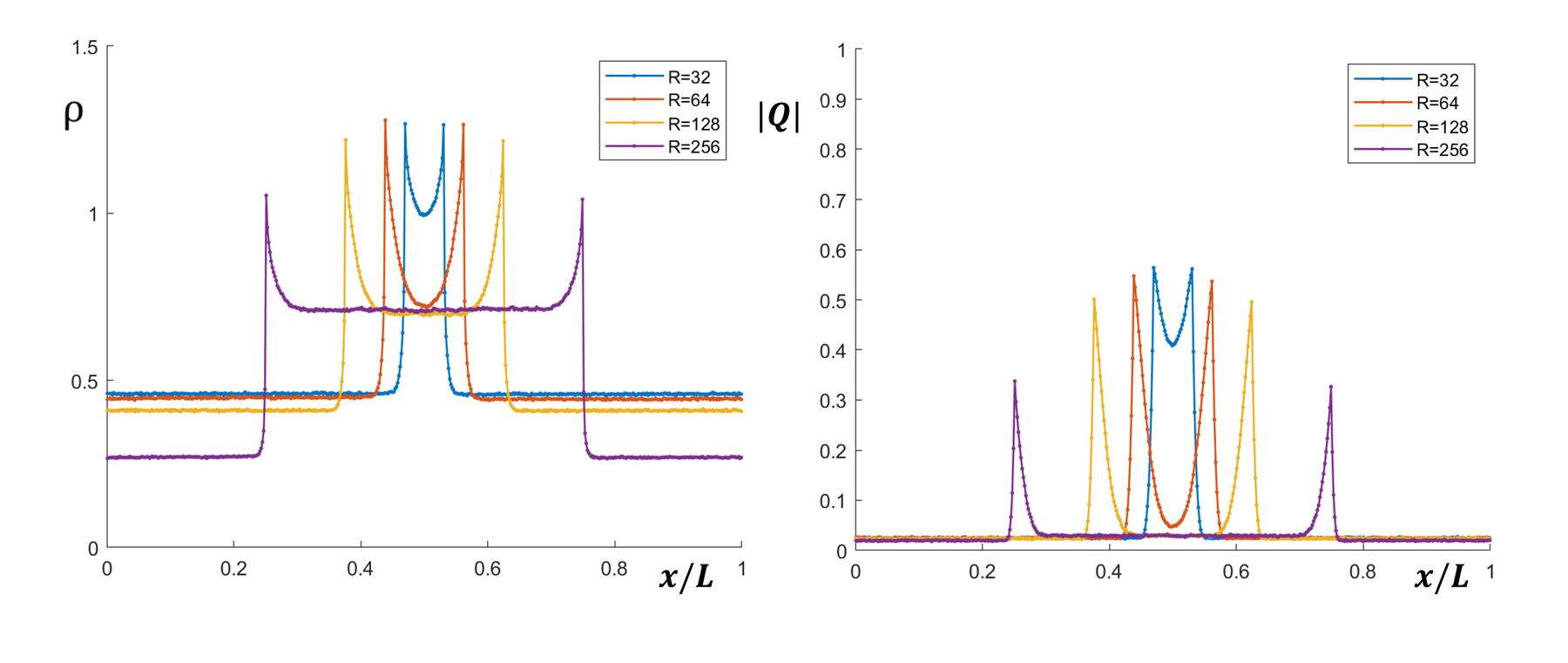}
    \caption{{\bf Finite size scaling of the density and nematic order in the presence fo repulsion} Left - Longitudinally averaged profiles of the density $\rho$ for different values of the wall separation $R = 32, 64, 128, 256$. Right - Same as left, but showing the amplitude of the nematic order parameter. Data is plotted against the rescaled distance $x/L$, where $x$ is the coordinate in the direction perpendicular to the walls, $L$ is system size. Here $\beta = 0.4$ and $L=512$.}
    \label{fig:finite_Size_scaling}
\end{figure}

Preliminary finite size scaling analysis suggests that in this case the extension of boundary layers is finite, as shown in Fig.\ref{fig:finite_Size_scaling}. Indeed, performing microscopic simulations for different wall separations $R$, at fixed system size $L$ one sees that the extension of the region where particles accumulate is limited to region close to the boundary and does not grow increasing wall separation, resulting in intensive accumulation of particles at the walls, at odds with what was found in Chapter \ref{confined} for confined polar flocks. Of course, changing the wall separation $R$ keeping the system size fixed can also influence the inner and outer density: we nevertheless believe that, as long as phase of the inner system is left unaltered, this should not invalidate our conclusions.

\subsection{Effective escape rate}
\begin{figure}[hbt!]
    \centering
    \includegraphics[width=0.9\linewidth]{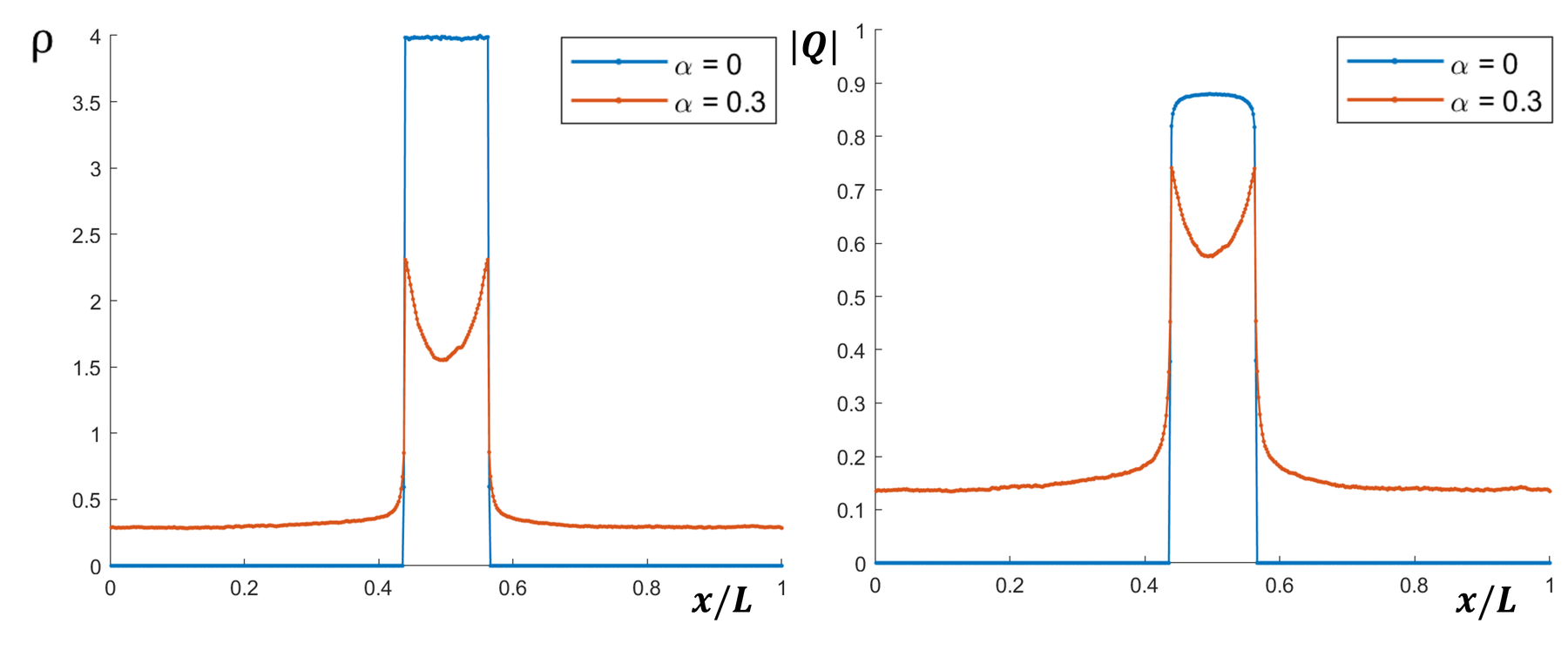}
    \caption{{\bf Profiles of density and nematic order in the absence repulsion, with and without escape rate} Left - Longitudinally averaged profiles of the density $\rho$ for different values of the escape probability $\alpha = 0, 0.3$, showing accumulation when particles have a finite probability to escape. Right - Same as left, showing the nematic order parameter. Data is plotted against the rescaled distance $x/L_x$, where $x$ is the coordinate in the direction perpendicular to the walls, $L_x$ is the size in the same direction.}
    \label{fig:profiles_slab_escape}
\end{figure}

When repulsion is present, the density in the outside region is non-zero (as previously discussed for the circular confinement), indicating that the confinement is not perfectly trapping and there is an exchange of particles from the inside and outside region. Indeed, particles outside will continue to fuel the inside structure and vice-versa: one might then think that this can be realized even in the absence inter-particle repulsion simply by adding and {\it escape probability} $\alpha$, that is, the probability that particles crossing from the illuminated to the dark region {\it do not} experience a reversal of their self propulsion direction, relaxing the perfect scoto-phobic behavior of Eq.\eqref{eq:micro_reflection}. Setting $\alpha = 1$ will make the virtual boundary disappear so that the escape probability $\alpha$ can be conveniently used to interpolate between perfect scoto-phobic behavior ($\alpha = 0$) and the free system ($\alpha = 1$).\\
Notably, we have verified that for escape probabilities $\alpha > \alpha_c\approx 0.2$, a simple model without repulsion $(\beta=0)$ is able to qualitatively recover the behavior of the model with repulsion, that is, accumulation and high nematic order at the virtual boundaries with the nematic orientation aligned longitudinally with respect to the boundaries, see Fig.\ref{fig:profiles_slab_escape}. In practice, the escape rate seem to play an analogous role to the one of inter-particle repulsion $\beta$ (see Fig. \ref{fig:profiles_slab_rep}) in reproducing the confined behavior.

This somehow surprising finding suggests that photo-tactic confinement by a slab geometry may be studied at the mesoscopic level considering a relatively simple system, that is the known equations for polar rods \cite{peshkov2012nonlinear} without repulsion in the presence of a proper boundary condition with a finite escape rate.

\subsection{Towards a simple mesoscopic description}
In this section we will show how it is possible to derive a boundary term describing photo-tactic confinement at the coarse-grained level starting from a simple lattice model, which is basically a variation of the AIM, seen in Section \ref{sec_aim}.\\
In the case of the slab geometry we can roughly map our model to a one dimensional lattice model, taking advantage of the invariance of the system in the longitudinal direction. Consider for instance a $d=1$ lattice model where the particles can take three orientations $\epsilon \in [-1, 0, 1]$ according to their self propelling direction ($\epsilon = 0$ takes into account both up and down orientation of the particles, which does not induce a bias in the diffusive motion). The particles will then hop to the left (right) neighboring site with probability $\frac{1-\epsilon}{2}$ $(\frac{1+\epsilon}{2})$ at a rate $u$, while they also undergo tumbles with rate $\lambda$.\\
Virtual boundaries are situated at site $k$ and site $k'$, with $k<k'$. Sites $i$ are illuminated for $k<i<k'$ and are dark otherwise. The particles will revert their polarity when crossing these boundaries from the illuminated to the dark region with rate $\gamma$ which directly controls the escape probability, i.e. $\gamma\propto 1-\alpha$.\\
Denoting the number densities as $n^+, n^-$ and $ n^0$ according to $\epsilon$ we can write the following evolution equations
\begin{subequations}\label{pm0}
    \begin{align}
    \dot{n}_i^+ &= u(n_{i-1}^+ - n_i^+) + \lambda\left[\frac{n_i^0+n_i^-}{2} - n_i^+\right] + \delta_{i,k}\gamma n_i^- - \delta_{i,k'}\gamma n_i^+,\\
    \dot{n}_i^- &= u(n_{i+1}^- - n_i^-) + \lambda\left[\frac{n_i^0+n_i^+}{2} - n_i^-\right] - \delta_{i,k}\gamma n_i^- + \delta_{i,k'}\gamma n_i^+,\\
    \dot{n}_i^0 &= u\left(\frac{1}{2}n_{i-1}^0 + \frac{1}{2} n_{i+1}^0 - n_i^0\right) + \lambda\left[\frac{n_i^+ + n_i^-}{2} - n_i^0\right].
    \end{align}
\end{subequations}
Taking the continuous limit and approximating the Kronecker delta with a normalized function strongly peaked at the boundary, $g(x)\approx\delta(x)$, we get 
\begin{subequations}
    \begin{align}
    \dot{n}^+ &= -v_0\partial_x n^+ + D\partial_{xx}^2 n^+ + \lambda\left[\frac{n^0+n^-}{2} - n^+\right] + g(x-x_L)\gamma n^- - g(x-x_R)\gamma n^+,\\
    \dot{n}^- &= v_0\partial_x n^- + D\partial_{xx}^2 n^- + \lambda\left[\frac{n^0+n^+}{2} - n^-\right]  - g(x-x_L)\gamma n^- + g(x-x_R)\gamma n^+,\\
    \dot{n}^0 &= D\partial_{xx}^2 n^0 + \lambda\left[\frac{n^++n^-}{2} - n^0\right] .
    \end{align}
\end{subequations}
where $v_0\equiv au$ and $D \equiv a^2d/2$ with $a$ being the lattice spacing. Also note that here we indicate with $x_L (x_R)$ the left (right) site of the interface.

We can now define the particle density $\rho$, polarity $p$ (along the $x$-direction) and nematic order $q$ as  
\begin{equation}
    \rho \equiv  n^+ + n^- + n^0, \;\;\;
    p \equiv n^+ - n^-, \;\;\;
    q\equiv n^+ + n^- - n^0.
\end{equation}
We can thus rewrite Eqs.\eqref{pm0} for these three variables as follows
\begin{subequations}
    \begin{align}
    \dot{\rho} &= -v_0\partial_x p + D\partial_{xx}^2 \rho,\\
    \dot{p} &= -\frac{v_0}{2}\partial_x (\rho + q) + D\partial_{xx}^2 p -\frac{3}{2} \lambda p +\frac{\gamma}{2} g(x-x_L)(\rho + q -2p) - \frac{\gamma}{2} g(x-x_R)(\rho + q + 2p),\\
    \dot{q} &= -v_0\partial_x p + D\partial_{xx}^2 q -\frac{3}{2}\lambda q.
    \end{align}
\end{subequations}
Far from the boundaries $x_L$ and $x_R$ one is left with the drift ($v_0\partial_x$ terms), diffusive ($D\partial^2_{xx}$) and tumbling ($\lambda$) terms. At the two boundaries, one also gets the boundary term modeling scoto-phobic reflecting boundaries. This has a simple, yet non-trivial, expression that is a combination of density polarity and nematic order and reads
\begin{equation}\label{eq:boundary_term}
    \frac{\gamma}{2} g(x-x_L)(\rho + q -2p) - \frac{\gamma}{2} g(x-x_R)(\rho + q + 2p)
\end{equation}
The derivation of this boundary term was done in the absence of alignment interactions between particles, which at this level of the description does not affect or modify the boundary term. Of course a more refined derivation of the boundary term should also consider the continuous rotational symmetry of the nematic angle, which is here instead treated as a discrete variable. Preliminary calculations of the boundary term, following a BGL approach to be reported elsewhere, show that the boundary term just derived is actually close to the correct one.\\
This boundary term can then be plugged into the mesoscopic equation for self-propelled rods of \cite{peshkov2012nonlinear} or by taking the zero repulsion (and zero spontaneous reversal) limit of the equations of \cite{patelli2019understanding} in order to model the effect of virtual confinement. These can be expressed in terms of the angular Fourier modes $f_k$, as seen in Sec.\ref{BGL}. The simple mesoscopic equations governing the lowest modes ($f_0 = \rho$, $f_1 = \rho\cdot(p_x, p_y)$ and $f_2 = \rho\cdot(Q_{xx}, Q_{xy})$) in the presence of photo-tactic confinement of polar active rods read 
\begin{equation}\label{eq:rods_f0}
        \partial_t\rho = - \frac{v_0}{2}(\nabla^*f_1 + \nabla f_1^*),
\end{equation}
\begin{equation}\label{eq:rods_f1}
\begin{split}
    \partial_t f_1 &= (\alpha[\rho] - \beta|f_2|^2) f_1 + \zeta f_1^* f_2 -\frac{v_0}{2}\nabla\rho - \frac{v_0}{2}\nabla^*f_2 + \gamma_1 f_2^*\nabla f_2\\
     &+\frac{\gamma}{2} g(x-x_L)(\rho + f_2 -2f_1) - \frac{\gamma}{2} g(x-x_R)(\rho + f_2 + 2f_1)
\end{split}
\end{equation}
\begin{equation}\label{eq:rods_f2}
    \partial_t f_2 = (\mu[\rho]+\tau|f_1|^2 -\xi|f_2|^2) f_2 + \omega f_1^2 - \frac{v_0}{2}\nabla f_1- \chi_1\nabla^* (f_1 f_2) + \kappa_1 f_1^* \nabla f_2,
\end{equation}
where $g(x)$ models the interface between light and dark regions and $\gamma \propto1-\alpha$ the escape rate. Moreover here $\nabla = \partial_x + i \partial_y$, $\nabla^* = \partial_x - i \partial_y$ and $\nabla^2 = \nabla\nabla^*$. The full expression of the coefficients can be found in \cite{patelli2019understanding}.

A detailed study of the effect of such boundary term at the mesoscopic level is still in progress but preliminary numerical simulations does not show a full success in reproducing the microscopic phenomenology: indeed the presence of the virtual boundary produces discontinuities between the inner and outer regions, that might require a more accurate mesoscopic description that takes into account the sharp density and nematic order profiles at the virtual boundaries, possibly suggesting a higher order closure of the mesoscopic equations. This is often the case when dealing with mechanical boundaries or obstacles, see for example \cite{codina2022small}, while fortunate cases where the gradients are more gently varying \cite{fava2024casimir}, are rather the exception. This mesoscopic analysis is the subject of ongoing research.
\section{Discussion}
To summarize, we have shown how it possible in a minimal Vicsek-style model to reproduce both accumulation at and alignment with virtual boundaries as found in \cite{kurjahn2024collective}. The formation of this ring structure depends on the inter particle repulsion, while it does not seem to depend on the phase of the fluid surrounding it: it is of course present in the case of disordered background, but perhaps more interestingly is the case of coexistence between the ring structure and a homogeneous background. In this case the homogeneous background interacting with ring creates a pair of $+1/2$ defects that are confined to the boundary of the circle.\\
In our microscopic point-wise model, accumulation and alignment are possible regardless of the confining shape and of the curvature, at odds with \cite{kurjahn2024collective} where a bend (for obtuse angles) to splay (for acute angles) transition is observed. To further simplify our setup, we studied a minimal setting for virtual confinement: a slab with two parallel and virtually reflecting walls.\\
The main effect of repulsion, at least in the case of the slab geometry, seems that of inducing an effective escape rate at the boundaries. Indeed we have shown how, even in the absence of repulsion, it is possible to obtain the same phenomenology (accumulation and alignment with the virtual boundaries) if a fraction $\alpha$ of the particles do not revert their direction at the boundaries. \\
Furthermore we have put forward a mesoscopic description based on this equivalence and derived a boundary interaction term from a simplified lattice model akin to the AIM for polar particles. The numerical and possibly analytical analysis (for instance at the linearized level) of these equations are however still at a preliminary stage and essentially left for future works.

Finally, it would be interesting to test if the equivalence between the escape rate and repulsion extends also to generic shapes. It would then be desirable to obtain a full mesoscopic description in the presence of repulsion, for instance considering the equations derived in \cite{patelli2019understanding} or exploring higher order closures of the relative kinetic level equations.
\pagestyle{plain}
\chapter{Conclusions \& Outlook}
Vicsek-style models constitute a simple yet powerful framework to investigate the effect of boundaries and bulk perturbations in vectorial active matter. At the microscopic level the implementation of the boundaries or of an external field is straightforward; this approach can then be complemented with a well-established field theoretical approach that allows to rationalize numerical results. This global conclusion aims to provide a summary of our finding and to point out possible additional issues that our work raises.

In Chapter \ref{directed} we have shown how explicit symmetry breaking, being it global (through a small homogeneous external field) or local (at the boundaries), affects the bulk system, leaving traces in measurable global observables. In the case of an external static and homogeneous field gently orienting the self-propulsion direction, both the mean flocking direction (for large enough times) or the static density correlations (for large enough systems) can be used to detect the presence of external fields or environmental cues. It would desirable to test these criteria to detect of confirm the directed nature of collective motion in {\it in vivo} experimental observations, for example in the process of wound healing. These criteria can also be extended to systems confined by two parallel and reflecting walls: we have shown that the behavior of a confined system is formally equivalent to a boundary free system perturbed by a small homogeneous external field.  

In Chapter \ref{confined} we focused our attention on the effects induced by confining a polar active fluid between two parallel and repelling walls. We have shown how the long-ranged correlations of fluctuations result in extensive boundary layers as opposed to the finite ones usually observed in scalar active matter, showing once again the far-reaching impact of boundaries in the context of active matter. Moreover, such fluctuations also produce an attractive Casimir-like force on the confining walls, which decays slowly and algebraically upon increasing wall separation. The Casimir exponent that controls the scaling of this force with wall separation is generic in the flocking regime and ultimately controlled by the scaling of bulk fluctuations. Both these results show a certain degree of universality, holding generically in the polar liquid phase and also w.r.t. the specific boundary interactions, as long as they are repelling (i.e. polar active particles tend to orient their self-propulsion direction away from the boundary interface). Our result also implies that in principle the bulk roughness exponent of the Toner \& Tu theory can be also measured by the finite-size scaling of the mechanical pressure exerted on the boundaries. Interestingly, we are also able to obtain theoretical predictions for both the velocity fluctuations and the density profile in the presence of confinement.\\
Numerical simulation of the collisional Vicsek model show that all of these results also hold beyond the dilute limit, paving the way to experimental test of confined flocking systems such as active colloids or hyper-confluent epithelial cells sheets. As interesting extension it would be also desirable to study the effect that such a confined polar active fluid has on a passive tracer.

In Chapter \ref{ch_nematic} we considered a more subtle confinement mechanism for nematic active matter inspired by scoto-phobic bacteria of Ref.\cite{kurjahn2024collective}. Physical boundaries that exert forces and torques on the active particles are replaced by a virtual light/dark interface which reverses the self-propulsion directions of particles that experience a decrease in illuminance. This interface has no effect on active particles when crossing from dark to light. Interestingly we observe that also such a virtual confinement can induce particle accumulation and alignment with such confining boundaries provided we include steric repulsion between our point-wise nematic particles: this result holds quite generically in a wide region of parameter space, regardless the phase of the free system (being it disordered or ordered). \\
The main effect of repulsion seems to consist in introducing an effective escape rate from the illuminated region. Indeed, introducing such an escape rate seems to be enough to reproduce qualitatively our results in the dilute, repulsion-less limit, at least in a slab geometry consisting in two infinite and parallel virtual light/dark interfaces. We complement our microscopic study deriving a set of mesoscopic field equation which is left for future studies.

We expect that many of or finding may be tested in experimental set ups (active colloids, epithelial cells \cite{giavazzi2017giant} moving in an anisotropic environment or under confinement, cyanobacteria, etc…), future direction for theoretical investigations may include the study of a passive tracers immersed in a confined polar active fluid, or perhaps the mechanical confinement of nematic active matter as possible further exploration of the results obtained in Ref.\cite{fava2024casimir}.

Finally we hope that these results have convinced the patient reader that a lot of interesting physics can be uncovered when vectorial active matter is perturbed and or confined. 
\appendix
\chapter{Appendix}
\section{Linearized structure factor: technical details}\label{linearized_SF}
In the following we derive the linearized structure factor \eqref{autocorr_eqt} from Eqs. ~\eqref{rho_lin}-\eqref{v_perp_lin}.
First, we rewrite Eqs. \eqref{rho_lin}-\eqref{v_perp_lin} in  Fourier space, according to \eqref{fourier_campi_TT} 
\begin{equation}   \label{rho_four_first}
    \begin{split}
         &[-i(\omega-v_2 q_{\parallel}) +D_{\rho_{\parallel}} q_{\parallel}^2 +D_{\rho_{\perp}} q_{\perp}^2-\phi q_{\parallel} \omega] \delta\hat{\rho} \,+\\
         & +[i \rho_0 q_{\perp} +D_{\rho v}q_{\perp} q_{\parallel}]\hat{v}_L=0,
    \end{split}
\end{equation}
\begin{equation}\label{v_L_first}
\begin{split}
        &\Big[ \frac{i c_o^2}{\rho_0}q_{\perp} -g_t q_{\perp}\omega +g_{\parallel} q_{\perp} q_{\parallel} \Big] \delta\hat{\rho}\,+\\
        &+[-i(\omega-\gamma q_{\parallel})+ D_L q_{\perp}^2 + D_{\parallel} q_{\parallel}^2 +h_v ]\hat{v}_L = \hat{f}_L
\end{split}
\end{equation}
where we have defined $\phi \equiv\rho_0 \mu_2$, $\gamma=\lambda_1 p_0(0)$ and $D_L \equiv D_B+D_T$. 
Moreover, $\hat{v}_L$ e $\hat{f}_L$ are the components along the longitudinal direction of $\hat{\bf v}_{\perp}$ and $\hat{\bf f}_{\perp}$ (the Fourier transformed transversal noise) defined as
\begin{equation}\label{A2}
    \hat{v}_L \equiv \hat{\bf v}_\perp \cdot \frac{\textbf{q}_{\perp}}{q_{\perp}} \ \ \text{and} \ \  \hat{f}_L \equiv \hat{\bf f}_\perp \cdot \frac{\textbf{q}_{\perp}}{q_{\perp}}.
\end{equation}
Notice that we have  omitted the equation for the $(d-2)$ transversal modes $\hat{\bf v}_T$, which are the components of $\hat{\bf v}_\perp$ orthogonal to ${\bf q_\perp}$: this is simply due to the fact that it is decoupled from Eqs. \eqref{rho_four_first}-\eqref{v_L_first} and it does not contribute to the longitudinal eigenmodes and the long-ranged behavior of density correlations \cite{toner2012reanalysis}.

We proceed to find the normal modes eigenfrequencies $\omega(\textbf{q})$ of Eqs.\eqref{rho_four_first}-\eqref{v_L_first}, that is, the complex frequencies at which non-zero solutions exist for zero noise, $\hat{f}_L=0$.
In the hydrodynamic limit ($q \rightarrow 0$) one obtains the complex conjugated eigenfrequencies
\begin{equation}
    \omega_{\pm}(\textbf{q}) = c_{\pm}(\theta_q)q-\epsilon_{\pm}(h, \textbf{q})\,.
    \label{norm_freq}
\end{equation}
Their real parts (the sound speeds) are unaffected by the external field and are given by
\begin{equation} \label{cpm}
    c_{\pm}(\theta_q)= \bigg( \frac{\gamma+v_2}{2}\bigg) \cos(\theta_q) \pm c_2(\theta_q);
\end{equation}
with
\begin{equation}\label{c2}
    c_2(\theta_q)\equiv \sqrt{\frac{(\gamma-v_2)^2\cos(\theta_q)^2}{4}+c_0^2\sin^2(\theta_q)}
\end{equation}
The only field-dependent terms are found to be the imaginary dampings $\epsilon_\pm(h, {\bf q})$ equal to
\begin{equation}
\epsilon_\pm (h, {\bf q}) = \epsilon_\pm (0, {\bf q})+ a_\pm(\theta_q) h \equiv \Tilde{\epsilon}_\pm(\theta_q) q^2 + a_\pm(\theta_q) h\,,
\end{equation}
where $a_\pm(\theta_q)$ is given by Eq. \eqref{eq:apm} and 
\begin{equation}\label{A3}
\Tilde{\epsilon}_\pm(\theta_q)=\frac{\mathrm{\Xi}_\pm(\theta_q)}{[2c_{\pm}(\theta_q)-(v_2+\gamma)\cos(\theta_q)]}\,.
\end{equation}
The numerator $\Xi_\pm$ of Eq. \eqref{A3} is rather complicated but only depends on the angle $\theta_q$,
\begin{equation}
\begin{split}
&\Xi_\pm(\theta_q) = 
-[D_L\sin^2(\theta_q) + D_\parallel\cos^2(\theta_q)]v_2\cos(\theta_q)+\\
&+[D_L \sin^2(\theta_q) + D_\parallel \cos^2(\theta_q) -\phi c_\pm(\theta_q)\,\cos(\theta_q)] 
c_\pm(\theta_q)+\\
&-[D_{\rho\parallel}\!\cos^2(\theta_q)\!+\!D_{\rho\perp}\!\sin^2(\theta_q)\!-\!\phi c_\pm(\theta_q)\!\cos(\theta_q)]\gamma\!\cos(\theta_q)
\\&+\frac{c_0^2}{\rho_0}D_{\rho v}\!\cos(\theta_q)\!\sin^2(\theta_q)\!-\![g_t c_\pm(\theta_q)\!+\! g_\parallel\!\cos(\theta_q)]\rho_0\!\sin^2(\theta_q).
\end{split}
\end{equation}

The solution of Eqs.\eqref{rho_four_first}-\eqref{v_L_first} can be now easily expressed in terms of the above eigenfrequencies $\omega_\pm$ (the zeros of the associated matrix determinant).  
In particular, in the hydrodynamic limit ($q\to0)$ we have
\begin{equation}
\delta\hat{\rho}(\omega, {\bf q})=\frac{\rho_{\rm 0}q\sin(\theta_q) \hat{f}_L}{(\omega-\omega_+({\bf q}))(\omega-\omega_-({\bf q}))}\,.  
\end{equation}
Correlating this solution pairwise we obtain 
\begin{equation} \label{auto_corr_dens}
        \langle \delta\hat{\rho}(\mathbf{q},\omega) \delta\hat{\rho}(-\mathbf{q},-\omega) \rangle
        = \frac{\rho_0^2 q^2 \sin^2(\theta_q) \Delta}{\{ [\omega-c_+(\theta_q) q]^2+[\epsilon_+(h, {\bf q})]^2\} \{[\omega-c_-(\theta_q) q]^2+ [\epsilon_-(h, {\bf q})]^2\}}.
\end{equation}
The equal time structure factor \eqref{autocorr_eqt} is finally recovered transforming back in real (equal) time by an integration over $\omega$.
\newpage

\section{Calculation of $S_{L_\perp}$}\label{calculation_SR}
\label{sm:sec-SR}
\begin{figure}[hbt!]
\centering
\includegraphics[width=0.35\textwidth]{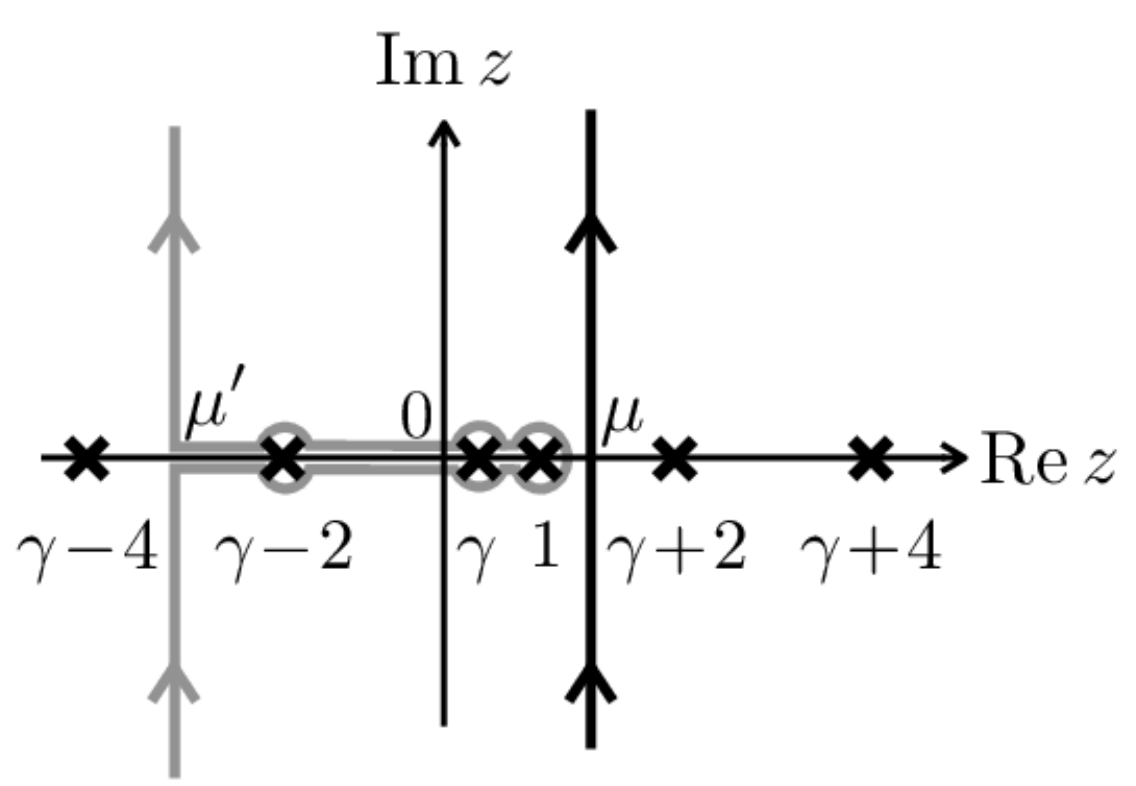}
\caption{Contours in the complex plane $z\in\mathbb{C}$ along which the integral in Eq.~\eqref{sm:eq-SR-Mell} is calculated for determining $S_{L_\perp}$. The black vertical contour with ${\rm Re}\, z = \mu$ is the one indicated in Eq.~\eqref{sm:eq-SR-Mell}, which can be deformed continuously into the equivalent grey contour used in order to determine the expansion of $S_{L_\perp}$ upon increasing $L_\perp$ in Eq.~\eqref{eq:sm-exp-SR-poles}. The crosses on the real axis indicate the poles of the integrand, which is otherwise an analytic function.}
\label{sm:fig-path}
\end{figure}

The sum $S_{L_\perp}$ in Eq.~\eqref{sm:eq-SR} can be conveniently evaluated by adapting to the present case the general strategy discussed, e.g., in Ref.~\cite{Gambassi03}. We first rewrite $S_{L_\perp}$ as
\begin{equation}
S_{L_\perp} = \frac{1}{L_\perp}\left(\frac{L_\perp}{\pi}\right)^{\gamma} \sum_{n=1}^\infty f(n),
\label{sm:eq-SR-2}
\end{equation}
in which we take into account the definition of $q_\perp(n)$ given after Eq.~\eqref{eq:v1} and we define
\begin{equation}
f(n) = \frac{1}{n^\gamma}\frac{1}{1 + (\Gamma \pi n/L_\perp)^2}.
\label{sm:eq-fn}
\end{equation}
Then, we introduce the Mellin transform $\hat f(z)$ of $f(n)$ in the complex plane $z \in \mathbb{C}$
\begin{equation}
\hat f(z) = \int_0^\infty \!\!dx\, f(x) x^{z-1} = \frac{\pi}{2}\frac{(\Gamma \pi/L_\perp)^{\gamma-z}}{\sin 
\left(\pi(z-\gamma)/2\right)}, 
\label{sm:eq-fnMellin}
\end{equation}
with $\gamma < \mbox{Re}\,z < 2+\gamma$. This condition on $\mbox{Re}\,z$ guarantees that the integral above is convergent and therefore that $\hat f(z)$ is an analytic function within that strip in the complex plane. The function $f(n)$ in Eq.~\eqref{sm:eq-SR-2} can be expressed in terms of $\hat f(s)$
\begin{equation}
    f(n) = \int_{\mu-i\infty}^{\mu+i\infty} \!\!\frac{dz}{2\pi i}\,\hat f(s) n^{-z},
\end{equation}
where $\mu$ is chosen within the domain of analiticity of $\hat f(z)$, i.e., $\gamma < \mu < 2+\gamma$. After this substitution, the remaining sum over the positive integer $n$ renders the Riemann function ${\bm\zeta}(z)$ provided that this sum converges, which additionally requires $\mu = \mbox{Re}\,z>1$. Accordingly, one can eventually write
\begin{equation}
S_{L_\perp} = \frac{1}{L_\perp}\left(\frac{L_\perp}{\pi}\right)^\gamma \int_{\mu-i\infty}^{\mu+i\infty} \!\!\frac{dz}{2\pi i}\,\hat f(z) {\bm\zeta}(z),
\label{sm:eq-SR-Mell}
\end{equation}
where $1 < \mu < 2+\gamma$. Note that the integration contour above consists of  a vertical line in the complex plane with fixed $\mbox{Re}\, z = \mu$, which is represented by the black path in Fig.~\ref{sm:fig-path}.

Noting that $\hat f(z) \propto L_\perp^{z-\gamma}$, the expansion of $S_{L_\perp}$ in decreasing powers of $L_\perp$ --- suitable for considering the case of large $L_\perp$ we are interested in --- can be obtained by shifting the integration contour in Eq.~\eqref{sm:eq-SR-Mell} towards smaller values of $z$, i.e., towards the left. In doing so, one has to account for the poles that the integrand has on the real axis. They are given by: $(i)$~the pole of ${\bm\zeta}(z)$ located at $z=1$, with residue ${\cal R}^* = 1$; $(ii)$~the poles of $\hat f(z)$ located at $\gamma - 2 m$, with $m\in \mathbb{Z}$ and residues ${\cal R}_m = (-1)^m(\Gamma\pi/L_\perp)^{2m}$. 

These poles are indicated by the crosses on the real axis in Fig.~\ref{sm:fig-path} and they are actually the only singularities of the integrand, which is otherwise an analytic function of $z$. Accordingly, the integral in Eq.~\eqref{sm:eq-SR-Mell} can be equivalently calculated, for example, along the grey contour depicted in Fig.~\ref{sm:fig-path}, in which the contributions of the horizontal portions of the path actually cancel each other in pairs. As a result, the integral along the grey contour is the same as that in Eq.~\eqref{sm:eq-SR-Mell} but with $\mu$ replaced by $\mu'$, plus the contributions of the portions of the paths encircling the poles, which --- due the orientation of the path and to Cauchy residue theorem --- render the residues at those poles. 
Accordingly, one finds
\begin{equation}
S_{L_\perp} = \frac{1}{L_\perp}\left(\frac{L_\perp}{\pi}\right)^\gamma \left[ \hat f(z=1){\cal R}^* + {\bm\zeta} (\gamma){\cal R}_0 + {\cal O}({\cal R}_1)\right],
\label{eq:sm-exp-SR-poles}
\end{equation}
where $\hat f(z=1) \propto L_\perp^{1-\gamma}$, ${\cal R}_0 = {\cal R}^* =1$, and  ${\cal R}_1 \sim L_\perp^{-2}$. Taking into account Eq.~\eqref{sm:eq-fnMellin}, this expansion yields Eq.~\eqref{eq:cas}, upon identifying the first term on the r.h.s.~with the $L_\perp$-independent constant $S_\infty$ in Eq.~\eqref{eq:inf} (see also Eqs.~\eqref{sm:eq-fn} and \eqref{sm:eq-fnMellin}) and the second term with the Casimir-like, leading finite-size contribution.

\newpage
\section{Derivation from full Toner \& Tu equations}\label{full_TT}
In order to determine the average density at the confining walls, we consider the TT equations for the mesoscopic density $\rho({\bf r}, t)$ and velocity ${\bf v} ({\bf r}, t)$ fields,
\begin{equation}
\partial_t \rho + \nabla \cdot (\rho{\bf v})=D_\rho \nabla^2 \rho,
\label{rho-evolution}
\end{equation}
\begin{equation}
\begin{split}
\partial_t{\bf v} = -\lambda_1({\bf v}&\cdot\nabla) {\bf v} - \lambda_2 {\bf v} (\nabla \cdot {\bf v}) - \lambda_3 \nabla |{\bf v}|^2 + (\mu - \kappa {\bf v}^2) {\bf v} -\sigma_1 \nabla \rho \\
&+D_1\nabla^2\mathbf{v} + D_v\nabla (\nabla \cdot {\bf v}) + D_2 (\nabla \cdot {\bf v})^2 {\bf v} + \mathbf{f},
\end{split}
\label{v-evolution}
\end{equation}
where we have included a nonzero density diffusion term for $\rho$, proportional to $D_\rho$, which in the TT equations is generated by renormalization \cite{TTR}. \\
The coefficients $D_i$, $\lambda_i$ ($i=1,2,3$) and $\sigma_i$ are related, respectively, to diffusion, advection and bulk pressure. We have also ignored nonlinear density contributions to the bulk pressure-like term $\sigma_1 \nabla \rho$. Spontaneous symmetry breaking in the ordered phase is induced by the familiar Ginzburg-Landau terms, which determine the modulus
\begin{equation}
  p_0\equiv|{\bf p}_0|=\sqrt{\frac{\mu(\rho_0)}{\kappa}}
\end{equation}
of the homogeneous solution $\rho({\bf r},t) = \rho_0$ velocity ${\bf v}={\bf p}_0$.\\
For simplicity, we have considered constant coefficients, only retaining the local density dependence of $\mu$ which plays a fundamental role in the onset of the phase separated regime, as we have seen in Section \ref{BGL}.

Due to confinement, ${\bf p}_0$ aligns with the unit vector ${\bf \hat{e}}_\parallel$ parallel to the walls. Assuming a highly ordered state, we proceed by spin-wave approximation \cite{Nishimori}, decomposing the velocity ${\bf v}$ into its longitudinal and transversal components, ${\bf v} = (p_0 + \delta v_\parallel) {\bf\hat{e}}_\parallel + {\bf v}_\perp $, and expanding in small velocity deviations from $ {\bf p}_0$. Discarding terms of order higher than $\delta v_\parallel \sim  {\bf v}_\perp^2$ (this ansatz will be justified a posteriori) and projecting in the longitudinal and transversal directions we obtain in two spatial dimensions 
\begin{equation}
  \partial_t \rho + p_0 \partial_\parallel \rho + \partial_\parallel (\rho \delta v_\parallel ) + \partial_\perp (\rho v_\perp) = D_\rho \left( \partial_\parallel^2 \rho + \partial_\perp^2 \rho \right)
\label{rhoSW}  
\end{equation}
\begin{equation}
\begin{split}
\partial_t \delta v_\parallel &=- 2 \Lambda p_0 \partial_\parallel \delta v_\parallel -
  \lambda_2 p_0 \partial_\perp  v_\perp -  \lambda_3 \partial_\parallel v_\perp^2 -2 \mu \delta v_\parallel - \kappa p_0 v_\perp^2 + \mu' p_0 \delta \rho\\
  &- \sigma_1 \partial_\parallel \rho + (D_1+D_v) \partial_\parallel^2 \delta v_\parallel +  D_v \partial_\parallel \partial_\perp v_\perp + D_1 \partial_\perp^2 \delta v_\parallel + D_2 p_0 (\partial_\perp v_\perp)^2 + f_\parallel
\end{split}
\label{parallelSW}  
\end{equation}
\begin{equation}
\begin{split}
\partial_t v_\perp &= - \lambda_1 p_0 \partial_\parallel v_\perp - 2 \lambda_3 p_0 \partial_\perp \delta v_\parallel -\Lambda \partial_\perp v_\perp^2 - \sigma_1 \partial_\perp \rho \\
&+ D_1 \partial_\parallel^2 v_\perp + D_v \partial_\parallel \partial_\perp \delta v_\parallel + (D_1 + D_v) \partial_\perp^2 v_\perp + f_\perp
\label{perpsSW}  
\end{split}
\end{equation}
where $v_\perp=|{\bf v}_\perp|$, $\mu' \equiv \partial_\rho\mu (\rho_0)$ and we have introduced the density deviation profile $\delta \rho = \rho - \rho_0$. Moreover, we have also introduced the new parameters $\Lambda = \lambda_3 + (\lambda_1+\lambda_2)/2$ and the notation $\partial_\parallel$ and $\partial_\perp$ to denote, respectively, spatial derivatives in the longitudinal and transversal direction.\\
In order to compute the average profiles we assume translational invariance w.r.t.~time and the longitudinal direction, i.e., $\partial_t\langle \cdot \rangle = \partial_\parallel \langle \cdot\rangle=0$. We are left with
\begin{equation}
\rho_0 \partial_\perp \langle v_\perp\rangle = D_\rho \partial_\perp^2 \langle \delta \rho \rangle
\label{rhoSWav}  
\end{equation}
\begin{equation}
\begin{split}
  \lambda_2 p_0 \partial_\perp  \langle v_\perp \rangle &= -2 \mu \langle \delta v_\parallel \rangle - \kappa p_0 \langle v_\perp^2 \rangle + \mu' p_0 \langle \delta \rho \rangle\\
  &+ D_1 \partial_\perp^2 \langle \delta v_\parallel \rangle +\frac{D_2p_0}{2} \partial_\perp^2 \langle v_\perp^2\rangle  
\end{split}
\label{parallelSWav}  
\end{equation}
\begin{equation}
 2 \lambda_3 p_0 \partial_\perp \langle \delta v_\parallel \rangle + \Lambda \partial_\perp \langle v_\perp^2 \rangle =
  - \sigma_1 \partial_\perp \langle \delta \rho \rangle +
 (D_1 + D_v) \partial_\perp^2 \langle v_\perp \rangle 
\label{perpsSWav}  
\end{equation}
where we have discarded higher-order contributions $\langle v_\perp \delta\rho\rangle$ and $\langle v_\perp \partial_\perp^2 v_\perp \rangle$ from Eqs.~(\ref{rhoSW}) and (\ref{parallelSWav}). \\
From Eq. (\ref{rhoSWav}) we obtain
\begin{equation}  \label{vperp_sm}
  \partial_\perp \langle {\bf v}_\perp \rangle \approx \frac{D_\rho}{\rho_0} \partial_\perp^2 \langle \delta\rho\rangle,
\end{equation}
which can be inserted in Eq. (\ref{parallelSWav}) to get
  \begin{equation}
  2 \mu \langle \delta v_\parallel \rangle = - \kappa p_0 \langle v_\perp^2 \rangle
  + \mu' p_0 \langle \delta \rho \rangle
  + D_1 \partial_\perp^2 \langle \delta v_\parallel \rangle - \lambda_2 \frac{D_\rho p_0}{\rho_0} \partial_\perp^2  \langle \delta \rho \rangle
  +\frac{D_2p_0}{2} \partial_\perp^2 \langle v_\perp^2\rangle  
  \label{parallelSWav2}
  \end{equation}
This is an equation which expresses the enslaving of $\langle \delta v_\parallel \rangle$ to $\langle\delta \rho \rangle$ and $\langle {\bf v}_\perp^2\rangle$. Following  \cite{TTR}, it can be solved iteratively at different orders in the transversal derivatives. To zeroth order in the derivatives it reads
\begin{equation}
  \langle \delta v_\parallel\rangle \approx \frac{\mu'(\rho_0)}{2\kappa p_0} \langle\delta \rho\rangle - \frac{\langle \mathbf{v}_\perp^2 \rangle}{2 p_0}\,.
  \label{enslave_app}
\end{equation}
Inserting Eq. (\ref{enslave_app}) in the r.h.s. of Eq. (\ref{parallelSWav2}) we obtain to second order in the derivatives
\begin{equation}
\langle \delta v_\parallel\rangle \approx \frac{\mu'(\rho_0)}{2\kappa p_0} \langle\delta \rho\rangle - \frac{\langle \mathbf{v}_\perp^2 \rangle}{2 p_0} + \left(\frac{D_1 \mu'}{4 \mu \kappa p_0}  - \frac{\lambda_2 D_\rho}{2 \rho_0 \kappa p_0} \right)\partial_\perp^2 \langle \delta \rho \rangle - \left[\frac{D_1}{4 \mu p_0} - \frac{D_2p_0}{4\mu}\right]  \partial_\perp^2 \langle v_\perp^2 \rangle
\label{enslave2}
\end{equation}
Finally, we insert Eqs. (\ref{vperp_sm}) and (\ref{enslave2}) into Eq. (\ref{perpsSWav}) to get, up to third order in transversal derivatives, the {\it profile equation}
\begin{equation}\label{eq:profile_app}
\begin{split}
\left[\frac{D_\rho}{\rho_0} \left(D_1+D_2+\frac{\lambda_2 \lambda_3}{\kappa} \right) - \frac{\lambda_3 D_1 \mu'}{2 \mu \kappa} \right]\, &\partial_\perp^3 \langle\delta \rho\rangle - \left( \sigma_1 + \frac{\lambda_3 \mu' }{\kappa}\right) \partial_\perp \langle\delta \rho\rangle =\\
\frac{\lambda_1+\lambda_2}{2} \partial_\perp \langle {\bf v}_\perp^2 \rangle &+ \left[\frac{\lambda_3 D_1}{2 \mu}-\frac{\lambda_3 D_2}{2\kappa} \right]\partial_\perp^3 \langle {\bf v}_\perp^2 \rangle 
\end{split}
\end{equation}
which should be compared with Eq.\eqref{eq:profile}, which was obtained from simplified equations. A few considerations are in order.\\
We added a diffusion term to the continuity equation, arguing it will be produced by renormalization: in any case, the $\partial_\perp^3 \langle \delta \rho \rangle$ term in Eq. (\ref{eq:profile_app})  also gets a contribution from $\mu'$, that is the density dependence of the linear Ginzburg-Landau coefficient. This third order derivative is a regularization term, as we have shown in Eq.\eqref{rho_sum} and in Fig.\ref{numerical profiles}. Note also that in the original TT expansion, third order derivatives are ignored, so in our treatment other contributions should appear, e.g. from terms in the velocity field dynamics $\sim \nabla (\nabla^2 \rho)$. Thus we expect a third order derivative of the density field to be present in the profile equation, acting as a regularizator (provided it is positive).\\
We also obtain a further contribution from the velocity fluctuations $\sim\partial_\perp^3\braket{\mathbf{v}_\perp^2}$ which can be integrated in our calculations for the density profile.\\
We ignored higher order terms in $\delta \rho$, which is less justified than the spin-wave expansion and can be problematic near the boundary where the density deviation diverge. Nevertheless, near the boundary (within an extrapolation length $\sim v_0$), the mesoscopic theory breaks down (see Fig.\ref{profile_boundary_layer}). We also ignored a somehow exotic anisotropic pressure term which may appear in TT equations, $\sim {\bf v} ( {\bf v} \cdot \nabla) \delta \rho$. This It simply adds a term $\sim \langle v_\perp \partial_\perp \delta \rho \rangle$ which we anyhow ignore here. Furthermore all coefficients do depend on $\rho$ and $|{\bf v}|^2$, but expanding around $p_0$ and $\rho_0$ we only get higher order contributions, with the only exception of an irrelevant contribution from the density dependence of $\kappa$, $-\kappa' p_0^3 \langle \delta \rho \rangle$ which normalizes the $\mu'$ coefficients, that is $\mu' \to \mu' - \kappa' p_0^2$.

\subsubsection{On the validity of a hydrodynamic approach}
One may wonder whether our use of the TT hydrodynamic theory, formally valid in the long wavelength limit can be justified in the presence of sharp boundaries where the slow fields deviations become large. In the following we justify its use.\\
First of all, the \textit{extensitivity} of the boundary layer in the thermodynamic limit means that $d_0\propto L_\perp$: this ensure that, asymptotically, the relative importance of higher order derivatives should become smaller and smaller since all the variations of the slow modes become more gentle.

\begin{figure}[hbt!]
    \centering
    \includegraphics[width=1\linewidth]{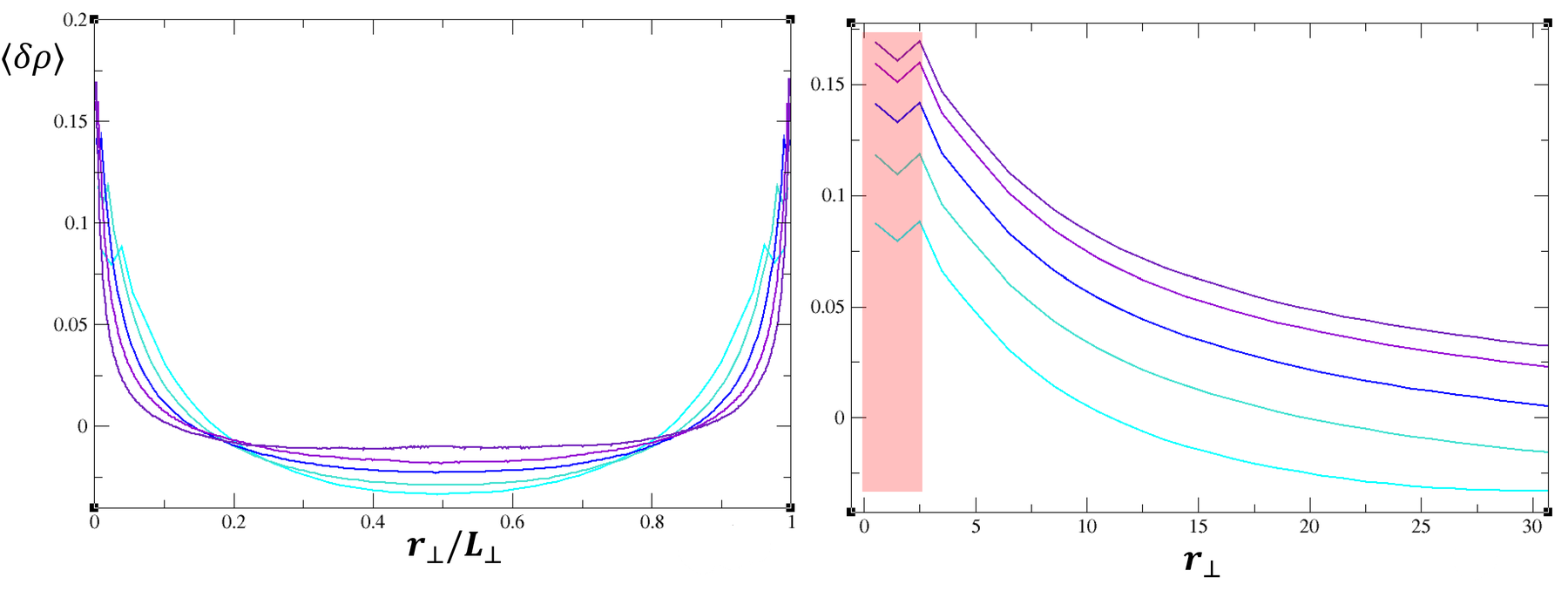}
    \caption{{\bf Average density deviation $\langle\delta\rho\rangle$ profiles at $v_0 = 2.5$ for different wall separations $L_\perp$} Left - Average density deviations profile $\langle\delta\rho\rangle$ at $v_0 = 2.5$ for different wall separations $L_\perp = 32, 64, 128, 256, 512$ (from cyan to purple) as a function of the rescaled transverse wall separation $r_\perp/L_\perp$. The density profiles show the same scaling behavior observed for $v_0 = 0.5$. Right - Zoom of data on the left side close to the left boundary as a function of the distance from the wall (not rescaled). The microscopic boundary layer (highlighted in red in figure), where we do not expect the hydrodynamic approach to be valid and thus where we observe deviations from it, is order of $\Lambda = v_0\Delta t/R_0 = 2.5$ here and thus negligible in the thermodynamic limit.}
    \label{profile_boundary_layer}
\end{figure} 

Secondly the details of the boundary interaction turn out to be relevant only in a very tiny region close to the wall of order $v_0$: we do not expect our predictions to be valid here (actually some of our calculations diverge, as we will show in the following) but this can be argued to be beyond an \textit{extrapolation length}, which is basically an offset that encapsulates the mismatch between microscopic and mesoscopic descriptions. This is already clear from Fig.\ref{fig2}a, where profiles with different $\beta$ turn out to coincide apart from one single point at the wall.

Moreover the influence of the wall can be compared to $\Lambda = v_0\Delta t/R_0$, with $\Delta t$ being the amplitude of the time-stepping in the discrete dynamics and $R_0$ the metric range of interaction, which is a dimensionless scale which measures the relative importance of self-propulsion over alignment.\\
Data reported in Fig.\ref{profile_boundary_layer} shows that the the details of the interaction with the wall, which result in a deviation from the hydrodynamic profile, are important in a region which is sub-extensive in the thermodynamic limit, comparable to $\Lambda$. This also explains why, in the case of $v_0 = 0.5$ different boundary interactions (tuned by $\kappa$) differ only for a single point close to the wall: simply because $v_0$ sets the width of the microscopic boundary to be less than unity.  

In conclusion, while the TT equations are a limitation when dealing with sharp boundaries (since they only retain derivatives up to second order), the extensitivity of the boundary layer guarantees that higher order derivatives are indeed less and less important. Nevertheless, we find it convenient to include third order derivatives in the profile equations as a regularization.

\sloppy
\printbibliography[heading=bibintoc, title=Bibliography]
\end{document}